  \documentclass[12pt,a4paper,doublespace,oneside]{book}

\usepackage{./styles/setspace}		
\usepackage{./styles/fancyhdr}		
\usepackage{ifthen}					
\usepackage{./styles/geometry}		
\usepackage{./styles/lgrind}
\usepackage{flafter}		

\usepackage[pdftex]{graphicx}
\usepackage{rotating}
\usepackage[T1]{fontenc}
\usepackage{subfig}
\usepackage{url}
\usepackage{amsmath}
\usepackage{amssymb}
\usepackage{fixltx2e}
\usepackage{authblk}
\usepackage{indentfirst}
\usepackage{stfloats}
\usepackage{array}
\usepackage{cite}
\usepackage{algorithmic}
\usepackage{algorithm}
\usepackage{color}
\usepackage{multicol}
\usepackage{multirow}
\usepackage{rotating}

\usepackage{nomencl}

\makenomenclature

\usepackage{makeidx}

\newcommand{\sign}{\operatorname{sign}}
\newcommand{\Rem}{\operatorname{Rem}}

\makeatletter
	\if@twoside
	\fi
\makeatother

\makeatletter%
	\def\cleardoublepage{%
		\clearpage%
		\if@twoside%
			\ifodd%
				\c@page%
			\else%
				\hbox{}%
				\thispagestyle{empty}%
				\newpage%
				\if@twocolumn%
					\hbox{}%
					\newpage%
				\fi%
			\fi%
		\fi%
	}%
\makeatother				

\fancypagestyle{plain}{%
	\fancyhf{} 
	\fancyfoot[CO,CE]{\thepage} 
	\renewcommand{\headrulewidth}{0pt} 
	\renewcommand{\footrulewidth}{0pt} 
}

\graphicspath{{Figures/}}

\begin{document}



\frontmatter

\renewcommand{\chaptermark}[1]{%
\markboth{#1}{#1}}

\begin{titlepage}

	\vspace*{\fill}

	\begin{center}
		\huge
		\textbf{Link Adaptation for Wireless Video Communication Systems}
	\end{center}
	\vfill
	
%
		
	\begin{center}
		\LARGE Husameldin Mukhtar
	\end{center}
	\vfill
	
	

	\begin{center}
		\Large December 2014 
	\end{center}

	\vfill
	
	\begin{center}
		\includegraphics[width=4.0in]{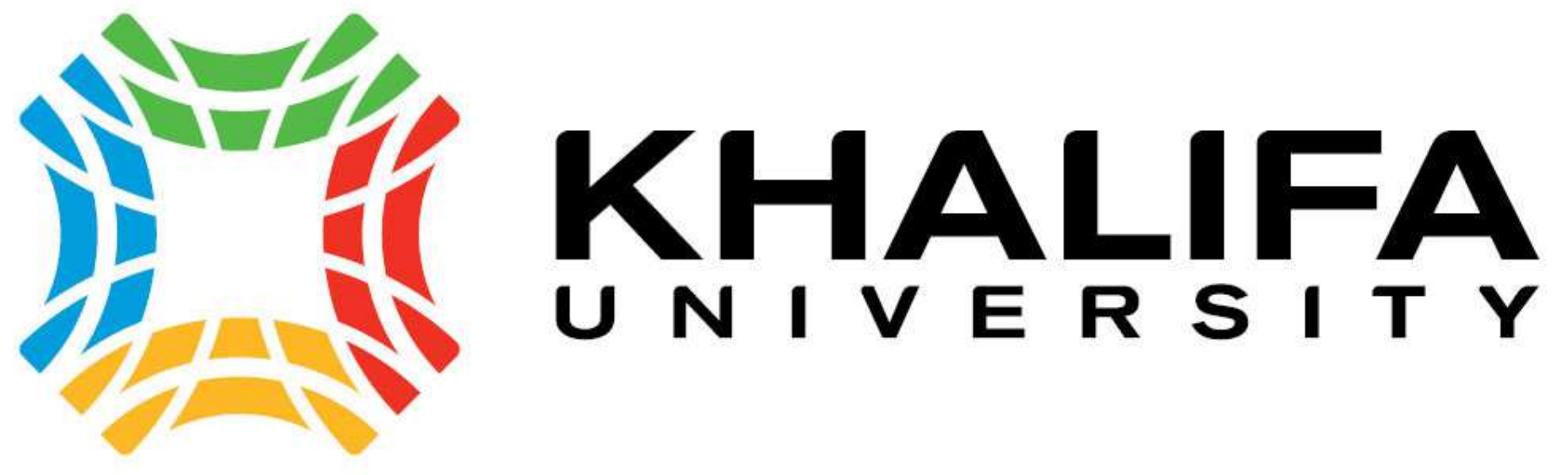}
	\end{center}

	\vfill

	{\Large		\noindent
	A thesis submitted to Khalifa University of Science, Technology, and Research in accordance with the requirements of the degree of PhD in Engineering in the Department of Electrical and Computer Engineering.
	}

\end{titlepage}

\addcontentsline{toc}{chapter}{Abstract}
%
%
%
%

\vspace{4cm}

\begin{center}
\begin{large}
{
LINK ADAPTATION FOR WIRELESS VIDEO COMMUNICATION SYSTEMS\\
}
\end{large}
\end{center}

\noindent
\begin{center}
Husameldin Mukhtar \\
PhD in Engineering\\
\end{center}

\noindent
\begin{center}
Department of Electrical and Computer Engineering\\
Khalifa University of Science, Technology and Research
\end{center}


\noindent
\begin{center}
\begin{large}
{
ABSTRACT\\
}
\end{large}
\end{center}
\vspace{1.5cm}

This PhD thesis considers the performance evaluation and enhancement of video communication over wireless channels. The system model considers hybrid automatic repeat request (HARQ) with Chase combining and turbo product codes (TPC). The thesis proposes algorithms and techniques to optimize the throughput, transmission power and complexity of HARQ-based wireless video communication. A semi-analytical solution is developed to model the performance of delay-constrained HARQ systems. The semi-analytical and Monte Carlo simulation results reveal that significant complexity reduction can be achieved by noting that the coding gain advantage of the soft over hard decoding is reduced when Chase combining is used, and it actually vanishes completely for particular codes. Moreover, the thesis proposes a novel power optimization algorithm that achieves a significant power saving of up to 80\%. Joint throughput maximization and complexity reduction is considered as well. A CRC (cyclic redundancy check)-free HARQ is proposed to improve the system throughput when short packets are transmitted. In addition, the computational complexity/delay is reduced when the packets transmitted are long. Finally, a content-aware and occupancy-based HARQ scheme is proposed to ensure minimum video quality distortion with continuous playback.\\

\noindent \textbf{Indexing Terms}: Hybrid automatic repeat request, turbo product codes, wireless channels, throughput, power optimization,  video quality.  

\chapter{Declaration and Copyright}
{
\section*{Declaration}
I declare that the work in this report was carried out in accordance with the regulations of Khalifa University of Science, Technology, and Research. The work is entirely my own except where indicated by special reference in the text. Any views expressed in the report are those of the author and in no way represent those of Khalifa University of Science, Technology, and Research. No part of the report has been presented to any other university for any degree.
\vspace*{1cm}

\noindent SIGNED: \dotfill \hspace{0.2cm}DATE: \dotfill
}
\vfill
\begin{singlespace}
{\section*{Copyright \copyright} 
No part of this report may be reproduced, stored in a retrieval system, or transmitted, in any
form or by any means, electronic, mechanical, photocopying, recording, scanning or otherwise, without prior written permission of the author. The thesis may be made available for consultation in Khalifa University of Science, Technology, and Research Library and for inter-library lending for use in another library and may be copied in full or in part for any bona fide library or research worker, on the understanding that users are made aware of their obligations under copyright, i.e. that no quotation and no information derived from it may be published without the author's prior consent.
}
\end{singlespace}

\chapter{Acknowledgments}

\noindent I would like to thank my PhD project supervisors, Dr. Arafat Al-Dweik and Dr. Mohammed Al-Mualla for their guidance and continuous support. They provided me with valuable technical knowledge and insightful research discussions throughout my PhD thesis work. I am also grateful to my supervisors for providing me with research growth opportunities such as supporting my visit to Western University in Canada and for their valuable advice in research dissemination methods. 

\noindent I also thank Dr. Abdallah Shami for inviting me to visit and present part of my PhD research work to his research group at Western University.
 
\noindent I would also like to thank Prof. Mahmoud Al-Qutayri for managing the PhD program and ensuring a healthy balance between our research and teaching duties. I am thankful to all my professors from the Electrical and Computer Engineering Department for their dedication and great teaching. In addition, I thank the College of Engineering at Khalifa University for offering me the Teaching Assistant Scholarship.

\noindent My thanks also go to my PhD co-students and friends with whom I enjoyed many great moments. It is really a rewarding experience to be part of such lively group of intellectuals and bright innovators. 

\noindent Finally, I would like to thank my family for their endless support. I gratefully acknowledge the love and devotion of my parents, brother and sisters who have always inspired me.	

\addcontentsline{toc}{chapter}{Contents}
\tableofcontents


\addcontentsline{toc}{chapter}{List of Figures}
\listoffigures

\clearpage      
\addcontentsline{toc}{chapter}{List of Tables}
\listoftables

\chapter{List of Acronyms}

\begin{singlespace}

%
%



\begin{list}{}
{
\setlength{\labelwidth}{2.5cm}
\setlength{\labelsep}{0.5cm}
\setlength{\leftmargin}{3cm}
}

  \item [{3GGP}]\begingroup 3rd Generation Partnership Project\nomeqref {2.0}
		\nompageref{27}
  \item [{A/D}]\begingroup Analog-to-Digital Conversion\nomeqref {2.0}
		\nompageref{18}
  \item [{ACK}]\begingroup Acknowledgment\nomeqref {2.0}\nompageref{12}
  \item [{ADL}]\begingroup Adaptive Data Loading\nomeqref {2.0}
		\nompageref{19}
  \item [{AMC}]\begingroup Adaptive Modulation and Channel Coding\nomeqref {2.0}
		\nompageref{24}
  \item [{ARQ}]\begingroup Automatic Repeat Request\nomeqref {1.0}
		\nompageref{2}
  \item [{AWGN}]\begingroup Additive White Gaussian Noise\nomeqref {1.0}
		\nompageref{3}
  \item [{BCH}]\begingroup Bose-Chaudhuri-Hocquenghem\nomeqref {2.0}
		\nompageref{30}
  \item [{BER}]\begingroup Bit Error Rate\nomeqref {1.0}\nompageref{3}
  \item [{BF}]\begingroup Brute-Force Method\nomeqref {5.8}
		\nompageref{70}
  \item [{BI}]\begingroup Bisection Method\nomeqref {5.8}
		\nompageref{70}
  \item [{bps}]\begingroup bits per second\nomeqref {3.28}
		\nompageref{49}
  \item [{BPSK}]\begingroup Binary Phase Shift Keying\nomeqref {2.0}
		\nompageref{15}
  \item [{BSC}]\begingroup Binary Symmetric Channel\nomeqref {7.4}
		\nompageref{100}
  \item [{CBR}]\begingroup Constant Bit Rate\nomeqref {4.0}
		\nompageref{57}
  \item [{CDF}]\begingroup Cumulative Density Function\nomeqref {3.23}
		\nompageref{48}
  \item [{CI/OFDM}]\begingroup Carrier Interferometry OFDM\nomeqref {2.0}
		\nompageref{19}
  \item [{CIF}]\begingroup Common Interchange Format\nomeqref {5.6}
		\nompageref{65}
  \item [{COFDM}]\begingroup Coded OFDM\nomeqref {2.0}\nompageref{19}
  \item [{CRC}]\begingroup Cyclic Redundancy Check\nomeqref {1.0}
		\nompageref{4}
  \item [{CSI}]\begingroup Channel State Information\nomeqref {6.0}
		\nompageref{83}
  \item [{CSMA/CA}]\begingroup Carrier Sense Multiple Access with Collision Avoidance\nomeqref {2.0}
		\nompageref{22}
  \item [{D/A}]\begingroup Digital-to-Analog Conversion\nomeqref {2.0}
		\nompageref{18}
  \item [{DCT}]\begingroup Discrete Cosine Transform\nomeqref {2.0}
		\nompageref{8}
  \item [{DFT}]\begingroup Discrete Fourier Transform\nomeqref {2.0}
		\nompageref{20}
  \item [{DL}]\begingroup Downlink\nomeqref {2.0}\nompageref{27}
  \item [{DLC}]\begingroup Data Link Control\nomeqref {7.0}
		\nompageref{94}
  \item [{DPSK}]\begingroup Differential Phase Shift Keying\nomeqref {2.0}
		\nompageref{17}
  \item [{DSL}]\begingroup Digital Subscriber Line\nomeqref {2.0}
		\nompageref{25}
  \item [{DSSS}]\begingroup Direct Sequence Spread Spectrum\nomeqref {2.0}
		\nompageref{21}
  \item [{DVB}]\begingroup Digital Video Broadcasting\nomeqref {2.0}
		\nompageref{29}
  \item [{DWT}]\begingroup Discrete Wavelet Transform \nomeqref {2.0}
		\nompageref{8}
  \item [{eBCH}]\begingroup Extended BCH\nomeqref {3.22}\nompageref{45}
  \item [{ECG}]\begingroup Equal Gain Combining\nomeqref {3.14}
		\nompageref{42}
  \item [{FAR}]\begingroup False Alarm Rate\nomeqref {7.12}
		\nompageref{104}
  \item [{FDD}]\begingroup Frequency Division Duplexing\nomeqref {2.0}
		\nompageref{26}
  \item [{FEC}]\begingroup Forward Error Correction\nomeqref {1.0}
		\nompageref{2}
  \item [{FFT}]\begingroup Fast Fourier Transform\nomeqref {2.0}
		\nompageref{18}
  \item [{FGS}]\begingroup Fine Granular Scalability\nomeqref {2.0}
		\nompageref{24}
  \item [{FHSS}]\begingroup Frequency Hopping Spread Spectrum\nomeqref {2.0}
		\nompageref{21}
  \item [{fps}]\begingroup frames per second\nomeqref {8.1}
		\nompageref{118}
  \item [{GBN}]\begingroup Go-back-N\nomeqref {2.0}\nompageref{12}
  \item [{GoB}]\begingroup Group of Blocks\nomeqref {2.0}
		\nompageref{14}
  \item [{GoP}]\begingroup Group of Pictures\nomeqref {2.0}
		\nompageref{8}
  \item [{GSM}]\begingroup Global System for Mobile Communications\nomeqref {2.0}
		\nompageref{28}
  \item [{HARQ}]\begingroup Hybrid Automatic Repeat Request\nomeqref {1.0}
		\nompageref{2}
  \item [{HARQ-CC}]\begingroup HARQ with Chase Combining\nomeqref {1.0}
		\nompageref{3}
  \item [{HARQ-IR}]\begingroup HARQ with Incremental Redundancy\nomeqref {1.0}
		\nompageref{3}
  \item [{HDD}]\begingroup Hard Decision Decoding\nomeqref {7.6}
		\nompageref{102}
  \item [{HDTV}]\begingroup High Definition Television\nomeqref {2.0}
		\nompageref{30}
  \item [{HIHO}]\begingroup Hard-Input Hard-Output\nomeqref {1.0}
		\nompageref{4}
  \item [{HIPERLAN}]\begingroup High Performance Radio LAN\nomeqref {2.0}
		\nompageref{21}
  \item [{HP}]\begingroup High Priority\nomeqref {2.0}\nompageref{16}
  \item [{HQAM}]\begingroup Hierarchical QAM\nomeqref {2.0}
		\nompageref{17}
  \item [{HSPA}]\begingroup High Speed Packet Access\nomeqref {2.0}
		\nompageref{24}
  \item [{ICI}]\begingroup Intercarrier Interference\nomeqref {2.0}
		\nompageref{19}
  \item [{IEEE}]\begingroup Institute of Electrical and Electronics Engineers\nomeqref {2.0}
		\nompageref{25}
  \item [{IFFT}]\begingroup Inverse FFT\nomeqref {2.0}\nompageref{18}
  \item [{ISI}]\begingroup Intersymbol Interference\nomeqref {2.0}
		\nompageref{18}
  \item [{ITU}]\begingroup International Telecommunication Union\nomeqref {2.0}
		\nompageref{25}
  \item [{JSCC}]\begingroup Joint Source Channel Coding\nomeqref {2.0}
		\nompageref{11}
  \item [{LDPC}]\begingroup Low Density Parity Check\nomeqref {2.0}
		\nompageref{30}
  \item [{LFSR}]\begingroup Linear Feedback Shift Register\nomeqref {7.0}
		\nompageref{94}
  \item [{LLR}]\begingroup Log-Likelihood Ratio\nomeqref {7.0}
		\nompageref{95}
  \item [{LOS}]\begingroup Line of Sight\nomeqref {2.0}\nompageref{25}
  \item [{LP}]\begingroup Low Priority\nomeqref {2.0}\nompageref{16}
  \item [{LPF}]\begingroup Low Pass Filter\nomeqref {2.0}
		\nompageref{18}
  \item [{LTE}]\begingroup Long Term Evolution\nomeqref {2.0}
		\nompageref{27}
  \item [{MAC}]\begingroup Medium Access Control\nomeqref {2.0}
		\nompageref{29}
  \item [{MBMS}]\begingroup Multimedia Broadcast and Multicast Services\nomeqref {2.0}
		\nompageref{29}
  \item [{MDC}]\begingroup Multiple Description Coding\nomeqref {2.0}
		\nompageref{9}
  \item [{MDR}]\begingroup Misdetection Rate\nomeqref {7.12}
		\nompageref{104}
  \item [{MIMO}]\begingroup Multiple-Input Multiple-Output\nomeqref {2.0}
		\nompageref{20}
  \item [{MISO}]\begingroup Multiple Input Single Output\nomeqref {2.0}
		\nompageref{31}
  \item [{MLD}]\begingroup Maximum Likelihood Decoding\nomeqref {2.0}
		\nompageref{12}
  \item [{MMSE}]\begingroup Minimum Mean Square Error\nomeqref {2.0}
		\nompageref{19}
  \item [{MOS}]\begingroup Mean Opinion Score\nomeqref {2.0}
		\nompageref{32}
  \item [{MRC}]\begingroup Maximal Ratio Combining\nomeqref {3.1}
		\nompageref{39}
  \item [{MSB}]\begingroup Most Significant Bit\nomeqref {2.0}
		\nompageref{16}
  \item [{NACK}]\begingroup Negative Acknowledgment\nomeqref {2.0}
		\nompageref{12}
  \item [{NLOS}]\begingroup Non-Line of Sight\nomeqref {2.0}
		\nompageref{25}
  \item [{OFDM}]\begingroup Orthogonal Frequency Division Multiplexing\nomeqref {2.0}
		\nompageref{17}
  \item [{OFDMA}]\begingroup Orthogonal Frequency Division Multiple Access\nomeqref {2.0}
		\nompageref{26}
  \item [{P/S}]\begingroup Parallel-to-Serial Conversion\nomeqref {2.0}
		\nompageref{18}
  \item [{PAPR}]\begingroup Peak-to-Average Power Ratio\nomeqref {2.0}
		\nompageref{18}
  \item [{PCCC}]\begingroup Parallel Concatenated Convolutional Codes\nomeqref {5.0}
		\nompageref{61}
  \item [{PDF}]\begingroup Probability Density Function\nomeqref {3.16}
		\nompageref{43}
  \item [{PDR}]\begingroup Packet Drop Rate\nomeqref {1.0}
		\nompageref{3}
  \item [{PED}]\begingroup Packet Error Detection\nomeqref {7.0}
		\nompageref{94}
  \item [{PER}]\begingroup Packet Error Rate\nomeqref {2.0}
		\nompageref{23}
  \item [{PHY}]\begingroup Physical Layer\nomeqref {7.0}\nompageref{94}
  \item [{PMF}]\begingroup Probability Mass Function\nomeqref {3.5}
		\nompageref{41}
  \item [{PSNR}]\begingroup Peak Signal to Noise Ratio\nomeqref {2.0}
		\nompageref{32}
  \item [{QAM}]\begingroup Quadrature Amplitude Modulation\nomeqref {2.0}
		\nompageref{15}
  \item [{QoE}]\begingroup Quality of Experience\nomeqref {1.0}
		\nompageref{4}
  \item [{QoS}]\begingroup Quality of Service\nomeqref {1.0}
		\nompageref{1}
  \item [{QPSK}]\begingroup Quadrature Phase Shift Keying\nomeqref {2.0}
		\nompageref{26}
  \item [{RLC}]\begingroup Radio Link Control\nomeqref {2.0}
		\nompageref{29}
  \item [{RTT}]\begingroup Round Trip Time\nomeqref {2.0}
		\nompageref{12}
  \item [{S/P}]\begingroup Serial-to-Parallel Conversion\nomeqref {2.0}
		\nompageref{18}
  \item [{SAS}]\begingroup Semi-Analytical Solution\nomeqref {1.0}
		\nompageref{3}
  \item [{SC-FDMA}]\begingroup Single Carrier Frequency Division Multiple Access\nomeqref {2.0}
		\nompageref{28}
  \item [{SISO}]\begingroup Soft-Input Soft-Output\nomeqref {1.0}
		\nompageref{3}
  \item [{SNR}]\begingroup Signal-to-Noise Ratio\nomeqref {1.0}
		\nompageref{4}
  \item [{SR}]\begingroup Selective Repeat\nomeqref {2.0}
		\nompageref{12}
  \item [{STD}]\begingroup Standard Deviation\nomeqref {7.0}
		\nompageref{94}
  \item [{SW}]\begingroup Stop-and-Wait\nomeqref {2.0}\nompageref{12}
  \item [{TDD}]\begingroup Time Division Duplexing\nomeqref {2.0}
		\nompageref{22}
  \item [{TDM}]\begingroup Time Division Multiplexing\nomeqref {2.0}
		\nompageref{32}
  \item [{TDMA}]\begingroup Time Division Multiple Access\nomeqref {2.0}
		\nompageref{22}
  \item [{TPC}]\begingroup Turbo Product Codes\nomeqref {1.0}
		\nompageref{3}
  \item [{TTI}]\begingroup Transmission Time Interval\nomeqref {2.0}
		\nompageref{24}
  \item [{UL}]\begingroup Uplink\nomeqref {2.0}\nompageref{27}
  \item [{UMTS}]\begingroup Universal Mobile Telecommunications System\nomeqref {2.0}
		\nompageref{24}
  \item [{UTRA}]\begingroup Universal Terrestrial Radio Access\nomeqref {2.0}
		\nompageref{24}
  \item [{VBR}]\begingroup Variable Bit Rate\nomeqref {8.1}
		\nompageref{120}
  \item [{VoD}]\begingroup Video on Demand\nomeqref {1.0}\nompageref{1}
  \item [{WCDMA}]\begingroup Wideband Code Division Multiple Access\nomeqref {2.0}
		\nompageref{24}
  \item [{WHT}]\begingroup Walsh-Hadamard Transform\nomeqref {2.0}
		\nompageref{20}
  \item [{WiMAX}]\begingroup Worldwide Interoperability for Microwave Access\nomeqref {2.0}
		\nompageref{26}
  \item [{WLAN}]\begingroup Wireless Local Area Network\nomeqref {2.0}
		\nompageref{21}
  \item [{XOR}]\begingroup Exclusive-OR\nomeqref {7.0}\nompageref{98}

\end{list}

\end{singlespace}

\chapter{List of Symbols}

\begin{singlespace}

%
%


\begin{list}{}
{
\setlength{\labelwidth}{2.5cm}
\setlength{\labelsep}{0.5cm}
\setlength{\leftmargin}{3cm}
}



  \item [{$(.)^{\ast }$}]\begingroup complex conjugation process\nomeqref {3.2}
		\nompageref{39}
  \item [{$\bar{\nu}$}]\begingroup number of ones in $\mathbf{\bar{g}}$\nomeqref {7.5}
		\nompageref{100}
  \item [{$\bar{g}(X)$}]\begingroup CRC generator polynomial\nomeqref {7.5}
		\nompageref{100}
  \item [{$\chi$}]\begingroup transmission data rate in bps\nomeqref {3.28}
		\nompageref{49}
  \item [{$\circ$}]\begingroup Hadamard product\nomeqref {3.1}
		\nompageref{38}
  \item [{$\delta$}]\begingroup search step size\nomeqref {5.8}
		\nompageref{71}
  \item [{$\epsilon$}]\begingroup ratio of $\delta$ to $\mathcal{\tilde{P}}_{0}$\nomeqref {5.8}
		\nompageref{71}
  \item [{$\eta$}]\begingroup throughput\nomeqref {3.2}\nompageref{40}
  \item [{$\gamma^{(l)}$}]\begingroup SNR during the $l$th transmission session\nomeqref {3.17}
		\nompageref{43}
  \item [{$\gamma_\text{C}$}]\begingroup combined SNR\nomeqref {3.17}
		\nompageref{43}
  \item [{$\hat{x}$}]\begingroup index of row in which first error is discovered\nomeqref {7.0}
		\nompageref{97}
  \item [{$\kappa$}]\begingroup number of information bits per subpacket\nomeqref {3.0}
		\nompageref{37}
  \item [{$\left\lceil .\right\rceil$}]\begingroup ceiling function\nomeqref {5.9}
		\nompageref{72}
  \item [{$\left\Vert . \right\Vert _{1}$}]\begingroup Manhattan norm\nomeqref {7.0}
		\nompageref{97}
  \item [{$\mathbb{C}$}]\begingroup percentage of concealed frames in decoded video\nomeqref {8.1}
		\nompageref{119}
  \item [{$\mathbb{E}\left\{.\right\}$}]\begingroup expected value of a random variable\nomeqref {3.2}
		\nompageref{40}
  \item [{$\mathbb{Q}$}]\begingroup average PSNR of the original transmitted video\nomeqref {8.1}
		\nompageref{119}
  \item [{$\mathbf{\bar{g}}$}]\begingroup vector representation of $\bar{g}(X)$\nomeqref {7.5}
		\nompageref{100}
  \item [{$\mathbf{\widehat{D}}$}]\begingroup decoded binary $n\times n$ matrix\nomeqref {7.0}
		\nompageref{95}
  \item [{$\mathbf{\widehat{d}}_{x}$}]\begingroup $x$th row in $\mathbf{\widehat{D}}$\nomeqref {7.0}
		\nompageref{96}
  \item [{$\mathbf{A}$}]\begingroup MRC weights\nomeqref {3.2}
		\nompageref{39}
  \item [{$\mathbf{C}$}]\begingroup TPC codeword matrix\nomeqref {3.0}
		\nompageref{37}
  \item [{$\mathbf{c}_{x}$}]\begingroup $x$th row in $\mathbf{C}$\nomeqref {7.0}
		\nompageref{96}
  \item [{$\mathbf{d}$}]\begingroup information bits sequence\nomeqref {3.0}
		\nompageref{37}
  \item [{$\mathbf{E}$}]\begingroup post-decoding error pattern matrix\nomeqref {7.0}
		\nompageref{96}
  \item [{$\mathbf{e}_{x}$}]\begingroup $x$th row in $\mathbf{E}$\nomeqref {7.0}
		\nompageref{96}
  \item [{$\mathbf{F}$}]\begingroup channel matrix\nomeqref {3.1}
		\nompageref{38}
  \item [{$\mathbf{g}$}]\begingroup vector representation of $g(X)$\nomeqref {7.0}
		\nompageref{97}
  \item [{$\mathbf{H}$}]\begingroup parity check matrix\nomeqref {7.0}
		\nompageref{96}
  \item [{$\mathbf{h}_{x}$}]\begingroup $x$th row in $\mathbf{H}$\nomeqref {7.0}
		\nompageref{97}
  \item [{$\mathbf{m}$}]\begingroup CRC encoded bits sequence\nomeqref {3.0}
		\nompageref{37}
  \item [{$\mathbf{R}$}]\begingroup received subpacket\nomeqref {3.1}
		\nompageref{38}
  \item [{$\mathbf{R}_\text{C}$}]\begingroup output of the Chase combiner\nomeqref {3.2}
		\nompageref{39}
  \item [{$\mathbf{S}$}]\begingroup syndrome matrix\nomeqref {7.0}
		\nompageref{96}
  \item [{$\mathbf{s}_{x}$}]\begingroup $x$th row in $\mathbf{S}$\nomeqref {7.0}
		\nompageref{96}
  \item [{$\mathbf{U}$}]\begingroup transmitted subpacket\nomeqref {3.1}
		\nompageref{38}
  \item [{$\mathbf{W}$}]\begingroup AWGN matrix\nomeqref {3.1}
		\nompageref{38}
  \item [{$\mathcal{\bar{P}}$}]\begingroup average transmission power to deliver an information bit\nomeqref {5.4}
		\nompageref{65}
  \item [{$\mathcal{\tilde{P}}_{0}$}]\begingroup initial searching point\nomeqref {5.9}
		\nompageref{72}
  \item [{$\mathcal{A}$}]\begingroup number of additions\nomeqref {5.9}
		\nompageref{72}
  \item [{$\mathcal{C}_\text{C}^\text{H}$}]\begingroup relative complexity of CRC detection to HIHO decoding\nomeqref {7.11}
		\nompageref{103}
  \item [{$\mathcal{C}_\text{C}^\text{S}$}]\begingroup relative complexity of CRC detection to SISO decoding\nomeqref {7.11}
		\nompageref{103}
  \item [{$\mathcal{C}_\text{S}$}]\begingroup relative complexity TPC self-detection to CRC detection\nomeqref {7.5}
		\nompageref{100}
  \item [{$\mathcal{I}_\text{BF}$}]\begingroup total number of iterations for Brute-Force\nomeqref {5.9}
		\nompageref{72}
  \item [{$\mathcal{I}_\text{BI}$}]\begingroup total number of iterations for Bisection\nomeqref {5.9}
		\nompageref{72}
  \item [{$\mathcal{M}$}]\begingroup number of multiplications\nomeqref {5.9}
		\nompageref{72}
  \item [{$\mathcal{N}$}]\begingroup length of packet\nomeqref {3.0}
		\nompageref{37}
  \item [{$\mathcal{P}^{\ast}$}]\begingroup optimal transmit power\nomeqref {5.7}
		\nompageref{69}
  \item [{$\mathcal{P}_{\max}$}]\begingroup maximum transmit power\nomeqref {5.7}
		\nompageref{69}
  \item [{$\mathcal{P}_{i}$}]\begingroup transmit power per bit during $i$th transmission round\nomeqref {5.4}
		\nompageref{65}
  \item [{$\mathcal{R}$}]\begingroup number of transmissions per packet\nomeqref {3.28}
		\nompageref{49}
  \item [{$\mathcal{T}$}]\begingroup transmission time of video sequence\nomeqref {5.6}
		\nompageref{66}
  \item [{$\mu$}]\begingroup throughput scaling factor required for power optimization\nomeqref {5.8}
		\nompageref{71}
  \item [{$\nu$}]\begingroup number of ones in $\mathbf{g}$\nomeqref {7.0}
		\nompageref{97}
  \item [{$\omega$}]\begingroup target bit rate in video encoder\nomeqref {5.6}
		\nompageref{67}
  \item [{$\Psi$}]\begingroup equivalent SNR\nomeqref {3.23}
		\nompageref{45}
  \item [{$\rho$}]\begingroup number of transmissions per subpacket\nomeqref {3.2}
		\nompageref{40}
  \item [{$\sigma_\text{f}^{2}$}]\begingroup variance of fading channel coefficients\nomeqref {3.1}
		\nompageref{38}
  \item [{$\sigma_\text{w}^{2}$}]\begingroup variance of AWGN\nomeqref {3.1}
		\nompageref{38}
  \item [{$\sign(.)$}]\begingroup signum function\nomeqref {7.0}
		\nompageref{95}
  \item [{$\tau$}]\begingroup transmission time using packet-based HARQ\nomeqref {3.28}
		\nompageref{49}
  \item [{$\tau_\text{s}$}]\begingroup transmission time using subpacket-based HARQ\nomeqref {3.28}
		\nompageref{49}
  \item [{$\tilde{\rho}$}]\begingroup unbounded number of transmissions per subpacket\nomeqref {3.2}
		\nompageref{40}
  \item [{$\widehat{\mathbb{Q}}$}]\begingroup average PSNR of received video\nomeqref {5.6}
		\nompageref{66}
  \item [{$\widehat{d}_{x}(X)$}]\begingroup polynomial representation of $\mathbf{\widehat{d}}_{x}$\nomeqref {7.0}
		\nompageref{96}
  \item [{$\widehat{Q}^{(i)}$}]\begingroup PSNR of $i$th frame in received video\nomeqref {5.6}
		\nompageref{66}
  \item [{$\zeta$}]\begingroup code rate\nomeqref {5.4}\nompageref{65}
  \item [{$\zeta_{\mathcal{P}}$}]\begingroup power efficiency\nomeqref {5.4}
		\nompageref{65}
  \item [{$\{.\}^{\top}$}]\begingroup matrix transpose operation\nomeqref {7.0}
		\nompageref{96}
  \item [{$a$}]\begingroup element of $\mathbf{A}$\nomeqref {3.2}
		\nompageref{39}
  \item [{$B_\text{C}$}]\begingroup playback buffer occupancy in frames\nomeqref {8.1}
		\nompageref{117}
  \item [{$B_\text{p}$}]\begingroup preroll buffer occupancy threshold\nomeqref {8.1}
		\nompageref{119}
  \item [{$B_\text{TH}$}]\begingroup buffer occupancy threshold for adapting ARQ limit\nomeqref {8.1}
		\nompageref{117}
  \item [{$B_\text{TH}^\text{(B)}$}]\begingroup $B_\text{TH}$ for B frame\nomeqref {8.1}
		\nompageref{117}
  \item [{$B_\text{TH}^\text{(I)}$}]\begingroup $B_\text{TH}$ for I frame\nomeqref {8.1}
		\nompageref{117}
  \item [{$B_\text{TH}^\text{(P)}$}]\begingroup $B_\text{TH}$ for P frame\nomeqref {8.1}
		\nompageref{117}
  \item [{$C$}]\begingroup product code\nomeqref {3.0}\nompageref{36}
  \item [{$C^{i}$}]\begingroup $i$th component code\nomeqref {3.0}
		\nompageref{36}
  \item [{$d_\text{min}$}]\begingroup minimum Hamming distance of component codeword\nomeqref {3.0}
		\nompageref{36}
  \item [{$E_\text{b}$}]\begingroup average energy per bit\nomeqref {3.23}
		\nompageref{45}
  \item [{$F$}]\begingroup number of frames in video sequence\nomeqref {5.6}
		\nompageref{66}
  \item [{$f$}]\begingroup element of $\mathbf{F}$\nomeqref {3.1}
		\nompageref{38}
  \item [{$f_\text{P}$}]\begingroup playback rate in fps\nomeqref {8.1}
		\nompageref{117}
  \item [{$g(X)$}]\begingroup generator polynomial\nomeqref {7.0}
		\nompageref{96}
  \item [{$K$}]\begingroup length of information bits sequence\nomeqref {3.0}
		\nompageref{37}
  \item [{$k$}]\begingroup number of information bits in component codeword\nomeqref {3.0}
		\nompageref{36}
  \item [{$L$}]\begingroup number of subpackets\nomeqref {3.0}
		\nompageref{37}
  \item [{$l_\text{crc}$}]\begingroup length of CRC bits\nomeqref {3.0}
		\nompageref{37}
  \item [{$l_\text{p}$}]\begingroup length of TPC parity bits\nomeqref {3.0}
		\nompageref{37}
  \item [{$M$}]\begingroup maximum number of allowed ARQ rounds\nomeqref {3.2}
		\nompageref{39}
  \item [{$N$}]\begingroup length of subpacket\nomeqref {3.0}
		\nompageref{37}
  \item [{$n$}]\begingroup length of component codeword\nomeqref {3.0}
		\nompageref{36}
  \item [{$N_\text{C}$}]\begingroup number of XOR operations for CRC detection\nomeqref {7.5}
		\nompageref{100}
  \item [{$N_\text{F}^{(i)}$}]\begingroup size of $i$th video frame\nomeqref {5.6}
		\nompageref{66}
	\item [{$N_\text{P}$}]\begingroup number of packets per video frame/picture\nomeqref {7.0}
		\nompageref{36}
  \item [{$N_\text{S}$}]\begingroup number of XOR operations in TPC self-detection\nomeqref {7.0}
		\nompageref{97}
  \item [{$N_\text{S}^{\text{lfsr}}$}]\begingroup $N_\text{S}$ using LFSR\nomeqref {7.0}
		\nompageref{97}
  \item [{$N_\text{S}^{\text{mtx}}$}]\begingroup $N_\text{S}$ using matrix multiplication\nomeqref {7.0}
		\nompageref{97}
  \item [{$N_{0}$}]\begingroup noise power spectral density\nomeqref {3.23}
		\nompageref{45}
  \item [{$N_{\text{HDD}}$}]\begingroup computational complexity for BCH HDD operation\nomeqref {7.9}
		\nompageref{101}
  \item [{$N_{\text{TPC}}^{\text{H}}$}]\begingroup HDD complexity of HIHO TPC per half iteration\nomeqref {7.9}
		\nompageref{101}
  \item [{$N_{\text{TPC}}^{\text{S}}$}]\begingroup HDD complexity of SISO TPC per half iteration\nomeqref {7.9}
		\nompageref{101}
  \item [{$p$}]\begingroup number of least reliable elements in Chase-II decoder\nomeqref {7.9}
		\nompageref{101}
  \item [{$P(.)$}]\begingroup probability of an event\nomeqref {3.2}
		\nompageref{40}
  \item [{$P_\text{D}$}]\begingroup subpacket drop rate\nomeqref {3.2}
		\nompageref{40}
  \item [{$p_\text{e}$}]\begingroup BER of BPSK in Rayleigh fading without combining\nomeqref {3.23}
		\nompageref{45}
  \item [{$P_\text{E}^{(i)}$}]\begingroup probability of subpacket error in $i$th ARQ round\nomeqref {3.2}
		\nompageref{40}
  \item [{$p_\text{e}^{\{\ell\}}$}]\begingroup BER of BPSK with $\ell$-order diversity in Rayleigh fading\nomeqref {3.23}
		\nompageref{45}
  \item [{$P_\text{F}$}]\begingroup false alarm probability\nomeqref {7.12}
		\nompageref{103}
  \item [{$P_\text{M}$}]\begingroup misdetection probability\nomeqref {7.12}
		\nompageref{103}
  \item [{$r$}]\begingroup element of $\mathbf{R}$\nomeqref {3.1}
		\nompageref{38}
  \item [{$s_{x}(X)$}]\begingroup syndrome polynomial\nomeqref {7.0}
		\nompageref{96}
  \item [{$t$}]\begingroup number of errors that can be corrected by a code\nomeqref {7.9}
		\nompageref{101}
  \item [{$T_\text{B}$}]\begingroup budget time for delivery of a video frame\nomeqref {8.1}
		\nompageref{117}
  \item [{$T_\text{F}$}]\begingroup delivery time of a video frame\nomeqref {8.1}
		\nompageref{117}
  \item [{$t_\text{p}$}]\begingroup propagation time\nomeqref {3.28}
		\nompageref{49}
  \item [{$u$}]\begingroup element of $\mathbf{U}$\nomeqref {3.1}
		\nompageref{38}
  \item [{$V$}]\begingroup number of transmitted subpackets\nomeqref {3.2}
		\nompageref{40}
  \item [{$w$}]\begingroup element of $\mathbf{W}$\nomeqref {3.1}
		\nompageref{38}
  \item [{$x$}]\begingroup row index\nomeqref {3.1}\nompageref{38}
  \item [{$y$}]\begingroup column index\nomeqref {3.1}\nompageref{38}
  \item [{$z_{i}$}]\begingroup random number that indicates if $i$th subpacket is dropped\nomeqref {3.2}
		\nompageref{40}

\end{list}

\end{singlespace}

\chapter{Publications Arising From This Research}

\begin{singlespace}


\begin{enumerate}

\section*{Journal Articles}

\item H. Mukhtar, A. Al-Dweik, M. Al-Mualla, and A. Shami, ``Adaptive hybrid ARQ system using turbo product codes with hard/soft decoding,'' \emph{IEEE Communications Letters}, vol. 17, no. 11, pp. 2132-2135, Nov. 2013.

\item H. Mukhtar, A. Al-Dweik, M. Al-Mualla, and A. Shami, ``Low complexity power optimization algorithm for multimedia transmission over wireless networks,'' \textit{IEEE Journal of Selected Topics in Signal Processing}, in press, DOI: 10.1109/JSTSP.2014.2331915, June 2014.

\item	H. Mukhtar, A. Al-Dweik and M. Al-Mualla, ``CRC-free hybrid ARQ system using turbo product codes,'' \textit{IEEE Transactions on Communications}, in press, DOI: 10.1109/TCOMM.2014.2366753, Nov. 2014.

\section*{Conference Papers}

\item	H. Mukhtar, A. Al-Dweik, and M. Al-Mualla, ``On the performance of adaptive HARQ with no channel state information feedback,'' accepted in \textit{IEEE Wireless Communication and Networking Conference (WCNC)}, New Orleans, LA, Sept. 2014.

\item	H. Mukhtar, ``Link adaptation for wireless video communication systems,'' in \textit{IEEE 20th International Conference on Electronics, Circuits, and Systems (ICECS)}, Abu Dhabi, UAE, Dec. 2013, pp. 66-67.

\section*{In Preparation}

\item H. Mukhtar, A. Al-Dweik, and M. Al-Mualla, ``Content-aware and occupancy-based adaptive hybrid ARQ for delay-sensitive video,'' Oct. 2014.

\item H. Mukhtar, A. Al-Dweik, and M. Al-Mualla, ``CRC-free hybrid ARQ system using a 2-D single parity error detection technique,'' Nov. 2014.

\section*{Others}

\item H. Mukhtar, A. Al-Dweik, and M. Al-Mualla, ``Hybrid ARQ with turbo product codes for wireless video communication,'' in \textit{Information and Communication Technology Research Forum (ICTRF)}, Abu Dhabi, UAE, May 2013. (Winner of best student oral-presentation)

\item H. Mukhtar, A. Al-Dweik, and M. Al-Mualla, ``Power optimization of hybrid ARQ with packet combining over Rayleigh fading channels,'' in \textit{ICTRF Graduate Engineering Research Symposium (GERS)}, Abu Dhabi, UAE, May 2014. 

\end{enumerate}


\end{singlespace}


\mainmatter

\renewcommand{\chaptermark}[1]{%
\markboth{\chaptername\ \thechapter.\ #1}{}}

\chapter{Introduction}
\label{ch:intro}

Delivery of digital video over wireless networks is becoming increasingly popular. Recent advances in wireless access networks provide a promising solution for the delivery of multimedia services to end-user premises. In contrast to wired networks, wireless networks not only offer a larger geographical coverage at lower deployment cost, but also support mobility. Video applications such as interactive video, live video, video on demand (VoD)\nomenclature{VoD}{Video on Demand}, and video surveillance will be available for users anytime, anywhere, and via any web-enabled device. Nevertheless, there are several challenges that currently attract the attention of a wide sector of the research community of wireless video communication. 

Due to the dynamic and erroneous nature of wireless channels, delivering video services with quality of service (QoS)\nomenclature{QoS}{Quality of Service} guarantees is a difficult task. Wireless channels have limited bandwidth and they introduce losses and errors due to noise, multipath fading and interference. On the other hand, video transmission has strict QoS requirements such as high data rates, bounded delay, and low packet drop rate. These QoS requirements are often traded-off with transmit power; however, the energy of mobile and hand-held wireless devices is limited by a small size battery which poses another challenge in wireless video communication.

To overcome these challenges many solutions have been proposed in the literature. These solutions can be categorized into video encoding techniques and wireless link adaptation techniques. Video encoding techniques have achieved significant improvements in compression rates to reduce bandwidth requirements. Compression techniques exploit video temporal redundancy where some video frames are predicted from selected reference frames. Therefore, encoded video is sensitive to packet losses where error might propagate for successive inter-dependent frames degrading the quality of the decoded video. Recent video encoding standards have introduced error resiliency techniques such as scalable video coding and periodic intra-coded frame refresh to improve the immunity of compressed video against packet loss and error propagation. These video encoding techniques alone are not enough to overcome bandwidth limitation and packet loss degradation. 

The other class of solutions which is used to further combat challenges in wireless transmission is link adaptation. Link adaptation is the process of dynamically changing transmission parameters such as modulation order, channel coding rate, and transmission power level based on the estimated condition of the wireless link for efficient utilization of system resources. Most modern wireless communication systems are equipped with link adaptation modules. These systems are mainly designed to operate as data-centric networks. As a result, their link adaptation is designed to maximize the spectral efficiency and transmission reliability often at the expense of increased latency. However, excessive latency and delay jitter cannot be tolerated for video applications especially for real-time video and interactive video. Moreover, in data networks, the transmitted data are treated with equal importance, whereas video content has unequal importance and the loss of some video packets has higher distortion impact on the received video quality compared to other less important packets. Therefore, the objective of this PhD project is to develop content-aware link adaptation solutions for minimizing video quality distortion in wireless systems with constraints in bandwidth, delay, and power.

Based on our literature search, we identified that the most effective link adaptation technique to minimize packet drop rate is hybrid automatic repeat request (HARQ)\nomenclature{HARQ}{Hybrid Automatic Repeat Request}. In HARQ, forward error correction (FEC)\nomenclature{FEC}{Forward Error Correction} is combined with automatic repeat request (ARQ)\nomenclature{ARQ}{Automatic Repeat Request} to achieve reliable transmission. HARQ employs retransmissions to minimize packet drop rate at the expense of increased delay. However, the maximum number of allowed retransmissions are limited to ensure bounded delay. There are mainly three types of HARQ which are Type-I HARQ, Type-II HARQ with Chase combining (HARQ-CC)\nomenclature{HARQ-CC}{HARQ with Chase Combining} and Type-II HARQ with incremental redundancy (HARQ-IR)\nomenclature{HARQ-IR}{HARQ with Incremental Redundancy} \cite{lin1983error,dahlman20114g}. Most modern communication standards employ Type-II HARQ for its superior performance when compared to Type-I HARQ. Moreover, HARQ-CC is preferred over HARQ-IR for its lower complexity. 

In this work, HARQ-CC is considered and implemented using turbo product codes (TPC) \nomenclature{TPC}{Turbo Product Codes} to achieve high coding gain performance. TPC are capacity-approaching FEC codes that can be implemented with reasonable complexity \cite{glavieux2007channel}. They support a wide range of codeword sizes and code rates. TPC are now included in some communication standards such as the IEEE-802.16 for fixed and mobile broadband wireless access systems \cite{5062485} and the IEEE-1901 for broadband power line networks \cite{5678772}. Nevertheless, TPC-based HARQ has not received enough attention in the literature. TPC systems are usually evaluated in terms of the bit error rate (BER)\nomenclature{BER}{Bit Error Rate} and packet drop rate (PDR)\nomenclature{PDR}{Packet Drop Rate}. However, other performance metrics such as throughput can be more informative in describing the efficiency and reliability of the transmission system.  

In this PhD project, we start by evaluating the performance of TPC-based HARQ in terms of throughput and delay. Truncated HARQ is adopted for bounded delay and a subpacket HARQ scheme is also employed for enhanced delay performance. A semi-analytical solution (SAS)\nomenclature{SAS}{Semi-Analytical Solution} is derived to obtain the system throughput and delay performance of HARQ-CC using the subpacket error probability of regular TPC systems. The SAS is derived for both additive white Gaussian noise (AWGN)\nomenclature{AWGN}{Additive White Gaussian Noise} and Rayleigh fading channels.

Video signals can be corrupted in the analog/pixel domain by different types of noise such as photon shot noise and camera noise \cite{ghazal2007real}. According to the central limit theorem, the aggregate noise is approximated as Gaussian noise. Source coding techniques and pre-processing or post-processing filters are used to reduce the effect of such noise. However, in this work, we are concerned with the development of link adaptation solutions at the Physical and Data Link layers for the transmission of digital video signals, which are modulated and transmitted over AWGN and Rayleigh wireless channels \cite{proakis1995digital}.

The TPC-based HARQ is evaluated for different codeword sizes and code rates using both extensive Monte Carlo simulations and the SAS. Based on the obtained results a link adaptation system is proposed where the subpacket/codeword size and code rate are dynamically adjusted based on the channel condition to maximize the HARQ throughput \cite{H-Mukhtar-2013}. Moreover, interesting observations are made based on the obtained throughput results. 

It is known that the ultimate coding gain of TPC is achieved by performing a number of soft-input soft-output (SISO)\nomenclature{SISO}{Soft-Input Soft-Output} iterative decoding processes which require considerable computational power. Hard-input hard-output (HIHO)\nomenclature{HIHO}{Hard-Input Hard-Output} decoding is considerably less computationally complex. However, in the literature, the high computational complexity of SISO decoding is justified by its high coding gain advantage over HIHO decoding. Nevertheless, the obtained throughput results for HARQ-CC reveal that the coding gain advantage of the SISO over HIHO decoding is significantly reduced and sometimes vanishes for some TPC codes when Chase combining is used \cite{H-Mukhtar-2013}.

In addition, the throughput results exhibit a staircase shape where the throughput remains fixed for a wide range of signal-to-noise ratios (SNRs)\nomenclature{SNR}{Signal-to-Noise Ratio}. Consequently, the transmit power can be reduced significantly while the throughput remains almost unchanged. The obtained results reveal that invoking power optimization algorithms can achieve a significant energy saving of about $80\%$ for particular scenarios \cite{mukhtar2014low}.

Moreover, TPC have error self-detection capabilities which have not been utilized in the literature for HARQ. TPC self-detection eliminates the need for additional redundancy codes such as cyclic redundancy check (CRC)\nomenclature{CRC}{Cyclic Redundancy Check} codes which are typically required for error detection in HARQ systems. Numerical and simulation results show that the CRC-free TPC-HARQ system consistently provides equivalent or higher throughput than CRC-based HARQ systems. The TPC self-detection also achieves lower computational complexity than CRC detection, especially for TPC with high code rates. In particular scenarios, the relative complexity of the self-detection approach with respect to popular CRC techniques is about $0.3\%$ \cite{mukhtar2014crc}.

All of these findings make TPC-based HARQ a suitable candidate to meet the strict QoS constraints of video transmission in terms of high data rates, bounded delay, reduced power consumption and low computational complexity. However, link adaptation techniques can still result in undesirable quality of experience (QoE)\nomenclature{QoE}{Quality of Experience}  if they are used without regard to the characteristics of video bitstreams. Video packets have unequal importance in terms of the impact on video quality distortion. The link adaptation techniques should consider this property in order to allocate the available system resources to the most important packets. In addition, the continuity of video playback is another important metric which affects QoE. Therefore, the video playback buffer occupancy should be monitored to ensure timely delivery of the most important packets within the available playback budget time. Hence, we propose a content-aware and occupancy-based adaptive HARQ scheme to ensure minimum quality distortion with continuous video playback.


The rest of the thesis is organized as follows. Chapter \ref{chap:2}~provides a literature review and background information about wireless video communication solutions. Chapter~\ref{chap:3} describes the TPC-HARQ system model and the semi-analytical solution. In Chapter \ref{chap:4}, an adaptive scheme is proposed to maximize the throughput of HARQ-CC using TPC with Hard/Soft decoding. In Chapter \ref{chap:5}, the SAS is used  to develop a low complexity power optimization algorithm which significantly reduces the transmit power without affecting the throughput. Chapter \ref{chap:6} studies the performance of the power optimization algorithm based on ARQ feedback. In Chapter~\ref{chap:7}, a CRC-free HARQ scheme based on TPC error correction and self-detection is proposed to reduce delay and complexity. In Chapter \ref{chap:8}, an occupancy-based and content-aware adaptive HARQ system is proposed for delay-sensitive video using the obtained findings and models from previous chapters. Finally, Chapter \ref{chap:9} summarizes the conclusions and future work.

\chapter{Literature Review}
\label{chap:2}
Different techniques have been proposed in the literature that constitute a solution space for the challenges in wireless video communications \cite{chou2007moi,hsu1999rate,atzori2007cycle,hassan2005paa,chuang2007content,li2006joint,zhai2003joint}. These techniques can be categorized into video coding, error control and physical layer techniques. Examples of video coding techniques are scalable video coding, bitstream switching, and transcoding. Error control techniques include error resilient coding, HARQ, and error concealment, whereas, physical layer techniques include adaptive modulation and power allocation. 



\section{Video Coding}
\begin{figure}[H]
\centering
\includegraphics[width=\textwidth]{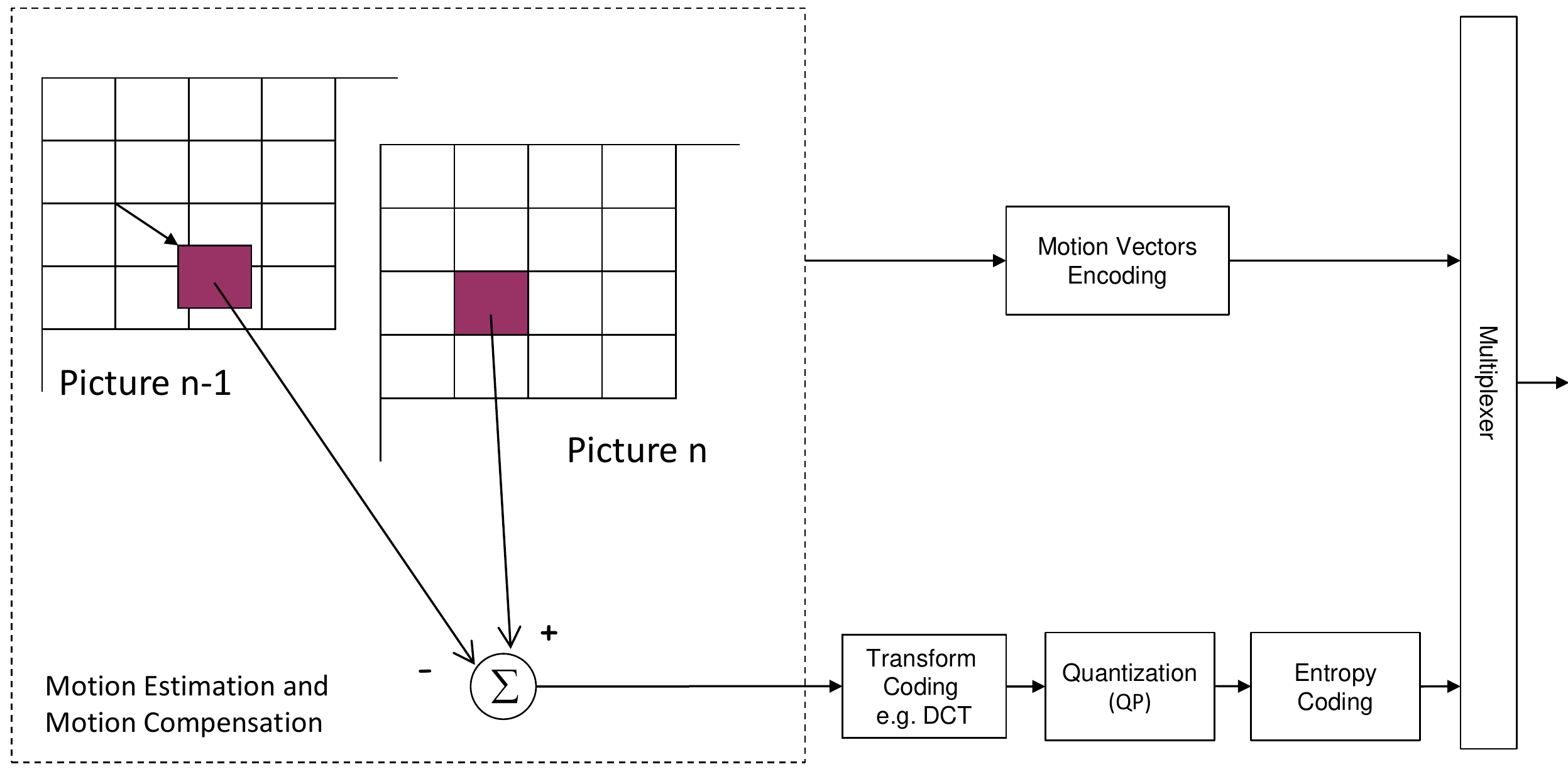}
\caption{Video compression stages.} 
\label{fig:vidcompr}
\end{figure}

Raw digital video contains an immense amount of data. It is composed of a time-ordered sequence of still images (frames). These images are required to be displayed at a certain rate so that objects' motion in a video sequence is perceived as continuous and natural by the human eye. Therefore, digital video transmission is considered one of the most bandwidth demanding data communication applications. Despite the recent advances in communication networks, channel bandwidth is still considered a scarce resource \cite{chou2007moi}. Hence, source coding and compression are essential in practical digital video communication. Video compression is composed of four main stages. These stages are motion estimation and compensation, transform coding, quantization and entropy coding as shown in Fig.~\ref{fig:vidcompr}.

\textbf{Motion estimation and compensation} are compression techniques which exploit video temporal redundancy \cite{al2002video}. In a video sequence, adjacent frames are very similar. Consequently, significant compression can be achieved by only encoding the differences between video frames. Motion estimation is the process of estimating the motion that occurred between a current frame and a reference frame. The estimated motion is represented by motion vectors which are then used to move the content of the reference frame to construct a motion-compensated prediction frame. This process is called motion compensation after which the encoder obtains the prediction error between the current frame and the motion-compensated frame.

\begin{figure}[H]
\centering
\includegraphics[width=4.5in]{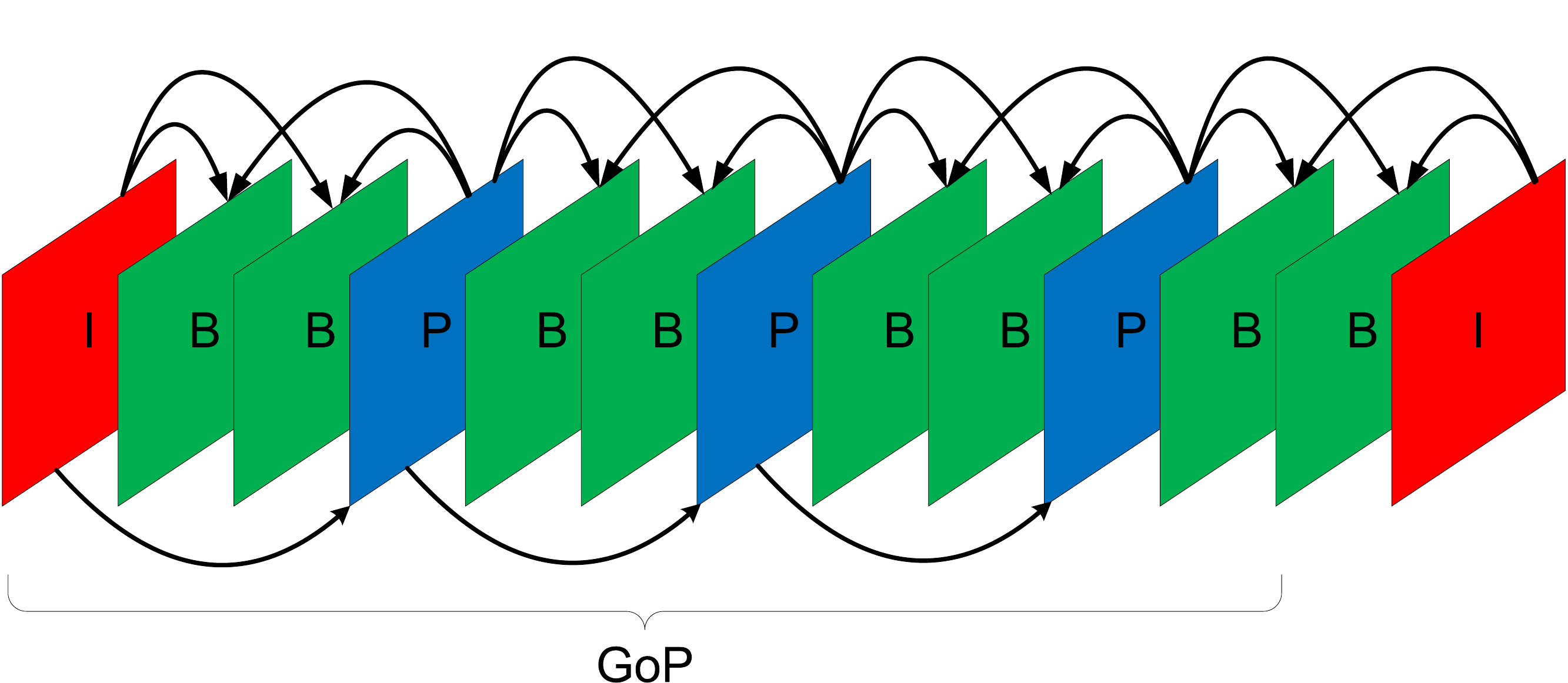}
\caption{Video frame types.} 
\label{fig:IPB}
\end{figure}

Most video codecs (e.g. MPEG-2, MPEG-4 \cite{watkinson2004mpeg}) implement motion estimation and motion compensation. A video sequence is encoded into 3 main frame types, namely, I, P, and B frames (see Fig.~\ref{fig:IPB}). I frames are 'intra-coded', independently of other frames using still image compression techniques (e.g. JPEG).  On the other hand, P and B frames are inter-coded based on previous or future encoded frames. P frames are 'predictively' coded based on previous I or P frames, while B frames are 'bi-directionally predicted' based on both previous and future I or P frames. B frames can also be used as a source of prediction as in the H.264 standard \cite{schwarz2007overview}. B frames achieve the highest compression level compared to other frame types. Nevertheless, I frames, which achieve low compression ratios, are introduced at regular intervals to help recover from transmission errors and limit error propagation among successive inter-coded frames. The frame distance between two consecutive I frames is known as the group of pictures (GoP)\nomenclature{GoP}{Group of Pictures} length.

\textbf{Transform coding} involves transforming video frames from the spatial domain into a more compact representation. Examples of transform coding are Discrete Cosine Transform (DCT)\nomenclature{DCT}{Discrete Cosine Transform} and Discrete Wavelet Transform (DWT)\nomenclature{DWT}{Discrete Wavelet Transform }. DCT is a transformation technique which is widely used in video compression. It provides transformation coefficients which represent video frames in the frequency domain. Unlike the data in the pixel domain, these coefficients are separable and with unequal importance. Knowing the fact that video frames (images) are low-frequency data by nature, compression can be achieved by considering the most important DCT coefficients which are the low-mid frequencies coefficients. Therefore, DCT is capable of reducing the spatial redundancy within a video frame by averaging out similar areas of color. On the other hand, DWT is a more sophisticated transformation with inherent scalability. In addition, DWT overcomes a drawback of block-based DCT known as blocking artifacts \cite{saha2000image}.

\textbf{Quantization} is a lossy compression technique where the transform coefficients are approximated by a discrete set of integer values. These quantized values are then represented in bits.

\textbf{Entropy coding} achieves additional compression by exploiting the redundancy in the bitstream. Entropy coding techniques such as run length coding, differential coding, and Huffman coding are applied to reduce this redundancy.

\subsection{Source Rate Control}
Source rate control is employed to adapt the source rate to the channel bandwidth variations with the objective of ensuring continuous playback \cite{atzori2007cycle}. This mechanism is implemented at the transmitter side. In certain schemes, the receiver monitors the channel condition and the state of decoder buffer \cite{hassan2007vso}. This information is fed back to the transmitter to adapt the source rate to match the available channel capacity. For example, when the available bandwidth decreases, the source rate will be reduced to avoid playback interruption by gracefully degrading the video quality. 

\subsection{Scalable Coding}
\begin{figure}[H]
\centering
\includegraphics[width=0.95\textwidth, angle=0]{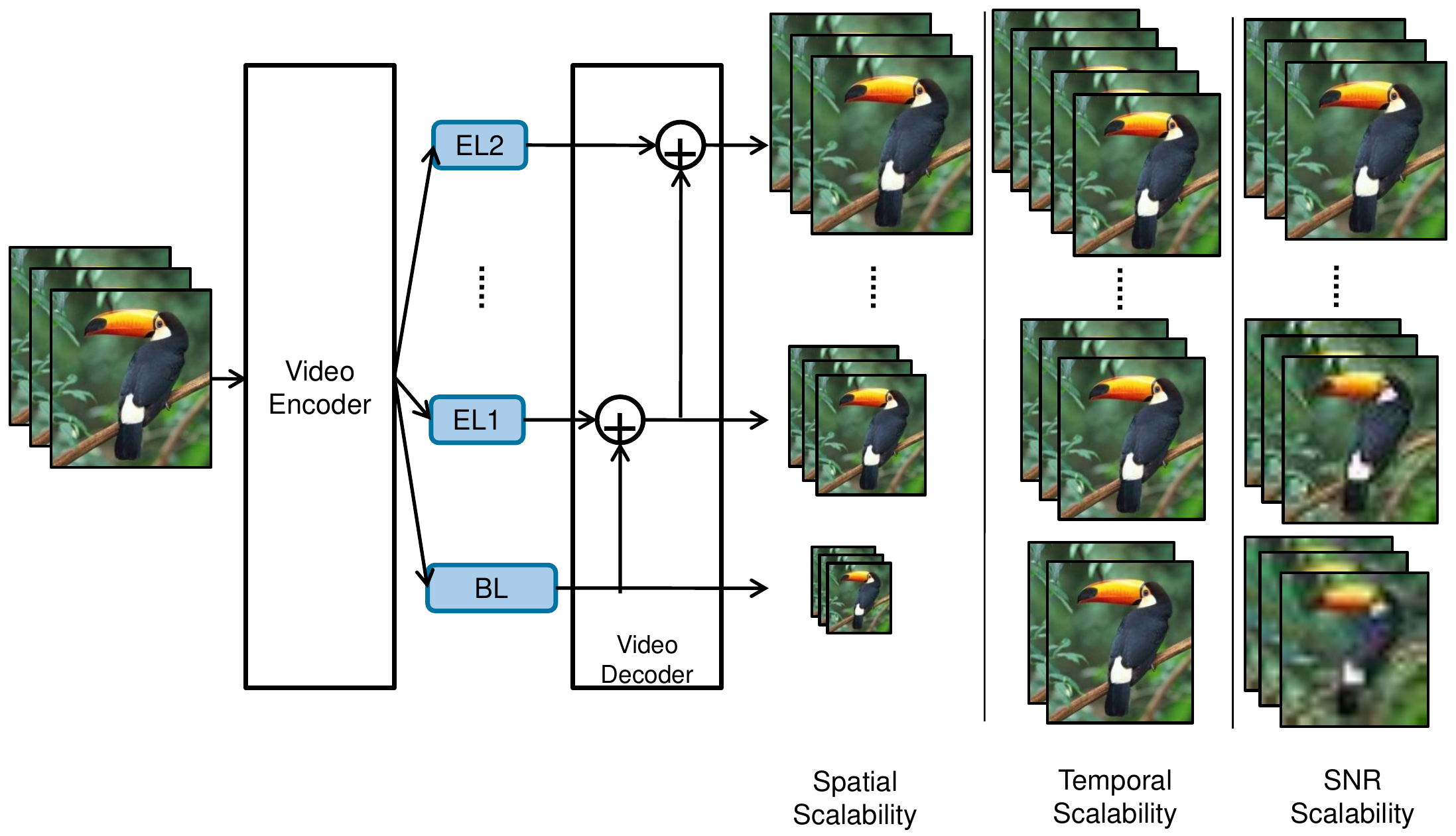}
\caption{Layered coding techniques.} 
\label{fig:layered}
\end{figure}

Scalable coding provides scalability to heterogeneous network links and video clients. Layered coding and multiple description coding (MDC)\nomenclature{MDC}{Multiple Description Coding} are the two classes of scalable coding. Layered coding encodes the video sequence into a base layer and multiple enhancement layers. The base layer provides a version of the original video sequence with minimum acceptable quality, whereas enhancement layers provide incremental improvement to the video quality when they are received. Nevertheless, enhancement layers can not be decoded without the base layer. Examples of layered coding techniques are spatial, temporal, and SNR layered coding. Fig. \ref{fig:layered} provides a general illustration of these layered coding techniques.

On the other hand, MDC encodes the video sequence into multiple descriptions (bitstreams). Unlike enhancement layers in layered coding, descriptions in MDC can be separately decoded. With each additional description, incremental improvement in video quality is achieved \cite{chou2007moi}.

\subsection{Bitstream Switching}
Bitstream switching is another adaptation technique which could be used when the other techniques do not guarantee the continuity of the playback. It requires the availability of several pre-encoded versions of the same video source with different encoding bit rates. Switching between the different bitstreams occurs according to the channel variations and decoder buffer occupancy. Usually, switching takes place at I frames to avoid the error drift problem \cite{sun2004sss}. Nevertheless, new types of encoded video frames (SI and SP) are introduced in \cite{karczewicz2003sas} to facilitate a more flexible drift-free switching. An example of how bitstream switching can be utilized to enhance the quality of the received video is explained in \cite{mukhtar2010multi}.

\subsection{Transcoding}
Transcoding relies on modifying the video encoding parameters to achieve a desired encoding bit rate. Examples of transcoding techniques are requantization and discarding high frequency DCT coefficients. A drawback of transcoding techniques is the drift problem. This problem arises when a reference frame is transcoded and subsequent dependent frames are predicted using this frame creating mismatch errors \cite{sun2004sss}.

\section{Error Control}
Error control mechanisms in video streaming includes forward error correction, automatic repeat request, error resilient coding, error concealment, and adaptive playback. These techniques are briefly described in the following subsections.

\subsection{Forward Error Correction}
FEC is a channel coding technique which adds redundancy to the bitstream. The introduced redundancy is structured in relation to the original data of the bitstream. An error in the received data will alter this structure and hence can be detected or even corrected. Hamming code, Reed-Solomon code, and convolutional codes are examples of FEC. FEC methods can improve throughput and can be static or adaptive. Adaptive FEC provides a more effective error control method where the FEC code rate is adapted to the channel state. In general, FEC introduces transmission latency due to the added redundancy. Nevertheless, this latency can be reduced by reducing the source rate to accommodate the FEC bits at the cost of slight reduction in video quality \cite{hassan2007vso}.  In addition, FEC can be jointly designed with the source coder to achieve effective video transmission \cite{argyriou2008erv}. This is often referred to by joint source channel coding (JSCC)\nomenclature{JSCC}{Joint Source Channel Coding} \cite{zhang2008joint} in which the channel coder provides different levels of protection based on the importance of source information.

\subsection{Automatic Repeat Request}
Another class of error control is retransmission or automatic repeat request. In this class, error detection techniques (e.g. parity check and CRC) are applied at the receiver to detect errors. The receiver sends acknowledgment (ACK)\nomenclature{ACK}{Acknowledgment} or negative acknowledgment (NACK)\nomenclature{NACK}{Negative Acknowledgment} messages to the transmitter to indicate whether a transmitted message was received correctly or not. In addition, the transmitter may initiate retransmission based on a timeout if ACK or NACK messages are delayed more than expected. This timeout is set relative to an estimated round trip time (RTT)\nomenclature{RTT}{Round Trip Time} which is the time between sending a message and receiving a positive ACK following the last successful retransmission of the same message. RTT can be estimated using a moving average of previously measured RTTs.

\begin{figure}[H]
\centering
\includegraphics[width=0.95\textwidth]{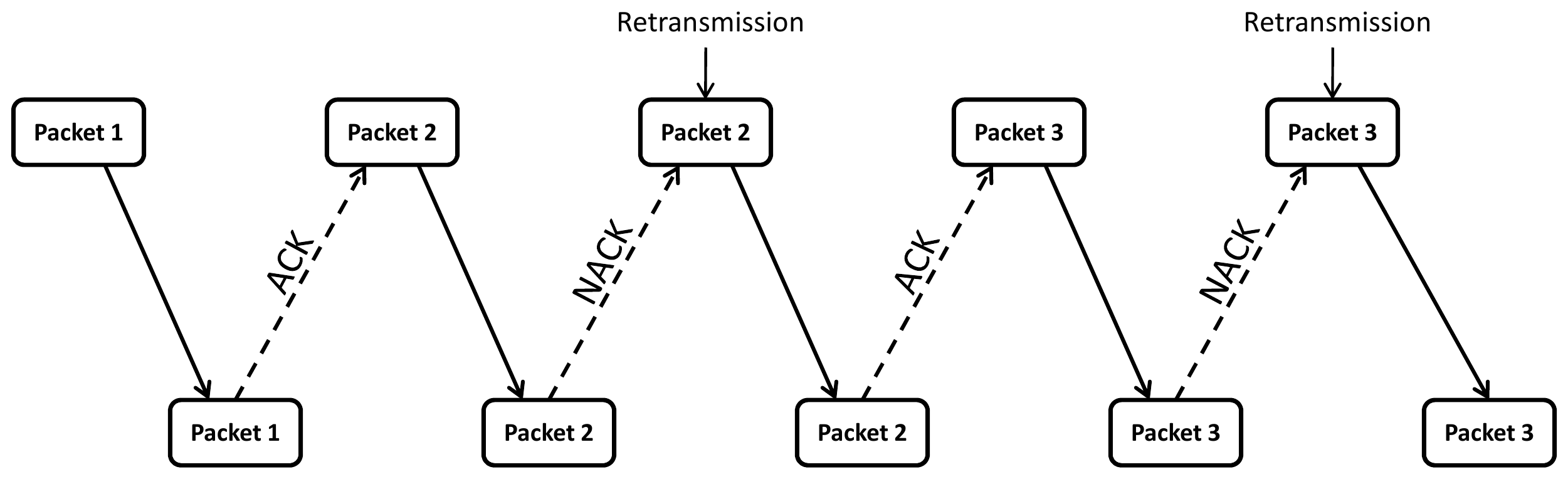}
\caption{Stop-and-Wait ARQ.}
\label{fig:SWARQ}
\end{figure}

\begin{figure}[H]
\centering
\includegraphics[width=0.95\textwidth]{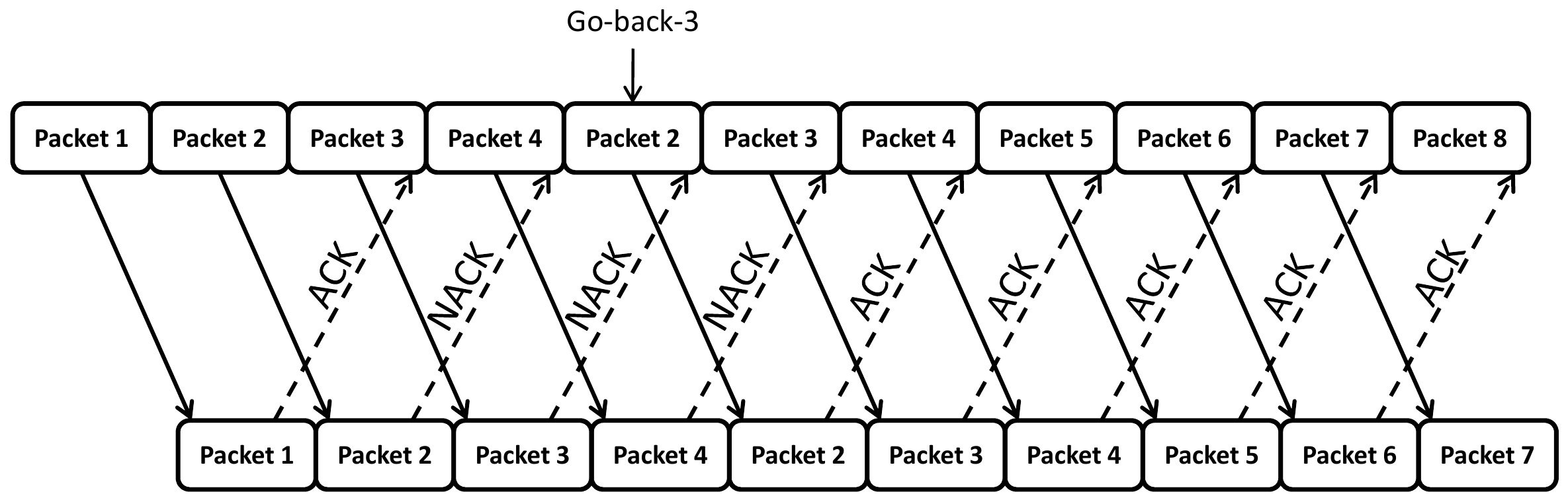}
\caption{Go-back-N ARQ.}
\label{fig:GBNARQ}
\end{figure}

\begin{figure}[H]
\centering
\includegraphics[width=0.95\textwidth]{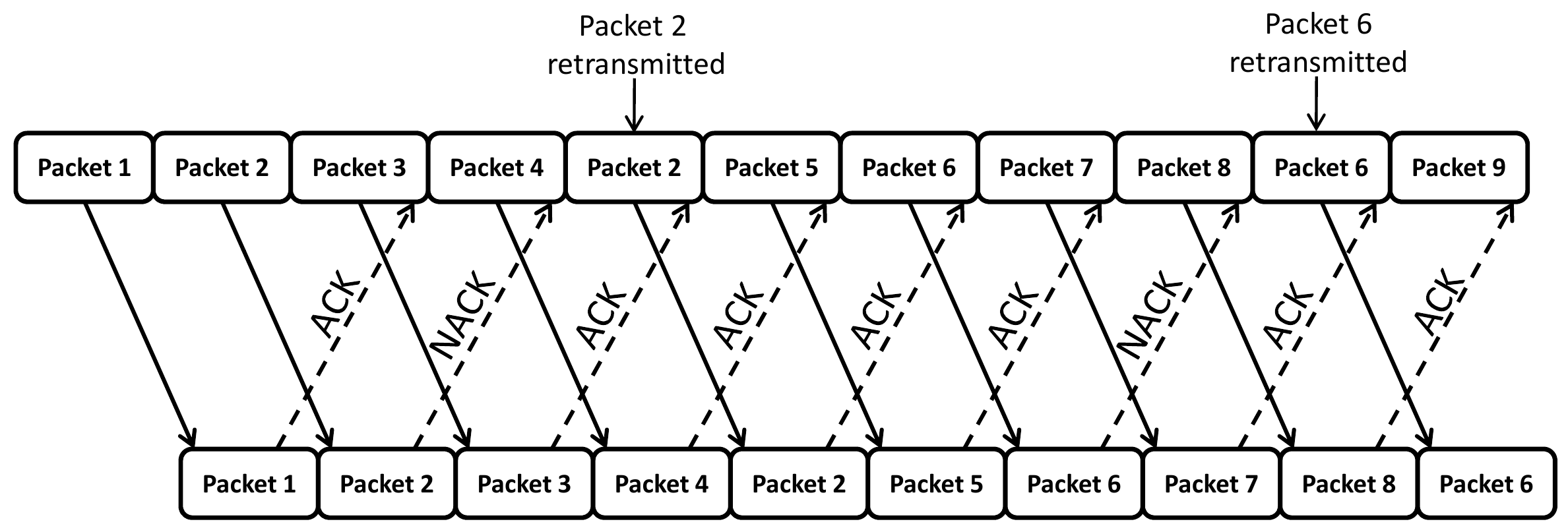}
\caption{Selective Repeat ARQ.}
\label{fig:SRARQ}
\end{figure}

There are 4 main types of ARQ protocols, namely, Stop-and-Wait (SW)\nomenclature{SW}{Stop-and-Wait}, Go-back-N (GBN)\nomenclature{GBN}{Go-back-N}, Selective Repeat (SR)\nomenclature{SR}{Selective Repeat}, and hybrid ARQ which is a combination of ARQ and FEC \cite{chou2007moi}. The operation of SW ARQ is depicted in Fig. \ref{fig:SWARQ}. The transmitter sends a packet and waits for its acknowledgment. SW ARQ is inefficient compared to GBN and SR because of the idle time spent waiting for an ACK or NACK. In GBN ARQ, the transmitter sends packets continuously. At the receiver, if a packet is received in error, it will be discarded and a NACK will be sent to the transmitter. The receiver continues to discard next packets until the originally discarded packet is received correctly. Upon receiving a NACK, the transmitter resends all packets that have not yet been positively acknowledged as shown in Fig. \ref{fig:GBNARQ}. SR ARQ is similar to GBN ARQ with the difference that only negatively acknowledged packets are retransmitted as shown in Fig. \ref{fig:SRARQ}. The transmission efficiency of SW, GBN, and SR is studied in \cite{mukhtar2011occupancy}.  


In hybrid ARQ, FEC is used to correct errors in the received packets. However, if errors can not be corrected, the receiver discards the erroneous packet and requests for retransmissions until the received packet can be decoded without errors or until the maximum number of allowed retransmissions is reached. In a modified operation of hybrid ARQ, erroneously received packets which the receiver fails to decode are not discarded. These erroneous packets are rather stored in a buffer memory and later combined with successive retransmissions to enhance the receiver decoding ability. This operation is referred to as hybrid ARQ with soft combining which is categorized into two classes, namely, Chase combining and incremental redundancy. In Chase combining the retransmissions contain the same coded bits as the original transmission \cite{chase1985code}, whereas, in incremental redundancy, the retransmissions are not identical to the original transmission \cite{metzner1979imp,le2012analytical}.      

\subsection{Error Resilient Coding}
Error resilient coding is a source error control method which improves the immunity of encoded video against errors or packet loss. Scalable coding, especially MDC, is considered a type of error resilient coding. When part of the bitstream (e.g. an enhancement layer) is corrupted by errors the remaining bitstream can still be decoded to reconstruct video frames with slight degradation in quality. Another example of error resilient coding is slice structured coding in which the video frame is spatially partitioned into groups of blocks (GoBs)\nomenclature{GoB}{Group of Blocks}. Each slice is then transmitted in a separate network packet introducing multiple synchronization points. In the event of a packet loss, the associated GoB is lost but the remaining parts of the frame can still be successfully decoded. In addition, data partitioning is another scheme which divides the different parts of a bitstream into groups according to their importance. For example high frequency transform coefficients are grouped together and considered of low importance. Data partitioning is usually combined with an unequal error protection scheme \cite{richardson2010h}.       

\subsection{Error Concealment}
Error concealment is another class of error control schemes which is implemented at the receiver with the objective to conceal data loss. Most error concealment techniques exploit spatial and/or temporal interpolation. Spatial interpolation approximates missing pixel values using neighboring pixel values. On the other hand, temporal interpolation approximates lost data from previous video frames \cite{wu2001svo}. 

\subsection{Adaptive Playback}
It is a common practice in most video streaming applications to prefetch some video frames in the decoder buffer before the start of playback to smooth out the variations in the end-to-end delay due to retransmissions and the variable bit rate nature of video streams. Generally speaking, it is required to match the arrival rate of video frames at the decoder buffer with the playback rate of the video player to avoid buffer starvation. 
Adaptive playback controls the video playback rate in an attempt to maintain a desired buffer occupancy at the video decoder. When the decoder buffer occupancy is below a predefined threshold, the playback rate is reduced to allow the buffer occupancy to increase. Conversely, when the decoder buffer occupancy is above the threshold reflecting good channel condition, the playback rate is increased to drain possible accumulation in the slow phase to prevent the video sequence from being desynchronized \cite{kalman2004amp}. 

\section{Physical Layer Techniques}
Advances in the physical layer techniques had always played a leading part in the evolution of wireless communication. These techniques enabled the development of wireless communication systems which offer improved transmission reliability, large network capacity, and high data rates. Examples of these physical layer techniques are briefly described in the following subsections.    

\subsection{Adaptive Modulation}
Adaptive modulation is a possible solution in which the modulation level is changed according to the channel condition and/or the buffer occupancy for effective bandwidth utilization and continuous playback. Increasing the level of modulation or the number of constellation points allows more bits per symbol, but at the same time increases the BER for a given SNR \cite{proakis1995digital}. Hence, when the channel is in a bad state, robust low-level modulation schemes such as binary phase shift keying (BPSK)\nomenclature{BPSK}{Binary Phase Shift Keying} can be used whereas when the channel is in a good state higher level modulation schemes such as 64-quadrature amplitude modulation (64-QAM)\nomenclature{QAM}{Quadrature Amplitude Modulation} can be used to achieve higher data rates. Adaptive modulation can be jointly designed with FEC while it could also be integrated with the source encoder to achieve effective video transmission \cite{argyriou2008erv,mukhtar2011occupancy}.

\subsection{Adaptive Power Allocation}
In adaptive power allocation the transmitter power is dynamically adjusted based on the channel condition to improve the spectral efficiency. The transmission power can be increased to achieve higher channel SNR and hence higher spectral efficiency. However, adaptive power allocation is usually constrained by a transmission power threshold which is usually dictated by telecommunication regulatory authorities. Moreover, adaptive power control can be used to enhance the power consumption efficiency by using the minimum required transmission power which satisfies a target BER. In multimedia application, adaptive power allocation can be used to unequally allocate power to different parts of the bitstream based on their importance to achieve higher quality of the received media. Examples of these power-adaptive multimedia transmission techniques are described in \cite{sabir2010unequal} and \cite{el2010joint}.

\subsection{Hierarchical Modulation}
\begin{figure}[H]
\centering
\subfloat[Hierarchical 16-QAM]{\label{fig:16QAMconst}\includegraphics[width=3.0in]{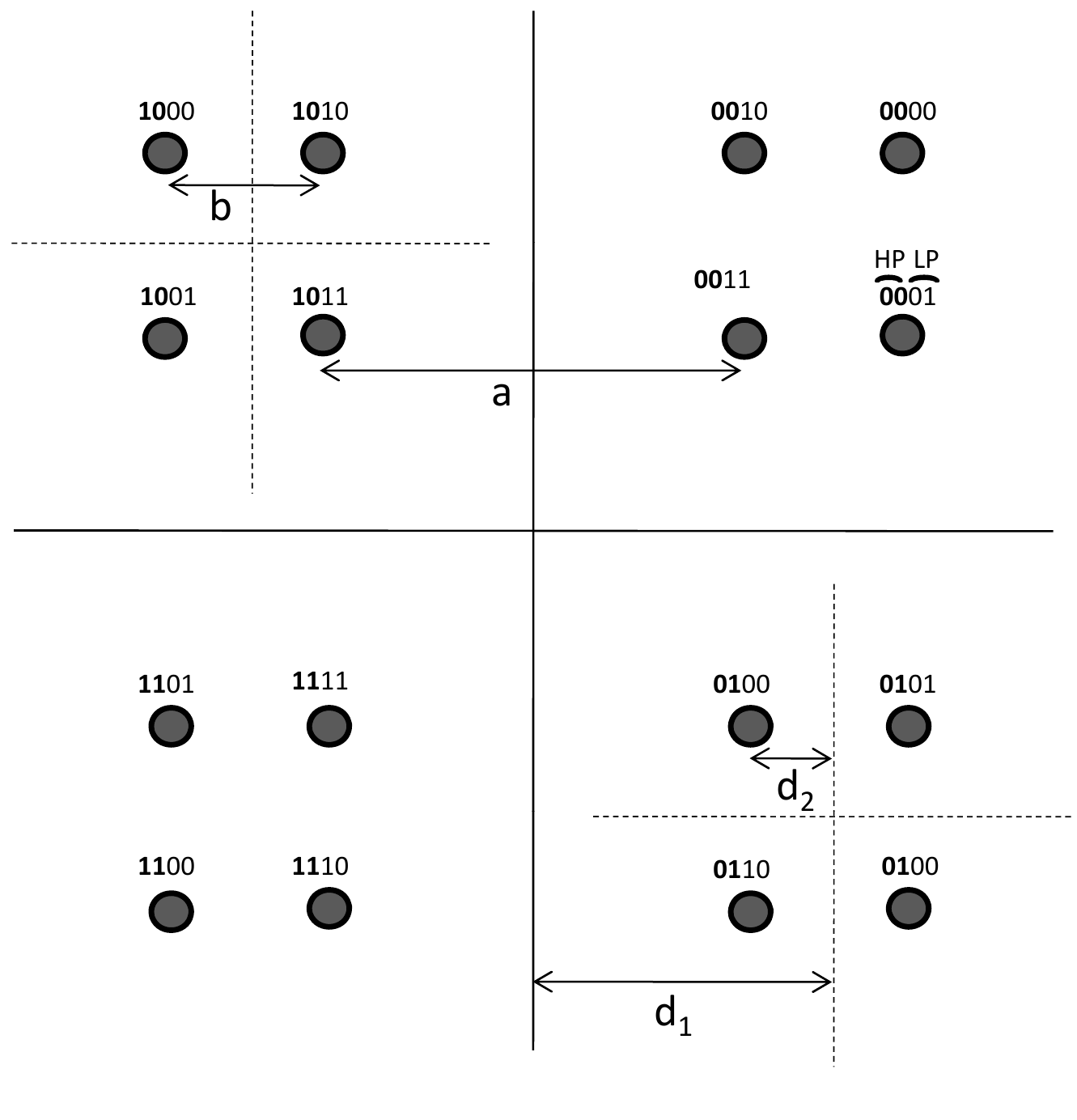}}
\subfloat[Hierarchical 64-QAM]{\label{fig:64QAMconst}\includegraphics[width=3.0in]{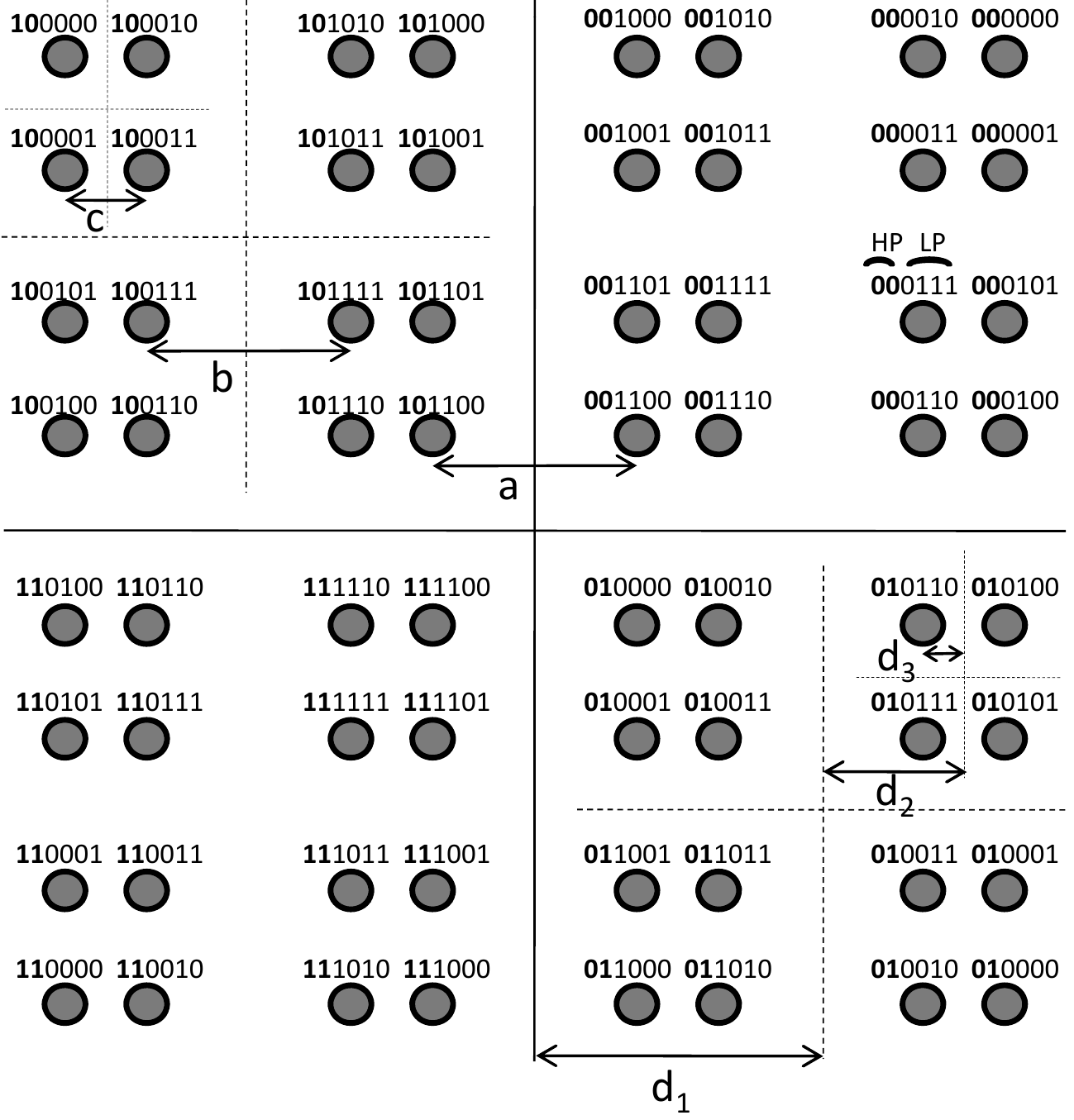}}
\caption{Hierarchical constellation for a)16-QAM, and b)64-QAM.}
\label{fig:const}
\end{figure}

Hierarchical modulation is an interesting variation of conventional modulation. It virtually divides a transmission channel into multiple sub-channels with unequal error protection without an increase in bandwidth \cite{barmada2005ptd}. A single bitstream can be separated into several multiplexed sub-streams with different levels of priority. The degree of protection of a sub-stream and the levels of hierarchy are controlled by the distances between constellation points (or regions) \cite{vitthaladevuni2003rae}. Fig.~\ref{fig:const} shows two examples of hierarchical constellations, one for 16-QAM and the other for 64-QAM. The highest priority (HP)\nomenclature{HP}{High Priority} sub-stream is transmitted using the most significant bits (MSBs)\nomenclature{MSB}{Most Significant Bit} while the lower priority (LP)\nomenclature{LP}{Low Priority} sub-streams are transmitted using the subsequent bits. 

Hierarchical QAM (HQAM)\nomenclature{HQAM}{Hierarchical QAM} is one of the popular hierarchical modulation schemes. It has already been incorporated in some digital video transmission standards such as DVB-T \cite{schertz2003hmt}. Hierarchical modulation can also be applied to other modulation schemes. An example of implementing hierarchical differential phase shift keying (DPSK)\nomenclature{DPSK}{Differential Phase Shift Keying} modulation is \cite{kwon2009higher}. A classical HQAM is the 64-HQAM where the conventional 64-QAM is transformed into three levels such that each level is associated with 2 bits. It is also possible to group two levels to be considered as one level and assign 4 bits to it. To control the relative degrees of protection between the levels, the ratios between the constellation distances are adapted \cite{6222293}.

\newpage

\subsection{Diversity Techniques for Fading Channels}
Diversity techniques are used to alleviate the adverse effects of multipath fading channels on the performance of wireless communication systems \cite{ghrayeb2007coding}. Diversity is achieved by transmitting replicas of the same message through multiple (idealy, independent) channels. Hence, the probability that all message components fade simultaneously is reduced. If the probability of error using one channel is $p_e$, then the probability of error when using $\ell$ channels is $p_e^{\ell}$. For Rayleigh fading channels, the probability of error is inversely proportional to the $\ell$th power of the average SNR \cite[p. 855]{proakis1995digital}. Examples of diversity methods are time diversity, frequency diversity and space diversity.        

In time diversity, the message replicas are transmitted during different time slots where the separation between consecutive time slots is equal to or greater than the coherence time of the fading channel. The coherence time is the time during which the channel remains almost unchanged. Time variation of fading channels can be classified into quasi-static fading, block fading and iid (independent and identically distributed) fast fading. In quasi-static fading channels, the time variation is very slow such that the channel remains almost fixed for the duration of multiple transmitted blocks or packets. In block fading channels, the channel coherence time is equal to the duration of a block of transmitted symbols, whereas, in iid fading channels, the channel fading coefficients independently change from one transmission symbol to another. This can be realized by employing interleavers to decorrelate the channel fading coefficients affecting adjacent symbols. Clearly, time diversity cannot be achieved in quasi-static channels but it is obtainable in block fading and iid fading channels.

In frequency diversity, the replicas are sent over multiple frequency bands where the separation between consecutive frequency bands is equal to or greater than the coherence bandwidth of the fading channel. The coherence bandwidth is the range of frequencies over which the fading can be considered constant. If the transmitted signal bandwidth is smaller than the channel coherence bandwidth, the fading channel is described as a flat fading channel. On the other hand, when the signal bandwidth is larger than the coherence bandwidth, the channel is said to be frequency selective.

Similarly, space diversity can be achieved by employing multiple antennas at the receiver, transmitter, or both. As a role of thumb, the separation between adjacent antennas in a uniform scattering environment should be more than half the wavelength of the transmitted signal so that the received replicas experience different channel fades. Most diversity methods employ combining techniques at the receiver to detect the transmitted signal from the received replicas. These combining techniques include selection combining, equal gain combining and maximal ratio combining.

It is worth noting that channel coding can be considered as a form of time diversity. For fully interleaved iid fading channels, optimum hard decision decoding provides a diversity order equal to half the minimum hamming distance of the code, whereas, optimum soft decision decoding provides diversity order equal to the minimum hamming distance \cite{ghrayeb2007coding}. Knowing that the probability of error is inversely proportional to the average SNR raised to the diversity order, soft decision decoding provides significant performance improvement when compared to hard decision decoding.

\subsection{Orthogonal Frequency Division Multiplexing (OFDM)}
\label{subsec:ofdm}
\begin{figure}[H]
\centering
\includegraphics[width=\textwidth]{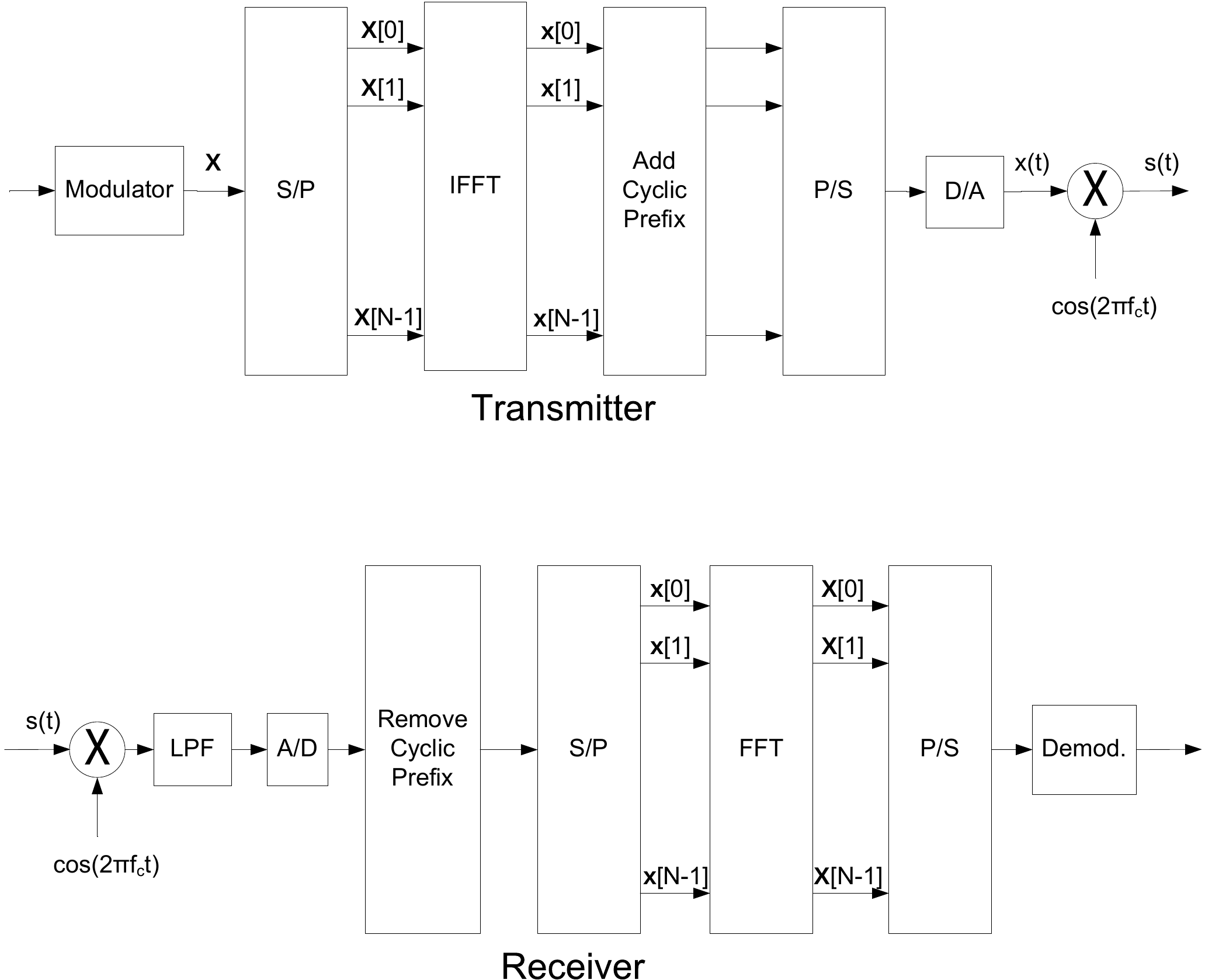}
\caption{OFDM system structure.}
\label{fig:OFDMTxRx}
\end{figure}

Orthogonal frequency division multiplexing (OFDM)\nomenclature{OFDM}{Orthogonal Frequency Division Multiplexing} is the technology of choice in many wireless communication systems \cite{andrews2007fundamentals}. OFDM is a spectrally efficient multicarrier modulation scheme with overlapping yet orthogonal subcarriers. OFDM is based on the idea of splitting a high bit rate data stream into multiple parallel lower bit rate streams, each transmitted over a separate subcarrier. Fig. \ref{fig:OFDMTxRx} describes the basic OFDM transmitter receiver structure where FFT is fast Fourier transform\nomenclature{FFT}{Fast Fourier Transform}, IFFT is inverse FFT\nomenclature{IFFT}{Inverse FFT}, S/P is serial-to-parallel conversion\nomenclature{S/P}{Serial-to-Parallel Conversion}, P/S is parallel-to-serial conversion\nomenclature{P/S}{Parallel-to-Serial Conversion}, A/D is analog-to-digital conversion\nomenclature{A/D}{Analog-to-Digital Conversion}, D/A is digital-to-analog conversion\nomenclature{D/A}{Digital-to-Analog Conversion}, and LPF is low pass filter\nomenclature{LPF}{Low Pass Filter}. 

In OFDM each symbol stream is transmitted over one narrowband subcarrier; therefore each symbol stream experiences a flat fade. This enables a simple OFDM receiver structure. In other words, dividing the input stream into many parallel lower bit rate streams increases the symbol duration of each stream. Thus, OFDM helps eliminate intersymbol interference (ISI)\nomenclature{ISI}{Intersymbol Interference}  by making the symbol duration large enough relative to the transmission channel delay spread. Moreover, OFDM offers more flexibility and granularity for resource allocation and management since subcarriers can be dynamically assigned to different users experiencing different channel conditions.

OFDM has two main disadvantages which are large peak-to-average power ratio (PAPR)\nomenclature{PAPR}{Peak-to-Average Power Ratio} and sensitivity to frequency offset caused by oscillators inaccuracies or Doppler shift due to mobility \cite{wang20094g,al2010carrier}. Large PAPRs hinder efficient utilization of the transmitter power amplifier since it will have a large backoff to ensure linear amplification of the transmitted signal. The frequency offset causes intercarrier interference (ICI)\nomenclature{ICI}{Intercarrier Interference} in OFDM. This requires reliable frequency offset estimation and ICI cancellation schemes, hence increasing the complexity of the receiver.

\begin{figure}[H]
\centering
\includegraphics[width=\textwidth]{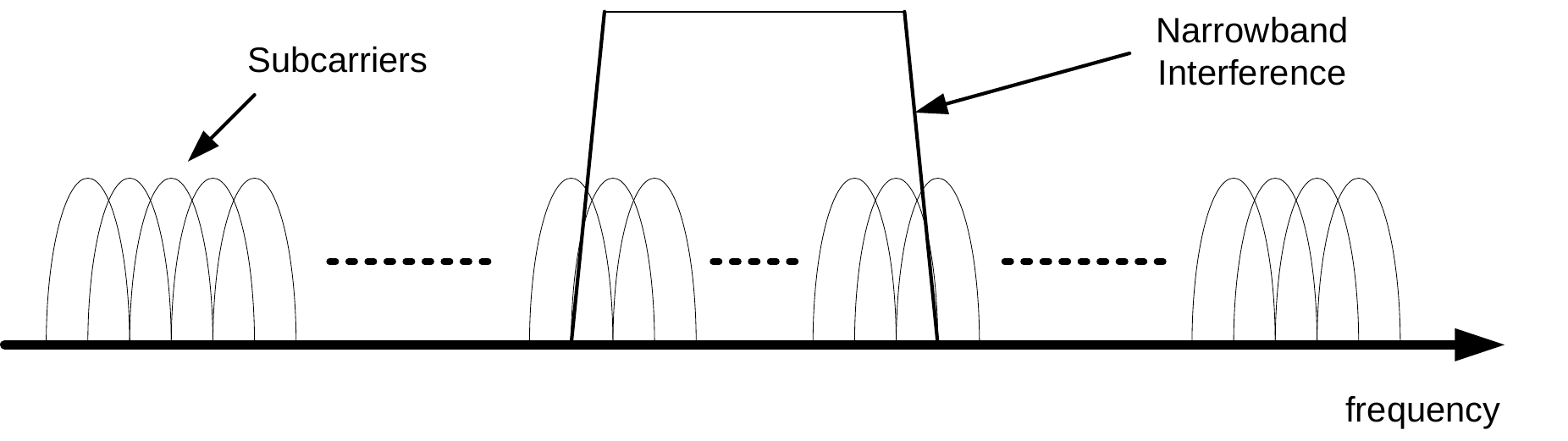}
\caption{Narrowband interference in OFDM.}
\label{fig:narrBand}
\end{figure}

Moreover, OFDM has an important drawback which is the sensitivity to narrowband interference/jamming. In OFDM, each information symbol is transmitted over a unique subcarrier. Therefore, in the presence of narrowband interference in one or more subcarriers the corresponding information symbols are likely to be lost. Narrowband interference is illustrated in Fig.~\ref{fig:narrBand}. Adaptive data loading (ADL)\nomenclature{ADL}{Adaptive Data Loading} and coded OFDM (COFDM)\nomenclature{COFDM}{Coded OFDM} can be used to alleviate this problem. However, ADL requires reliable feedback and COFDM can significantly reduce the good throughput \cite{wu2005narrowband}.

\subsubsection{OFDM with Spreading Codes}
A modified OFDM architecture which is referred to as carrier interferometry OFDM (CI/OFDM)\nomenclature{CI/OFDM}{Carrier Interferometry OFDM} is proposed in \cite{wu2005narrowband} to combat narrowband interference by exploiting frequency diversity. The proposed architecture in \cite{wu2005narrowband} improves the BER performance without any throughput loss and without the need for feedback from the receiver to the transmitter. Unlike conventional OFDM, CI/OFDM spreads each symbol into all subcarriers using orthogonal spreading codes, so that information in a symbol stream will not be completely lost in case of narrowband interference. A minimum mean square error (MMSE)\nomenclature{MMSE}{Minimum Mean Square Error} combiner is employed at the receiver to recover the symbol streams from the subcarriers which are not under narrowband interference. The proposed OFDM architecture in \cite{wu2005narrowband} uses discrete Fourier transform (DFT)\nomenclature{DFT}{Discrete Fourier Transform} as the spreading transform after the IFFT block in the OFDM transmitter. However, other spreading transforms can also be used instead such as the Walsh-Hadamard transform (WHT)\nomenclature{WHT}{Walsh-Hadamard Transform} \cite{ahmed2010ofdm,wang2010walsh}.

\subsection{Multiple-Input Multiple-Output (MIMO)}
Multiple-input multiple-output (MIMO)\nomenclature{MIMO}{Multiple-Input Multiple-Output} systems utilize multiple antennas at the transmitter and/or the receiver to increase the transmission data rate through spatial multiplexing or to improve transmission reliability through spatial diversity. MIMO has been adopted by many commercial wireless systems such as IEEE 802.11n, mobile WiMAX, and LTE. MIMO significantly improves the transmission performance without an increase in the transmission bandwidth or power \cite{ahmadi2010mobile}. In spatial multiplexing, multiple independent data streams are multiplexed and transmitted from separate antennas. In good channel conditions, the receiver can detect the different data streams, achieving a significant improvement in the system capacity. In spatial diversity, the same data stream is transmitted from each antenna. At the receiver, the multiple signal paths are combined to improve the transmission reliability. Another MIMO technique is beamforming in which the strength of the transmitted or received signal can be adjusted based on its direction. This is done by assigning to the antenna elements different weights which are appropriately selected to direct the transmitted signal toward the intended receiver and away from interference \cite{andrews2007fundamentals}.

\section{Link Adaptation in Wireless Communication Standards}
Link adaptation is the process of dynamically changing transmission parameters such as modulation order, channel coding rate, and transmission power level based on the estimated condition of the wireless link for efficient utilization of system resources and enhanced quality of service for end-users \cite{correia2006mobile,foukalas2008cross}.

In MIMO-OFDM systems, link adaptation is extended into the frequency and spatial domains. The adaptation is not based on only temporal channel variations but also variations in the frequency and spatial domains. Adaptive coding, bit loading, and power loading can be performed on a tone-by-tone basis. Moreover, MIMO mode switching and adaptive beamforming can be performed based on the channel condition to improve transmission reliability and spectral efficiency.

Hybrid ARQ with soft combining is also considered an implicit link adaptation technique because the employed retransmissions in hybrid ARQ compensates for channel variations in an implicit way \cite{dahlman20114g}. Moreover, in hybrid ARQ the coding rate is adjusted based on the result of the decoding at the receiver and, therefore, the adaptation is implicitly dependent on the channel condition. This can be considered as an advantage over explicit link adaptation techniques which require explicit estimates of the channel condition. However, hybrid ARQ results into an increased end-to-end delay when compared to explicit link adaptation schemes.

\subsection{Wireless Local Area Networks (WLANs)}
The two main wireless local area network (WLAN)\nomenclature{WLAN}{Wireless Local Area Network} standards are the IEEE 802.11 and the High Performance Radio LAN (HIPERLAN)\nomenclature{HIPERLAN}{High Performance Radio LAN}. Both standards supported link adaptation since their inception by defining different operating modes. The IEEE 802.11 is more popular and had undergone several amendments, whereas the HIPERLAN has only two versions, namely, HIPERLAN Type 1 (HIPERLAN/1) and HIPERLAN Type 2 (HIPERLAN/2). 

The first version of the IEEE 802.11 (Legacy 802.11) was released in 1997 where two operating modes were defined with two different transmission rates in the 2.4 GHz band \cite{correia2006mobile}. The legacy standard is based on the frequency hopping spread spectrum (FHSS)\nomenclature{FHSS}{Frequency Hopping Spread Spectrum} and the direct sequence spread spectrum (DSSS)\nomenclature{DSSS}{Direct Sequence Spread Spectrum}. Legacy 802.11 was shortly enhanced in the 802.11b which only uses DSSS and offers higher data rates due to the use of complementary code keying as the modulation scheme. Along with 802.11b, another version which is known as 802.11a was released with a more flexible design. The 802.11a operates in the 5 GHz band, uses OFDM, and offers 8 transmission rates from 6 to 54 Mbps. Another version is the 802.11g which operates in the 2.4 GHz band, uses both OFDM and DSSS, and also provides transmission rates up to 54 Mbps. A recent version of the standard is the 802.11n which operates in both bands, supports MIMO, and provides 77 operating modes with data rates up to 600 Mbps \cite{802.11n}. 

The introduction of MIMO in 802.11n, with up to 4 streams, enabled the standard to support this large number of operating modes. When one stream is used, 8 transmission modes are supported. Additional 24 modes are supported when 2, 3 or 4 spatial streams are used with the same modulation scheme in all streams. Moreover, 802.11n provides the option to use different modulation schemes for the different MIMO streams allowing for more transmission modes.   

HIPERLAN/1 supports two transmission rates, whereas, HIPERLAN/2 provides seven operating modes with data rates up to 54 Mbps \cite{pahlavan2001principles,korowajczuk2011lte}. Both types operate in the 5 GHz band. The IEEE 802.11 family and HIPERLAN/1 use the carrier sense multiple access with collision avoidance (CSMA/CA)\nomenclature{CSMA/CA}{Carrier Sense Multiple Access with Collision Avoidance} as the medium access and medium sharing technique. On the other hand, HIPERLAN/2 uses time division multiple access and time division duplexing (TDMA/TDD)\nomenclature{TDMA/TDD}{Time Division Multiple Access}\nomenclature{TDD}{Time Division Duplexing}. Moreover, in HIPERLAN/2, Selective Repeat ARQ is used for error control \cite{doufexi2002comparison}, whereas, Stop-and-Wait ARQ is used in the IEEE 802.11. Table \ref{wlantable} compares the different WLAN standards and summarizes their main characteristics.     

\begin{table}[H]
\caption{Summary of WLAN specifications and features.}
\label{wlantable}
\vspace{\abovecaptionskip}
\includegraphics[width=\textwidth]{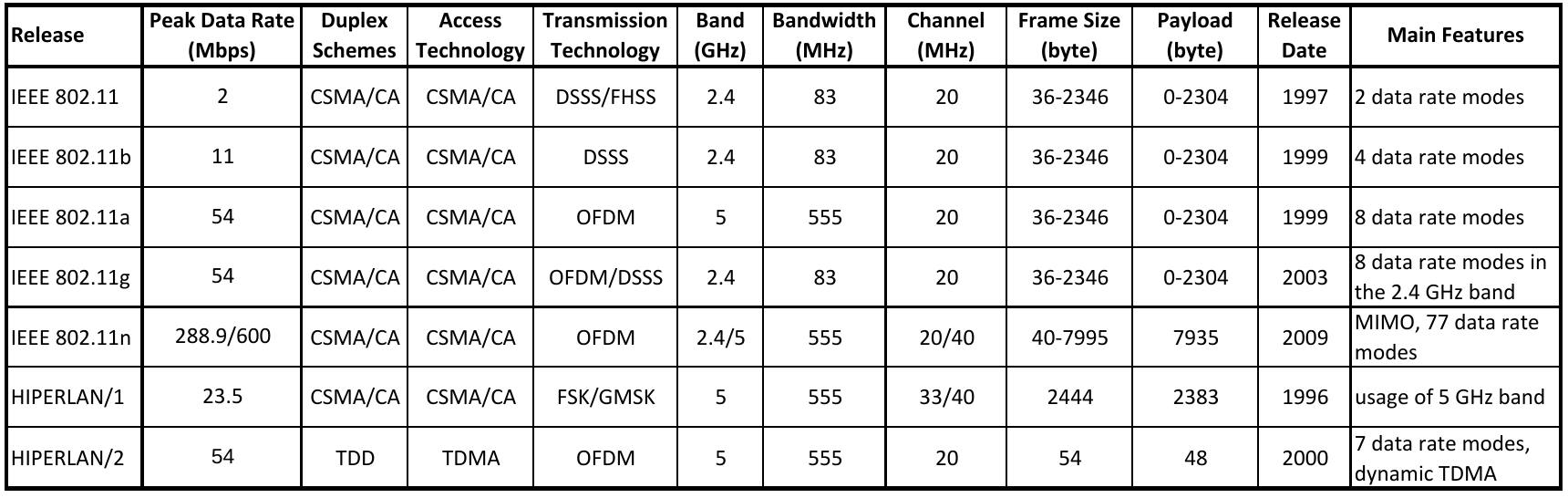}
\end{table}          

A link adaptation scheme for video transmission over OFDM-based WLANs was proposed in \cite{ferre2003link}. The objective of the proposed scheme in \cite{ferre2003link} was to improve the QoS for video transmission in terms of the overall received video quality (i.e., PSNR), rather than maximizing the error-free throughput. The scheme searches offline for the transmission mode which results in the maximum PSNR for a given video sequence transmission over a range of channel SNR values. Based on the exhaustive search, several SNR regions and switching points are identified and a look-up table is constructed to show each region and its corresponding optimum transmission mode. 

The PSNR-based link adaptation scheme in \cite{ferre2003link} is not suitable for real-time video since the PSNR computation and the exhaustive search are done offline. Therefore, the authors proposed another link adaptation scheme in \cite{ferré2006video} which uses packet error rate (PER)\nomenclature{PER}{Packet Error Rate} thresholds. The authors argue that PER is closely related to the PSNR and hence can be used instead as a more practical decision metric. Empirical results were provided in \cite{ferré2006video} to show the impact of PER thresholds selection on the PSNR of the received video. However, the authors ignored the effect of video content, error concealment, and ARQ on the switching thresholds. Moreover, they did not provide an algorithm to obtain optimum PER thresholds.

In \cite{pierre2008distortion}, a distortion-based link adaptation scheme for video transmission over WLANs was proposed. An estimate of the video distortion was used as the decision metric for adapting the link speed to the channel condition. The estimation model depends on the PER and operates in a GoP basis. Moreover, the distortion model takes into consideration the impact of error propagation and error concealment but does not account for retransmissions. The distortion model assumes a fixed PER within a GoP. This assumption makes the distortion model inaccurate especially for fast varying channels where the channel condition may change within a GoP. A more sophisticated distortion model may be needed to enhance the performance of the link adaptation scheme proposed in \cite{pierre2008distortion}.

A cross-layer link adaptation algorithm for the transmission of video over WLANs was proposed in \cite{loiacono2010cross}. The authors argue that switching to a lower link speed improves SNR-BER performance but increases channel contention (i.e., collision probability). They introduce a distortion model which takes into account packet losses due to collision as well as channel errors. The adaptation algorithm operates in a GoP basis similar to the algorithm described in \cite{pierre2008distortion}. The authors in \cite{loiacono2010cross} implemented their algorithm in Axis wireless cameras which uses IEEE 802.11b to demonstrate the performance gains of their proposed algorithm through experimental results. Nevertheless, they ignored the playback buffer occupancy in their results which is important to evaluate the continuity of video playback.

The authors in \cite {fallah2008link} proposed a link adaptation scheme for efficient transmission of H.264 scalable video over multirate WLANS. They introduced a distortion model for H.264 scalable video with fine granular scalability (FGS)\nomenclature{FGS}{Fine Granular Scalability}. The authors claim that conventional enhancement layer dropping techniques are not optimal. They proposed a link adaptation scheme to maximize the quality of the received video by adjusting the transmission mode/rate for each layer based on its relative importance. They utilized MIMO in addition to adaptive modulation and channel coding (AMC)\nomenclature{AMC}{Adaptive Modulation and Channel Coding}. The playback buffer dynamics are also ignored in this paper.

       
\subsection{High Speed Packet Access (HSPA)}
High Speed Packet Access (HSPA)\nomenclature{HSPA}{High Speed Packet Access} is an evolution of the radio access technology in the Universal Mobile Telecommunications System (UMTS)\nomenclature{UMTS}{Universal Mobile Telecommunications System} which is known as Universal Terrestrial Radio Access (UTRA)\nomenclature{UTRA}{Universal Terrestrial Radio Access} or Wideband Code Division Multiple Access (WCDMA)\nomenclature{WCDMA}{Wideband Code Division Multiple Access} \cite{correia2006mobile}. HSPA has evolved through several releases providing incremental performance improvements over time \cite{dahlman20114g}. The first release of HSPA was in Release 5 of the UMTS standard in 2002. This release contained link adaptation features in the downlink, namely, adaptive modulation and coding, channel dependent scheduling and hybrid ARQ. In Release 6, link adaptation features were added for uplink transmissions. The adaptation decisions can be taken every 2 ms which is the transmission time interval (TTI)\nomenclature{TTI}{Transmission Time Interval} in HSPA. 
Spatial multiplexing of two layers in the downlink, as well as downlink 64-QAM and uplink 16-QAM were introduced in Release 7. However, simultaneous usage of spatial multiplexing and 64-QAM in the downlink was later enabled in Release 8. Moreover, carrier aggregation was introduced in Release 9 to increase the supported bandwidth from 5 MHz to 10 MHz in the downlink and the uplink by using two carriers. In Release 10, the supported bandwidth was further increased in the downlink by using 4 carriers. A multi-process Stop-and-Wait hybrid ARQ is used in parallel in HSPA for error control \cite{hspaPHY2001}. Table \ref{hspatable} outlines the evolution of HSPA and summarizes the main features and specifications of its different releases.




\begin{table}[H]
\caption{Summary of HSPA specifications and features.}
\label{hspatable}
\vspace{\abovecaptionskip}
\includegraphics[width=\textwidth]{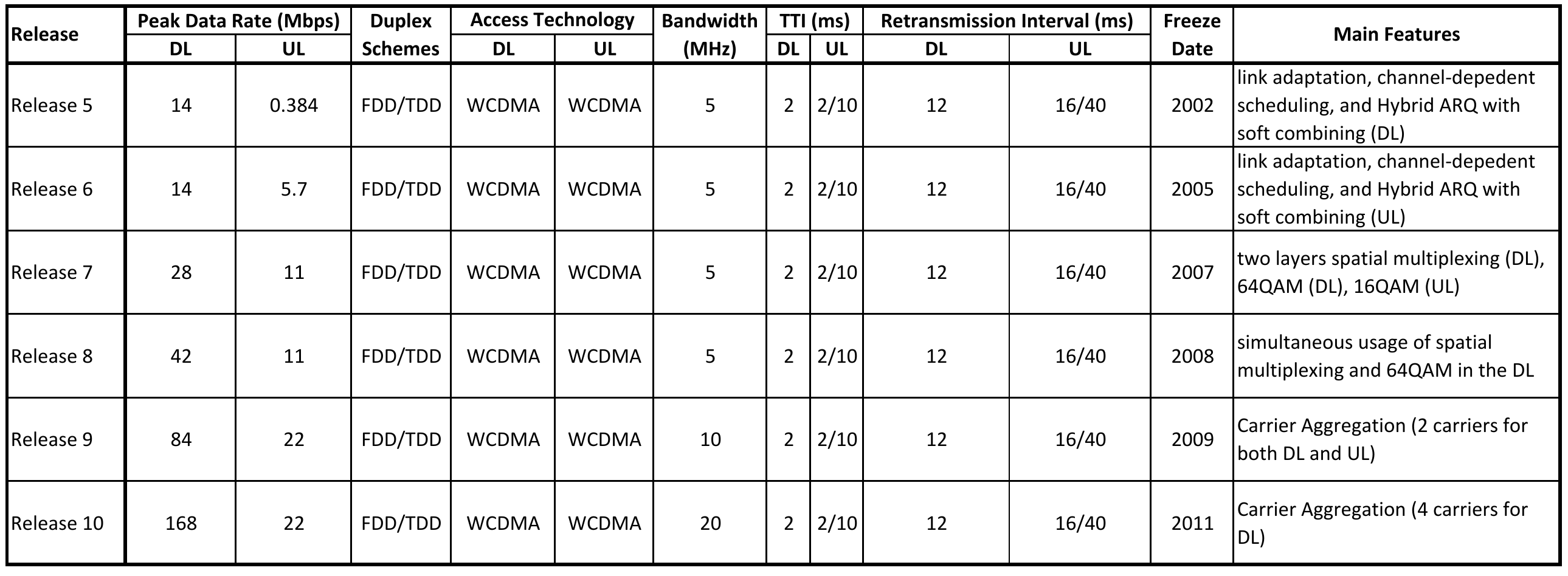}
\end{table} 

\subsection{IEEE 802.16}
IEEE 802.16 is a wireless broadband access technology also developed by the Institute of Electrical and Electronics Engineers (IEEE)\nomenclature{IEEE}{Institute of Electrical and Electronics Engineers} \cite{marks2004ieee}. The standardization was initiated in 1998, but the first standard was published on 2002 and then several versions of the standard were developed. 802.16 was mainly intended to provide last mile broadband access services as an alternative to cable and Digital Subscriber Line (DSL)\nomenclature{DSL}{Digital Subscriber Line} networks. The initial version of the standard was designed for line of sight (LOS)\nomenclature{LOS}{Line of Sight} communication in the 10-66 GHz band. The next version, 802.16a, was also intended for fixed wireless access application. However, it provided support for non-line of sight (NLOS)\nomenclature{NLOS}{Non-Line of Sight} communication in the 2-11 GHz band. In 2005, another version of the standard, 802.16e, was introduced to support mobility. A recent version of the standard is the 802.16m which was approved by the International Telecommunication Union (ITU)\nomenclature{ITU}{International Telecommunication Union}, in 2010, as an IMT-Advanced-compliant (i.e. 4G) technology \cite{dahlman20114g}.

The IEEE 802.16 specifications contain an extremely large set of alternative features and options, hindering practical implementation of the standard. Therefore, an industry-led cooperation, known as the WiMAX (Worldwide Interoperability for Microwave Access)\nomenclature{WiMAX}{Worldwide Interoperability for Microwave Access} forum, was established to enable and promote practical implementation of the standard. The WiMAX forum selects a subset of features from the standard to construct an implementable specification, known as a WiMAX System Profile. For example, System Profile Release 2.0 corresponds to the 802.16m specifications. Moreover, 802.16e is the basis for many commercial 802.16-based products. The system profiles for the 802.16e are usually referred to as WiMAX or Mobile WiMAX.

Mobile WiMAX is an OFDM-based technology which supports the 5 and 10 MHz bandwidth with a carrier spacing of 10.94 kHz. Although Frequency Division Duplexing (FDD)\nomenclature{FDD}{Frequency Division Duplexing} and Time Division Duplexing (TDD) are both supported in 802.16e, Mobile WiMAX is focused on TDD operation. Mobile WiMAX defines a 5 ms frame which contains 48 OFDM symbols divided into a downlink part and an uplink part. Control signaling for link adaptation such as hybrid ARQ and scheduling is sent at the beginning of the downlink part. Therefore, the adaptation can be provided once every 5 ms. Moreover, Mobile WiMAX supports quadrature phase shift keying (QPSK)\nomenclature{QPSK}{Quadrature Phase Shift Keying}, 16-QAM and 64-QAM along with Turbo codes for channel coding. It also supports MIMO techniques (in the downlink and uplink) such as open loop spatial multiplexing.

\begin{figure}[H]
\centering
\includegraphics[width=\textwidth]{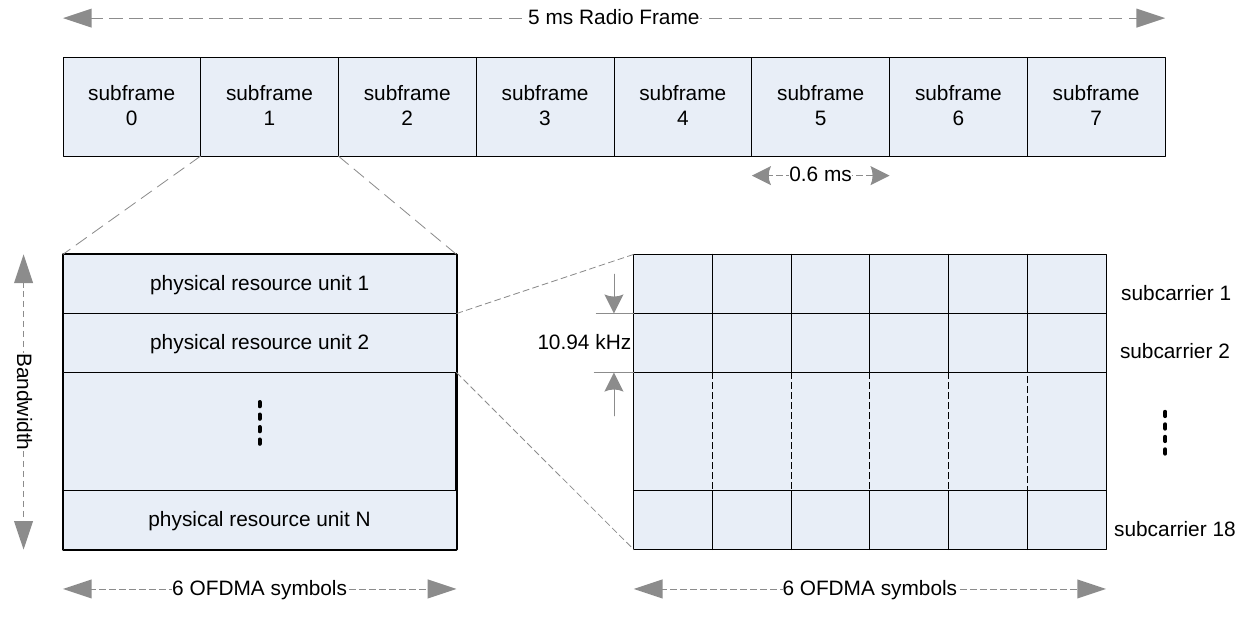}
\caption{Example physical structure for an IEEE 802.16m radio frame.}
\label{fig:wimaxFrame}
\end{figure}

802.16m was initiated to meet the IMT-Advanced requirements defined by the ITU. New complementary features were added in 802.16m to extend the capabilities of the 802.16e radio access technology. The prominent new feature which allowed for higher data rates was the carrier aggregation for bandwidths beyond 20 MHz. 802.16m also achieved reduced latency by introducing a shorter subframe with 0.6 ms of length which reduced the transmission time interval. Moreover, 802.16m uses OFDMA (Orthogonal Frequency Division Multiple Access)\nomenclature{OFDMA}{Orthogonal Frequency Division Multiple Access} as the multiple access scheme in both the downlink and the uplink. \textit{Physical resource units}, composed of a number of contiguous frequency subcarriers, were defined in 802.16m. Each resource unit contains 18 subcarriers by six contiguous OFDMA symbols occupying a total bandwidth of $(18-1)\times10.94k\approx180~kHz$. The basic unit for resource allocation is referred to as a \textit{logical resource unit} which is equal in size to the physical resource unit, but can be distributed or localized. Distributed resource units are used to achieve frequency diversity gain, whereas, localized resource units are used for frequency selective scheduling \cite{ahmadi2009overview}. Fig. \ref{fig:wimaxFrame} shows an example physical structure for an IEEE 802.16m radio frame. It should be noted that the IEEE 802.16m supports other partially different numerologies with different subcarrier spacing and subframe sizes.      

In addition, IEEE 802.16m uses adaptive asynchronous HARQ in the downlink (DL)\nomenclature{DL}{Downlink}, whereas non-adaptive synchronous HARQ is used in the uplink (UL)\nomenclature{UL}{Uplink} \cite{ahmadi2009overview}. The HARQ in both utilizes a multi-process Stop-and-Wait protocol. The DL asynchronous HARQ offers the flexibility to adapt the resource allocation and transmission format for the HARQ retransmissions based on the air interface conditions. On the other hand, in the UL non-adaptive synchronous HARQ, the parameters and resource allocation for the retransmissions are pre-defined; hence, no explicit signaling is required to inform the receiver about the retransmission schedule.




\subsection{Long Term Evolution (LTE)}
LTE stands for Long Term Evolution\nomenclature{LTE}{Long Term Evolution} which is one of the most recent mobile broadband access technologies developed by 3GGP (3rd Generation Partnership Project)\nomenclature{3GGP}{3rd Generation Partnership Project}. LTE development was initiated to provide reduced latency, higher user data rates, and improved system capacity and coverage while maintaining compatibility/interaction with other 3GGP technologies such as HSPA and GSM (Global System for Mobile Communications)\nomenclature{GSM}{Global System for Mobile Communications} \cite{wang20094g}. LTE has a flat radio access network architecture unlike previous 3GPP technologies which require a centralized network component \cite{pedersen2009overview}. LTE provides spectrum flexibility by supporting bandwidths from 1.4 MHz up to 20 MHz. Moreover, LTE supports both TDD and FDD with OFDMA in the downlink and SC-FDMA (Single Carrier Frequency Division Multiple Access)\nomenclature{SC-FDMA}{Single Carrier Frequency Division Multiple Access} in the uplink because the transmit signal of SC-FDMA has a lower PAPR when compared to the OFDM/OFDMA signal, allowing for improved power efficiency in mobile terminals.

\begin{figure}[H]
\centering
\includegraphics[width=\textwidth]{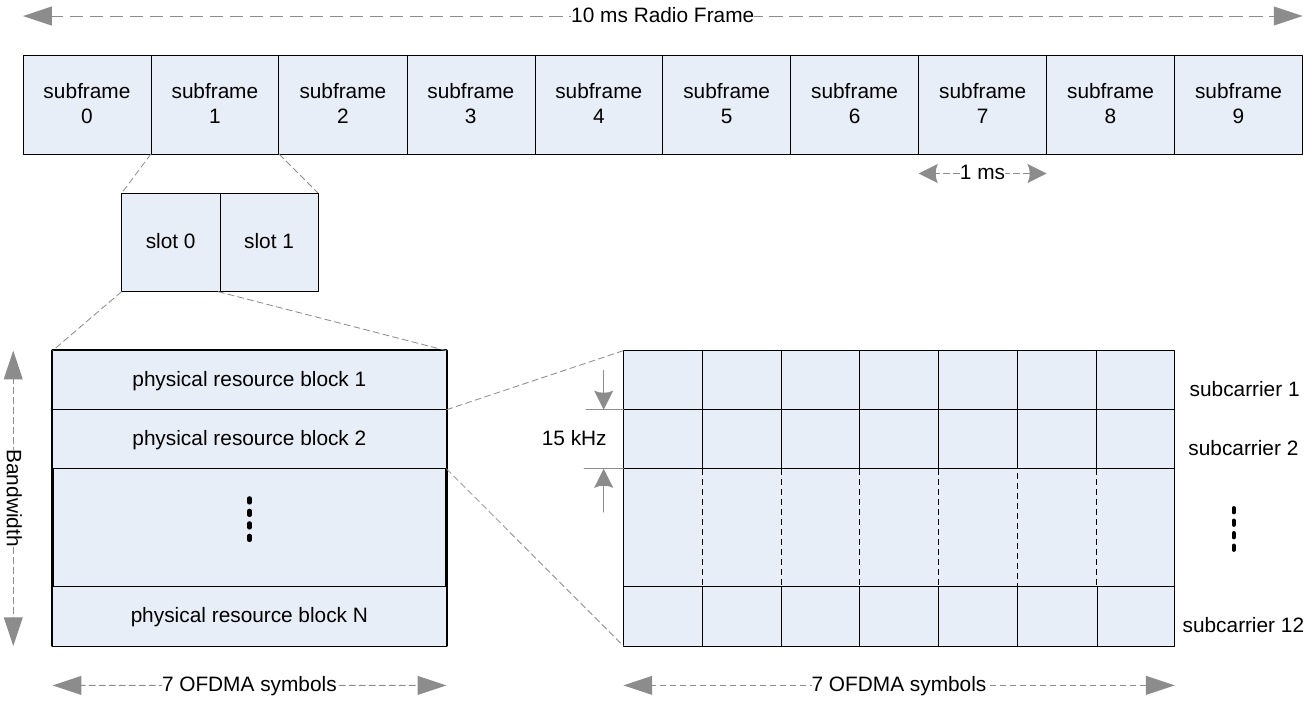}
\caption{LTE radio frame physical structure.}
\label{fig:lteFrame}
\end{figure}

In the downlink, LTE defines a 10 ms radio frame which includes 10 subframes of 1 ms each. Each subframe is further divided into two slots with a length of 0.5 ms. Each slot contains 7 OFDMA symbols and a variable number of subcarriers depending on the available bandwidth. The physical resource block in LTE includes 12 of these subcarriers per slot. The subcarrier spacing in LTE is 15 kHz; hence, one physical resource block occupies a 180 kHz bandwidth similar to the physical resource block bandwidth in IEEE 802.16m. Fig. \ref{fig:lteFrame} describes the LTE radio frame physical structure. 

Link adaptation, including AMC and flexible resource scheduling, is also supported in LTE. The supported modulation schemes are QPSK, 16-QAM, and 64-QAM. The subframe interval corresponds to the TTI which is 1 ms. The first 1-3 OFDM symbols in a TTI are used to transmit control information for link adaptation. This short TTI enables faster link adaptation when compared to other mobile broadband standards. In addition, MIMO operation is also supported for enhancing either user throughput or cell throughput.         
In addition, LTE uses a two-level ARQ design, HARQ at the medium access control (MAC)\nomenclature{MAC}{Medium Access Control} layer and an outer ARQ at the radio link control (RLC)\nomenclature{RLC}{Radio Link Control} layer \cite{larmo2009lte}. The HARQ mechanism at the MAC layer is similar to the solution adopted in IEEE 802.16m. A multi-process Stop-and-Wait HARQ is used with asynchronous retransmission in the DL and synchronous retransmissions in the UL. The MAC layer HARQ provides small delay, simplicity, and low control overhead. Nevertheless, the outer RLC ARQ is required to provide additional reliability by handling residual errors that are not detected by the lightweight HARQ. The outer ARQ of the RLC layer employs a window-based Selective Repeat protocol.

The first release of LTE specifications, known as release 8, was completed in 2008. Another release, with additional features such as the multimedia broadcast and multicast services (MBMS)\nomenclature{MBMS}{Multimedia Broadcast and Multicast Services} mode, was introduced in 2009. LTE release 10 or LTE-Advanced was introduced in 2010 to ensure fulfillment of the IMT-Advanced requirements. Additional features were also added in release 10 such as carrier aggregation and extended multi-antenna transmission \cite{dahlman20114g}.  



\subsection{Digital Video Broadcasting (DVB)}
Digital Video Broadcasting (DVB)\nomenclature{DVB}{Digital Video Broadcasting} is a European project created at the beginning of the 1990s \cite{reimers2006dvb,fischer2010digital}. The technical specifications which are being developed in the DVB project are mainly related to the transmission of digital television over satellite, cable, or via terrestrial transmitters. The standard for digital video broadcasting by satellite is referred to as DVB-S which was adopted in 1994. Reliable modulation schemes such as QPSK and 8-PSK are used in DVB-S. DVB-S operates in the 11-13 GHz band and the 14-19 GHz band for the downlink and the uplink, respectively. A satellite channel bandwidth of 26-36 MHz are usually used in DVB-S. Two forward error correction codes are used in DVB-S, namely a Reed-Solomon block code followed by a convolutional code. In DVB-S, the video transport packet is 188 bytes, expanded to 204 bytes by the Reed-Solomon code. The packet size is further expanded depending on the used code rate of the convolutional code. A data rate of 38 Mbps is typically achieved when QPSK is used with a 3/4 convolutional code. Theoretically, DVB-S can achieve a maximum data rate of 66.5 Mbps when 8-PSK is used with a 7/8 code rate.

In 2003, DVB-S2 was introduced as a second version for video broadcasting over satellite channels with new modulation and coding methods. DVB-S2 was mainly developed to support high definition television (HDTV)\nomenclature{HDTV}{High Definition Television} coverage. In addition to QPSK and 8-PSK, other modulation schemes such as amplitude phase shift keying (16-APSK and 32-APSK) and hierarchical 8-PSK were introduced in DVB-S2. Moreover, new error protection mechanisms were adopted such as BCH (Bose-Chaudhuri-Hocquenghem)\nomenclature{BCH}{Bose-Chaudhuri-Hocquenghem} and LDPC (low density parity check)\nomenclature{LDPC}{Low Density Parity Check}. Depending on the code rate of the BCH coding followed by the LDPC coding and by means of padding, the DVB-S2 packet has a length of 8100 or 2025 bytes. The FEC packet is then divided into multiple slots to comprise a physical layer frame with 90 symbols in each slot. A data rate of 49 Mbps is typically achieved when QPSK is used with a 9/10 LDPC code. Theoretically, DVB-S2 can achieve a maximum data rate of 122.46 Mbps when 32-APSK is used with a 9/10 code rate.

DVB-T is another standard which was defined in 1995 for terrestrial transmission of digital television. Terrestrial television coverage are needed when satellite reception or cable reception are infeasible or inadequate. Terrestrial broadcasting may also be required to provide portable or mobile TV reception and local TV services. DVB-T is an OFDM-based system which operates in the VHF (30-300 MHz) and the UHF (300-3000 MHz) bands. Moreover, DVB-T defines a channel bandwidth of 6, 7 or 8 MHz. The modulation schemes used in DVB-T are QPSK, 16-QAM and 64-QAM as well as hierarchical 16-QAM and hierarchical 64-QAM. The compressed video transport packet and the FEC mechanism are the same as in the the DVB-S standard. In addition, DVB-T defines a physical frame composed of 68 OFDM symbols and every four frames are grouped into a super frame. A maximum data rate of 31.67 Mbps can be achieved in DVB-T when 64-QAM is used with a 7/8 code rate and a channel bandwidth of 8 MHz. DVB-T supports link adaptation such as adaptive modulation and FEC coding.    

In 2008, DVB-T2 was introduced as a new terrestrial video broadcasting standard which is capable of terrestrial HDTV coverage and enhanced mobile TV reception. DVB-T2, similar to DVB-T, is an OFDM-based system but it supports larger number of subcarriers and additional channel bandwidths, namely, 1.7, 5, 6, 7, 8 and 10 MHz. DVB-T2 also operates in the VHF and UHF bands. It supports 256-QAM in addition to the modulation schemes which are used in DVB-T. Moreover, rotated Q-delayed constellations were introduced in DVB-T2 to enhance the modulation schemes reliability. DVB-T2 uses the same error protection mechanism and FEC frame structure as in the DVB-S2. However, DVB-T2 defines a different physical frame structure which enable a more flexible link adaptation \cite{vangelista2009key}. A maximum data rate of 50.32 Mbps can be achieved by DVB-T2 when 256-QAM is used with a 5/6 code rate and an 8 MHz channel bandwidth. DVB-T2 also supports multiple input single output (MISO)\nomenclature{MISO}{Multiple Input Single Output} smart antenna techniques as an option to improve the system coverage.

The DVB project also created a standard for digital video broadcasting for handheld mobile terminals which is abbreviated as DVB-H. This standard was introduced in 2003 to provide convergence between mobile radio networks (e.g. UMTS) and broadcasting networks (e.g. DVB-T). DVB-H provides a framework for a modified DVB-T which combines the bi-directionality advantage of mobile radio networks with the high data rate advantage of broadcasting networks. For example, when services are requested by a single user via the mobile radio return channel, the requested stream is transmitted via the downlink of the mobile radio network. However, when the same service is request by a large number of users at the same time via the mobile radio network, the requested stream is broadcasted to these users using the terrestrial broadcast network. The DVB-H introduces some additional operation modes to the DVB-T such as a 5 MHz channel and an OFDM mode with 4096 carriers.

Another DVB standard, derived from the DVB-T, is the DVB-SH which stands for digital video broadcast via satellite and terrestrial path. DVB-SH combines features from DVB-T/H and DVB-S2. It uses terrestrial broadcasting to provide coverage for populated areas while it uses broadcasting via satellite to provide coverage for rural areas. DVB-SH is OFDM-based, but it also supports single carrier time division multiplexing (TDM)\nomenclature{TDM}{Time Division Multiplexing} in the satellite to terminal path. Finally, ARQ is not included in the DVB standard; however, an alternative retransmission protocol is used. This protocol is referred to as DVB-RET \cite{dvbiptv}. Table~\ref{dvbtable} provides a summary of DVB specifications and features.

\begin{table}[H]
\caption{Summary of DVB specifications and features.}
\label{dvbtable}
\vspace{\abovecaptionskip}
\includegraphics[width=\textwidth]{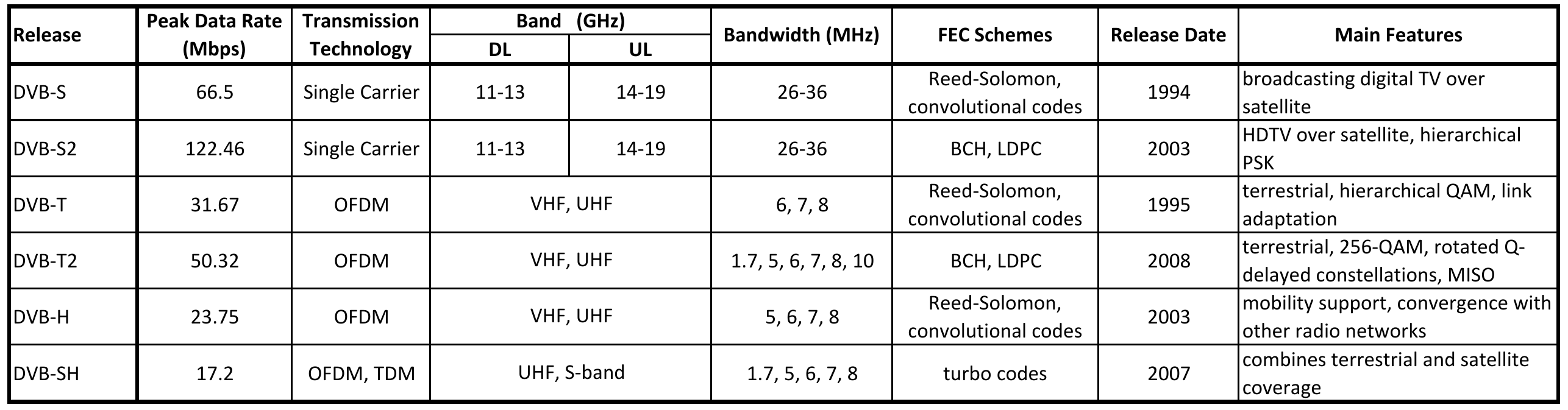}
\end{table} 

\section{Video Quality Metrics}
Various video quality assessment techniques are proposed in the literature~\cite{Lambrecht96,Pinson04,Wang2004,damera2000image}. Assessment techniques in which quality metrics are mainly based on mathematical quantification are classified as objective approaches. Other assessment techniques that rely on  viewers perception of the video quality are classified as subjective. In general, video quality has two aspects: spatial and temporal.  Spatial video quality is typically measured using peak signal to noise ratio (PSNR)\nomenclature{PSNR}{Peak Signal to Noise Ratio} metric. Temporal quality pertains to the viewer perception of the screen changes with time. It is usually measured using a subjective approach such as the mean opinion score (MOS)\nomenclature{MOS}{Mean Opinion Score}. 

\begin{figure}[H]
\centering
\subfloat[Concealed Frame by Freezing Previous Frame]{\label{fig:Irand}\includegraphics[width=\textwidth]{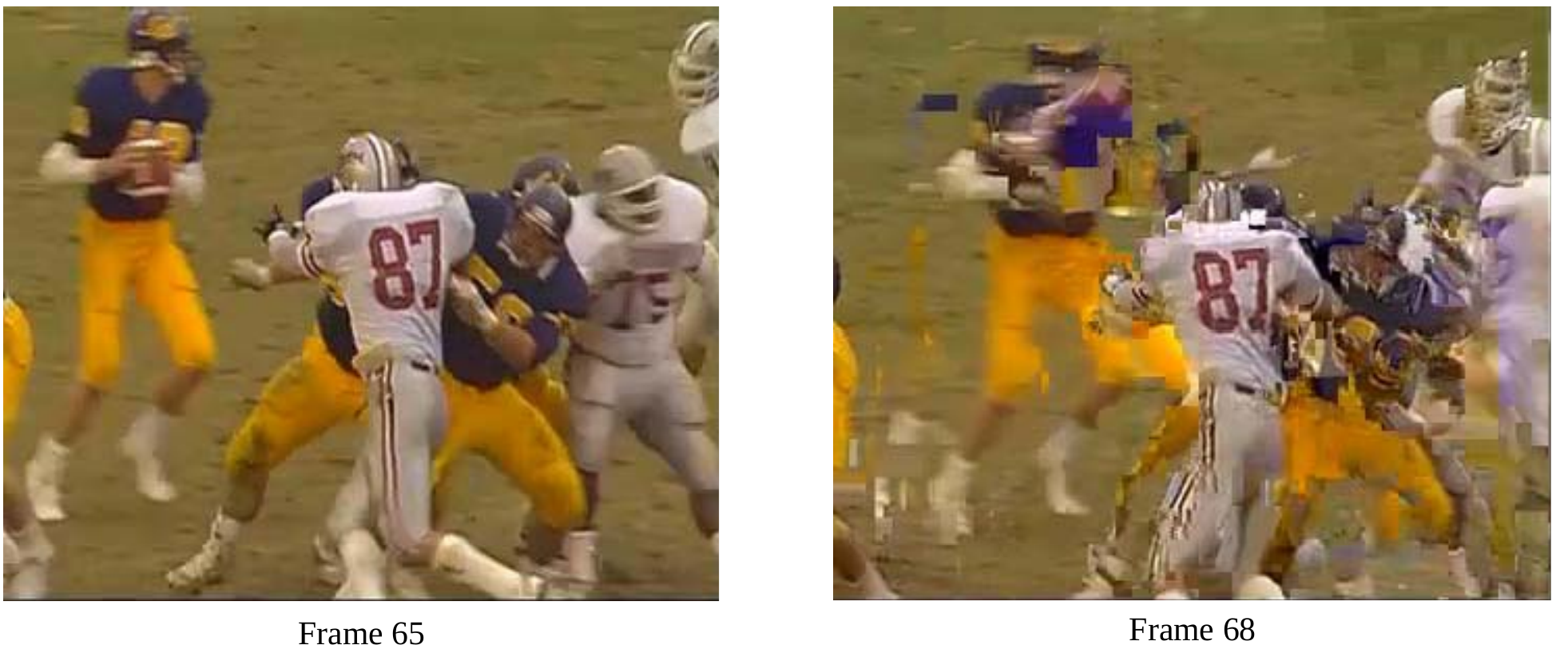}}\\
\subfloat[Original Frame]{\label{fig:Isyscontig1}\includegraphics[width=\textwidth]{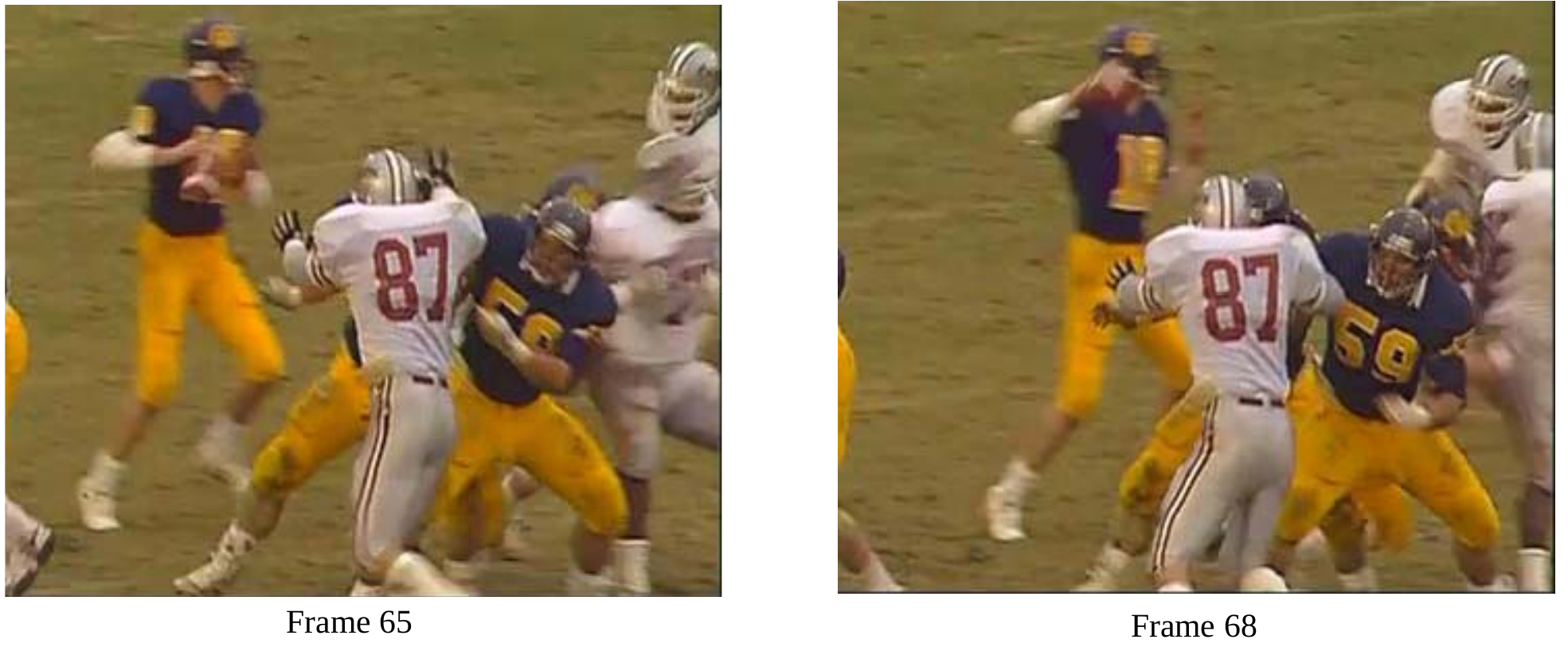}}\\
\subfloat[Concealed Frame by Motion Copy]{\label{fig:Isyscontig2}\includegraphics[width=\textwidth]{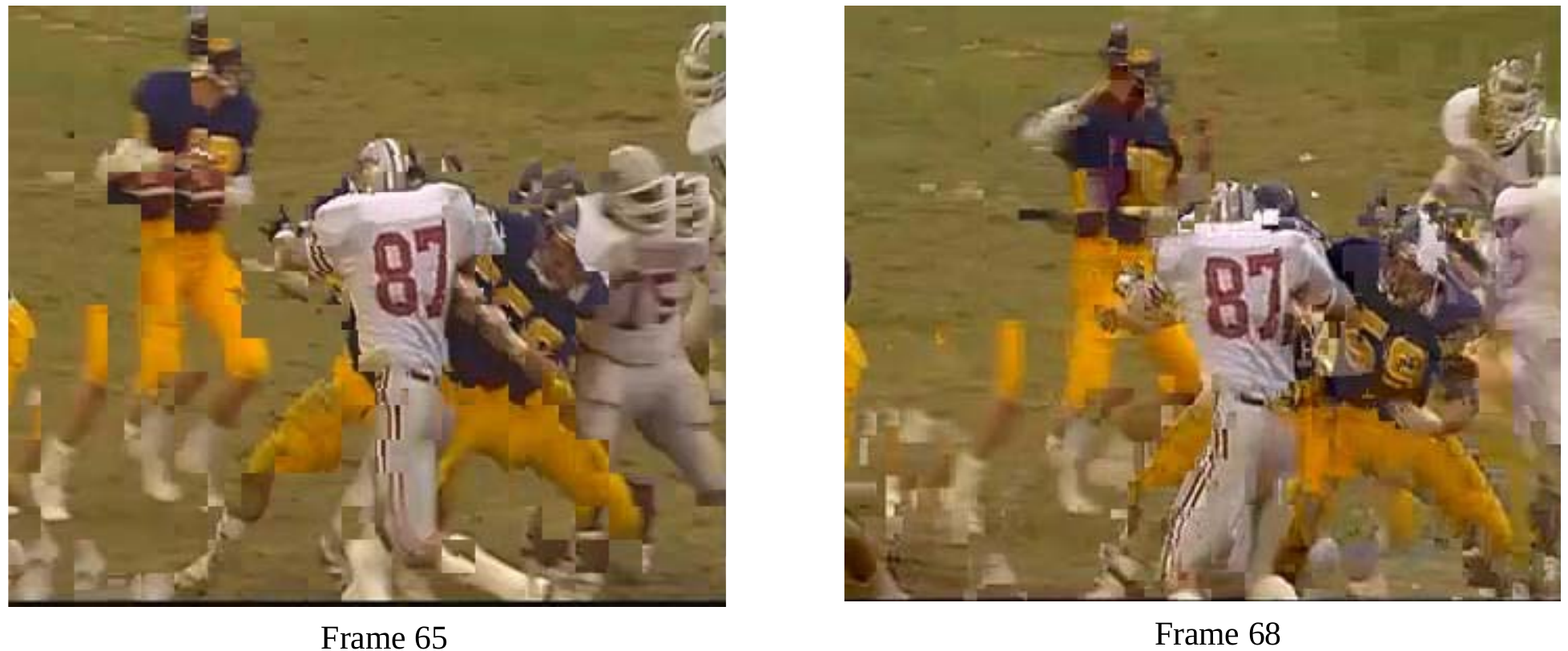}}
\caption{Frames 65 and 68 of the``football" sequence.}
\label{fig:frame64}
\end{figure}

Each of the two commonly used measures (PSNR or MOS) has its own drawbacks. MOS assessment is time consuming, slow, and expensive. PSNR is a full-reference metric that requires a priori knowledge of the original video sequence which is typically not available at the client side. In addition, it is known that PSNR values are not necessarily correlated with perceptual quality. For example, consider Fig.~\ref{fig:frame64}. This figure shows frames 65 and 68 of the ``Football" video sequence that was encoded using the H.264/AVC JM encoder \cite{website:JM}. The original two frames are shown in Fig.~\ref{fig:frame64}(b) for the sake of comparison. The transmission process was intentionally disturbed to result in the loss of frame 65. 

In Fig.~\ref{fig:frame64}(a), the loss of this frame was concealed by freezing the previous frame. Frame 68 of Fig.~\ref{fig:frame64}(a) shows the impact of error propagation when frame copy is used as the concealment method. On the other hand, in Fig.~\ref{fig:frame64}(c) the loss of frame 65 was concealed by motion copy~\cite{wu2006error}. Similarly, frame 68 of Fig.~\ref{fig:frame64}(c) shows the impact of error propagation when motion copy is used. Clearly, frame 65 in Fig.~\ref{fig:frame64}(a) is of a better perceptual quality when compared to that in Fig.~\ref{fig:frame64}(c). Nevertheless, the PSNR in Fig.~\ref{fig:frame64}(c) is $2$ dB higher than that of Fig.~\ref{fig:frame64}(a). However, this may not be the case for future frames that might reference this concealed frame. As a result, a high MOS value could be associated with a relatively low PSNR. Therefore, one can argue that PSNR alone is not enough to assess the video quality in the presence of transmission errors. These errors could lead to playback buffer starvation which in turn degrades the temporal quality.

\begin{figure}[H]
\centering
\includegraphics[width=0.8\textwidth]{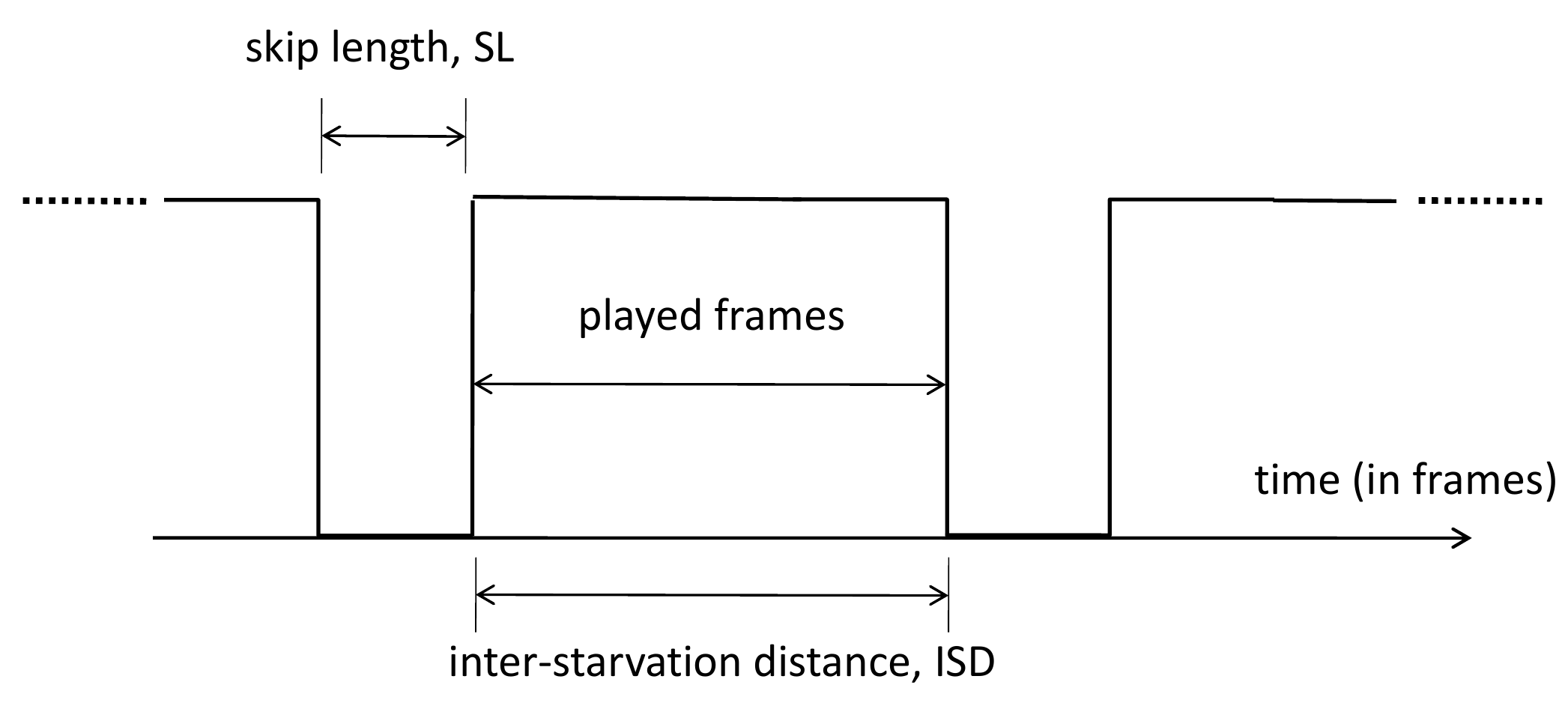}
\caption{Definitions of skip length and inter-starvation distance metrics.}
\label{fig:Skip-Def}
\end{figure}

To better quantify the effect of losing frames due to transmission errors, a temporal measure that complements PSNR was proposed in \cite{hassan2010skip}. This measure is called the skip length which can also be utilized to predict the PSNR. On the occurrence of any starvation instant, the skip length indicates how long (in frames) this starvation will last on average. The rationale behind skip length as a metric for temporal quality is the fact that it is better for the human eye to watch a continuously played back video at a lower quality rather than watching a higher quality video sequence that is frequently interrupted. An additional temporal quality metric that emanates from the skip length is the inter-starvation distance. It is the distance in frames that separates successive starvation instants. This metric complements the skip length in the sense that if the latter is small but very frequent then the quality of the played back video would be degraded. Therefore, large inter-starvation distances in conjunction with small skip lengths would result in a better played back video quality. Fig.~\ref{fig:Skip-Def} illustrates the definitions of these two metrics.

\chapter{TPC HARQ System Model}
\label{chap:3}

In this work, we consider unicast HARQ-based video transmission using TPC. For digital video transmission, video frames/pictures are compressed and encoded into binary sequences which are then packetized. Each transmitted packet has a fixed size $\mathcal{N}$ and each packet is composed of $L$ subpackets with size $N$ such that $\mathcal{N}=L\times N$. A subpacket is a TPC codeword which contains $\kappa$ information bits. Therefore, the number of packets required to contain a video frame of size $K$ bits is estimated as $N_\text{P}=\dfrac{K}{L\kappa}$. 

\begin{figure}[H]
\captionsetup{format=hang}
\centering
\includegraphics[]{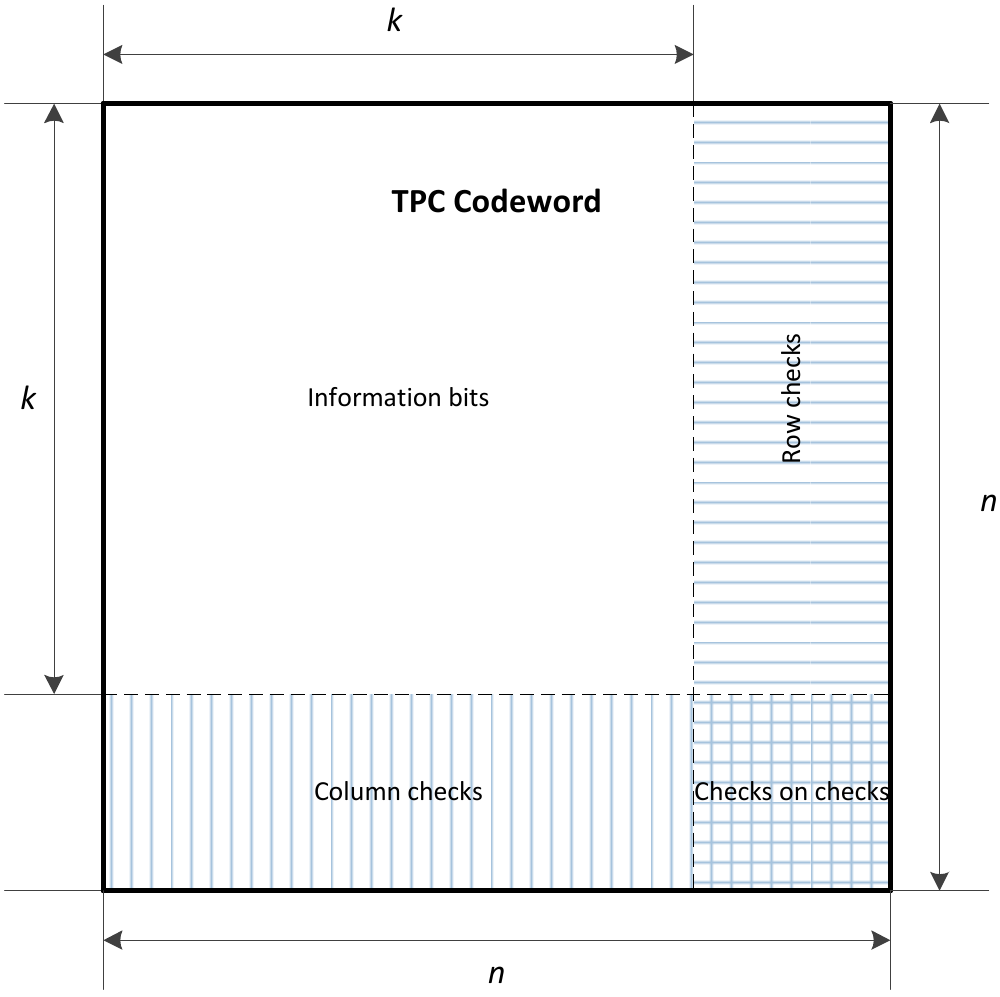}
\caption{Square TPC codeword.}
\label{fig:TPCcodeword}
\end{figure}

TPC are two dimensional FEC codes constructed by
serially concatenating two linear block codes $C^{i}$ ($i=1,2$). The two
component codes $C^{i}$ have the parameters ($n_{i},k_{i},d_\text{min}^{(i)}$)
which describe the codeword length, number of information bits, and minimum
Hamming distance, respectively \cite{al2011closed}. To build a product code,
$k_{1}\times k_{2}$ information bits are placed in a matrix of $k_{1}$ rows
and $k_{2}$ columns. The $k_{1}$ rows are encoded by code $C^{1}$ and a matrix
of size $k_{1}\times n_{1}$ is generated. Then, the $n_{1}$ columns are
encoded by the $C^{2}$ code and a two-dimensional codeword of size
$n_{2}\times n_{1}$ is obtained. The parameters of the product code $C$ are
($n_{1}\times n_{2},k_{1}\times k_{2},d_\text{min}^{(1)}\times d_\text{min}^{(2)}$).
Without loss of generality, we consider square TPC in this paper where
$n_{1}=n_{2}\triangleq n$, $k_{1}=k_{2}\triangleq k$ and $d_\text{min}%
^{(1)}=d_\text{min}^{(2)}\triangleq d_\text{min}$, and hence, the product code will be
denoted as $(n,k,d_\text{min})^{2}$. Fig.~\ref{fig:TPCcodeword} shows an illustration of a square TPC codeword.

\nomenclature{$C^{i}$}{$i$th component code}
\nomenclature{$C$}{product code}
\nomenclature{$n$}{length of component codeword}
\nomenclature{$k$}{number of information bits in component codeword}
\nomenclature{$d_\text{min}$}{minimum Hamming distance of component codeword}

Consider that the information bits sequence $\mathbf{d=[}d_{1}$, $d_{2}$,$%
\cdots $, $d_{K}\mathbf{]}$ is to be transmitted over a TPC-HARQ system, $%
d_{i}\in \left\{ 0\text{, }1\right\}$. For ease of exposition, let us for now assume that $N_\text{P}=1$. Then the sequence $\mathbf{d}$ is
divided into $L$ equal and independent parts $\mathbf{d}=\left[ \mathbf{d}%
^{(1)}\text{, }\mathbf{d}^{(2)}\text{, }\cdots \mathbf{d}^{(L)}\right] $,
where $\mathbf{d}^{(i)}\mathbf{=}\left[ d_{1}^{(i)}\text{, }d_{2}^{(i)}\text{%
,}\cdots d_{\kappa}^{(i)}\right] $ and $\kappa=K/L$. Each of the $%
\mathbf{d}^{(i)}$ sequences is applied to a CRC encoder where $l_\text{crc}$ bits
are appended to $\mathbf{d}^{(i)}$ for error detection purposes at the
receiver side. The CRC encoder output can be written as $\mathbf{m}^{(i)}%
\mathbf{=}\left[ d_{1}^{(i)}\text{, }d_{2}^{(i)}\text{,}\cdots \text{, }%
d_{\kappa}^{(i)}\text{, }m_{1}^{(i)}\text{, }m_{2}^{(i)}\text{, }\cdots \text{, }%
m_{l_\text{crc}}^{(i)}\right] $, which is then applied to a TPC encoder that
appends $l_\text{p}=n^{2}-k^{2}$ bits to $\mathbf{m}^{(i)}$ as described in \cite%
{pyndiah1998iterative} to form a TPC codeword matrix $\mathbf{C}^{(i)}$, $%
i\in \{1$, $2$, $\cdots $, $L\}$. It is worth noting that the TPC encoder
requires $k^{2}=\kappa+l_\text{crc}$ bits to form a TPC codeword with size $n^{2}$ bits.

\nomenclature{$\mathbf{d}$}{information bits sequence}
\nomenclature{$\mathbf{m}$}{CRC encoded bits sequence}
\nomenclature{$\mathbf{C}$}{TPC codeword matrix}
\nomenclature{$K$}{length of information bits sequence}
\nomenclature{$L$}{number of subpackets}
\nomenclature{$\kappa$}{number of information bits per subpacket}
\nomenclature{$l_\text{crc}$}{length of CRC bits}
\nomenclature{$l_\text{p}$}{length of TPC parity bits}

\begin{figure}[H]
\captionsetup{format=hang}
\centering
\includegraphics[]{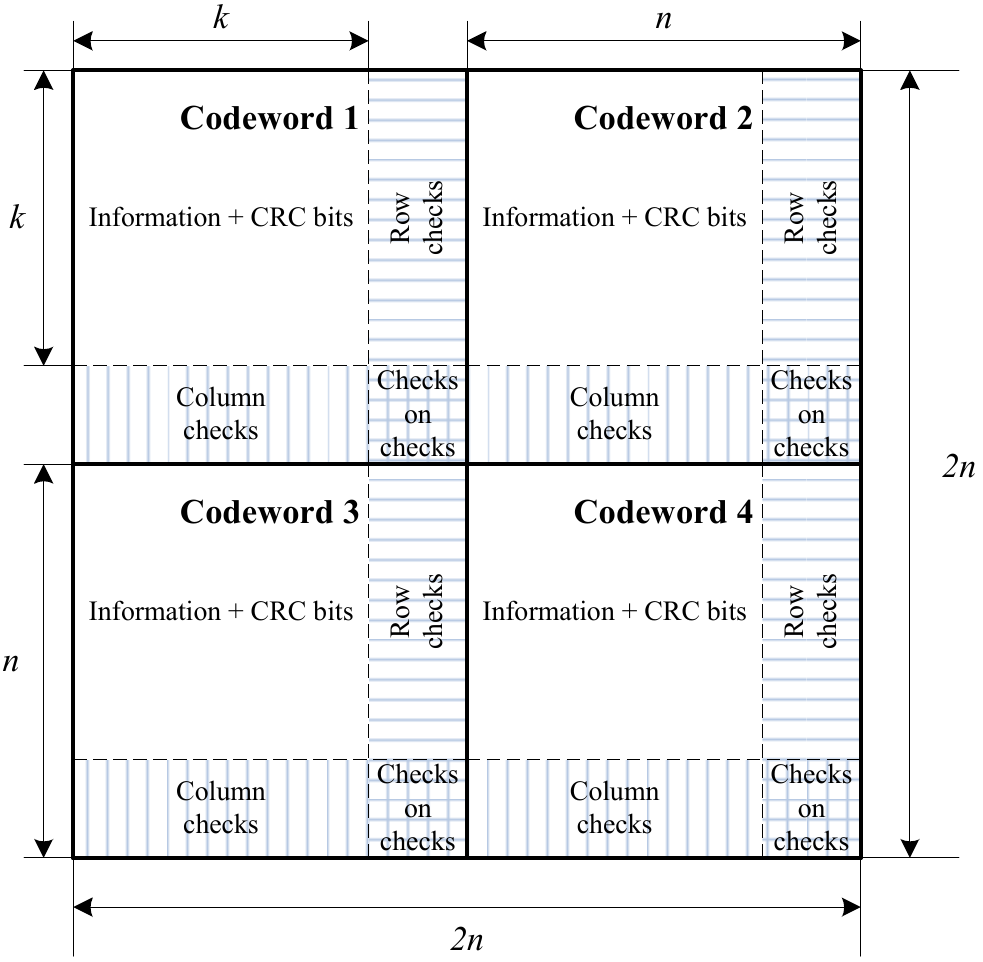}
\caption{Example of TPC-HARQ packet with $L=4$.}
\label{fig:codeword}
\end{figure}

In this work, we consider an HARQ system where each transmitted packet has
a fixed size $\mathcal{N}$, which is an integer multiple of the TPC codeword size $%
n^{2}\triangleq N=\kappa+l_{p}+l_\text{crc}$. Moreover, $\mathcal{N}$ is considered to be
fixed regardless of $\kappa$, $K$ or $L$ \cite{linadaptive2012}. Consequently,
each packet will be composed of $L$ TPC codewords. An example of the TPC-HARQ packet with $\mathcal{N}=4\times N$ is given in Fig.~\ref{fig:codeword}.

\nomenclature{$\mathcal{N}$}{length of packet}
\nomenclature{$N$}{length of subpacket} 

Splitting the data packet into smaller subpackets is necessary to
improve the system throughput and delay performance\cite{zhou2006optimum}. The TPC encoder output is interleaved to decorrelate the channel fading
effects. Most error correction codes are not effective in correcting burst errors which occur in deep fading. Therefore, ineterleavers are typically used in practice to transform the bursty error channel into a channel with independent random errors at the cost of some additional delay \cite{ghrayeb2007coding}. The required length of the interleaver is related to the coherence time of the fading channel. However, small coherence time values are typically observed in practice where wireless channels are modeled as block fading channels. Hence, an interleaver with a reasonable length will suffice without introducing a significant delay.

The interleaved bits are then modulated using BPSK modulation and transmitted through a Rayleigh fading channel.
After deinterleaving, the received subpackets can be expressed as%
\begin{equation}
\mathbf{R}^{(i)}=\mathbf{F}^{(i)}\circ \mathbf{U}+\mathbf{W}^{(i)}
\label{E-R1}
\end{equation}%
where $\mathbf{F}$ is the channel matrix, $\mathbf{U}$ is the transmitted
subpacket, the symbol `$\circ$' denotes the Hadamard product and $\mathbf{W}$
is the AWGN. Each of the matrices $\mathbf{F}$, $\mathbf{U}$, and $\mathbf{W}
$ in (\ref{E-R1}) consists of $n\times n$ elements denoted as $f_{x,y}$, $%
u_{x,y}$ and $w_{x,y}$ where $x$ and $y$ denote the row and column indices,
respectively. The channel and AWGN components are iid zero-mean complex Gaussian random variables
with variance $\sigma_\text{f}^{2}$ and $\sigma_\text{w}^{2}$, respectively.

\nomenclature{$\mathbf{R}$}{received subpacket}
\nomenclature{$\mathbf{F}$}{channel matrix}
\nomenclature{$\mathbf{U}$}{transmitted subpacket}
\nomenclature{$\mathbf{W}$}{AWGN matrix}
\nomenclature{$r$}{element of $\mathbf{R}$}
\nomenclature{$f$}{element of $\mathbf{F}$}
\nomenclature{$u$}{element of $\mathbf{U}$}
\nomenclature{$w$}{element of $\mathbf{W}$}
\nomenclature{$x$}{row index}
\nomenclature{$y$}{column index}
\nomenclature{$\sigma_\text{f}^{2}$}{variance of fading channel coefficients}
\nomenclature{$\sigma_\text{w}^{2}$}{variance of AWGN}  
\nomenclature{$\circ$}{Hadamard product}

\begin{figure}[H]
\centering
\includegraphics[width=\textwidth]{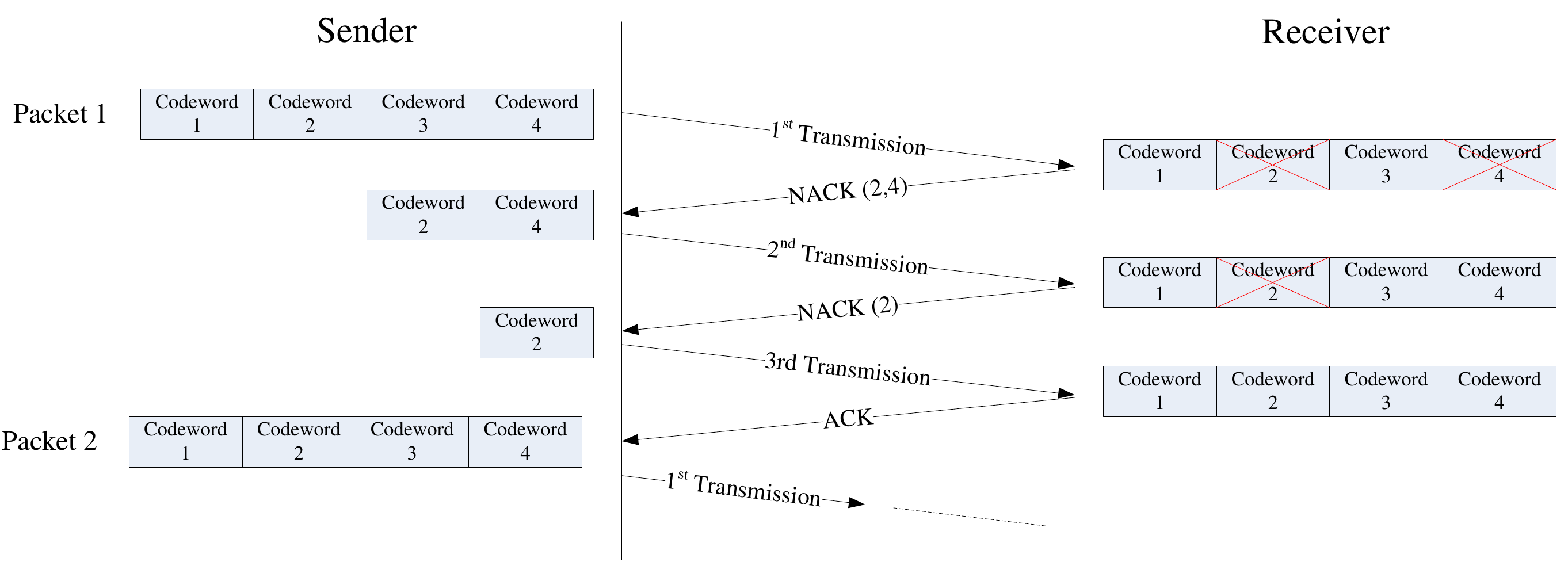}
\caption{TPC-HARQ with subpacket retransmissions.}
\label{fig:Fig1}
\end{figure}

At the receiver side, the received packet is split into $L$ subpackets that
will be decoded independently. In an adaptive system where $L$ may change, the transmitter should inform the receiver of the used $L$ in order to be able to perform decoding successfully. After decoding, the $L$ subpackets are checked for errors independently. If all subpackets are error free, an ACK is sent to the transmitter to proceed
with the transmission of the next packet. Otherwise, the soft versions of
the erroneous subpackets are stored and a NACK is
sent to instruct the transmitter to retransmit the erroneous subpackets as illustrated in Fig.~\ref{fig:Fig1}. This process is similar to the parallel Stop-and-Wait in LTE and WiMAX \cite{larmo2009lte,cipriano2010over}. Parallel Stop-and-Wait provides comparable delay performance to window-based Selective Repeat with lower complexity \cite{larmo2009lte}. 

In this work, we assume error-free feedback channel. HARQ feedback messages (ACK/NACK) are typically very small and can be well-protected \cite{zhai2006rate}. Hence, they can be transmitted with negligible loss probability. Nevertheless, imperfect feedback is an important problem to be considered. There are several solutions proposed in the literature which consider this problem such as in \cite{meyer2006arq} and \cite[p. 207]{kurose2003computer}.

Once a subpacket is retransmitted, the new received version and the stored versions
of that subpacket can be combined to enhance the system performance, which
is known as Chase combining. The subpackets combining is usually performed
using maximal ratio combining (MRC)\nomenclature{MRC}{Maximal Ratio Combining}. It is worth noting that conventional MRC is not optimal in HARQ systems because it ignores the failure of previous transmissions \cite{6129543}. In conventional diversity techniques, MRC was proven optimal for independent fading channels. Nevertheless, in HARQ systems, transmission replicas are not independent. The failure of preceding transmissions is the condition for subsequent transmissions. 

\begin{figure}[H]
\centering
\includegraphics[]{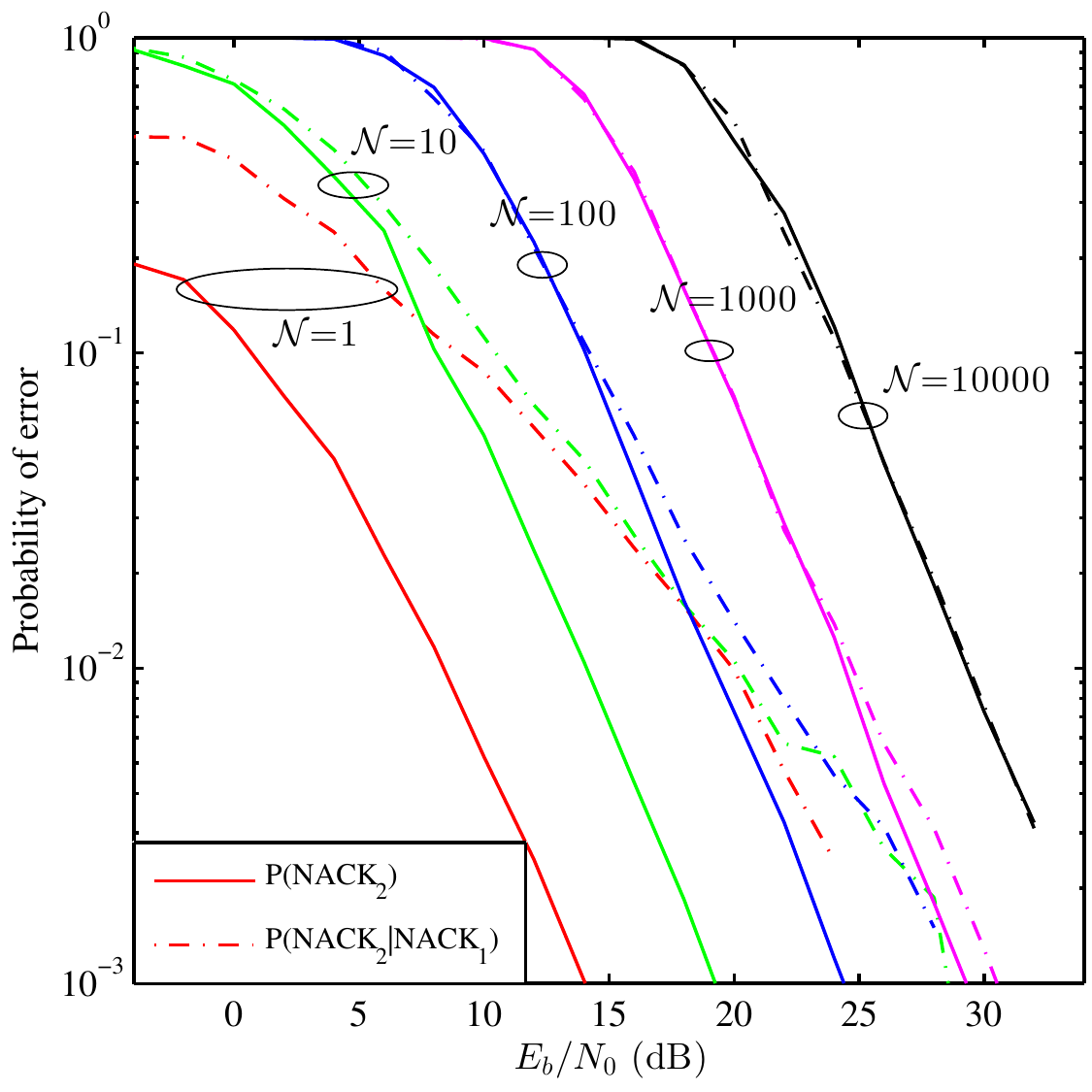}
\caption{Probability of having a NACK in the second transmission when assuming dependent and independent transmissions.}
\label{fig:perConditional}
\end{figure}

Ignoring the dependency between transmission rounds may result in overoptimistic prediction of the performance of HARQ systems. To demonstrate this argument, we simulated an ARQ system and obtained the probability of having a NACK in the second transmission while assuming dependent and independent transmissions denoted as $P(\text{NACK}_2|\text{NACK}_1)$ and $P(\text{NACK}_2)$, respectively. Fig.~\ref{fig:perConditional} shows the failure probabilities in the second transmission as a function of the SNR per bit ($E_\text{b}/N_0$) for different block sizes. 

The figure shows that for small block sizes, the error probability when assuming dependent transmissions is higher than the error probability when assuming independent transmissions. However, for block sizes greater than 100 bits, the difference between $P(\text{NACK}_2|\text{NACK}_1)$ and $P(\text{NACK}_2)$ is negligible. Moreover, the results reported in \cite{6129543} show that the optimal combiner outperforms the MRC only for very short block lengths. Otherwise, the MRC and optimal combiner provide roughly the same throughput. Practically speaking, a typical block length is much larger than 100 symbols. Consequently, we adopt MRC to implement the subpacket combining.


The output of the Chase combiner is equal to%
\begin{equation}
\mathbf{R}_\text{C}=\sum_{i=1}^{\ell }\mathbf{A}^{(i)}\circ \mathbf{R}^{(i)}
\end{equation}%
where $\mathbf{A}^{(i)}$ is the MRC weights matrix, $a_{x,y}^{(i)}=\left(
f_{x,y}^{(i)}\right) ^{\ast }\left[ \sum_{i=1}^{\ell }\left\vert
f_{x,y}^{(i)}\right\vert ^{2}\right] ^{-1}$ and $(.)^{\ast }$ denotes the
complex conjugation process \cite%
{proakis1995digital}. Based on the decoder implementation, the combiner
output $\mathbf{R}_\text{C}$ might be demodulated and converted to binary bits if
HIHO decoding is desired; otherwise, $\mathbf{R}_\text{C}$ is fed directly to the
TPC decoder for SISO decoding \cite{pyndiah1998iterative}. The remaining
processes are identical to the first transmission session. The
retransmission process is repeated until all subpackets are error free, or
the maximum number of transmissions $M$ is reached. Therefore, the NACK requires $L$ bits to indicate the locations of the erroneous subpackets. If the number of transmission sessions is equal to M, the erroneous subpackets are dropped. 

\nomenclature{$\mathbf{R}_\text{C}$}{output of the Chase combiner}
\nomenclature{$\mathbf{A}$}{MRC weights}
\nomenclature{$a$}{element of $\mathbf{A}$}
\nomenclature{$(.)^{\ast }$}{complex conjugation process}
\nomenclature{$M$}{maximum number of allowed ARQ rounds}

Turbo product codes are powerful FEC codes that can provide high coding
gain. Nevertheless, the complexity of TPC decoders can be very high when maximum likelihood
decoding (MLD)\nomenclature{MLD}{Maximum Likelihood Decoding} is used. Therefore, sub-optimum iterative decoding methods
are alternatively used to reduce the complexity while providing satisfactory
performance \cite{pyndiah1998iterative}. 

In the iterative decoding of TPC, all rows of the code matrix are decoded
sequentially followed by column decoding. A full iteration corresponds to
the decoding of all set of rows and columns while a half iteration
corresponds to the decoding of either all the rows or all columns. All
row/column component codewords are decoded independently. The row/column
decoding in every iteration is performed using MLD when HIHO is considered,
and it is performed as described in \cite{pyndiah1998iterative} when SISO is
considered. The decoding process is terminated if the maximum number of
iterations is reached, or if all rows and columns are valid codewords of
their respective elementary codes.

\section{HARQ System Throughput}

\nomenclature{$\eta$}{throughput}
\nomenclature{$V$}{number of transmitted subpackets}
\nomenclature{$z_{i}$}{random number that indicates if $i$th subpacket is dropped}
\nomenclature{$P_\text{D}$}{subpacket drop rate}
\nomenclature{$P_\text{E}^{(i)}$}{probability of subpacket error in $i$th ARQ round}
\nomenclature{$\mathbb{E}\left\{.\right\}$}{expected value of a random variable}
\nomenclature{$\tilde{\rho}$}{unbounded number of transmissions per subpacket}
\nomenclature{$\rho$}{number of transmissions per subpacket}
\nomenclature{$P(.)$}{probability of an event}

The transmission efficiency or throughput $\eta$, is defined as the ratio of
the number of information bits received successfully to the total number of
transmitted bits \cite[p. 461]{lin1983error}. Given that $V$ subpackets are transmitted, then%
\begin{align}
\eta &  =\frac{\kappa z_{1}+\kappa z_{2}+\cdots+\kappa z_{V}}{N\rho_{1}+N\rho_{2}%
+\cdots+N\rho_{V}}\nonumber\label{E-Throu-1}\\
&  =\frac{\kappa\sum_{i=1}^{V}z_{i}}{N\sum_{i=1}^{V}\rho_{i}}%
\end{align}
where $1\leq\rho_{i}<M$ is a random number that represents the total number of
transmissions per subpacket and $z_{i}$ is a random number that indicates if
the subpacket is dropped; $z_{i}=0$ if the $i$th subpacket is dropped, and $1$
otherwise. However, because $z_{i}\in\left\{  0,1\right\}  $, then$\frac{1}%
{V}\sum_{i=1}^{V}z_{i}$ is just the ratio of the non-zero elements to the
total number of transmitted subpackets, i.e., the complement of the subpacket
drop rate $P_\text{D}$. Given that $V\rightarrow\infty$, using the law of large
numbers $\frac{1}{V}\sum_{i}\rho_{i}\rightarrow\mathbb{E}\left\{
\rho\right\}  $ and $\frac{1}{V}\sum_{i=1}^{V}z_{i}\rightarrow(1-P_\text{D})$,
where $\mathbb{E}\left\{  .\right\}  $ denotes the expected value. Therefore
(\ref{E-Throu-1}) can be written as
\begin{equation}
\eta=\frac{1}{\mathbb{E}\left\{  \rho\right\}  }\frac{\kappa}{N}(1-P_\text{D})
\label{eq:eta}%
\end{equation}
where $P_\text{D}$ can be computed by noting that a subpacket is dropped if the $M$
transmissions fail,%
\begin{equation}
P_\text{D}=\prod\limits_{i=1}^{M}P_\text{E}^{(i)} \label{eq:Pd}%
\end{equation}
where $P_\text{E}^{(i)}$ is the probability that a subpacket is declared erroneous
during the $i$th transmission round. In other words, $P_\text{E}^{(i)}$ can be expressed as
\begin{equation}
P_\text{E}^{(i)}=
\begin{cases}
P(\text{NACK}_1)\text{,} & \text{for}~i=1 \\ 
P(\text{NACK}_i\left|\text{NACK}_1,\cdots,\text{NACK}_{i-1}\right.)\text{,} & \text{for}~i>1
\end{cases}%
\end{equation}
The probability $P_\text{E}^{(i)}$ can be accurately estimated as $P(\text{NACK}_i)$ for block sizes larger than 100 bits where the dependency assumption is dropped \cite{zhou2006optimum,long2009perf}.
It is worth noting that the throughput analysis can be obtained using the renewal reward theory as well 
\cite{Zorzi96,Caire01}.

To compute $\mathbb{E}\left\{
\rho\right\}$, we initially assume that the number of transmissions per
subpacket is unbounded, and the erroneous subpackets are immediately discarded
by the receiver. Consequently, the unbounded number of transmissions per
subpacket $\tilde{\rho}$ has a Geometric probability mass function (PMF)\nomenclature{PMF}{Probability Mass Function} that
is given by
\begin{equation}
P(\tilde{\rho}=\ell)=P_\text{E}^{\ell-1}(1-P_\text{E})\text{, }1\leq\tilde{\rho}<\infty
\end{equation}
where $P_\text{E}$ is the probability that a subpacket is declared erroneous when
combining is not used and it is the same for all ARQ rounds. Hence, the
expected value of $\tilde{\rho}$ is given by
\begin{equation}
\mathbb{E}\left\{  \tilde{\rho}\right\}  =\sum_{i=1}^{\infty}iP(\tilde{\rho
}=i)=\frac{1}{1-P_\text{E}}.
\end{equation}
However, with subpacket combining, the value of $P_\text{E}$ changes as a function
of the transmission round index. Hence,%
\begin{align}
P\left(  \tilde{\rho}=\ell\right)   &  =P_\text{E}^{(1)}P_\text{E}^{(2)}\cdots
P_\text{E}^{(\ell-1)}\left[  1-P_\text{E}^{(\ell)}\right] \nonumber\\
&  =\left[  1-P_\text{E}^{(\ell)}\right]  \prod\limits_{i=1}^{\ell-1}P_\text{E}^{(i)}.
\end{align}
And%
\begin{equation}
\mathbb{E}\left\{  \tilde{\rho}\right\}  =\sum_{i=1}^{\infty}\text{ }i\text{
}\left[  1-P_\text{E}^{(i)}\right]  \prod\limits_{j=1}^{i-1}P_\text{E}^{(j)}.
\end{equation}
However, in practical HARQ systems, the number of transmissions is limited to
$M$, which results in a truncated HARQ. The PMF of the truncated $\rho$
becomes
\begin{equation}
P(\rho=\ell)=%
\begin{cases}
P(\tilde{\rho}=\ell)\text{,} & \ell\text{$\in$}\left\{  \text{$1$, $2$%
,$\cdots$, $M-1$}\right\} \\
1-P(\tilde{\rho}<M)\text{, \ } & \ell=M\\
0\text{,} & \text{otherwise}
\end{cases}
\end{equation}
where $P(\tilde{\rho}<M)=\sum_{i=1}^{M-1}P\left(  \tilde{\rho}=i\right)  $.
After some straightforward manipulations, $\mathbb{E}\left\{  \rho\right\}  $
can be expressed as,
\begin{equation}
\mathbb{E}\left\{  \rho\right\}  =M-\sum_{i=1}^{M-1}\left[  M-i\right]
\left[  1-P_\text{E}^{(i)}\right]  \prod\limits_{j=1}^{i-1}P_\text{E}^{(j)}.
\label{eq:avgRs}%
\end{equation}
Finally, the throughput $\eta$ can be computed by substituting (\ref{eq:Pd})
and (\ref{eq:avgRs}) into (\ref{eq:eta}),%
\begin{equation}
\eta=\frac{\kappa}{N}\frac{1-\prod\nolimits_{i=1}^{M}P_\text{E}^{(i)}}{M-\sum
_{i=1}^{M-1}\left[  M-i\right]  \left[  1-P_\text{E}^{(i)}\right]  \prod
\nolimits_{j=1}^{i-1}P_\text{E}^{(j)}}. \label{E-eta-1}%
\end{equation}

Alternatively, since the number of retransmissions
takes only non-negative integer values, $\mathbb{E}\left\{ \rho\right\} 
$ can also be written as \cite{Caire01, garcia2009probability} 
\begin{equation}
\mathbb{E}\left\{ \rho\right\} =\sum_{i=1}^{M}P\left( \rho\geq
i\right) =1+\sum_{i=1}^{M-1}\prod_{j=1}^{i}P_\text{E}^{(j)}.  \label{eq:avgRs2}
\end{equation}%
Therefore, $\eta $ is given by 
\begin{equation}
\eta =\frac{\kappa }{N}\frac{1-\prod\nolimits_{i=1}^{M}P_\text{E}^{(i)}%
}{1+\sum_{i=1}^{M-1}\prod_{j=1}^{i}P_\text{E}^{(j)}}.  \label{E-eta-2}
\end{equation}%

Unfortunately, computing $P_\text{E}^{(\ell)}$analytically is not feasible because
TPC error correction capability depends on the error pattern rather than the
number of errors \cite{al2011closed}. Moreover, using Monte Carlo simulation
to evaluate $\eta$ could be tedious as well due to the retransmission process
inherent in HARQ systems. Consequently, developing a semi-analytical solution is indispensable in such scenarios. In sections \ref{sec:awgn} and \ref{sec:fading},
the SAS is derived for both AWGN and fading channels, respectively.

\section{Semi-Analytical Solution, AWGN Channels}
\label{sec:awgn}

As it can be noted from (\ref{E-eta-1}), computing $\eta$ requires the
knowledge of $P_\text{E}^{(\ell)}$ for $\ell=1$, $2$, $\ldots$ , $M$, where
$P_\text{E}^{(\ell)}$ is the subpacket error rate during the $\ell$th transmission
session. In AWGN channels, MRC is equivalent to equal gain combining (ECG)\nomenclature{ECG}{Equal Gain Combining},
which only provides array gain. Therefore, the packet combining effect is
limited to the enhancement of the overall SNR, which can be computed as the
superposition of all SNRs over $\ell$ transmission sessions,%
\begin{equation}
\gamma_\text{C}=\sum_{l=1}^{\ell}\gamma^{(l)}\text{, }\ell\in\left\{  1\text{,
}2\text{, }\ldots\text{ , }M\right\}
\end{equation}
where $\gamma^{(l)}$ is the SNR during the $l$th transmission session. Assuming that the channel statistics does not change over successive transmissions of the same packet, i.e., $\gamma^{(1)}=\gamma^{(2)}=\cdots\gamma^{(\ell)}=\gamma$, then $\gamma_\text{C}=\ell\gamma$. Therefore, the
subpacket error rate given that $\ell$ transmissions have been combined can be
computed as%
\begin{equation}
P_\text{E}^{(\ell )}|_{\text{SNR}=\gamma }=P_\text{E}|_{\text{SNR}=\ell \gamma }
\label{eq:PpAWGN}
\end{equation}%
where $P_\text{E}^{(\ell )}|_{\text{SNR}=\gamma }$ is $P_{E}^{(\ell )}$ computed
at SNR=$\gamma $ and $P_\text{E}|_{\text{SNR}=\ell \gamma }$ is $P_\text{E}$ computed
at SNR=$\ell \gamma $. Therefore, $P_\text{E}^{(\ell)}$ is equal to $P_\text{E}$ except that the SNR is
replaced by $\gamma_\text{C}$. Therefore, the SAS can be used to compute $\eta$
without the need to simulate the complete TPC-HARQ\ system. Instead, only
$P_\text{E}$ is needed, which is just the packet error probability of the regular
TPC system without ARQ. Consequently, the simulation time can be reduced remarkably.

\nomenclature{$\gamma_\text{C}$}{combined SNR}
\nomenclature{$\gamma^{(l)}$}{SNR during the $l$th transmission session}

\section{Semi-Analytical Solution, Fading Channels}
\label{sec:fading}

Unlike the AWGN channel case, MRC in fading channels provides both array and
diversity gains where the average SNR is improved and the statistics of the
fading channel is changed. In particular, given that the $\ell$ successive
transmissions of a particular subpacket experience iid Rayleigh fading, the
envelope probability density function (PDF)\nomenclature{PDF}{Probability Density Function} of the combiner output will
be Nakagami-$\ell$, the combiner output SNR $\gamma_\text{C}=\sum_{i=1}^{\ell
}\gamma^{(i)}$ and the PDF of $\gamma_\text{C}$ is Chi-squared with $2\ell$ degrees
of freedom \cite[p. 859]{proakis1995digital}.

\begin{figure}[H]%
\centering
\includegraphics[
]%
{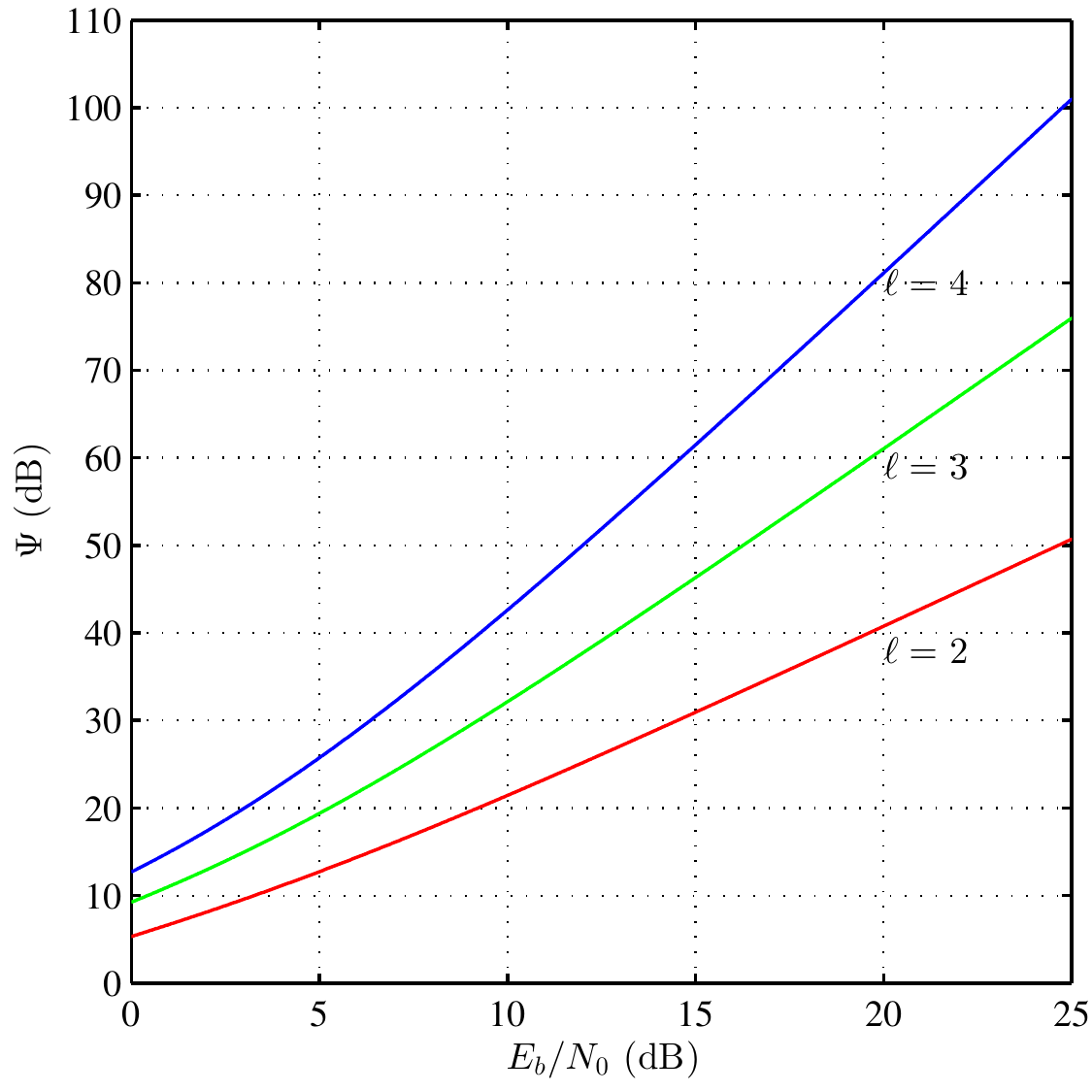}%
\caption{{Equivalent SNR $\Psi$ after combining for diversity order
$\ell=2,3,4$ in Rayleigh fading channels.}}%
\label{fig:snrGain1}%
\end{figure}

As it can be noted from the discussion given in the previous paragraph,
deriving the SAS for fading channels following the same approach used for the
AWGN case is not feasible because it ignores the diversity gain of the MRC
process. To resolve this problem, the closed-form expression for the uncoded
BER with $\ell$-order diversity is invoked where the BER
difference between the single and multiple transmissions is mapped to an SNR
improvement, which is then used in a similar way as for the AWGN channel case.
Towards this goal, note that the BER for uncoded BPSK with $\ell$-order
diversity in Rayleigh fading channels is given by \cite{proakis1995digital}
\begin{equation}
p_\text{e}^{\{\ell\}}=\left[  \frac{1}{2}\left(  1-\sqrt{\Lambda}\right)  \right]
^{\ell}\sum_{i=0}^{\ell-1}\binom{\ell-1+i}{i}\left[  \frac{1}{2}+\frac{1}%
{2}\sqrt{\Lambda}\right]  ^{i} \label{eq:peM}%
\end{equation}
where $\Lambda=\dfrac{\gamma}{1+\gamma}$. The equivalent BER of BPSK over
Rayleigh fading channels without combining can be described as
\begin{equation}
p_\text{e}=\frac{1}{2}\left(  1-\sqrt{\dfrac{\Psi}{1+\Psi}}\right)
\label{E-BER-Rayleigh}%
\end{equation}
where $\Psi$ denotes the equivalent SNR. Therefore,
\begin{equation}
p_\text{e}^{\{\ell\}}|_{SNR=\gamma}=p_\text{e}|_{SNR=\Psi}.
\end{equation}
By equating (\ref{eq:peM}) with (\ref{E-BER-Rayleigh}) and solving for $\Psi$
we obtain
\begin{equation}
\Psi=\frac{\beta_{\ell}^{2}}{1-\beta_{\ell}^{2}} \label{eq:alpha}%
\end{equation}
where
\begin{equation}
\beta_{\ell}{=1-2\left[  \frac{1}{2}\left(  1-\sqrt{\Lambda}\right)  \right]
^{\ell}\sum_{i=0}^{\ell-1}\binom{\ell-1+i}{i}\left[  \frac{1}{2}\left(
1+\sqrt{\Lambda}\right)  \right]  ^{i}.} \label{eq:A}%
\end{equation}
Fig.~\ref{fig:snrGain1} shows $\Psi$ as a function of ${\gamma}=E_\text{b}/N_{0}%
$\ where $E_\text{b}$\ is the average energy per bit and $N_{0}$\ is the noise
power spectral density.

Therefore, $P_\text{E}^{(\ell)}$ can be computed as
\begin{equation}
P_\text{E}^{(\ell)}|_{SNR=\gamma}=P_\text{E}|_{SNR=\Psi}\text{, }\ell\in\left\{
1,\text{ }2\text{, }\ldots\text{ , }M\right\}  \text{.}%
\end{equation}

\nomenclature{$p_\text{e}^{\{\ell\}}$}{BER of BPSK with $\ell$-order diversity in Rayleigh fading}
\nomenclature{$p_\text{e}$}{BER of BPSK in Rayleigh fading without combining}
\nomenclature{$\Psi$}{equivalent SNR}
\nomenclature{$E_\text{b}$}{average energy per bit}
\nomenclature{$N_{0}$}{noise power spectral density}

\begin{figure}[H]%
\centering
\includegraphics[width=4.4in]%
{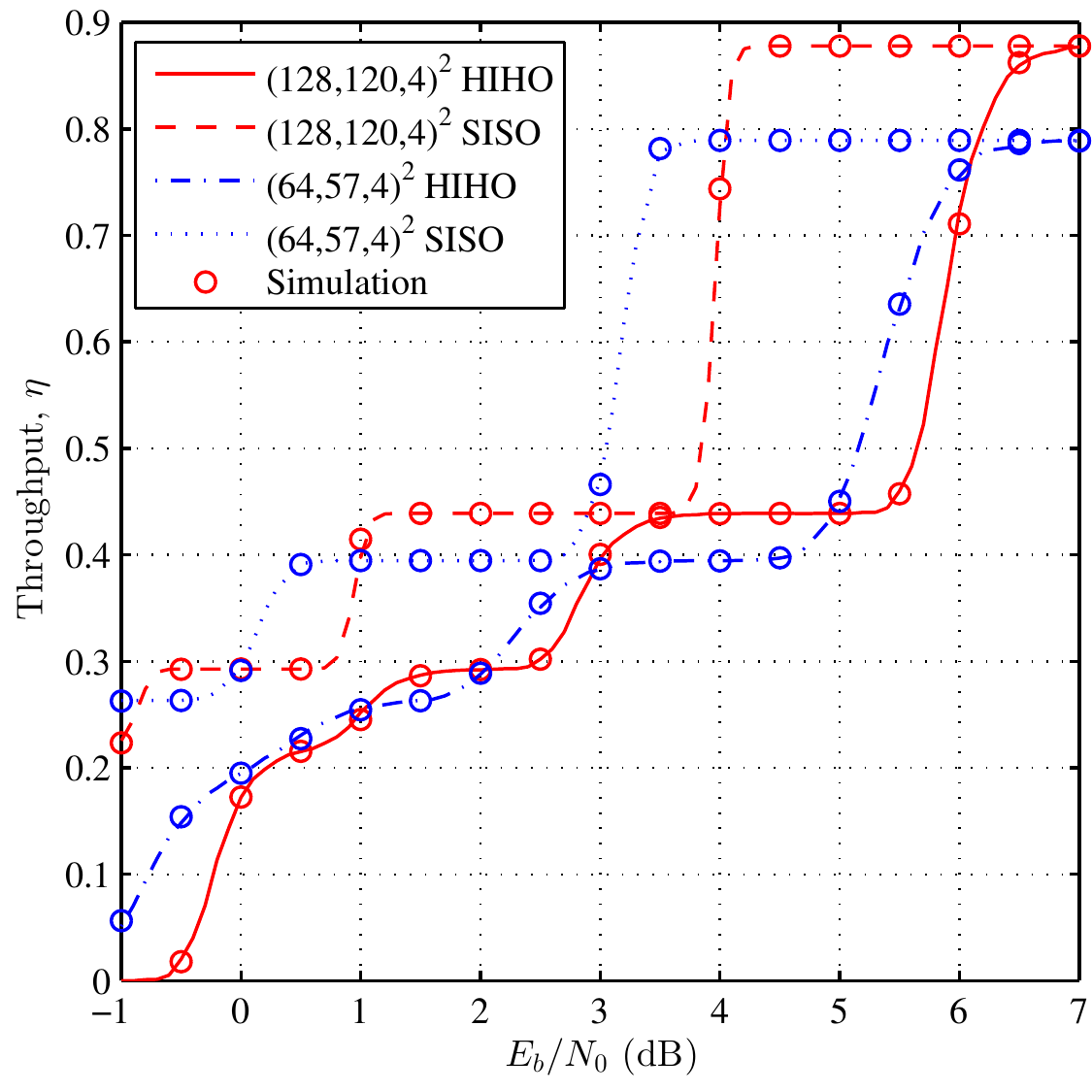}%
\caption{{Throughput simulation and semi-analytical results in AWGN
channels.}}%
\label{fig:thrAWGN1}%
\end{figure}

\begin{figure}[H]%
\centering
\includegraphics[width=4.4in]%
{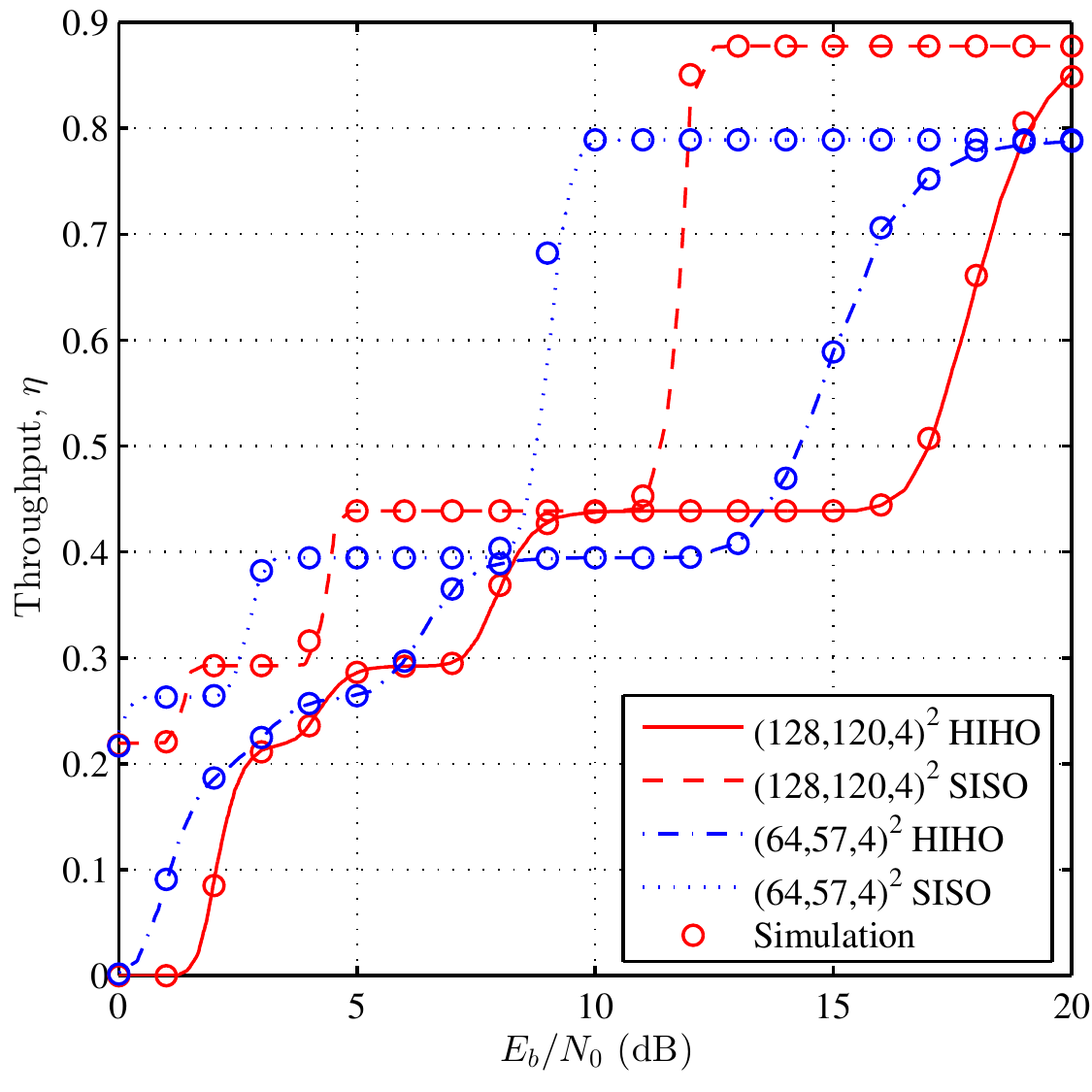}%
\caption{{Throughput simulation and semi-analytical results in Rayleigh fading
channels.}}%
\label{fig:thrFading1}%
\end{figure}

To verify the semi-analytical solution, a TPC-HARQ system is simulated for two
different TPC where the extended Bose-Chandhuri-Hocquenghen (eBCH)\nomenclature{eBCH}{Extended BCH} is used as
the component code. The two TPC are the eBCH$(128,120,4)^{2}$ and
eBCH$(64,57,4)^{2}$. The packet size for both cases is fixed at $\mathcal{N}%
=128^{2}=16$,$384$~bits and hence, $L=1$ for the first case and $L=4$ for the
latter case. A 16-bit CRC code is used before encoding with TPC for error
detection. Each simulation run consists of $1000$\ packets and the maximum
number of ARQ rounds allowed is $M=4$. The SISO and HIHO decoders are
configured to perform a maximum of four iterations. The number of reliability
bits for the Chase decoder is set to $4$ bits \cite{pyndiah1998iterative},
\cite{chase1972class}. As it can be noted from Fig.~\ref{fig:thrAWGN1} and
\ref{fig:thrFading1}, the semi-analytical results closely match the simulation results.

As it can be noted from the results in Fig.~\ref{fig:thrAWGN1}, the throughput
for the two cases has several crossovers. This can be explained by noting that
the eBCH$(64,57,4)^{2}$ has lower code rate but stronger error correction
capability \cite{pyndiah1998iterative}. For example, the throughput of the
SISO eBCH$(128,120,4)^{2}$ is larger than the SISO eBCH$(64,57,4)^{2}$ at high
$E_\text{b}/N_{0}\gtrsim4$ dB where both systems have an average number of
transmission sessions of one. At $E_\text{b}/N_{0}<4$ dB, the eBCH$(128,120,4)^{2}$
fails to correct most of the packets in the first transmission session, and
hence a second transmission is required, which increases the average number of
transmissions to two and reduces the system throughput by $50\%$. Due to its
stronger error correction capability, the eBCH$(64,57,4)^{2}$ retains the same
throughput for lower $E_\text{b}/N_{0}$ $(\sim3.5$ dB$)$, which creates the first
crossover. Moreover, it can be noted from the figure that the slope of the
throughput during the transition periods for the $L=4$ case is lower than the
$L=1$ case, because the subpacket retransmission has less effect on the
throughput as compared to the full packet retransmission. The same behavior
can be observed for the fading channel throughput given in Fig.
\ref{fig:thrFading1}.
   
\begin{figure}[H]
\begin{center}
\includegraphics[
]%
{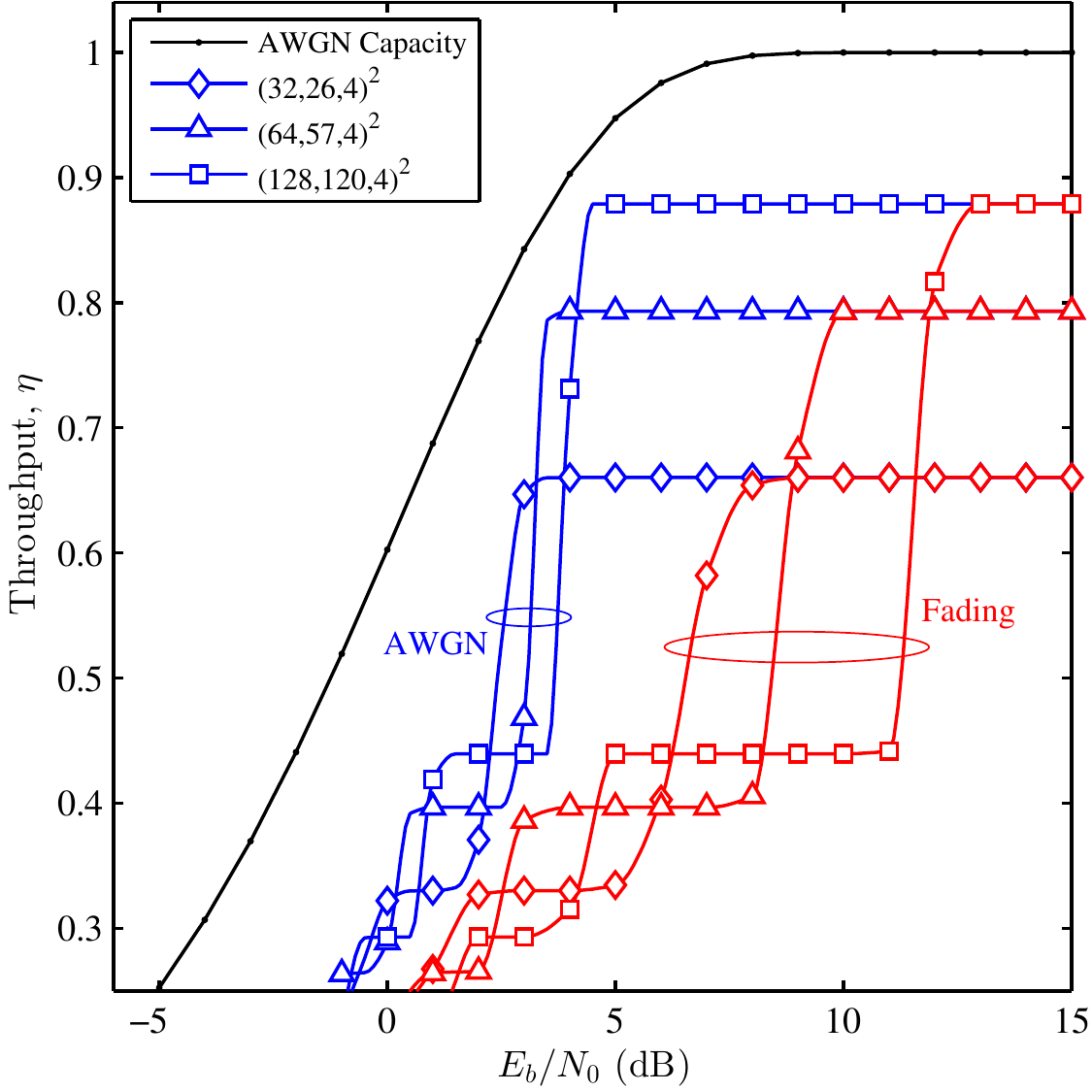}%
\caption{TPC-HARQ throughput versus channel capacity when BPSK and SISO decoding are used in AWGN and Rayleigh fading channels.}%
\label{fig:capacity}%
\end{center}
\end{figure}

Fig.~\ref{fig:capacity} compares the
performance of TPC-HARQ when combining and SISO decoding are used to the
theoretical limit of reliable communication in AWGN channels for BPSK. In
this work, the TPC encoder output is interleaved making the individual
symbols experience independent channel fading. Therefore, the effect of the
channel would be equivalent to fast fading where the standard capacity can
be used to describe the limits of reliable communications \cite[p. 905]%
{proakis1995digital}. The capacity of AWGN channel is shown in the figure as
an upper bound for the throughput of TPC-HARQ in both AWGN and Rayleigh
fading channels.

 \section{HARQ Delay Analysis}
For delay analysis of HARQ systems, computing the average number of transmission
sessions per \textbf{packet} is required in addition to the average number of transmission
sessions per \textbf{subpacket}. The total
number of transmission sessions per packet $\mathcal{R}$ is equal to the number
of NACKs sent to the transmitter, in addition to the initial transmission
session. Consequently, $\mathcal{R}$ is equal to the maximum number of
transmissions for each of the $L$ subpackets,%
\begin{equation}
\mathcal{R}=\max\left\{  \rho^{\{1\}}\text{, }\rho^{\{2\}}\text{, }\cdots\text{,
}\rho^{\{L\}}\right\}  \text{, }1\leq\rho^{\{i\}}<M.
\end{equation}
Therefore, the cumulative density function (CDF)\nomenclature{CDF}{Cumulative Density Function} of the number of transmission
sessions in untruncated HARQ is
\begin{align}
F_{\tilde{\mathcal{R}}}(\ell)  &  =P(\tilde{\mathcal{R}}\leq\ell)\nonumber\\
&  =P(\tilde{\rho}^{\{1\}}\leq\ell)\cdot P(\tilde{\rho}^{\{2\}}\leq
\ell)\cdots P(\tilde{\rho}^{\{L\}}\leq\ell)
\end{align}
where $\tilde{\rho}^{\{i\}}$ denotes the unbounded number of transmissions for
the $i$th subpacket. Subsequently, the PMF of the maximum of $L$ subpacket
transmissions without truncation can be obtained as
\begin{equation}
P(\tilde{\mathcal{R}}=\ell)=\left[  1-\prod\limits_{i=1}^{\ell}%
P_\text{E}^{(i)}\right]  ^{L}-\left[  1-\prod\limits_{i=1}^{\ell-1}P_\text{E}%
^{(i)}\right]  ^{L}%
\end{equation}
and the PMF of packet transmissions in truncated HARQ is
\begin{equation}
P(\mathcal{R}=\ell)=%
\begin{cases}
P(\tilde{\mathcal{R}}=\ell)\text{,} & \text{for $\ell=1,2,\cdots,M-1$}\\
1-P(\tilde{\mathcal{R}}<M)\text{,} & \text{for $\ell=M$}%
\end{cases}
. \label{eq:pmf4}%
\end{equation}
Therefore, the average number of packet transmissions in truncated HARQ with
subpacket fragmentation is
\begin{equation}
\mathbb{E}\left\{  \mathcal{R}\right\}  =\sum_{i=1}^{M}i\cdot P(\mathcal{R}=i)
\label{eq:avgR1}%
\end{equation}

\nomenclature{$\mathcal{R}$}{number of transmissions per packet}
\nomenclature{$\tau_\text{s}$}{transmission time using subpacket-based HARQ}
\nomenclature{$\tau$}{transmission time using packet-based HARQ}
\nomenclature{$\chi$}{transmission data rate in bps}
\nomenclature{$t_\text{p}$}{propagation time}

Let $\tau_\text{s}$ be the time (in seconds) required to send $K$ information bits available at the source side using subpacket-based HARQ. Based on the average numbers of transmissions
$\mathbb{E}\left\{  \rho\right\}  $ and $\mathbb{E}\left\{  \mathcal{R}\right\}
$, $\tau_\text{s}$ can be estimated by
\begin{equation}
\tau_\text{s}=\left(\mathbb{E}\left\{  \rho\right\}  \dfrac
{L\,N}{\chi}+\mathbb{E}\left\{  \mathcal{R}\right\} 2t_\text{p}\right)\dfrac{K}{L\kappa}
\label{eq:taus}%
\end{equation}
where $\chi$ is the data rate in bits per second (bps)\nomenclature{bps}{bits per second} and $t_\text{p}$ is the
propagation time from source to destination. On the other hand, the transmission time in HARQ with no subpacketization is given by
\begin{equation}
\tau=\mathbb{E}\left\{  \rho\right\}\left(  \dfrac
{N}{\chi}+ 2t_\text{p}\right)\dfrac{K}{\kappa}
\label{eq:tau}%
\end{equation}

\begin{figure}[H]%
\centering
\includegraphics[
]%
{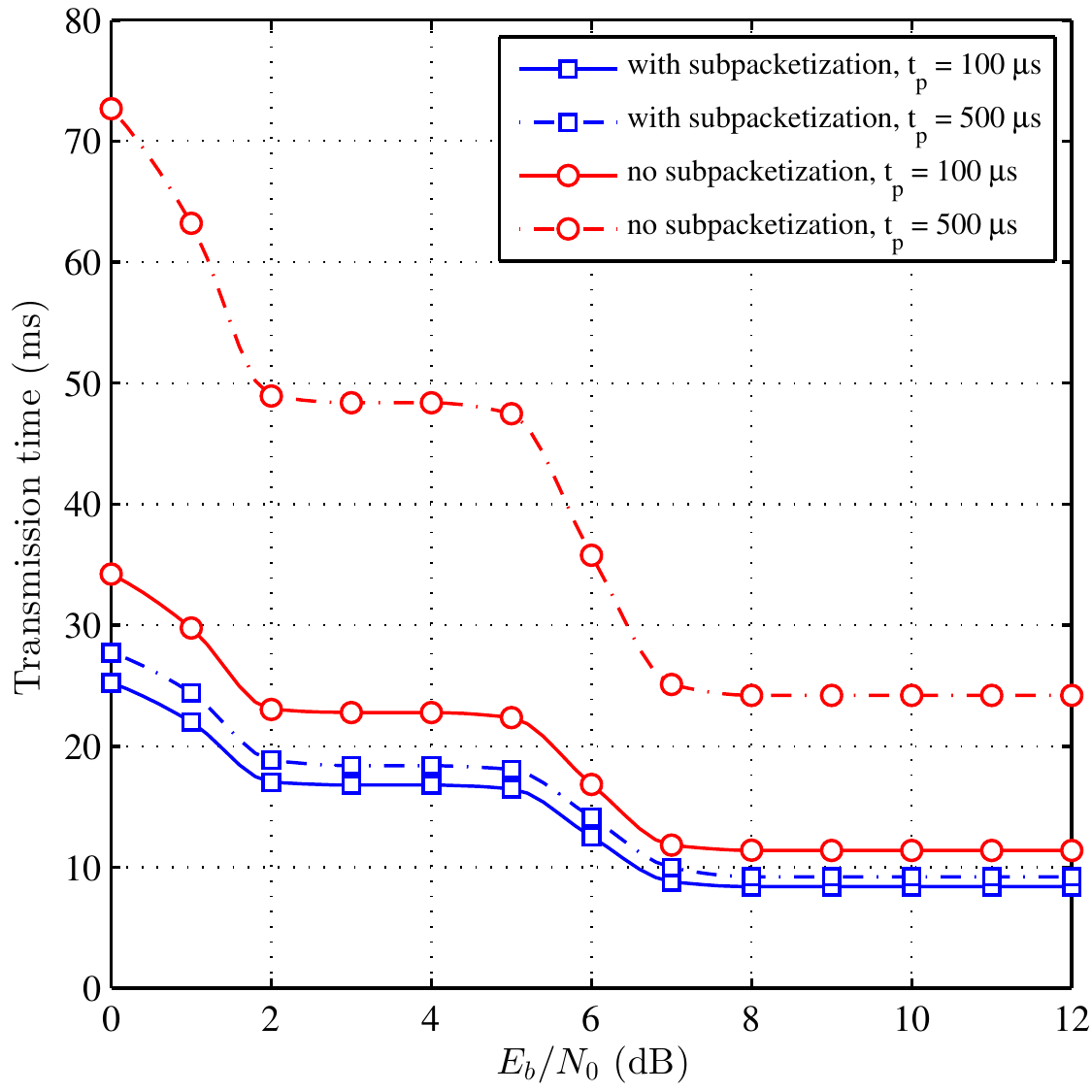}%
\caption{{Delay performance of subpacket and packet-based HARQ for eBCH $(32,26,4)^2$ with SISO decoding in Rayleigh fading channels.}}%
\label{fig:subDelay}%
\end{figure}

Fig.~\ref{fig:subDelay} shows that subpacket-based HARQ significantly reduces the transmission delay when compared to HARQ with no subpacketization. Both systems are simulated using TPC code eBCH $(32,26,4)^2$ with SISO decoding in Rayleigh fading channels. For subpacket-based HARQ, $N=32^2=1,024$~bits, $\mathcal{N}=128^2=16,384$~bits and $L=\mathcal{N}/N=16$, whereas, for the regular HARQ, $N=\mathcal{N}=1,024$~bits. Moreover, for both systems, $l_\text{crc}=16$~bits, $\kappa=26^2-l_\text{crc}=660$~bits, $K=10,560$~bits, and $\chi=2$~Mbps. The results in the figure are obtained for $t_\text{p}=100$ and $500\,\mu$s. The advantage of subpacket-based HARQ over regular HARQ increases as $t_\text{p}$ increases. We also see that the subpacket HARQ system with $t_\text{p}=500\,\mu$s still achieves lower delay than the packet-based HARQ with $t_\text{p}=100\,\mu$s.

\chapter{Adaptive Hybrid ARQ System using Turbo Product Codes with Hard/Soft Decoding}
\label{chap:4}

The BER performance of TPC has been
considered extensively in the literature. However, other performance metrics
such as throughput can be more informative in particular systems. In this
chapter, the throughput performance of HARQ
is considered using TPC with iterative hard and soft
decision decoding. Monte Carlo simulation and semi-analytical solutions are
used to evaluate the throughput of HARQ-TPC system for a wide selection
of codes. The obtained results reveal that the coding gain advantage of SISO over HIHO decoding is reduced significantly when the throughput is adopted
as the performance metric, and it actually vanishes completely for some
codes. When adaptive coding is used, the soft decoding advantage is limited
to about $1.4$ dB.



The ultimate coding gain of TPC is achieved by performing a number of
SISO iterative decoding processes that are applied
to each row and column in the codeword matrix, which requires considerable
computational power \cite{al2011closed}. Consequently, reducing the
computational complexity of TPC has received significant attention in the
literature as reported in \cite{al2011closed} and \cite{chen2007efficient}. The computational complexity constraint of TPC
becomes even more severe for systems that employ ARQ protocol because particular packets have to be retransmitted, and
hence decoded several times. 

\section{Related Work}

In the literature, Al-Dweik \textit{et al.} \cite{al2011closed}, \cite%
{al2009hybrid}, \cite{al2009non} proposed new techniques to
reduce the SISO decoders complexity by improving the BER
performance of the less complex TPC decoders, namely the HIHO decoders. Although such techniques managed to reduce the
BER gap between SISO and HIHO decoders, the SISO BER remains considerably
smaller. For example, the SISO eBCH(32, 21, 6)$^{2}$ and eBCH(64, 51, 6)$%
^{2} $ have a coding gain advantage of more than $2$ dB over the HIHO ones.
The gap becomes larger with higher code rates as in the case of the eBCH(32,
26, 4)$^{2}$ and eBCH(128, 120, 4)$^{2}$, where the coding gain difference
surges to about $4$ dB \cite{al2011closed}. Consequently, the low
complexity might not be sufficient to justify adopting HIHO decoding for
practical systems due to the high coding gain penalty.

In general, most of the work considered in the literature aimed at
minimizing the computational complexity under fixed BER constraint \cite%
{al2011closed}, \cite{dave2001efficient}. However, the BER is not necessarily
sufficient to describe the QoS for systems that
incorporate HARQ, where the throughput \cite{linadaptive2012} or delay \cite{5439329}
are more desired performance metrics. Therefore, this work considers the
throughput evaluation of TPC HARQ systems using SISO and HIHO decoding.
Extensive Monte Carlo simulation and semi-analytical results are produced
for various TPC codeword sizes and rates. Surprisingly, the obtained results
reveal that HIHO can offer throughput that is equivalent to SISO decoding
for particular codes and SNRs. The comparison is
then extended to include adaptive HARQ systems with the objective of
maximizing the system throughput.

\section{Adaptive TPC HARQ System Model}

The HARQ system considered in this chapter is similar to the one described in Chapter~\ref{chap:3}. However, we assume in this chapter that the transmission channel introduces only AWGN.
The decoding of each subpacket is performed using SISO or HIHO decoding \cite%
{pyndiah1998iterative}, \cite{al2009hybrid}, \cite{al2009non}. The
adaptation process is performed to maximize the system throughput by
selecting a particular $L$ and TPC codeword size based on the channel condition, which is
the SNR for the considered system.

\section{Numerical and Simulation Results}

The TPC HARQ system is evaluated using extensive simulations with values of $n=128$, $64$, $32$, and $16$, and using all possible values of $k$ that gives code rates larger
than $0.25$. The maximum number of transmissions allowed is $M=4$. For each simulation run $1000$ packets are transmitted. The SISO and HIHO decoders are configured to
perform a maximum of four iterations. Moreover, the number of reliability bits is set to 4 in the SISO decoder. 

\begin{figure}[H]
\centering
\includegraphics[]{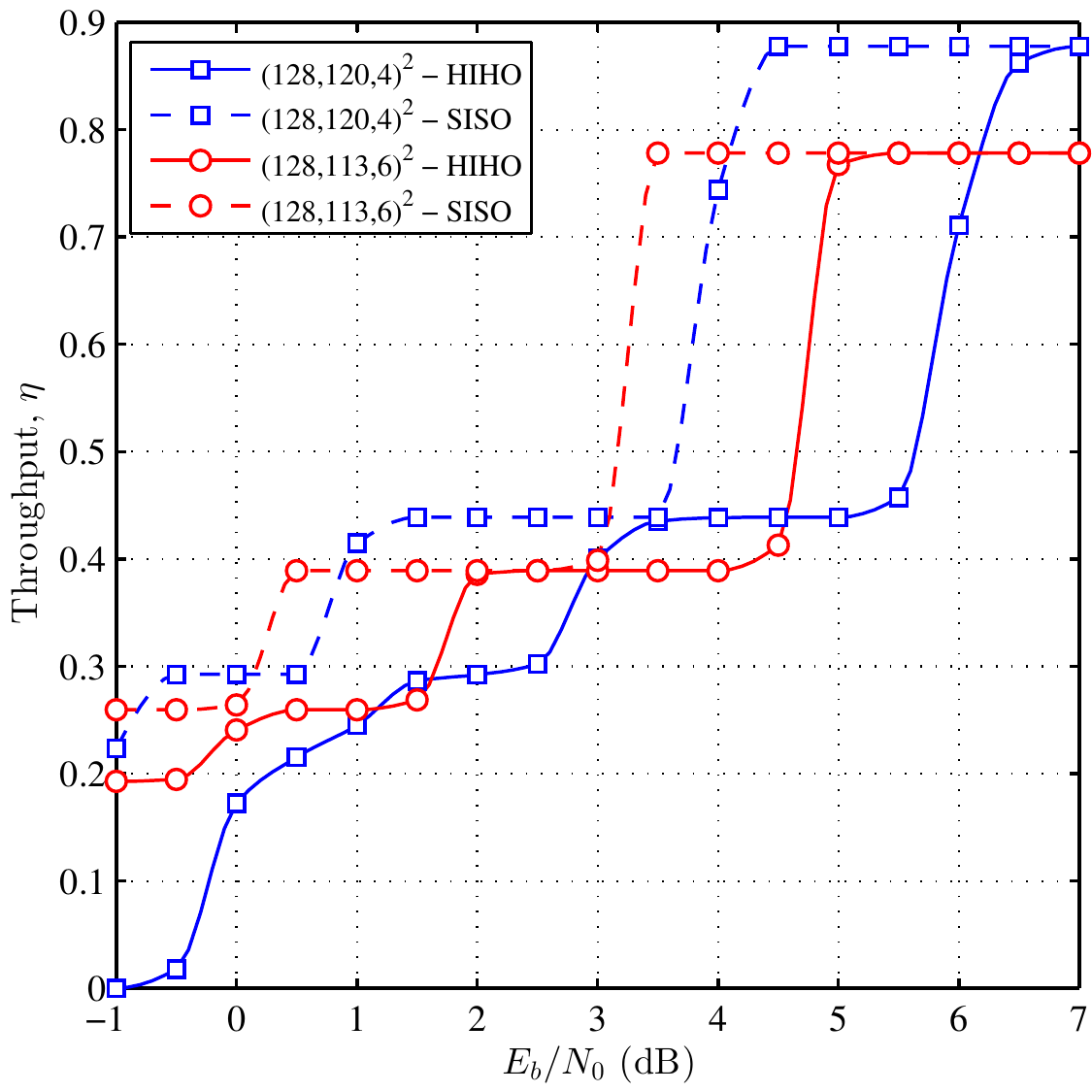}
\caption{Throughput of TPC HARQ with MRC for $(128,120,4)^{2}$ and $(128,113,6)^{2}$ in AWGN channels.}
\label{fig:SISOvsHIHO113}
\end{figure}

The throughput results for the TPC HARQ using the eBCH $(128,120,4)^{2}$ and $(128,113,6)^{2}$ are given in Fig.~\ref{fig:SISOvsHIHO113}. As it can be noted from the figure, the throughput of the SISO and HIHO decoding for the $(128,113,6)^{2}$ is approximately equal for $E_\text{b}/N_{0}\gtrsim 5$~dB, and in the range from $2$ to $3$ dB, which is remarkably different from the BER performance for these codes. The $(128,120,4)^{2}$ exhibits a similar behavior except that it is for a smaller range of $E_\text{b}/N_{0}$.

\begin{figure}[H]
\centering
\subfloat[HIHO]{\label{fig:etaAllHard}\includegraphics[width=0.5\textwidth]{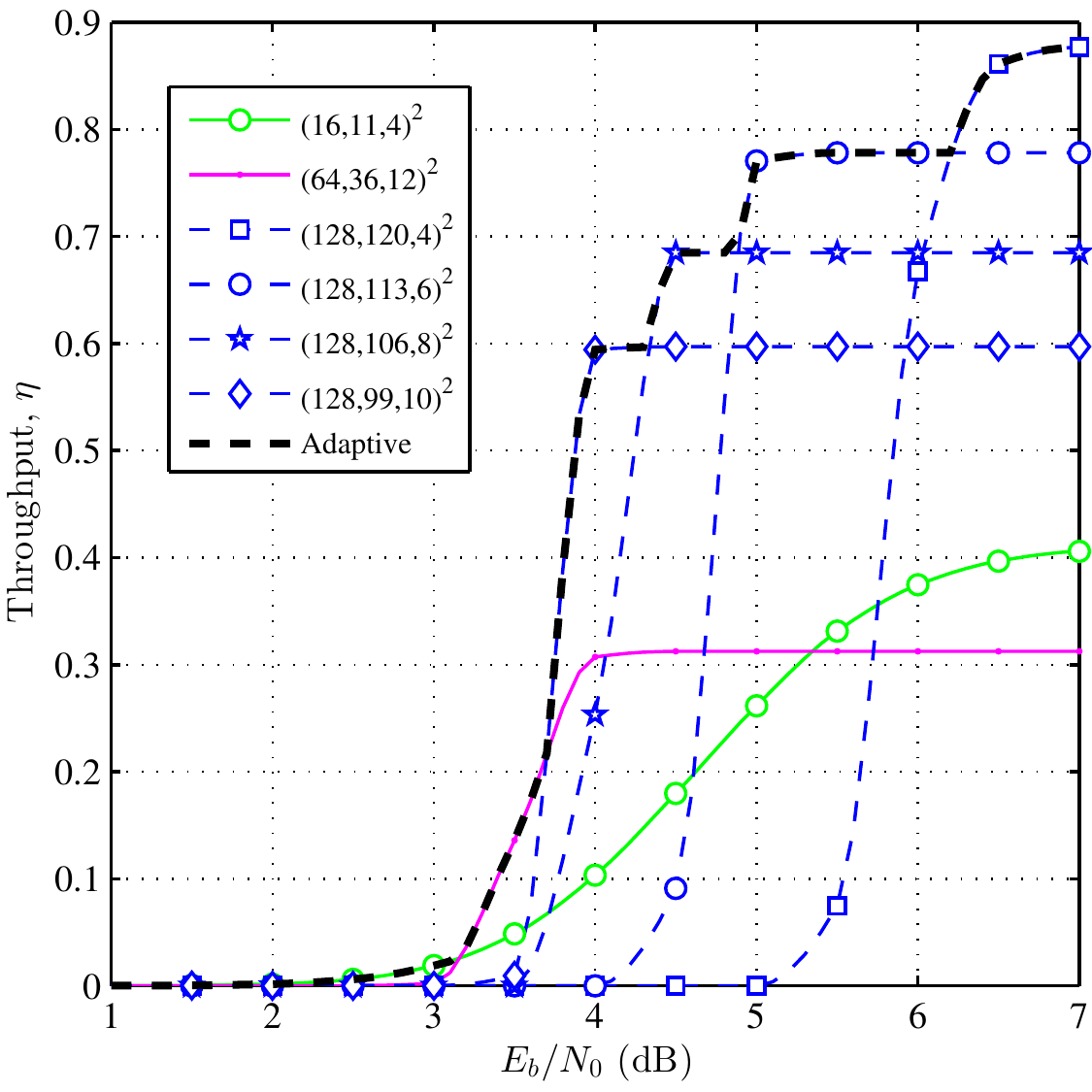}}
\subfloat[SISO]{\label{fig:etaAllSoft}\includegraphics[width=0.5\textwidth]{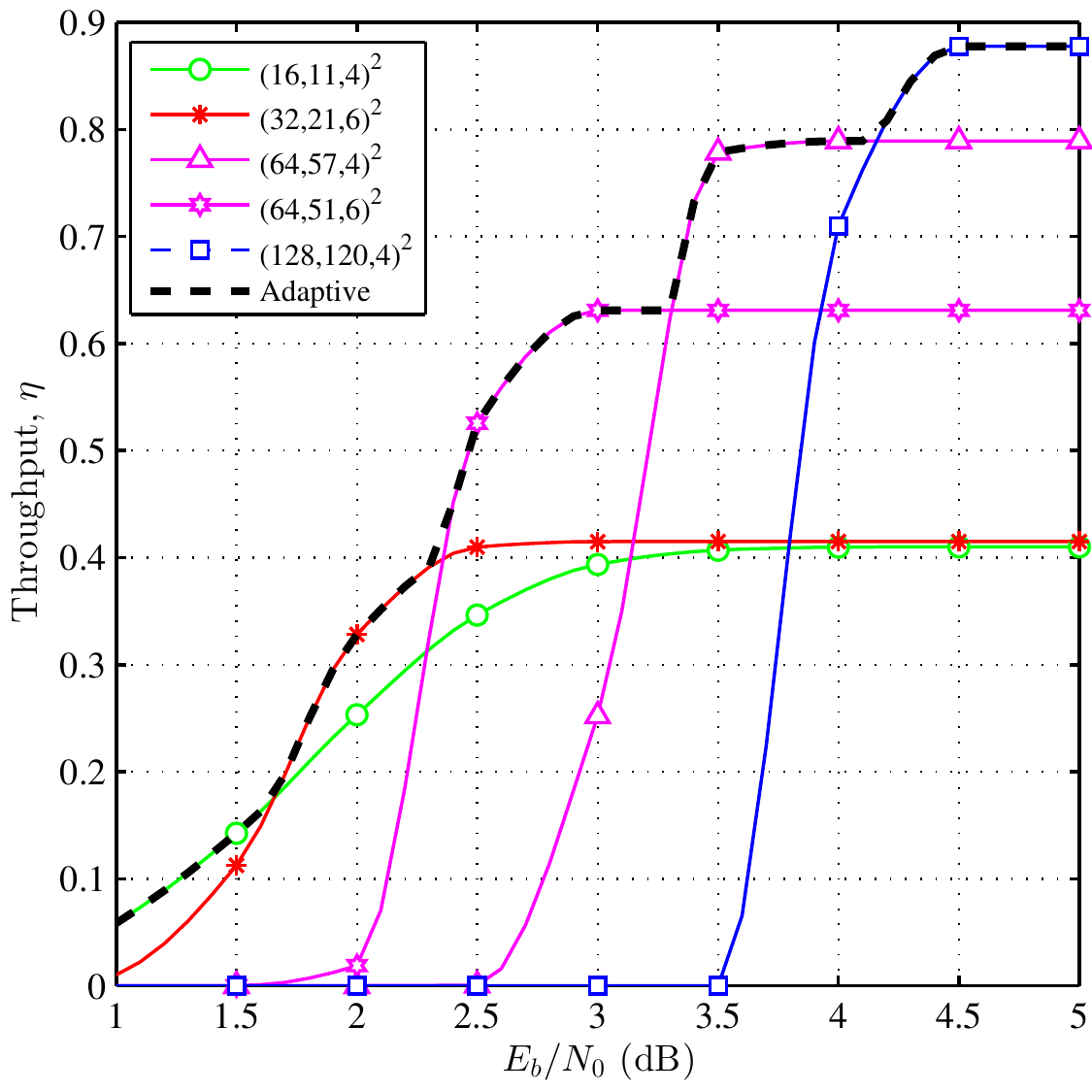}}
\caption{Throughput of TPC HARQ in AWGN channels when subpacket combining is not used.}
\label{fig:etaAll}
\end{figure}

\begin{figure}[H]
\centering
\subfloat[HIHO]{\label{fig:etaAllHardm}\includegraphics[width=0.5\textwidth]{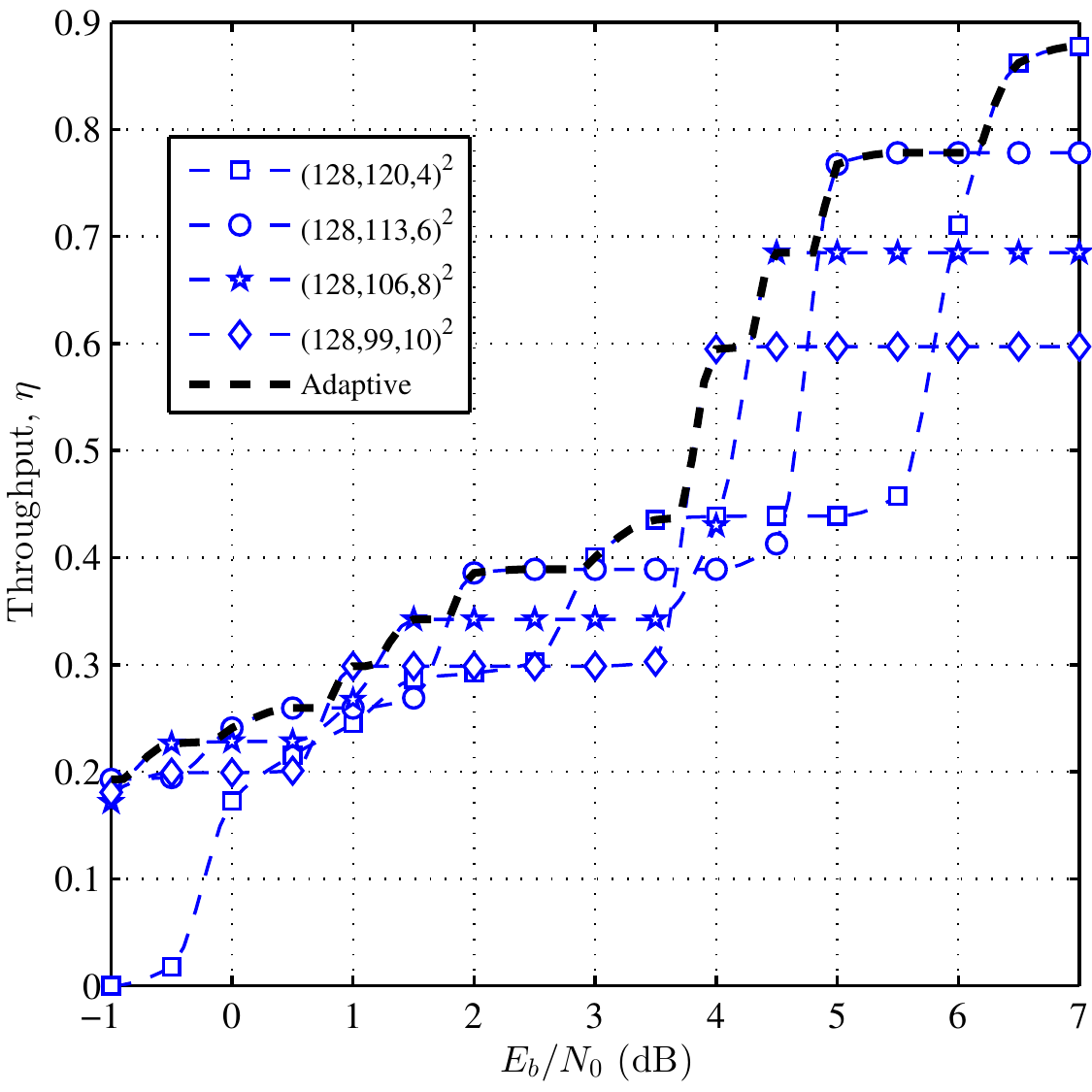}}
\subfloat[SISO]{\label{fig:etaAllSoftm}\includegraphics[width=0.5\textwidth]{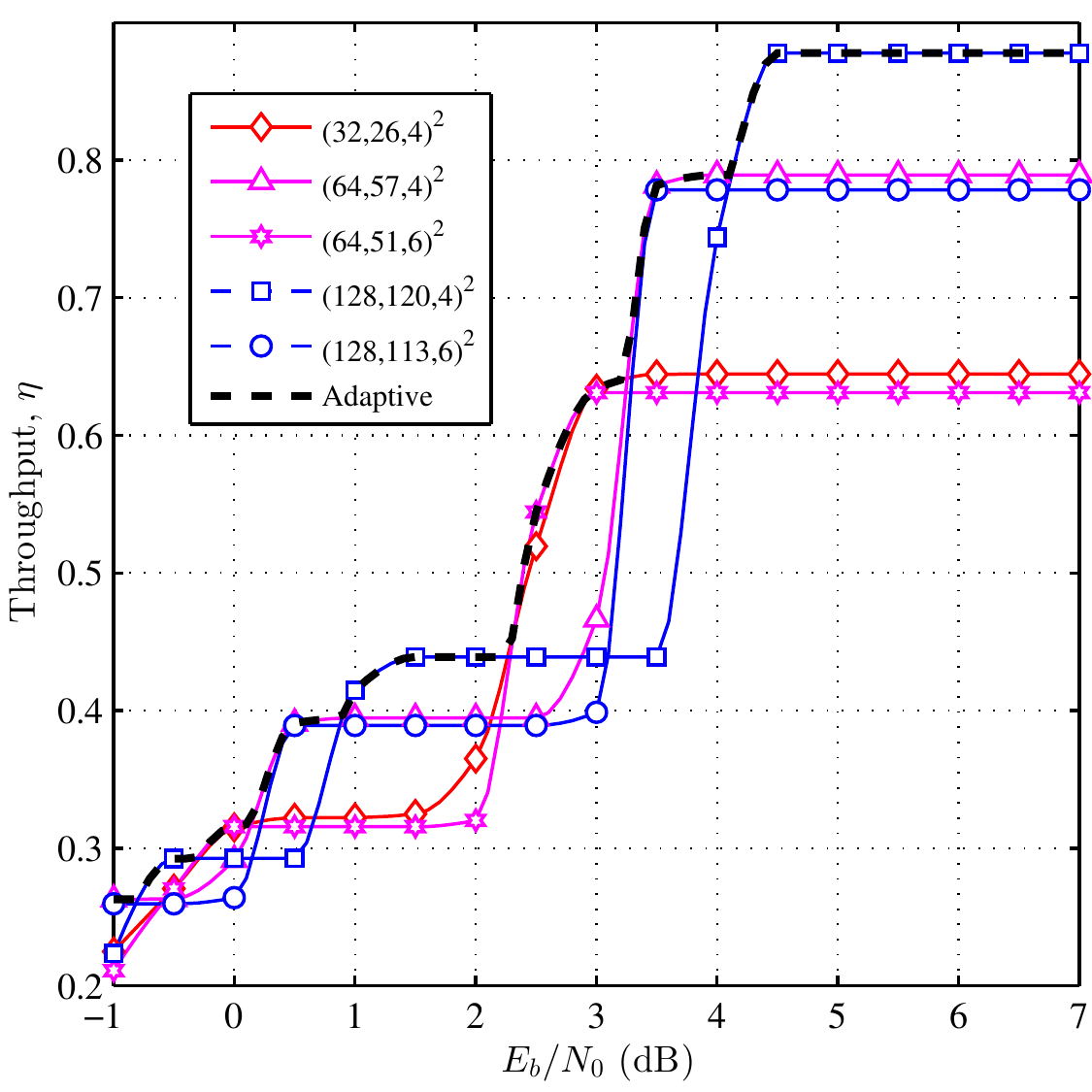}}
\caption{Throughput of TPC HARQ with MRC in AWGN channels.}
\label{fig:etaAll2}
\end{figure}

Fig.~\ref{fig:etaAll} shows the throughput of the adaptive TPC HARQ system. The codes contributing to the maximum throughput are only shown in the figure. It can be noticed from Fig.~\ref{fig:etaAll} that for a varying channel condition, the subpacket size and code rate should be adaptively changed to achieve the maximum throughput. 
The throughput of the adaptive TPC HARQ when chase combining is used is shown in Fig.~\ref{fig:etaAll2}.

\begin{figure}[H]
\centering
\includegraphics[]{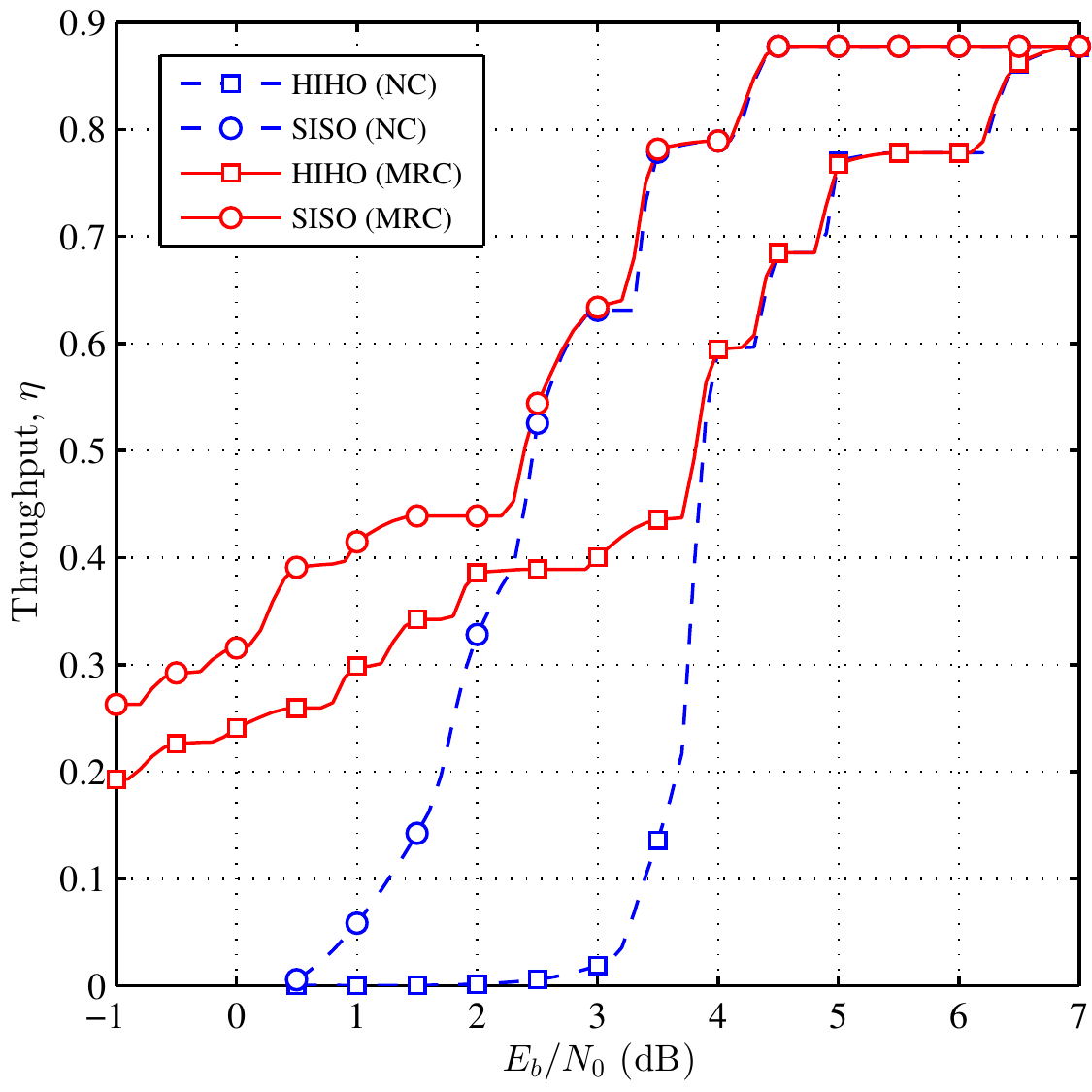}
\caption{Throughput comparison of adaptive TPC HARQ system with HIHO and SISO in AWGN channels.}
\label{fig:etaSfvsHd}
\end{figure}

Moreover, the adaptive TPC HARQ performance with HIHO decoding is compared with SISO decoding in Fig.~\ref{fig:etaSfvsHd}. MRC in the figures legend refers to maximal ratio combining and NC refers to no combining. SISO decoding achieves better performance when compared to HIHO decoding. However, the gab between SISO and HIHO performances diminishes when Chase combining (i.e. MRC) is used.  
 
\begin{figure}[H]
\centering
\includegraphics[]{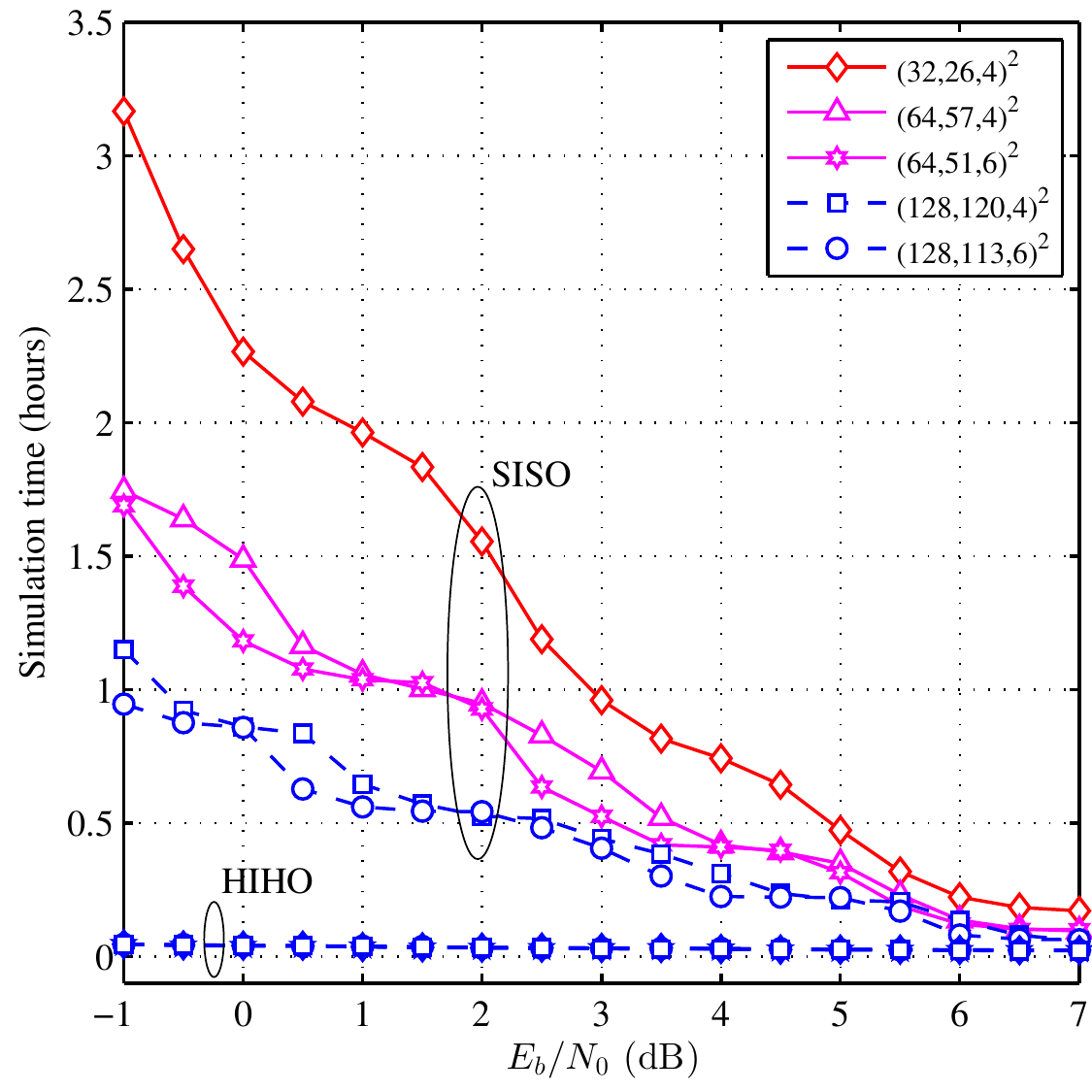}
\caption{The simulation time of SISO and HIHO TPC HARQ.}
\label{fig:SimTime}
\end{figure}

To compare the complexity of the TPC HARQ system with HIHO and SISO decoding, the simulation time is measured for both systems under identical operation conditions. The simulation time in hours is given in Fig. \ref{fig:SimTime} as a function of $E_\text{b}/N_{0}$ for the codes contributing to the maximum throughput in the adaptive TPC HARQ with MRC. The TPC decoder stops whenever the decoder converges to the correct codeword as described in \cite{al2011closed}; otherwise it completes $4$ full iterations. As it can be noted from the figure, increasing the number of subpackets $L$ for the SISO TPC HARQ increases the complexity significantly, which is due to the fact the complexity of SISO TPC decoders is dominated by the number of soft decision decoding operations performed for each row and column rather than the size of the component codeword. The figure also shows that the complexity of the HIHO-TPC is substantially smaller than the SISO for low and moderate SNRs. For high SNR the difference shrinks as the SISO decoder stops mainly after the first half iteration \cite{al2011closed}. However, the SISO still requires much longer simulation time, which ranges from $2$ to $7$ times that of the HIHO based on the code used.

In the literature, the high computational complexity of SISO decoding is justified by its high coding gain advantage over HIHO decoding as demonstrated by the BER curves in \cite{al2011closed}. However, by comparing the throughput of the adaptive HIHO and SISO systems with MRC in Fig.~\ref{fig:etaSfvsHd}, it can be observed that the average $E_\text{b}/N_{0}$ gab between the two curves is only about $1.4$ dB. Therefore, the penalty of adopting HIHO TPC HARQ is not as high as suggested by the BER results. Moreover, the processing power required by the SISO decoder is much higher than the HIHO decoder. Consequently, it can be argued that HIHO is a more practical solution to be used in an HARQ scheme to significantly reduce the complexity at the expense of slight reduction in the system throughput.


The slight reduction in throughput caused by adopting HIHO instead of SISO TPC HARQ is not significant for particular systems such as video transmission. To demonstrate this, a video transmission system is simulated using both HIHO and SISO TPC HARQ assuming a channel condition with $E_\text{b}/N_{0}=4$~dB. The error free channel speed of the transmission system is $1.88$~Mbps and therefore the QoS transmission bit rate when SISO HARQ is used is $0.8\times1.88=1.5$~Mbps. Note that $0.8$ is the transmission efficiency (i.e. throughput) of SISO TPC HARQ at $E_\text{b}/N_{0}=4$~dB as shown in Fig.~\ref{fig:etaSfvsHd}. Similarly, the QoS transmission bit rate when HIHO HARQ is used is $0.6\times1.88=1.13$~Mbps. Initially, the Football video sequence is pre-encoded as a constant bit rate (CBR)\nomenclature{CBR}{Constant Bit Rate} stream with a target bit rate of $1.4$~Mbps which slightly less than the QoS offered by SISO TPC HARQ to ensure continuous playback and to provide a a margin for variations in the video frame sizes (in bits). 

\begin{figure}[H]
\centering
\subfloat[\footnotesize{high source bit rate - SISO}]{\label{fig:buffS}\includegraphics[width=0.33\textwidth]{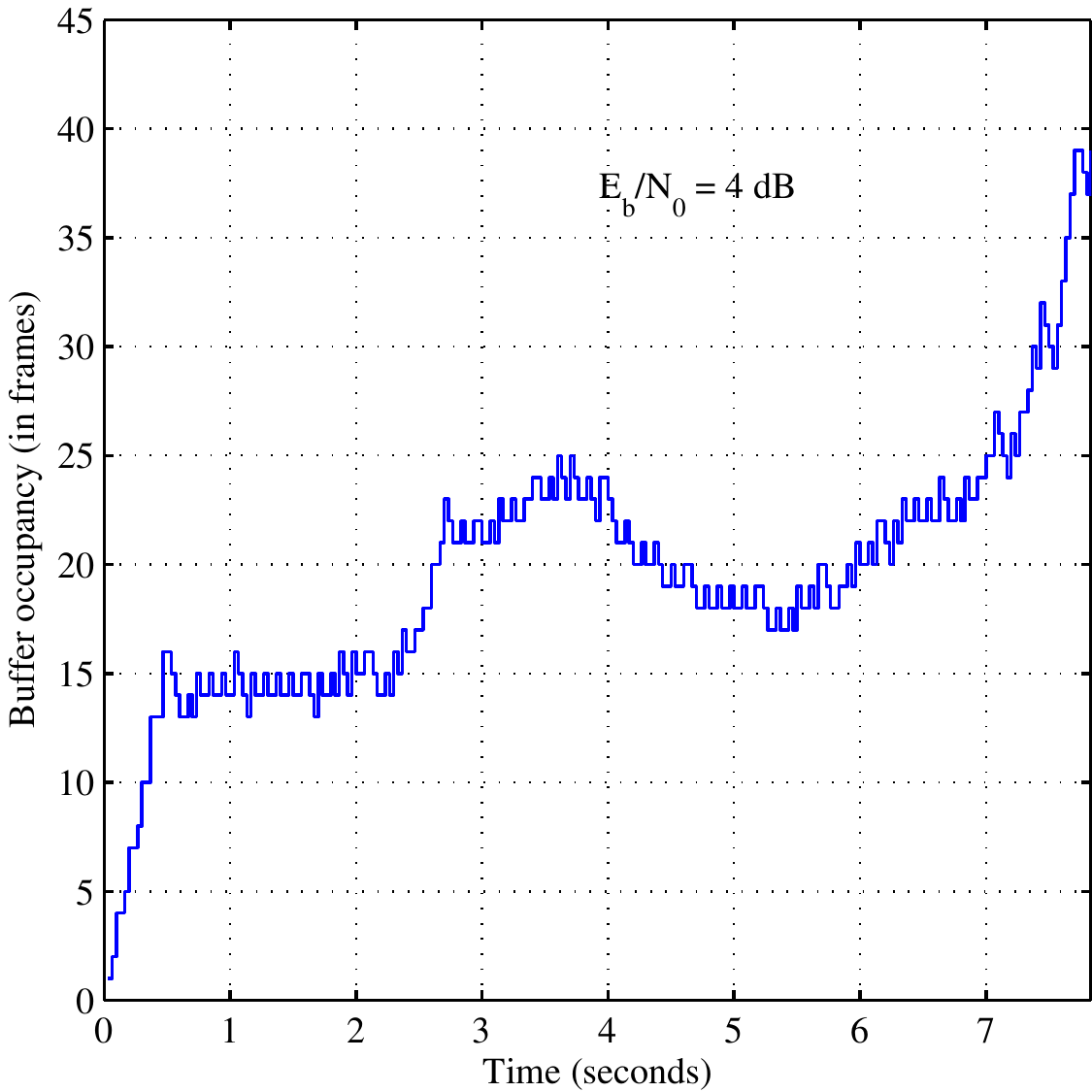}}
\subfloat[\footnotesize{high source bit rate - HIHO}]{\label{fig:buffH}\includegraphics[width=0.33\textwidth]{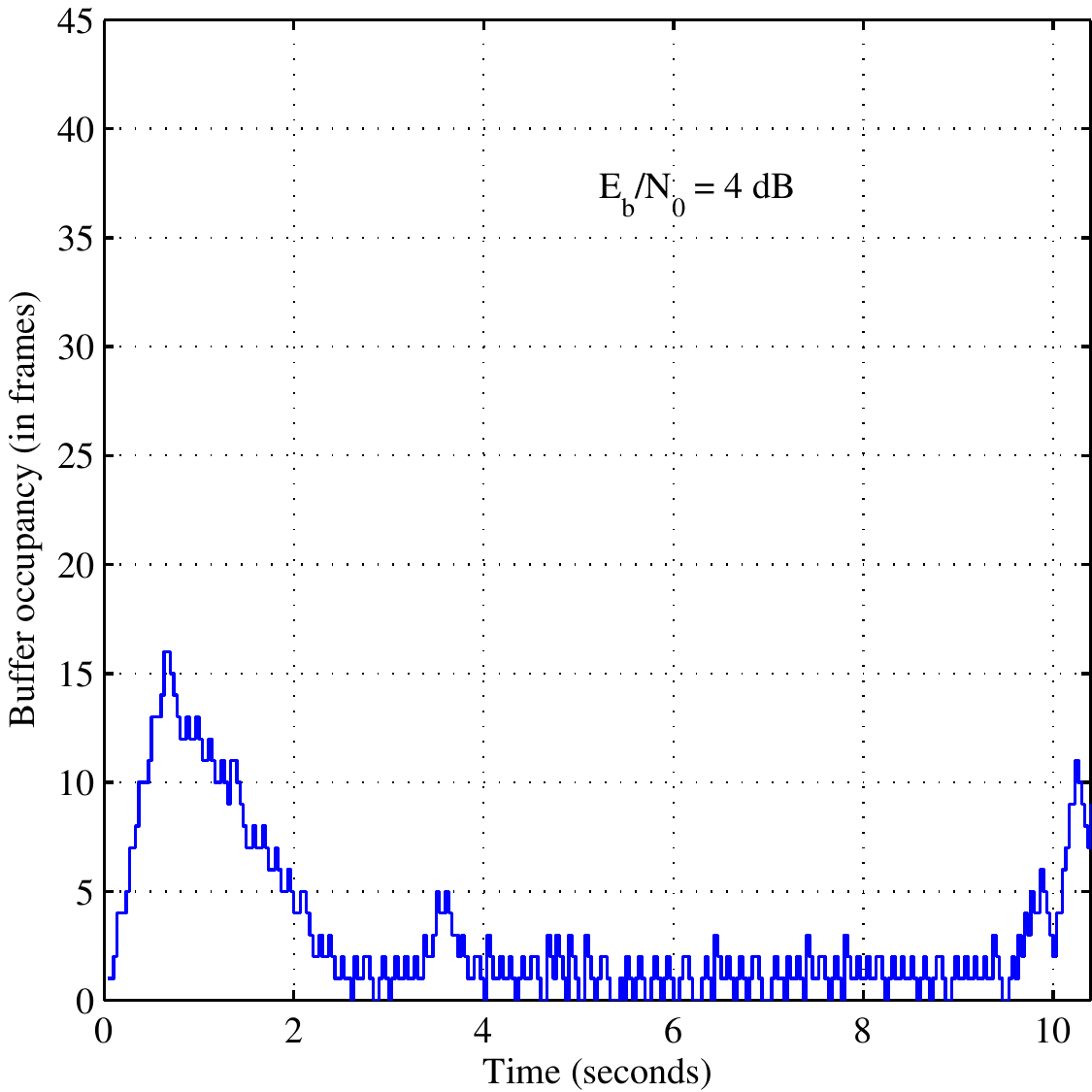}}
\subfloat[\footnotesize{lower bit rate - HIHO}]{\label{fig:buffH2}\includegraphics[width=0.33\textwidth]{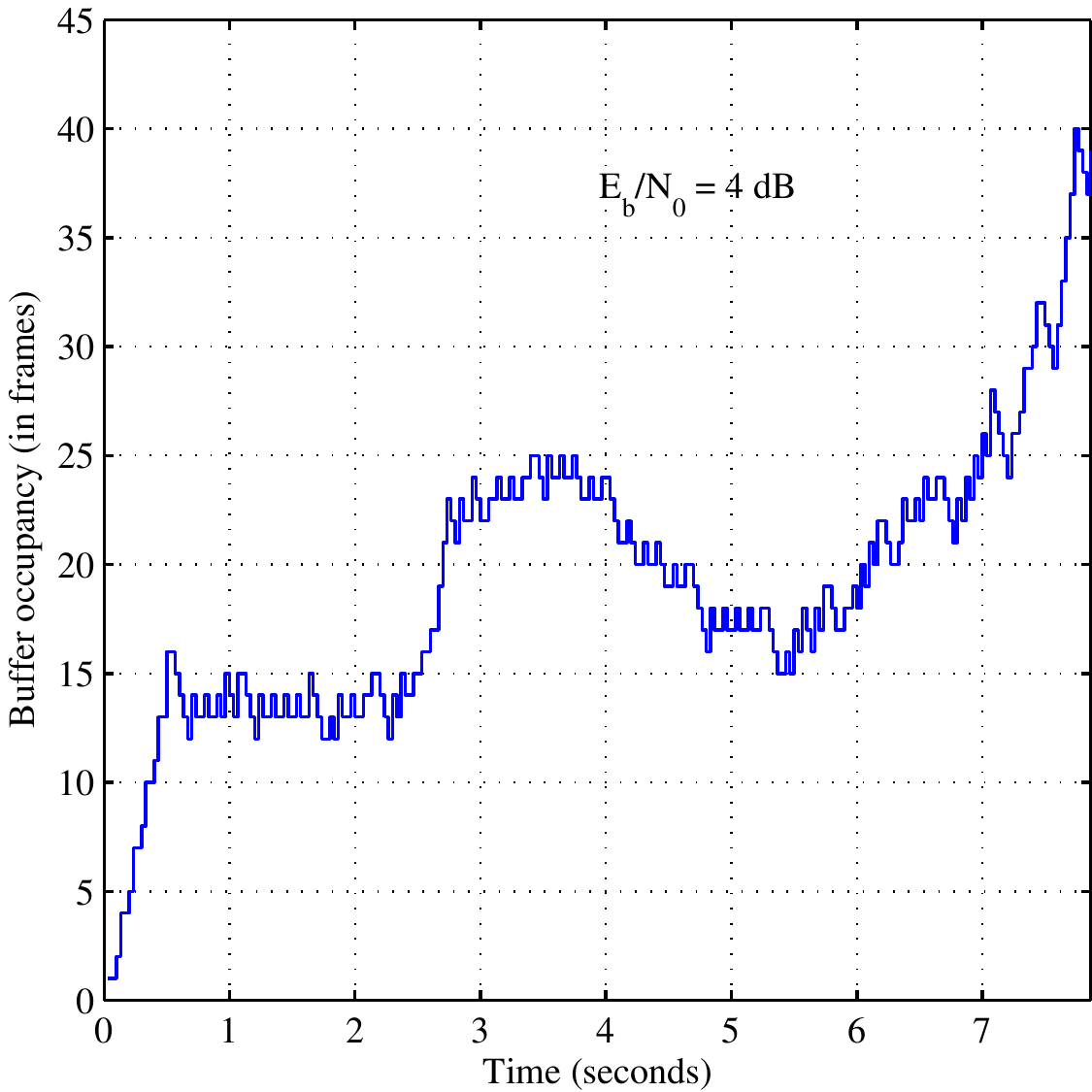}}
\caption{Video playback buffer occupancy when using adaptive TPC HARQ in AWGN channels with $E_\text{b}/N_{0}=4$~dB.}
\label{fig:buff}
\end{figure}

\begin{figure}[H]
\centering
\includegraphics[]{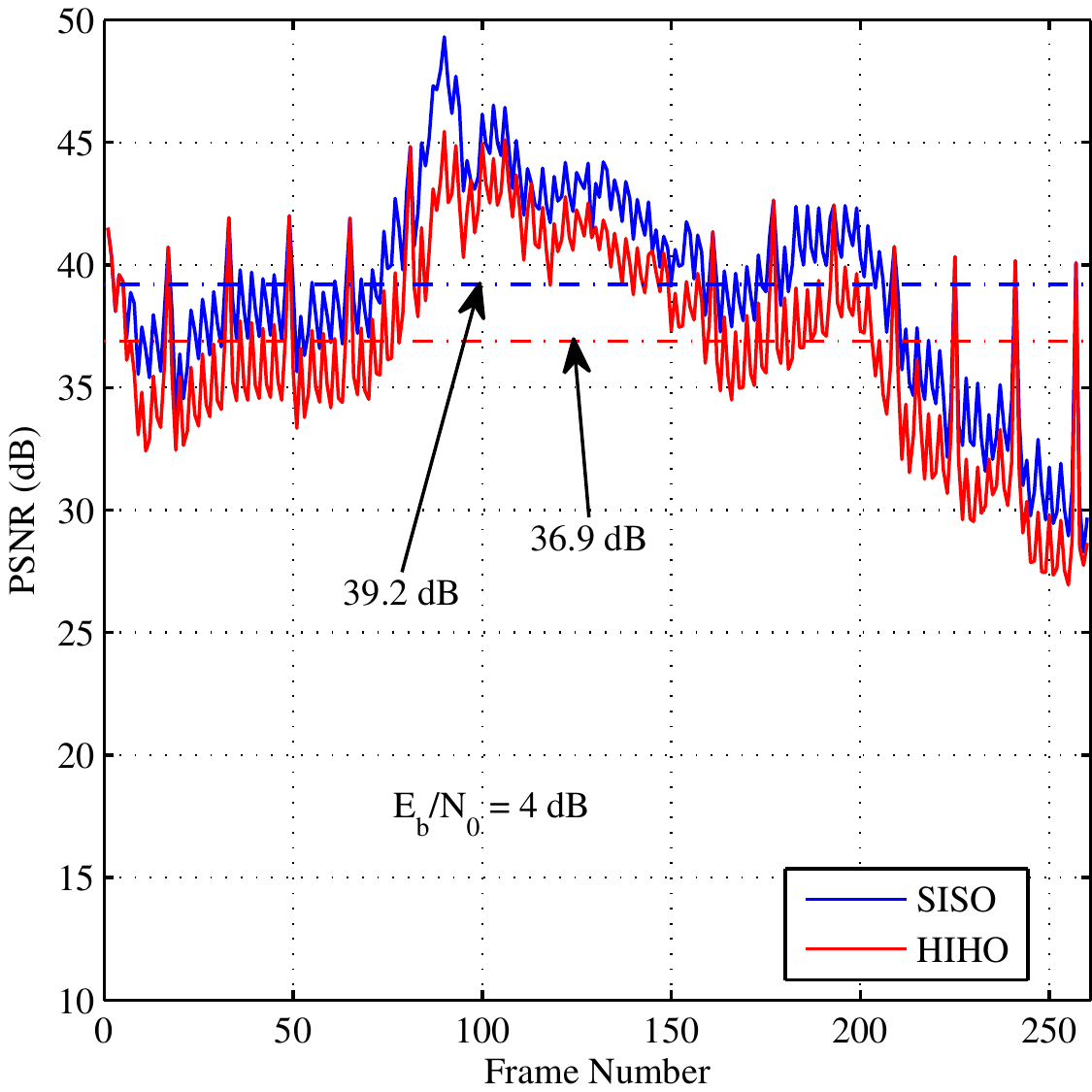}
\caption{PSNR quality of the received high source bit rate video and the reduced source bit rate video.}
\label{fig:psnrTPC}
\end{figure}

\begin{figure}[H]
\centering
\subfloat[\footnotesize{Frame 227 - SISO}]{\label{fig:227_S}\includegraphics[width=0.33\textwidth]{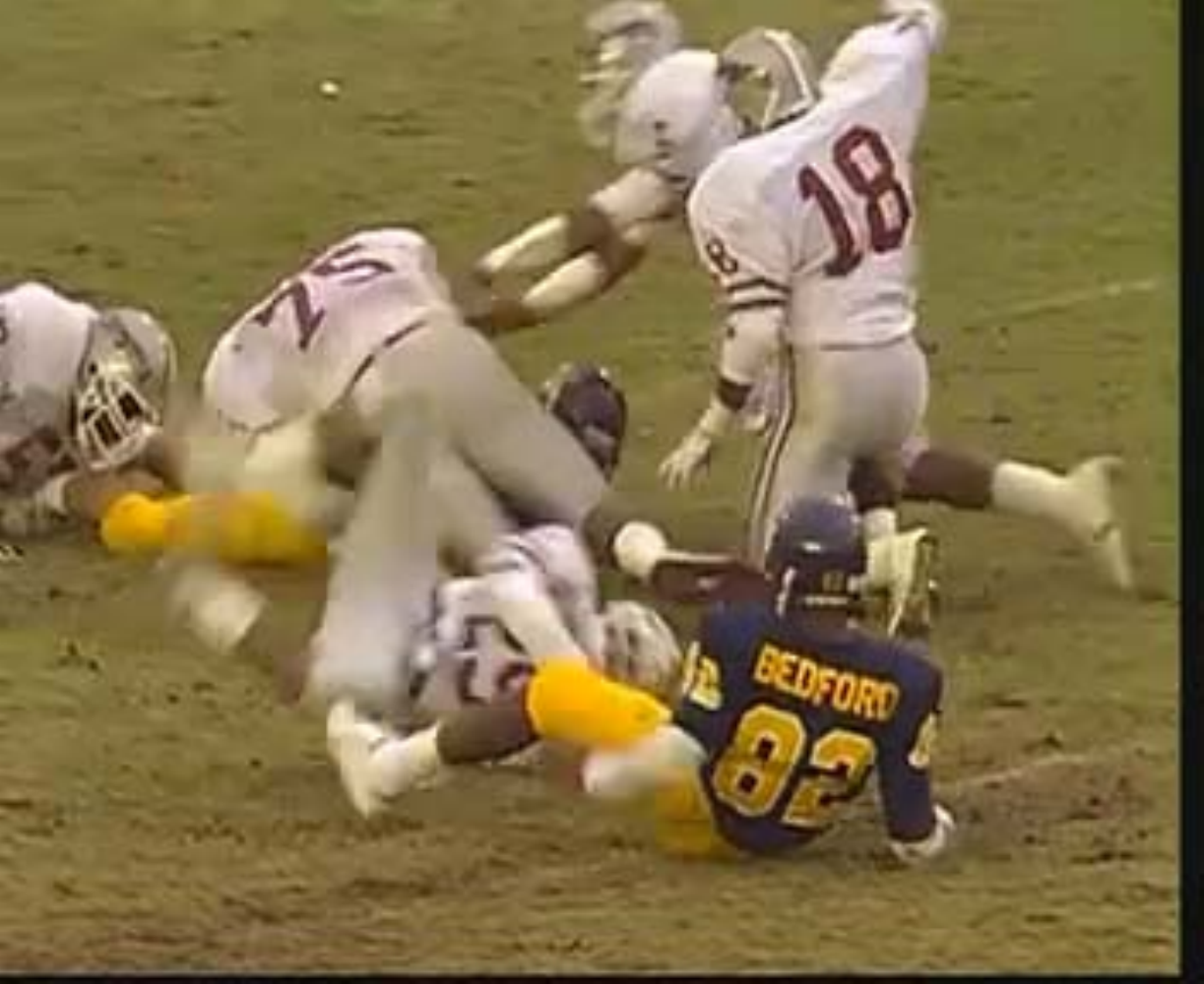}}
\subfloat[\footnotesize{Frame 230 - SISO}]{\label{fig:230_S}\includegraphics[width=0.33\textwidth]{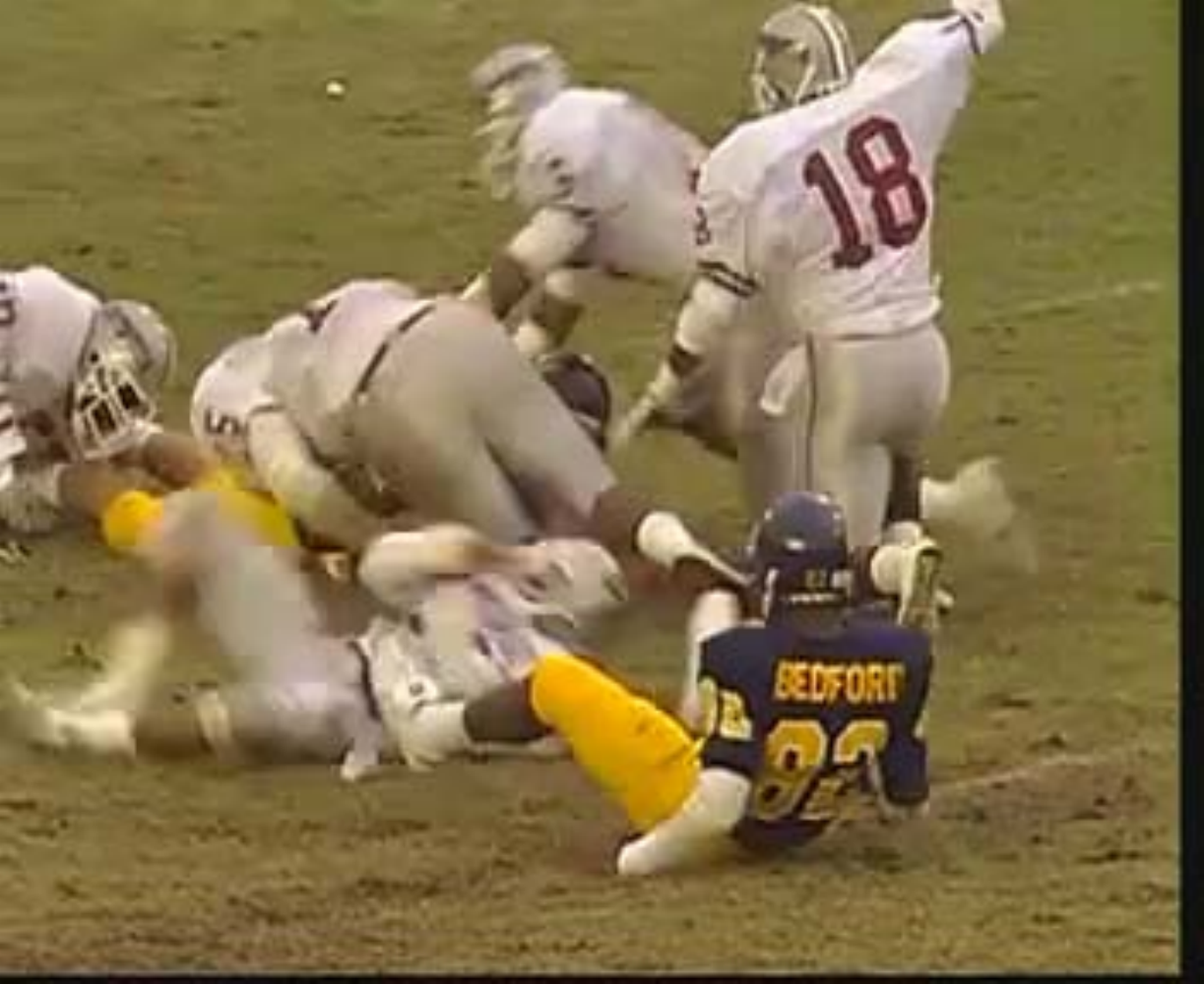}}
\subfloat[\footnotesize{Frame 260 - SISO}]{\label{fig:260_S}\includegraphics[width=0.33\textwidth]{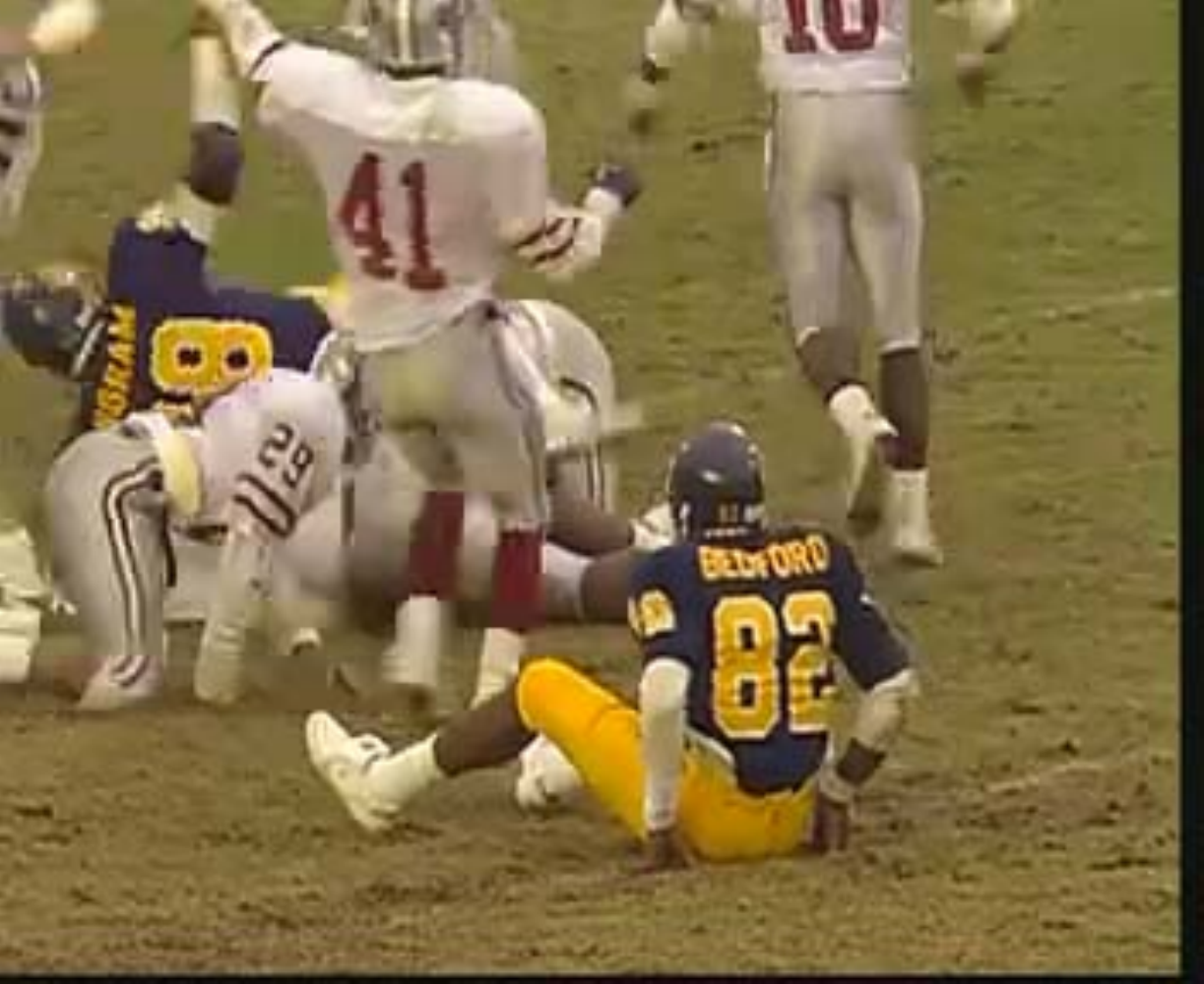}}\\
\subfloat[\footnotesize{Frame 227 - HIHO}]{\label{fig:227_H}\includegraphics[width=0.33\textwidth]{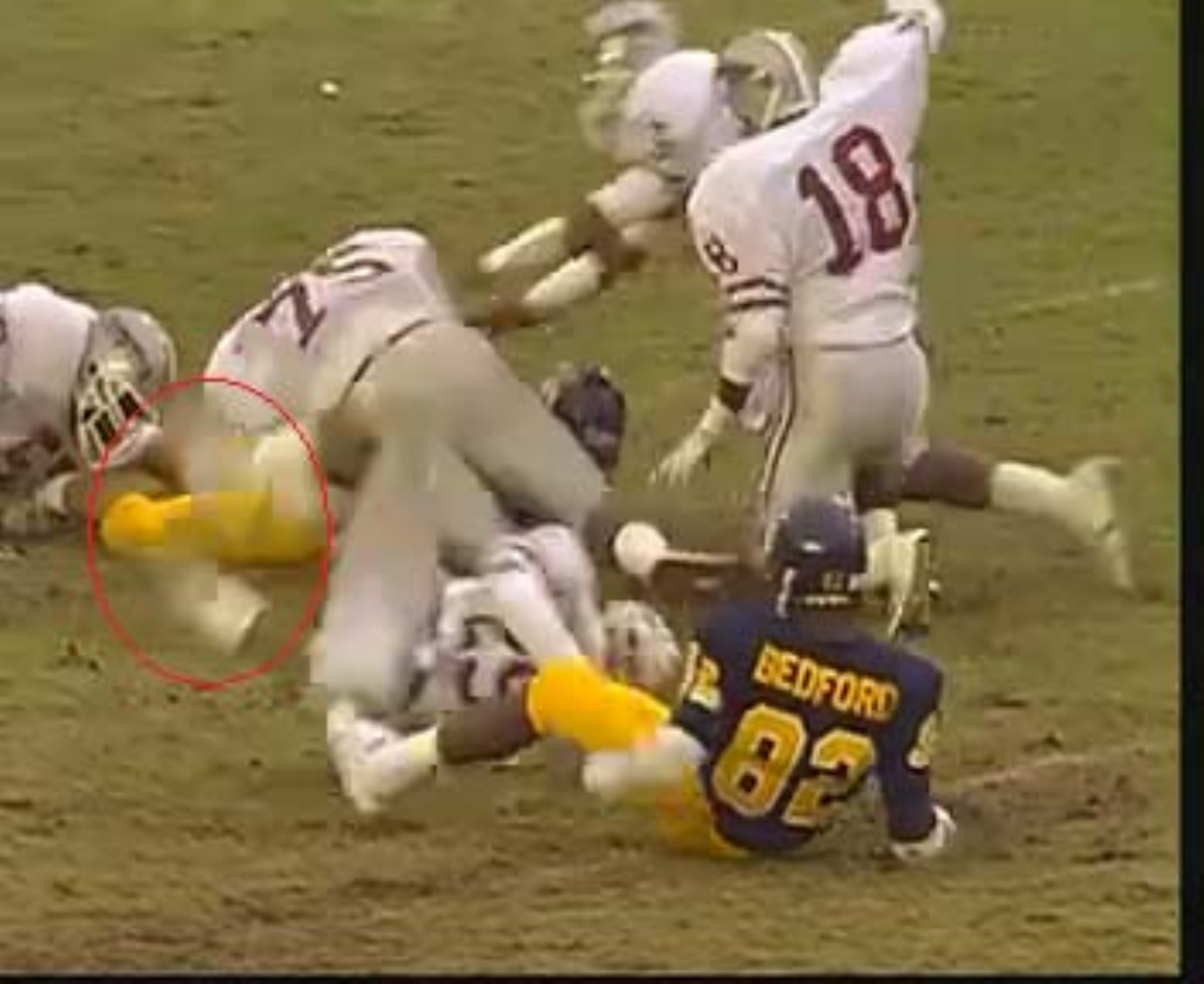}}
\subfloat[\footnotesize{Frame 230 - HIHO}]{\label{fig:230_H}\includegraphics[width=0.33\textwidth]{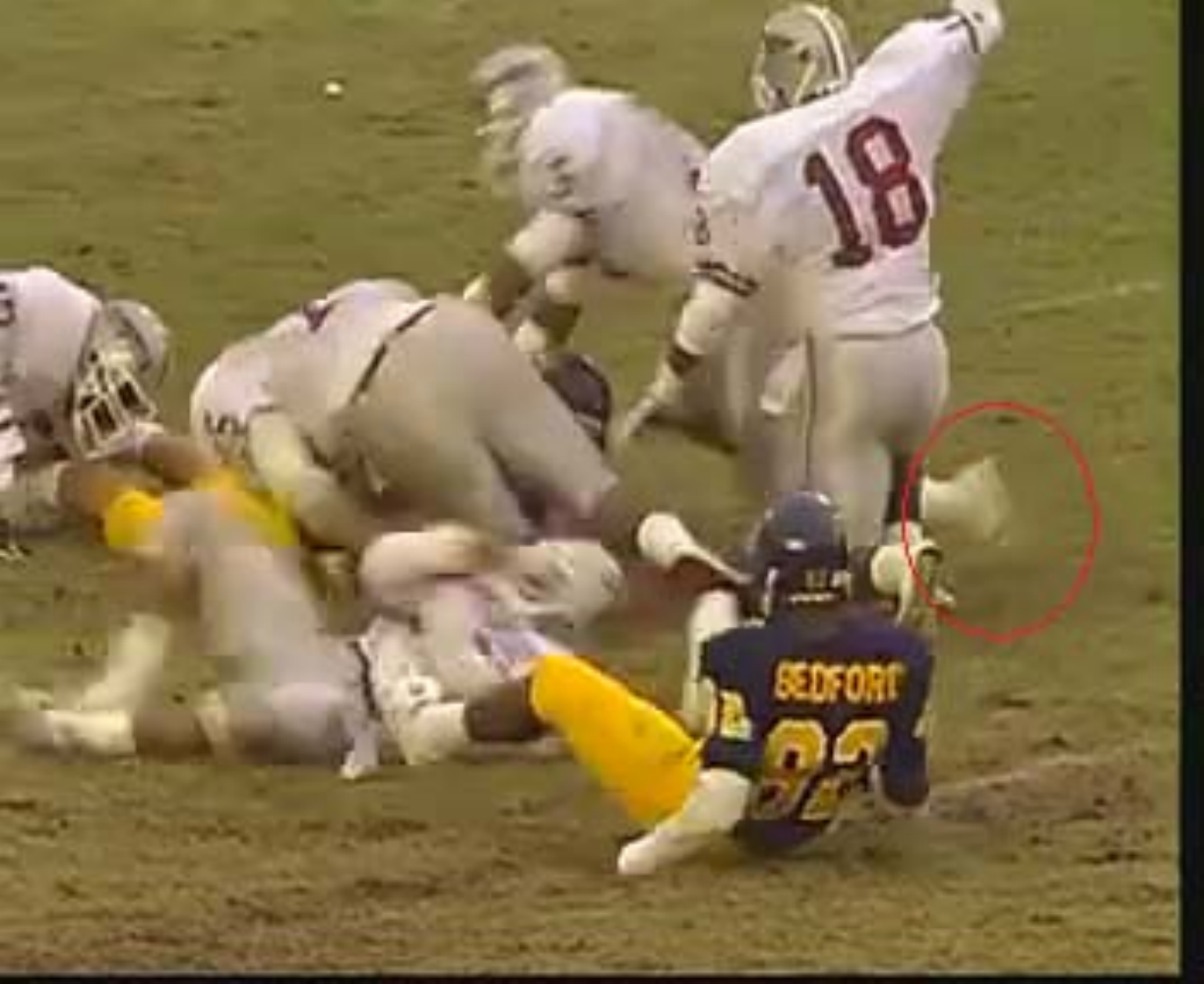}}
\subfloat[\footnotesize{Frame 260 - HIHO}]{\label{fig:260_H}\includegraphics[width=0.33\textwidth]{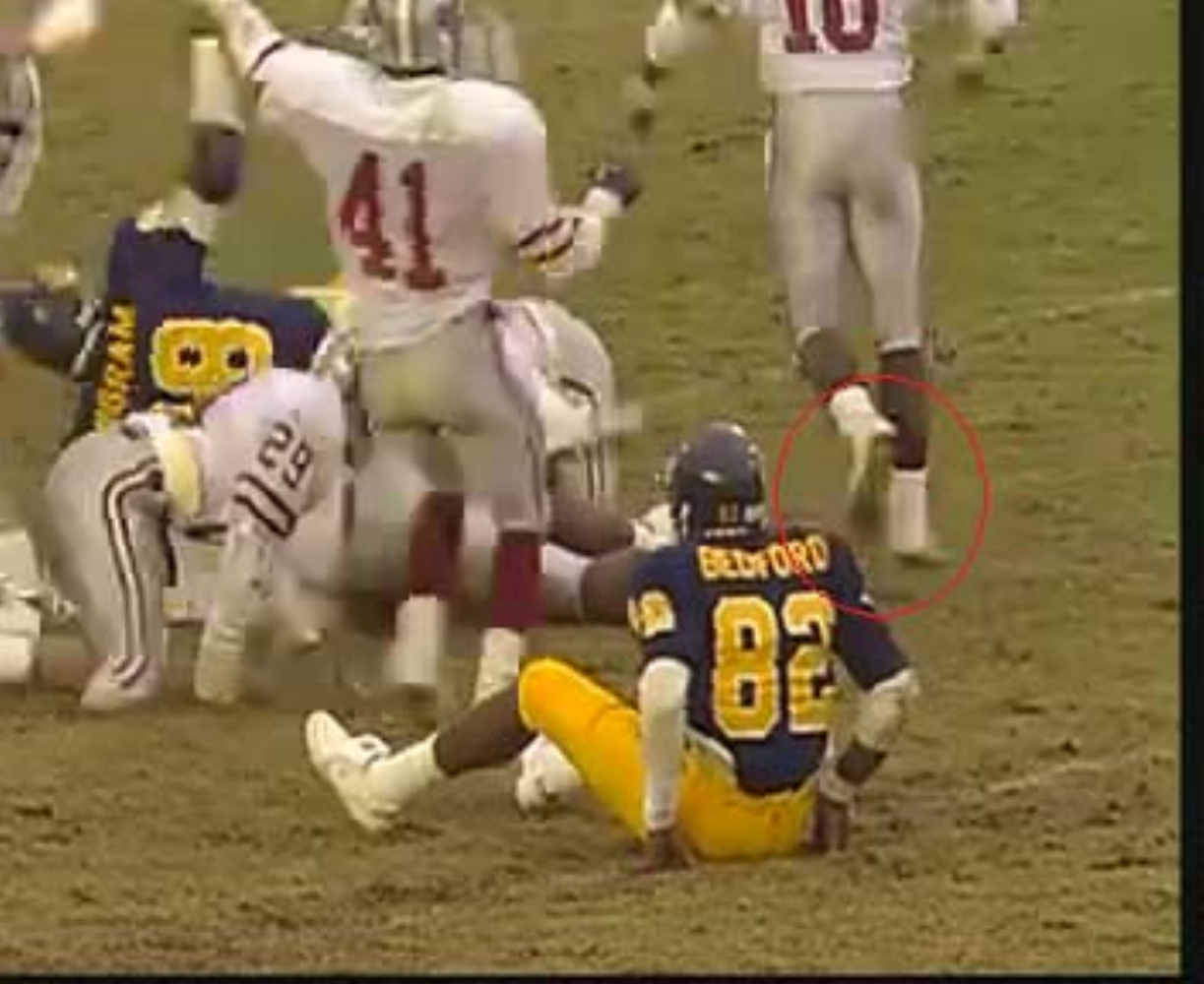}}
\caption{Frame by frame quality comparison.}
\label{fig:frmbyfrm}
\end{figure}

Fig.~\ref{fig:buffS} shows the playback buffer occupancy at the receiver when the SISO TPC HARQ is used. It can be seen that continuous playback is maintained without buffer starvation. However, when HIHO TPC HARQ is used the offered QoS ($1.13$~Mbps) is less than the bit rate of the transmitted video stream ($1.4$~Mbps). Therefore, playback buffer starvation occurs as shown in Fig.~\ref{fig:buffH}. Playback buffer starvation corresponds to interruption in the video playback. To make the video stream suitable for the QoS offered by HIHO TPC HARQ, the video sequence was re-encoded with a bit rate equal to ($1.04$~Mbps). The playback buffer starvation was completely avoided as shown in Fig.~\ref{fig:buffH2}, however, at the expense of slight reduction in the PSNR quality of the received video as shown in Fig.~\ref{fig:psnrTPC}. This small reduction in the PSNR is unnoticeable during video playback. Even when comparing the two video streams frame by frame, the slight reduction in the quality of the HIHO video is not easily observed (see Fig.~\ref{fig:frmbyfrm}).          

\section{Conclusions}
\label{sec:Conclusions4}
An adaptive subpacket HARQ scheme is proposed in this
chapter. The proposed HARQ is based on TPC with HIHO
and SISO decoding. The adaptation process is performed to
maximize the system throughput by changing the subpacket
size and code rate based on the channel SNR. Extensive
Monte Carlo simulation and semi-analytical results are used
to compare the performance of HIHO and SISO systems with
and without adaptation. The obtained results demonstrate
that the throughput performance of HIHO and SISO decoding
is drastically different as compared to BER performance
where the throughput exhibited a staircase shape. The staircase
throughput implies that power adaptation should be used as
it can reduce the power consumption while maintaining the
throughput unchanged. 
The obtained results show that the HIHO is only
about 1.4 dB less than the SISO when the throughput is used as
the performance metric for HARQ with Chase combining. It is also shown that this advantage may not be significant for particular systems such as video transmission.

\chapter{Low Complexity Power Optimization Algorithm for Multimedia Transmission over Wireless Networks}
\label{chap:5}


This chapter considers the problem of transmit power optimization for multimedia
applications in continuous high speed transmission over wireless networks. The
power optimization process is developed by noting that some performance
metrics such as throughput, delay and PSNR for
particular systems with HARQ may exhibit a
staircase behavior. In such scenarios, the corresponding metric remains fixed
for a wide range of SNRs. Consequently, the transmit
power can be reduced significantly while the relevant metric remains almost
unchanged. The obtained results reveal that invoking power optimization
algorithms can achieve a significant power saving of about $80\%$ for
particular scenarios. The system considered in this work is a truncated HARQ
with TPC and parallel concatenated convolutional codes
(PCCC)\nomenclature{PCCC}{Parallel Concatenated Convolutional Codes}. Chase combining is also used to combine the retransmitted packets with
the original transmission. The semi-analytical solution presented in Chapter~\ref{chap:3} is exploited to
obtain the system throughput in AWGN and
Rayleigh fading channels. The obtained results also show that using the
throughput, delay or PSNR as performance metrics provides equivalent power
saving results.



%

Wireless networks are becoming increasingly popular for
various communication applications with pivotal advantages such as mobility
support and larger geographical coverage at lower deployment cost when
compared to wired networks. Nevertheless, wireless channels have limited bandwidth and
they introduce losses and errors due to noise, multipath fading and
interference. Therefore, providing the desired QoS for
some applications is a challenging task in wireless communication, and QoS is
often traded-off with transmit power \cite{Shami-2011}. However, the energy of
mobile and hand-held devices is limited by a small size battery. Hence, it
should be efficiently managed and utilized.

Moreover, the enabling technologies of pervasive and ubiquitous computing have
been growing rapidly along with their energy consumption requirements.
However, the energy capabilities of the batteries in mobile devices is growing
at a much lower rate due to the constraints on size, weight, and recharging
frequency. Consequently, energy minimization subject to a desired QoS is
crucial to sustain the growth of the wireless communication sector
\cite{lan2003resource,cui2005energy}.


\section{Related Work}

Several solutions have been proposed in the literature to optimize ARQ or HARQ
for low transmit energy subject to a desired QoS. The minimum average
transmission energy for a truncated ARQ protocol is investigated in
\cite{seo2007optimal} under a reliability constraint defined by the frame drop
rate. A simple Stop-and-Wait ARQ protocol without packet combining is used
along with a parallel concatenated convolutional turbo code with rate $1/4$.
Independent block fading is assumed such that the channel varies independently
at every transmission round. In \cite{stanojev2009energy}, the total average
energy is minimized for truncated HARQ which uses soft combining and
convolutional codes. The optimal transmission energy per bit and packet
duration are determined under a fixed outage probability constraint. The
energy per bit is the same for all retransmissions and independent block
fading is also assumed. An optimal power assignment scheme is proposed for
truncated ARQ with Chase combining in \cite{su2011optimal} to minimize the
total transmission power for a target outage probability and total power
budget. However, the peak transmit power is unbounded. In addition, a slowly
fading channel is assumed where the channel does not change during
retransmissions of the same packet. In \cite{chaitanya2011outage} and
\cite{chaitanya2013optimal}, the average transmit power for truncated HARQ
with incremental redundancy and Chase combining is minimized under a specified
packet drop rate with constraints on both total power budget and peak transmit
power. The proposed scheme is evaluated for independent block fading using
numerical models but no simulation results are provided.

As it can be noted from the aforementioned paragraph, most HARQ power
optimization schemes proposed in the literature minimize the transmit energy
for a given packet drop rate, and hence they are not directly optimized to
minimize the delay or maximize the throughput. Consequently, one of the main
objectives of this work is to reduce the transmit energy under throughput
constraints. The main motivation for adopting this approach is the unique
behavior of several HARQ systems where the throughput varies in a staircase
manner as a function of the SNR
\cite{kallel1990generalized,W-Yafeng-VTC2003,qian2007improving,linadaptive2012,le2012analytical,aoun2012throughput,H-Mukhtar-2013}%
. The staircase behavior of the throughput implies that the transmit power can
be controlled eruptively to minimize the total transmit energy while
maintaining the throughput unchanged. Such approach can lead to considerable
power saving without sacrificing the system throughput performance. To the
best of our knowledge, there is no work reported in the literature that
exploited the flat regions in the throughput performance to optimize the
HARQ systems' energy.

In this chapter, we minimize the total transmit energy required to achieve a
particular throughput, or equivalently a delay, in truncated HARQ with Chase
combining. A semi-analytical model is developed to obtain the system
throughput for truncated HARQ in AWGN and Rayleigh fading channels. The
semi-analytical model is then exploited to find the minimum transmit power
required to achieve a particular throughput, delay or PSNR. The HARQ is implemented using TPC for
error correction and Stop-and-Wait for the ARQ. The TPC are considered with
hard and soft iterative decoding. Moreover, the used TPC-based HARQ implements
subpacket fragmentation \cite{zhou2006optimum} for effective adaptation to the
channel dynamics and reduced transmission delay when compared to conventional
Stop-and-Wait protocol. The HARQ is also considered using PCCC.

%
%

\section{Power Efficiency of HARQ Systems}

\label{sec:analysis5}

An important metric to assess ARQ systems is the power efficiency
$\zeta_{\mathcal{P}}$, which can be defined as the ratio of the minimum
possible transmit power to the average transmit power. However, the average
transmit power definition should consider the fact that certain subpackets can
be dropped at the receiver side. Therefore, we define $\mathcal{\bar{P}}$ as
average power required to transmit an information bit without error,
\begin{equation}
\mathcal{\bar{P}=}\frac{N\rho_{1}\mathcal{P}_{1}+N\rho_{2}%
\mathcal{P}_{2}+\cdots+N\rho_{V}\mathcal{P}_{V}}{\kappa z_{1}+\kappa z_{2}%
+\cdots+\kappa z_{V}}. \label{E-Pav-1}%
\end{equation}
where $\mathcal{P}_{i}$ is the transmit power per bit during the $i$th
transmission session. Consequently $\zeta_{\mathcal{P}}$ can be computed as
\begin{equation}
\zeta_{\mathcal{P}}=\frac{\mathcal{\bar{P}}_{\min}}{\mathcal{\bar{P}}}
\label{E-eta-P}%
\end{equation}
where $\mathcal{\bar{P}}_{\min}$ is the minimum average power required to transmit an information bit without error. Assuming equal power assignment for all transmission rounds $\mathcal{P}_{1}=\mathcal{P}%
_{2}=\cdots=\mathcal{P}$, then the
average power in (\ref{E-Pav-1}) can be written as
\begin{equation}
\mathcal{\bar{P}}=\mathcal{P}\frac{N}{\kappa}\frac{\sum_{i=1}^{V}\rho_{i}}%
{\sum_{i=1}^{V}z_{i}}.
\end{equation}
Under the same conditions and assumptions used to derive $\eta$, the average
power can be expressed as
\begin{equation}
\mathcal{\bar{P}}=\frac{\mathcal{P}}{\eta}\text{, \ \ \ \ }\frac{\mathcal{P}%
}{\zeta}\leq\mathcal{\bar{P}<\infty} \label{E-P_Av}%
\end{equation}
where $\zeta=\frac{\kappa}{N}$ is the code rate. However, since $\mathcal{\bar{P}}_{\min}$
is a constant in (\ref{E-eta-P}), then maximizing $\zeta_{\mathcal{P}}$ is
equivalent to minimizing $\mathcal{\bar{P}}$.

\nomenclature{$\mathcal{\bar{P}}$}{average transmission power to deliver an information bit}
\nomenclature{$\zeta_{\mathcal{P}}$}{power efficiency}
\nomenclature{$\mathcal{P}_{i}$}{transmit power per bit during $i$th transmission round}
\nomenclature{$\zeta$}{code rate}

For video communication systems, metrics such as PSNR and transmission delay
are commonly used to evaluate the system performance. The PSNR is used as a
measure of video spatial quality, whereas transmission delay is used to
evaluate the continuity of video playback at the receiver.

Let $\mathcal{T}$ be the time (in seconds) required to send all video frames
available at the source side. Based on (\ref{eq:taus}), $\mathcal{T}$ can be estimated by
\begin{equation}
\mathcal{T}=\sum_{i=1}^{F}\left(  \mathbb{E}\left\{  \rho\right\}  \dfrac
{L\,N}{\chi}+\mathbb{E}\left\{  \mathcal{R}\right\}  2t_\text{p}\right)
\dfrac{N_\text{F}^{(i)}}{\kappa\times L} \label{eq:Ttot}%
\end{equation}
where $N_\text{F}^{(i)}$ is the size of the $i$th video frame in (bits) and $F$ is the number of video frames to be
transmitted. Moreover, the average PSNR is defined as
\begin{equation}
\widehat{\mathbb{Q}}=\dfrac{1}{F}\sum_{i=1}^{F}\widehat{Q}^{(i)}
\label{eq:psnrAvg}%
\end{equation}
where $\widehat{Q}^{(i)}$ is the PSNR of the $i$th frame in the received video.

\nomenclature{$\mathcal{T}$}{transmission time of video sequence}
\nomenclature{$N_\text{F}^{(i)}$}{size of $i$th video frame}
\nomenclature{$\widehat{\mathbb{Q}}$}{average PSNR of received video}
\nomenclature{$\widehat{Q}^{(i)}$}{PSNR of $i$th frame in received video}
\nomenclature{$F$}{number of frames in video sequence}

To visualize the performance of the HARQ system as a video transmission
system, the popular Football video sequence with Common Interchange Format
(CIF)\nomenclature{CIF}{Common Interchange Format} is considered. The H.264 JM reference software \cite{website:JM} is used
to encode $F=260$ frames with constant bit rate. Moreover, the video
sequence is encoded with $\{\text{IBBPBBP}\cdots\}$ GoP
structure with a length of 16 frames. The target encoding bit rate is
initially set at $\omega=\left(  120^{2}/128^{2}\right)  \chi$ where $\chi$ is
assumed to be $2$ Mbps.

\begin{figure}[H]%
\centering
\includegraphics[
]%
{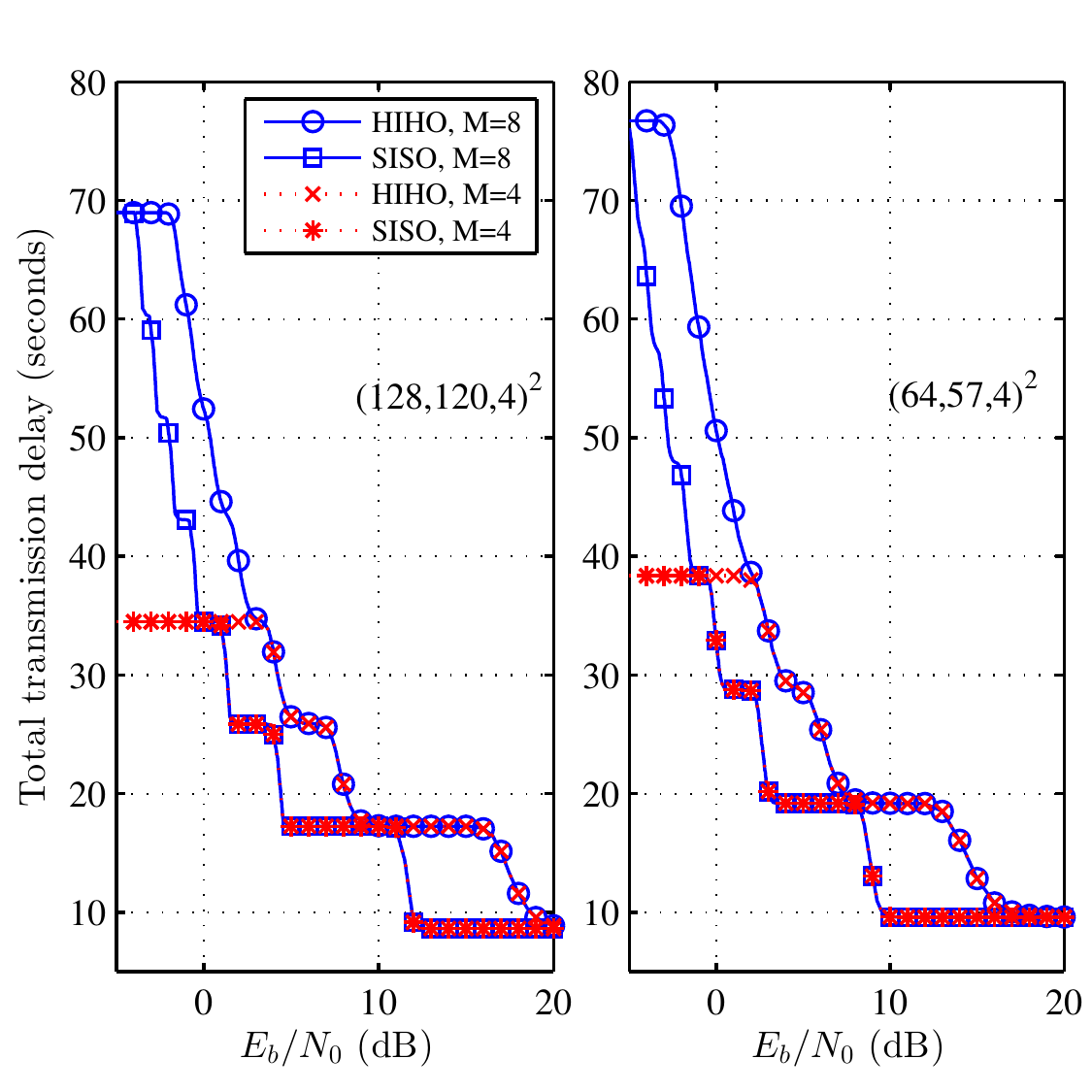}%
\caption{{Total transmission delay over Rayleigh fading channels for the Football sequence assuming a fixed
encoding bit rate.}}%
\label{fig:Ttot}%
\end{figure}

Fig.~\ref{fig:Ttot} shows the total transmission delay when
eBCH $(128,120,4)^{2}$ and eBCH $(64,57,4)^{2}$ are used with hard and soft
decoding using two truncation factor values of $M=4$ and $8$. Rayleigh fading and propagation
time $t_\text{p}=10~\mu\text{s}$ are assumed. As it can be noted from the
figure, the total transmission delay increases as the channel SNR decreases;
however, similar to the throughput, the total transmission delay exhibits a
staircase behavior where $\mathcal{T}$ stays unchanged for a range of SNR values. For
the $M=4$ case, the delay reaches the maximum when at low $E_\text{b}/N_{0}$ where
the number of transmissions per packet is equal to $M$. In such scenarios,
most of the subpackets are dropped because the maximum number of transmissions
is reached while these subpackets still have errors. For the $M=8$ case, the
transmission delay values are equivalent to the $M=4$ case at high and
moderate $E_\text{b}/N_{0}$ values. However, the performance of the system changes
drastically at low $E_\text{b}/N_{0}$ values where the value of $\mathcal{T}$ increases
sharply and the staircase phenomenon almost vanishes.

\begin{figure}[H]%
\centering
\includegraphics[
]%
{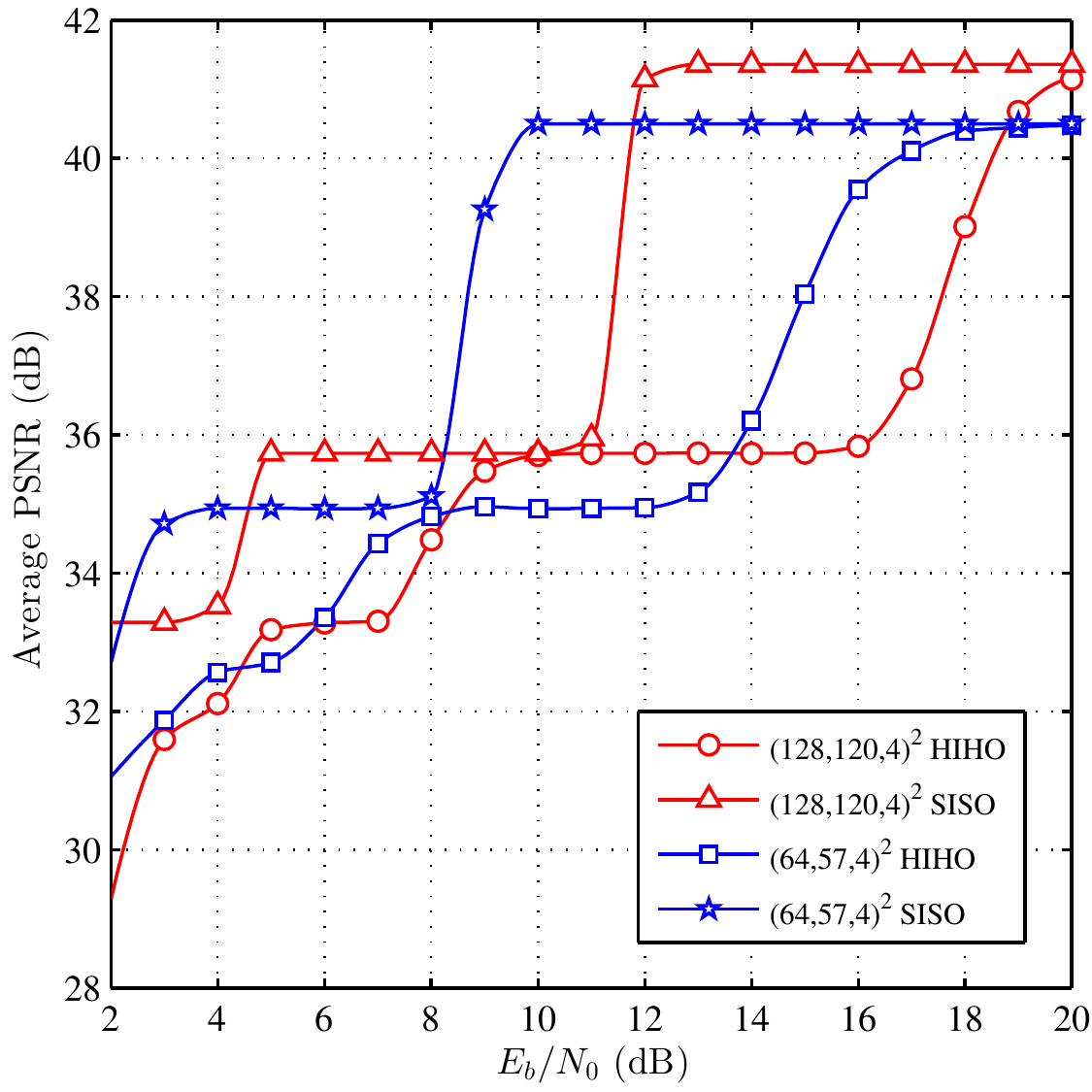}%
\caption{Average PSNR of the received video over Rayleigh fading channels when the source encoding bit rate
is adjusted to match the offered system throughput.}%
\label{fig:psnrAvg}%
\end{figure}

For particular video communication systems
\cite{mukhtar2011occupancy,mukhtar2010multi}, the source encoding bit rate is
adjusted to match the offered system throughput in an attempt to ensure timely
delivery of video frames at the receiver. Adjusting the source encoding bit
rate in accordance to the offered system throughput affects the average PSNR.
Fig.~\ref{fig:psnrAvg} shows the average PSNR of the received video when
eBCH$(128,120,4)^{2}$ and eBCH$(64,57,4)^{2}$ are used with hard and soft
decoding. The target bit rate in the video encoder is set to $\omega=\eta\chi$
to match the offered system throughput. Similar to the throughput $\eta$, the
average PSNR decreases as the channel SNR decreases, but stays unchanged for
certain ranges of SNR.

\nomenclature{$\omega$}{target bit rate in video encoder}


\section{Power Optimization Problem Formulation}

\label{sec:algorithm5} Although it cannot be concluded directly from
(\ref{E-eta-1}), the graphical representation of $\eta$ given in
Fig.~\ref{fig:thrAWGN1} and Fig.~\ref{fig:thrFading1} implies that $\eta$ is
not strictly increasing/decreasing as a function of $E_\text{b}/N_{0}$, or
equivalently the transmit power $\mathcal{P}$. Therefore, the instantaneous
transmit power $\mathcal{P}$ in (\ref{E-Pav-1}) can be changed in particular
scenarios to minimize the average transmit power $\mathcal{\bar{P}}$ while
maximizing $\eta$. For example, assuming that the results in
Fig.~\ref{fig:thrAWGN1} and Fig.~\ref{fig:thrFading1} are generated using the
maximum possible transmit power ($\mathcal{P}_{\max}$) in each ARQ round, the
throughput achieved by the HARQ system will be maximum as well, which is
denoted as $\eta(\mathcal{P}_{\max})$. However, as depicted by the throughput
results in the figures, there is no single unique value of $\mathcal{P}$ that
corresponds to $\eta(\mathcal{P}_{\max})$. Instead, $\eta(\mathcal{P}_{\max})$
can be achieved for a wide range of $\mathcal{P\leq P}_{\max}$. Therefore,
based on the SNR, the value of $\mathcal{P}$ can be reduced substantially
without degrading the throughput. Based on this critical observation, we can
apply a power optimization algorithm to minimize the average transmission power
per information bit $\mathcal{\bar{P}}$ by using the minimum possible
transmission power in each ARQ round ($\mathcal{P}^{\ast}$) while maximizing
the throughput. Therefore, the optimum power can be computed as
\begin{equation}
\mathcal{P}^{\ast}=\left\{
\begin{array}
[c]{c}%
\arg\min\limits_{\mathcal{P}}{\ }\mathcal{\bar{P}}\\
\text{subject to: }\eta(\mathcal{P}^{\ast})\geq\mu\eta(\mathcal{P}_{\max
})\text{, }\mathcal{P}^{\ast}\leq\mathcal{P}_{\max}%
\end{array}
\right.  \label{E-opt-1}%
\end{equation}
where $\mu$ is a design parameter that allows the designer to trade-off power
with throughput. It is also required to avoid converging to a local minimum as
it will be explained later in this section.

\nomenclature{$\mathcal{P}_{\max}$}{maximum transmit power}
\nomenclature{$\mathcal{P}^{\ast}$}{optimal transmit power}

The optimal solution for (\ref{E-opt-1}) can be obtained by computing the
vector $\mathbb{P}\mathcal{=[{P}}_{1}$, $\mathcal{{P}}_{2}$, $...\mathcal{]}$
that satisfies $\frac{\partial}{\partial\mathcal{P}}\mathcal{\bar{P}=}0$, and
then select%
\begin{equation}
\mathcal{P}^{\ast}=\min\left(  \mathbb{P}\right)  |_{\left\{  \eta
(\mathcal{P}^{\ast})\geq\mu\eta(\mathcal{P}_{\max})\text{, }\mathcal{P}^{\ast
}\leq\mathcal{P}_{\max}\right\}  }.
\end{equation}
However, the subpacket probability $P_\text{E}$ that is required to compute
$\mathcal{\bar{P}}$ does not have a closed-form representation to allow for
the optimum solution. Moreover, by noting the relation between $\eta$ and
$\mathcal{P}$ or $E_\text{b}/N_{0}$, it is straightforward to conclude that
minimizing $\mathcal{\bar{P}}$ for fixed $\eta$ is equivalent to selecting the
minimum $\mathcal{P}$ or $E_\text{b}/N_{0}$ in the flat regions in the throughput
curves. Consequently, computing $\mathcal{P}^{\ast}$ can be performed by
computing the derivative of (\ref{E-eta-1}) numerically, and then use a
particular search algorithm to find $\mathcal{P}^{\ast}$. Consequently, the
solution becomes suboptimal and the difference from the optimal value depends
on the search step size.

\begin{figure}[H]%
\centering
\includegraphics[
]%
{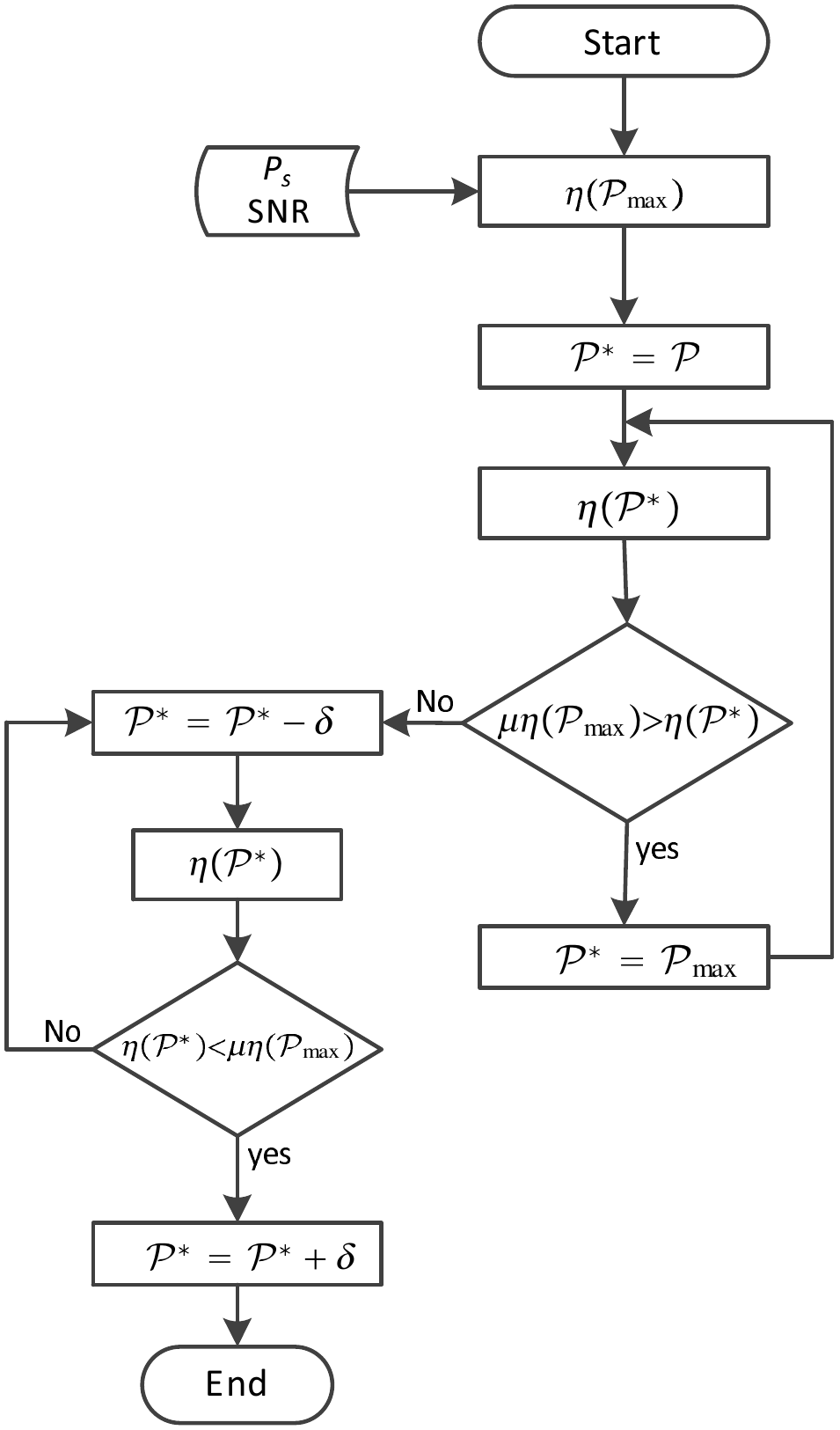}%
\caption{Power optimization algorithm flowchart.}%
\label{flowchart}%
\end{figure}

To simplify the presentation of the proposed system, we assume initially that
brute-force search is the method used to find $\mathcal{P}^{\ast}$ as described
in the optimization algorithm given in Table~\ref{T-Opt}. The algorithm is
also shown in a flowchart diagram in Fig.~\ref{flowchart}.

\begin{table}[tbp]%
\renewcommand{\arraystretch}{0.6}
\caption{Power optimization algorithm.}%
\begin{tabular}
[c]{lllll}%
1. & compute $\eta(\mathcal{P}_{\max})$ &  &  & \\
2. & set $\mathcal{P}^{\ast}=\mathcal{P}$ &  &  & \\
3. & compute $\eta(\mathcal{P}^{\ast})$ &  &  & \\
4. & if $\mu\eta(\mathcal{P}_{\max})>\eta(\mathcal{P}^{\ast})$ &  &  & \\
5. & \ \ \ \ set $\mathcal{P}^{\ast}=\mathcal{P}_{\max}$ &  &  & \\
6. & \ \ \ \ got to 3 &  &  & \\
7. & else &  &  & \\
8. & set $\mathcal{P}^{\ast}=\mathcal{P}^{\ast}-\delta$ &  &  & \\
9. & compute $\eta(\mathcal{P}^{\ast})$ &  &  & \\
10. & if $\eta(\mathcal{P}^{\ast})<\mu\eta(\mathcal{P}_{\max})$ &  &  & \\
11. & $\ \ \ \ \mathcal{P}^{\ast}=\mathcal{P}^{\ast}+\delta$ &  &  & \\
12. & else: go to 8 &  &  & \\
13. & end &  &  &
\end{tabular}
\label{T-Opt}%
\end{table}%

Although brute-force search might be of substantial complexity for particular
systems, it is not the case for this problem as the computation required in
every iteration is small. Nevertheless, more efficient search algorithms can
still be invoked if higher efficiency searching is desired. In this work, we
use the Bisection method to demonstrate the additional complexity reduction
gained by adopting efficient search algorithms. The search process starts by
verifying if increasing the transmit power to $\mathcal{P}_{\max}$ will
increase the throughput, and if it is the case, the algorithm starts the
search at $\mathcal{P}^{\ast}=\mathcal{P}_{\max}-\delta$, where $\delta$ is
the search step size. The search process is repeated until the condition
$\eta(\mathcal{P}^{\ast})<\mu\eta(\mathcal{P}_{\max})$ is satisfied. If
increasing the transmit power does not increase the throughput, the search
starts at $\mathcal{P}^{\ast}=\mathcal{P}-\delta$, and then evolves
iteratively until the condition $\eta(\mathcal{P}^{\ast})<\mu\eta
(\mathcal{P}_{\max})$ is satisfied. The optimization process requires the
knowledge of SNR at the transmitter side in order to estimate $P_\text{E}$ and
$\eta$. Moreover, assuming that the channel is linear, then the SNR is
linearly proportional to $\mathcal{P}$, and hence the $P_\text{E}$ and $\eta$ can
be computed as a function of $\mathcal{P}$. The value of $P_\text{E}$ can also be
obtained from the measurements documented by particular communication
standards \cite{LTE,6297413}. In video communications with ARQ, a large number
of subpackets is transmitted in short time intervals; hence, $P_\text{E}$ can be
accurately estimated at the transmitter side by exploiting the NACKs sent from
the receiver.

\nomenclature{$\delta$}{search step size}
\nomenclature{$\mu$}{throughput scaling factor required for power optimization}
\nomenclature{$\epsilon$}{ratio of $\delta$ to $\mathcal{\tilde{P}}_{0}$}

\subsection{Computational Complexity}

Since the optimization algorithm proposed in (\ref{E-opt-1}) does not have a
closed form analytical solution, then its complexity depends on the particular
search method used and the computations performed per iteration, which is
dominated by (\ref{E-eta-1}) and (\ref{eq:alpha}). As it can be noted from
(\ref{E-eta-1}) and (\ref{eq:alpha}), the total number of real additions
$(\mathcal{A})$ and multiplications $(\mathcal{M})$ for each iteration is
$\mathcal{A}=\tfrac{M}{2}\left(  M+7\right)  +1$ and $\mathcal{M}=\tfrac{M}%
{2}\left(  M+11\right)  -3$, respectively. In this work we consider two
searching methods, namely, the Brute-Force (BF)\nomenclature{BF}{Brute-Force Method} and Bisection (BI)\nomenclature{BI}{Bisection Method}. The
Brute-Force algorithm uses linear search to find $\mathcal{P}^{\ast}$ and hence
its complexity depends on the search step size $\delta$ and the initial SNR.
Before starting the backward search, the optimization algorithm first assures
that the initial searching point $\mathcal{\tilde{P}}_{0}$ is larger than the
optimal transmission power $\mathcal{P}^{\ast}$, which requires two
computations of $\eta$. Then, the algorithm starts the backward search at step
8 in Table~\ref{T-Opt}. The search step size $\delta$ is a function of
$\mathcal{\tilde{P}}_{0}$ where $\delta=\epsilon\mathcal{\tilde{P}}_{0}$.
Therefore, the total number of iterations is given by
\begin{align}
\mathcal{I}_\text{BF}  &  =\left\lceil \frac{1}{\delta}\left(  \mathcal{\tilde{P}%
}_{0}-\mathcal{P}^{\ast}\right)  \right\rceil +2\nonumber\\
&  =\left\lceil \Gamma\left(  1-\frac{\mathcal{P}^{\ast}}{\mathcal{\tilde{P}%
}_{0}}\right)  \right\rceil +2 \label{E-BF}%
\end{align}
where $\Gamma=1/\epsilon$ and $\left\lceil .\right\rceil $ is the ceiling
function. Consequently, (\ref{E-BF}) implies that $2\leq\mathcal{I}_\text{BF}\leq\left\lceil \Gamma\right\rceil +2$.

\nomenclature{$\mathcal{A}$}{number of additions}
\nomenclature{$\mathcal{M}$}{number of multiplications}
\nomenclature{$\mathcal{\tilde{P}}_{0}$}{initial searching point}
\nomenclature{$\left\lceil .\right\rceil$}{ceiling function}
\nomenclature{$\mathcal{I}_\text{BF}$}{total number of iterations for Brute-Force}
\nomenclature{$\mathcal{I}_\text{BI}$}{total number of iterations for Bisection}

For the Bisection method, the number of iterations $\mathcal{I}_\text{BI}$ required
to have an error less than $\epsilon\mathcal{\tilde{P}}_{0}$ is given by
\begin{align}
\mathcal{I}_\text{BI}  &  =\left\lceil \log_{2}\left(  \frac{\mathcal{\tilde{P}%
}_{0}-\mathcal{P}_{\min}}{\epsilon\mathcal{\tilde{P}}_{0}}\right)
\right\rceil +2\nonumber\\
&  =\left\lceil \log_{2}\left(  \Gamma\right)  \right\rceil +2 \label{E-BiSec}%
\end{align}
As it can be noted from (\ref{E-BiSec}), setting $\mathcal{P}_{\min}=0$ makes
$\mathcal{I}_\text{BI}$ a function of the desired tolerance $\epsilon$ while it is
independent of $\mathcal{\tilde{P}}_{0}$ and $\mathcal{P}^{\ast}$. Because the
number of iterations $\mathcal{I}_\text{BI}$ is inversely proportional to the
logarithm of $\epsilon$, changing the search tolerance will not have
significant effect on $\mathcal{I}_\text{BI}$. For example, $\mathcal{I}_\text{BI}|_{\epsilon=0.1}=6$ and $\mathcal{I}_\text{BI}|_{\epsilon=0.01}=9$. Therefore,
the number of iterations performed is very small when efficient search methods
are employed. In addition, most communication standards employ delay
constrained HARQ with small values of $M$, and hence, the computations
required in every iteration is small. Consequently, the proposed algorithm has
low complexity even for a small search step or tolerance.

\section{Numerical and Simulation Results}

\label{sec:Results5} In this section, simulation and numerical results are
presented to evaluate the performance of the proposed power optimization
algorithm. The results are obtained for both AWGN and Rayleigh fading
channels. The interleaving and deinterleaving process are assumed to be ideal;
hence, the envelope of all bits in a particular subpacket are assumed to be
iid Rayleigh distributed random variables. The HARQ system is constructed
using TPC and PCCC. The considered TPC are the eBCH$(128,120,4)^{2}$ and
eBCH$(64,57,4)^{2}$ with SISO and HIHO decoding. The PCCC considered in this
work are described at the end of this section . The subpacket error
probability $P_\text{E}$ which is required to compute the SAS is obtained via Monte
Carlo simulation using the parameters described in Section \ref{sec:analysis5}
except that $M=1$.

\begin{figure}[H]%
\centering
\includegraphics[
width=4.3in]%
{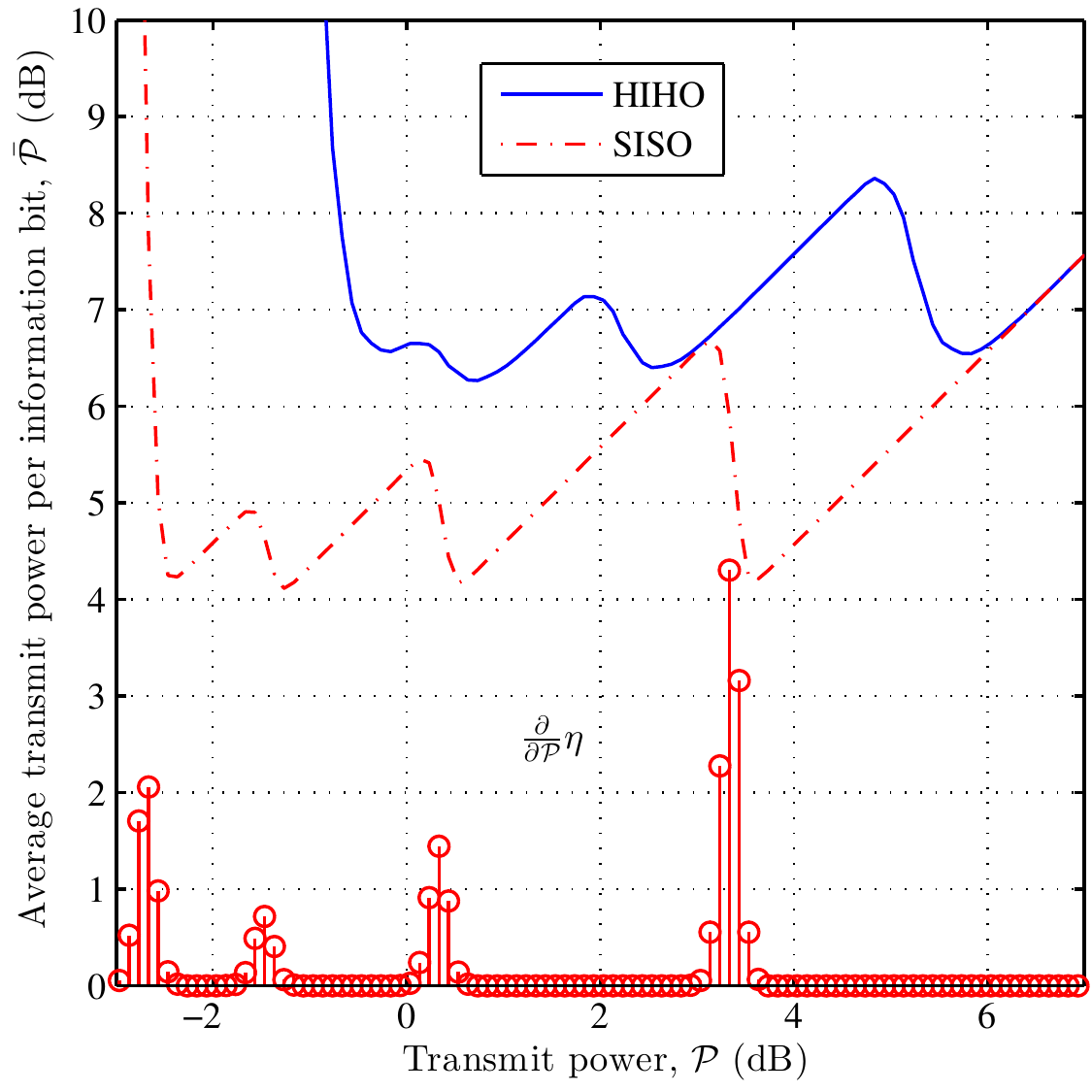}%
\caption{{Average transmit power $\mathcal{\bar{P}}$ with respect to transmit
power $\mathcal{P}$ for HARQ with eBCH$(128,120,4)^{2}$ and $N_{0}=1$ in AWGN
channels.}}%
\label{fig:totP_Pt_128AWGN1}%
\end{figure}

\begin{figure}[H]%
\centering
\includegraphics[
width=4.3in]%
{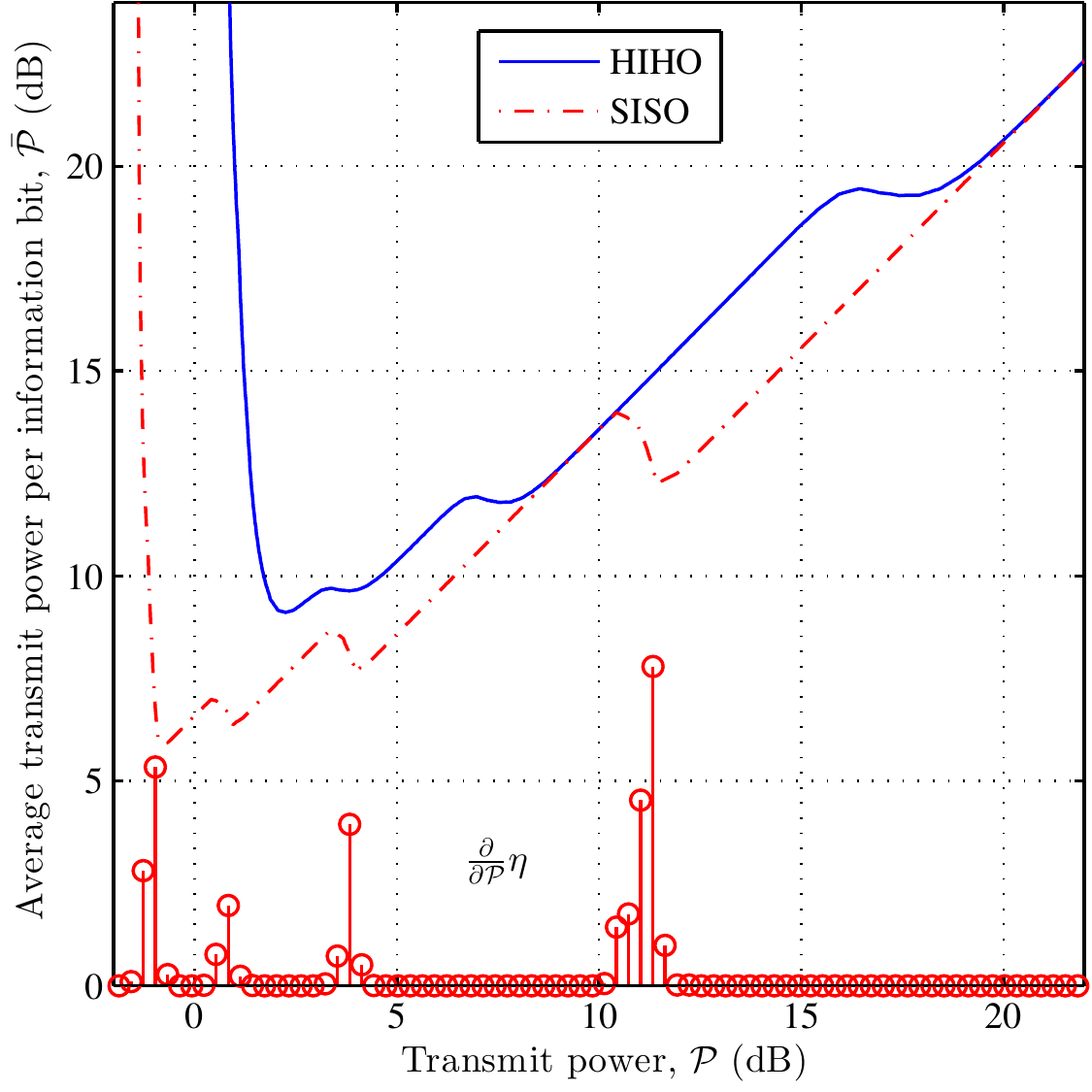}%
\caption{{Average transmit power $\mathcal{\bar{P}}$ with respect to transmit
power $\mathcal{P}$ for HARQ with eBCH$(128,120,4)^{2}$ and $N_{0}=1$ in
Rayleigh fading channels.}}%
\label{fig:totP_Pt_128Fading1}%
\end{figure}

\begin{figure}[H]%
\centering
\includegraphics[
]%
{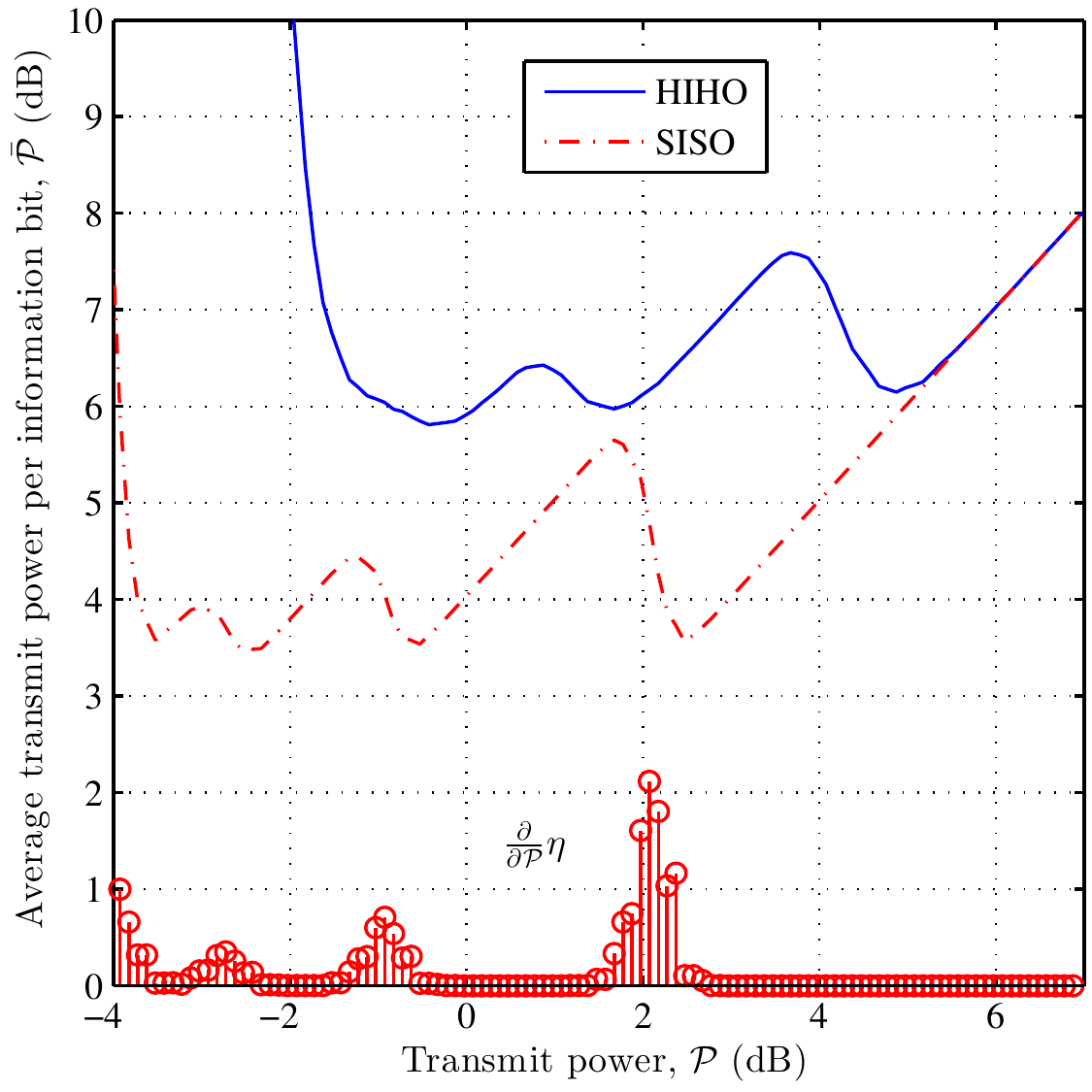}%
\caption{{Average transmit power $\mathcal{\bar{P}}$ with respect to transmit
power $\mathcal{P}$ for HARQ with eBCH$(64,57,4)^{2}$ and $N_{0}=1$ in AWGN
channels.}}%
\label{fig:totP_Pt_64AWGN1}%
\end{figure}

\begin{figure}[H]%
\centering
\includegraphics[
]%
{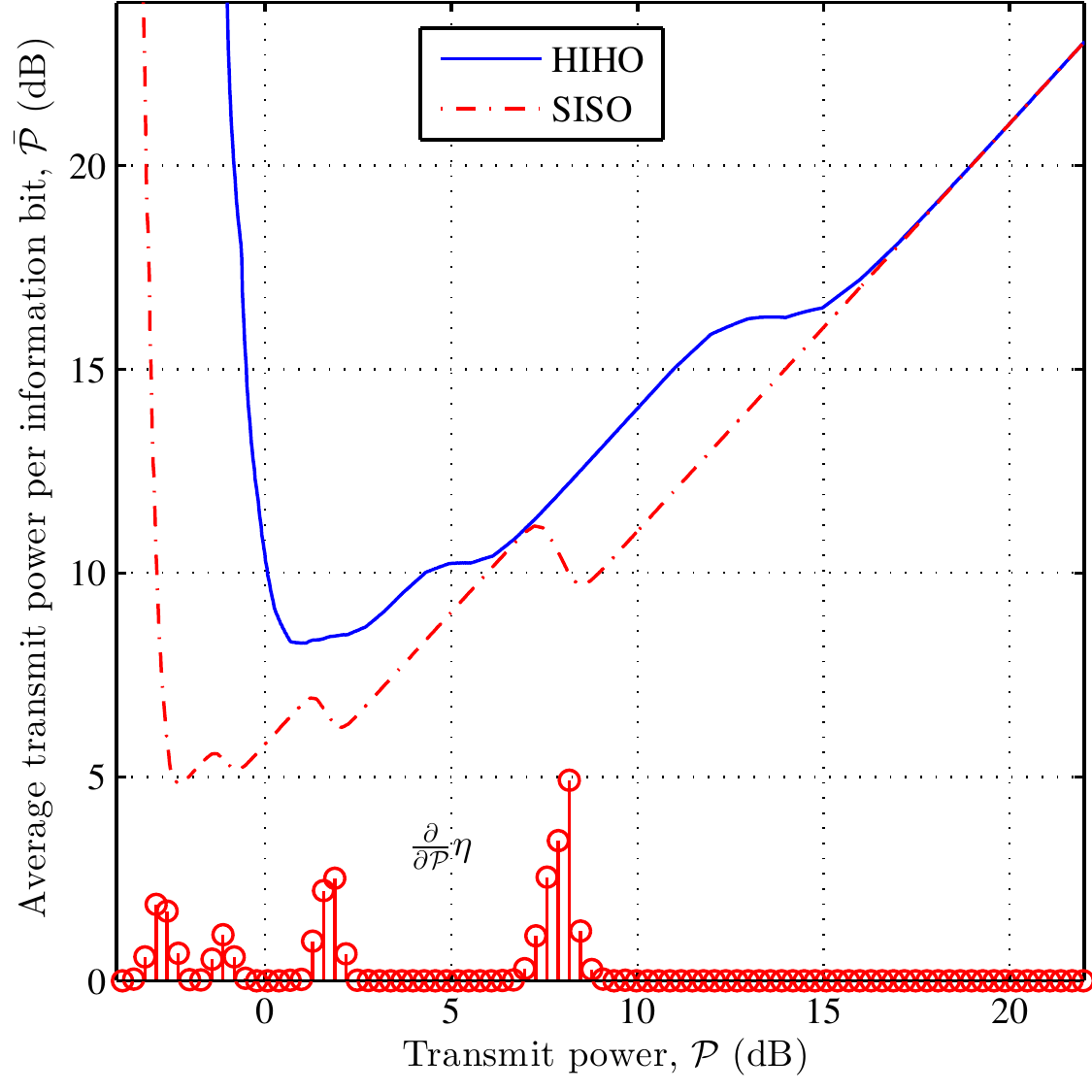}%
\caption{{Average transmit power $\mathcal{\bar{P}}$ with respect to transmit
power $\mathcal{P}$ for HARQ with eBCH$(64,57,4)^{2}$ and $N_{0}=1$ in
Rayleigh fading.}}%
\label{fig:totP_Pt_64Fading1}%
\end{figure}

Figures \ref{fig:totP_Pt_128AWGN1} to \ref{fig:totP_Pt_64Fading1} present
$\mathcal{\bar{P}}$ versus $\mathcal{P}$ for HARQ with eBCH$(128,120,4)^{2}$
and eBCH$(64,57,4)^{2}$ in AWGN and Rayleigh fading channels. The results also present
the numerical derivative of $\eta$ for the SISO decoding results. As it can be
noted from the figures, there exists an optimum value of $\mathcal{P}$ that
minimizes $\mathcal{\bar{P}}$; hence, it is a global minimum. Therefore, given
that the system can tolerate some time delay, then using such value of
$\mathcal{P}$ leads to substantial power saving. The location of the global
minimum is more apparent in fading channels due to the impact of diversity.
However, since the number of transmissions per subpacket is bounded by $M$,
the value of $\mathcal{\bar{P}}$ increases drastically at low SNR as the
number of dropped subpackets increases. Eventually, $\mathcal{\bar{P}}$
becomes infinite when no reliable transmission is possible. However, if there
are constraints on the delay, then the optimum power should be selected such
that the throughput is maximized as well. Hence, the problem becomes a max-min
optimization where the optimum value of $\mathcal{P}$ is the maximum of all
minima values. As it can be noted from Figures \ref{fig:totP_Pt_128AWGN1} to
\ref{fig:totP_Pt_64Fading1}, the derivative of the throughput is almost zero
for a wide range of $\mathcal{P}$, which indicates that significant power
saving will be achieved because increasing $\mathcal{P}$ will directly
increase $\mathcal{\bar{P}}$ while $\eta$ is fixed in these regions. Moreover,
the figures show that SISO decoding reduces $\mathcal{\bar{P}}$ when compared
to HIHO decoding, however, at the expense of increased complexity.
Consequently, the transmit power and the computational power at the receiver
are actually conflicting parameters in HARQ systems.

For very low transmit power devices, the computational power represents a
non-negligible component in the device power budget. Therefore, minimizing the
computational power at the receiver by controlling the transmit power could be
crucial for such low power devices.

Another interesting property that can be observed from Figures
\ref{fig:totP_Pt_128AWGN1} to \ref{fig:totP_Pt_64Fading1} is that increasing
the transmit power $\mathcal{P}$ does not necessarily increase the average
transmit power $\mathcal{\bar{P}}$. For example, it can be noted from Fig.
\ref{fig:totP_Pt_128AWGN1} that increasing the value of $\mathcal{P}$ (when
SISO decoding is used) from ${3.1}$ to ${3.6}$ dB decreases the value of
$\mathcal{\bar{P}}$ by about ${44}\%$. Such behavior is achieved because the
addition of $0.5$ dB to $\mathcal{P}$ almost eliminates the probability of
having a second transmission round, and hence the average power $\mathcal{\bar
{P}}$ decreases by about $1-10^{0.05}/2\approx0.44$.

\begin{figure}[H]%
\centering
\includegraphics[
width=4.0in]%
{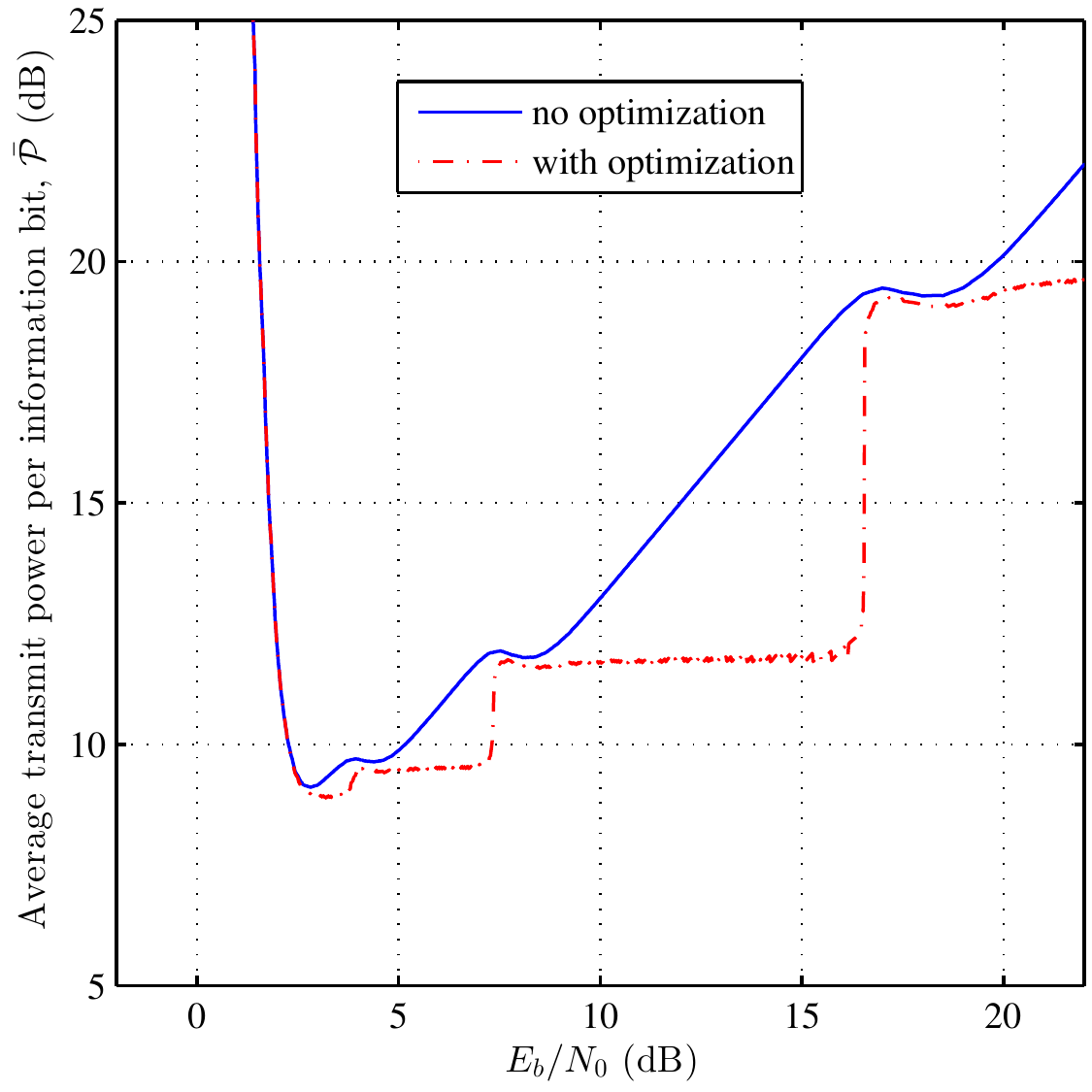}%
\caption{{Average transmit power with and without optimization with respect to
$E_\text{b}/N_{0}$ for HARQ with HIHO eBCH$(128,120,4)^{2}$, $\mu=0.95$, and
$\epsilon=0.01$ in Rayleigh fading channels.}}%
\label{fig:totP_snr_H_128Fading}%
\end{figure}

\begin{figure}[H]%
\centering
\includegraphics[
width=4.0in]%
{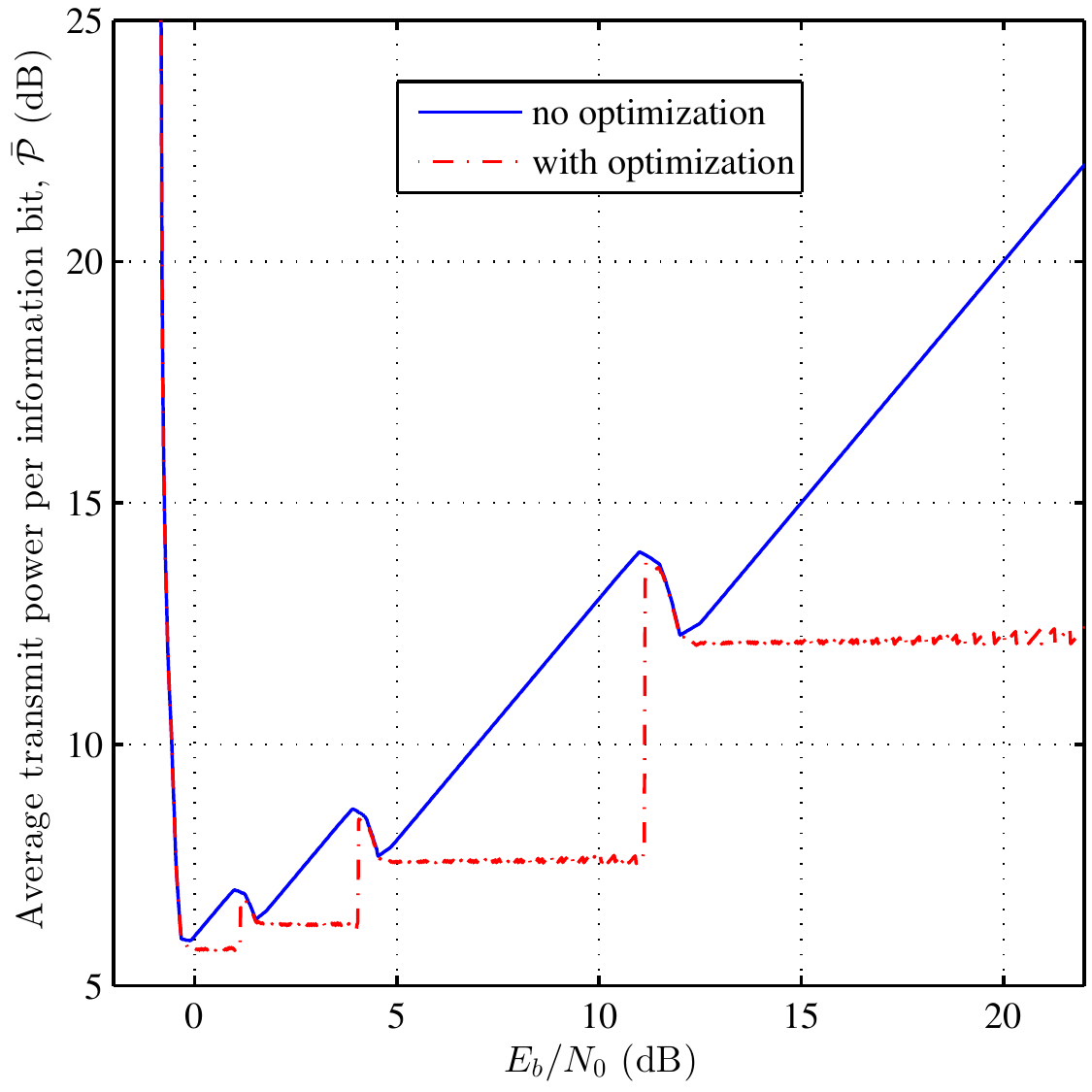}%
\caption{{Average transmit power with and without optimization with respect to
$E_\text{b}/N_{0}$ for HARQ with SISO eBCH$(128,120,4)^{2}$, $\mu=0.95$, and
$\epsilon=0.01$ in Rayleigh fading channels.}}%
\label{fig:totP_snr_S_128Fading}%
\end{figure}

Fig.~\ref{fig:totP_snr_H_128Fading} and Fig.~\ref{fig:totP_snr_S_128Fading}
compare the average transmit power of the TPC-HARQ with and without
optimization for HIHO eBCH$(128,120,4)^{2}$ and SISO eBCH$(128,120,4)^{2}$ in
Rayleigh fading channels. The values of $E_\text{b}/N_{0}$ in
Fig.~\ref{fig:totP_snr_H_128Fading} and Fig.~\ref{fig:totP_snr_S_128Fading}
corresponds to the initial channel SNR per bit when the optimization algorithm
is not used and peak transmit power is used. As it can be noted from the
figures, the proposed power optimization algorithm reduces $\mathcal{\bar{P}}$ significantly.

\begin{figure}[H]%
\centering
\includegraphics[
]%
{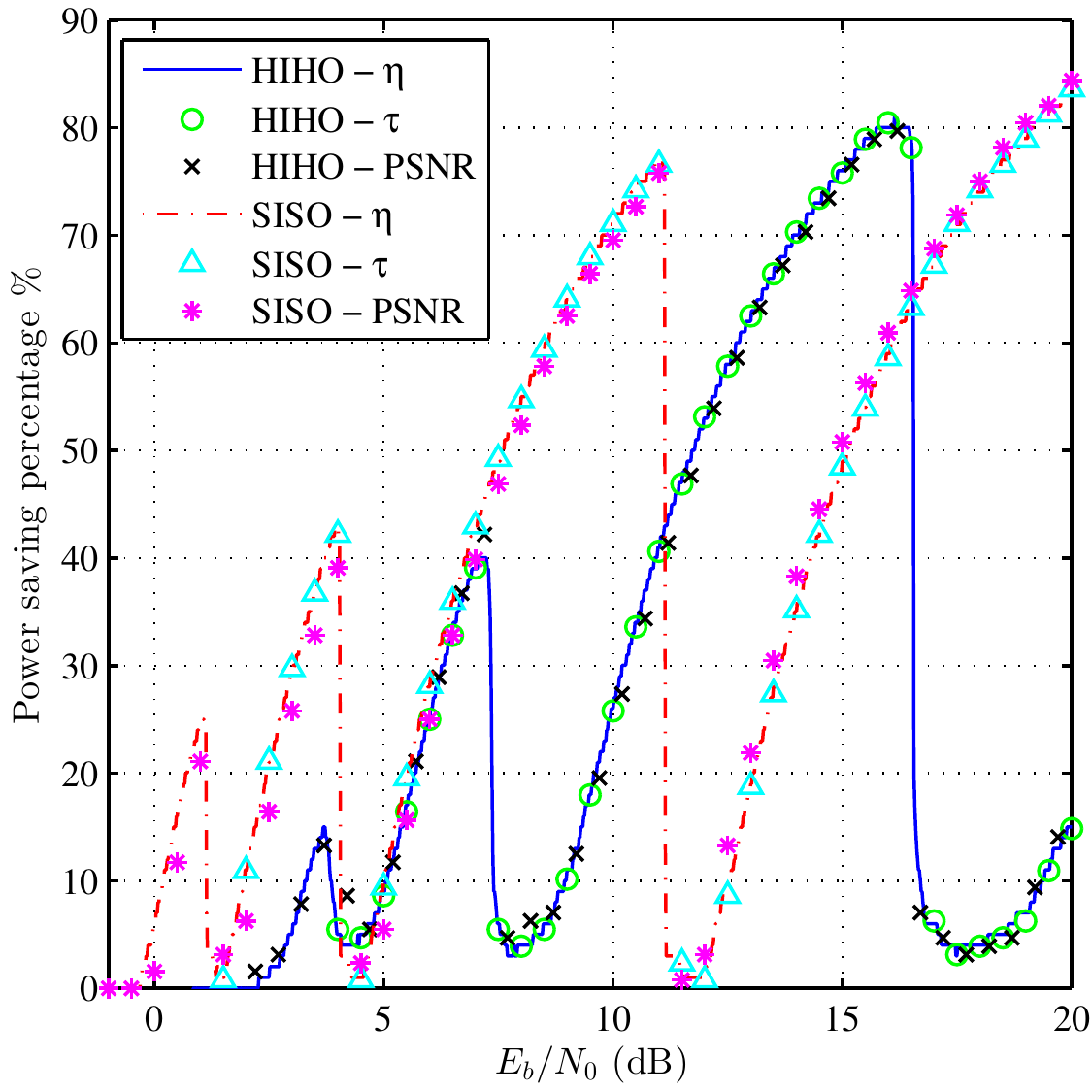}%
\caption{{Power saving percentage due to optimization with small search step
}${(\epsilon=0.01)}${ for HARQ with eBCH$(128,120,4)^{2}$ in Rayleigh fading
channels.}}%
\label{fig:powerSaving128Fading1}%
\end{figure}

\begin{figure}[H]%
\centering
\includegraphics[
]%
{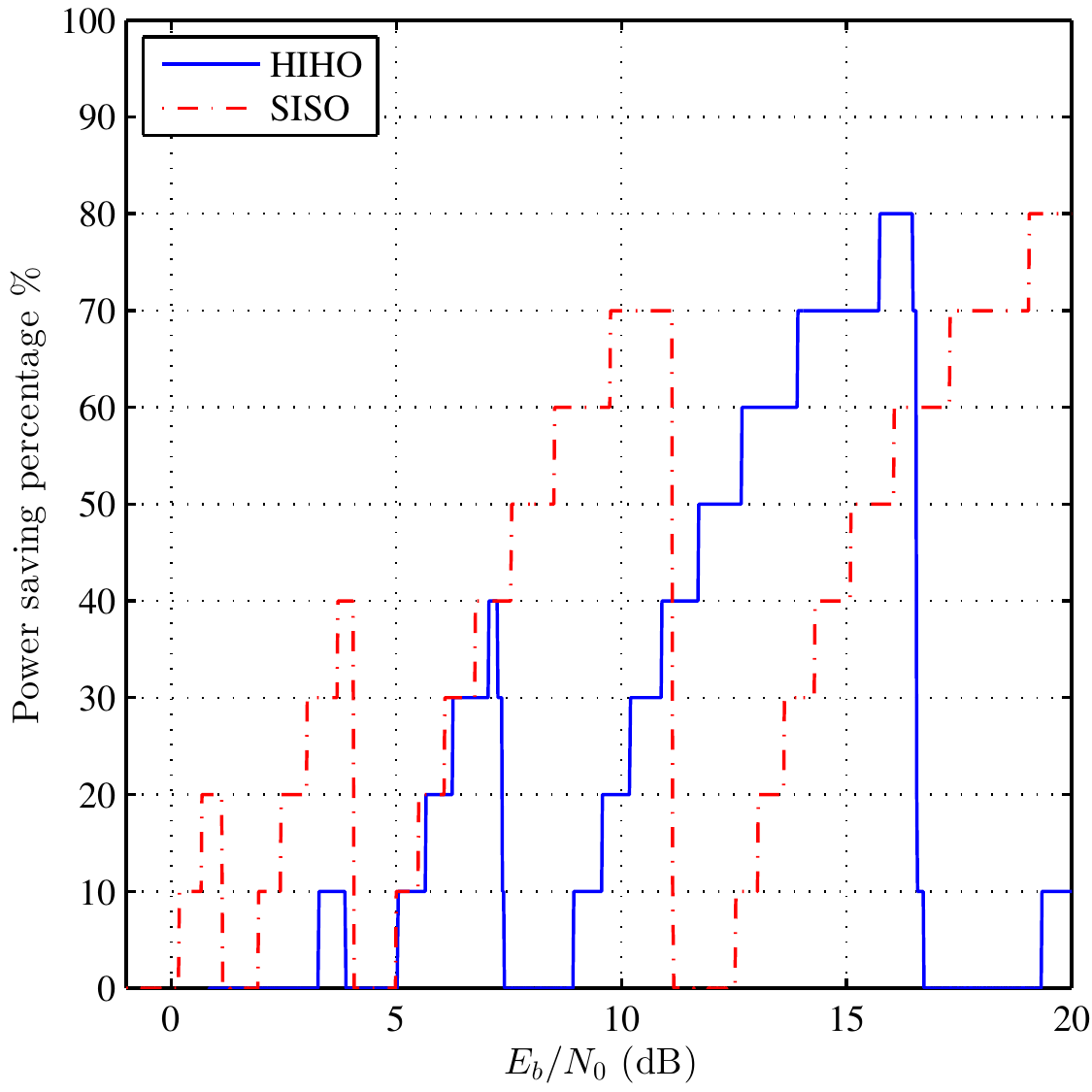}%
\caption{{Power saving percentage due to optimization with large search step
($\epsilon=0.1$) for HARQ with eBCH$(128,120,4)^{2}$ in Rayleigh fading.}}%
\label{fig:powerSaving128Fading21}%
\end{figure}

Fig.~\ref{fig:powerSaving128Fading1} shows the power saving percentage defined
as
\begin{equation}
\text{Power saving $\%$}=\dfrac{\mathcal{\bar{P}}-\mathcal{\bar{P}}^{\ast}%
}{\mathcal{\bar{P}}}\times100
\label{eq:pwrSav}
\end{equation}
$\mathcal{\bar{P}}^{\ast}$ is the average transmit power when $\mathcal{P}%
^{\ast}$ is used. As shown in the figure, a remarkable power saving of about
$80\%$ can be achieved without sacrificing the throughput. The algorithm
parameters are set to $\epsilon=0.01$ and $\mu=0.95$. Increasing the search
parameter $\epsilon$ can reduce the search time for the suboptimal
transmission power $\mathcal{P}^{\ast}$; however, at the expense of power
saving granularity. Fig.~\ref{fig:powerSaving128Fading21} shows the power
saving percentage achieved by the optimization algorithm when $\epsilon=0.1$.

The proposed optimization algorithm is also applied to the delay and PSNR
results obtained for the eBCH$(128,120,4)^{2}$, which are given in
Fig.~\ref{fig:Ttot} and Fig.~\ref{fig:psnrAvg} using $M=4$. The algorithm
tolerance is set to $5\%$ for the delay and $1\%$ for the average PSNR;
however, the search step is kept the same for all cases where $\epsilon=0.01$.
Fig.~\ref{fig:powerSaving128Fading1} compares the power saving achieved for
each performance metric. The figure demonstrates that using the throughput as
a constraint is equivalent to using the PSNR or transmission delay as
constraints for power optimization in video communication systems.
Nevertheless, it can be argued that using the throughput as the optimization
constraint is more practical. That is because the throughput can be obtained
semi-analytically with less computational complexity when compared to the PSNR
and transmission delay.

\begin{figure}[H]%
\centering
\includegraphics[
]%
{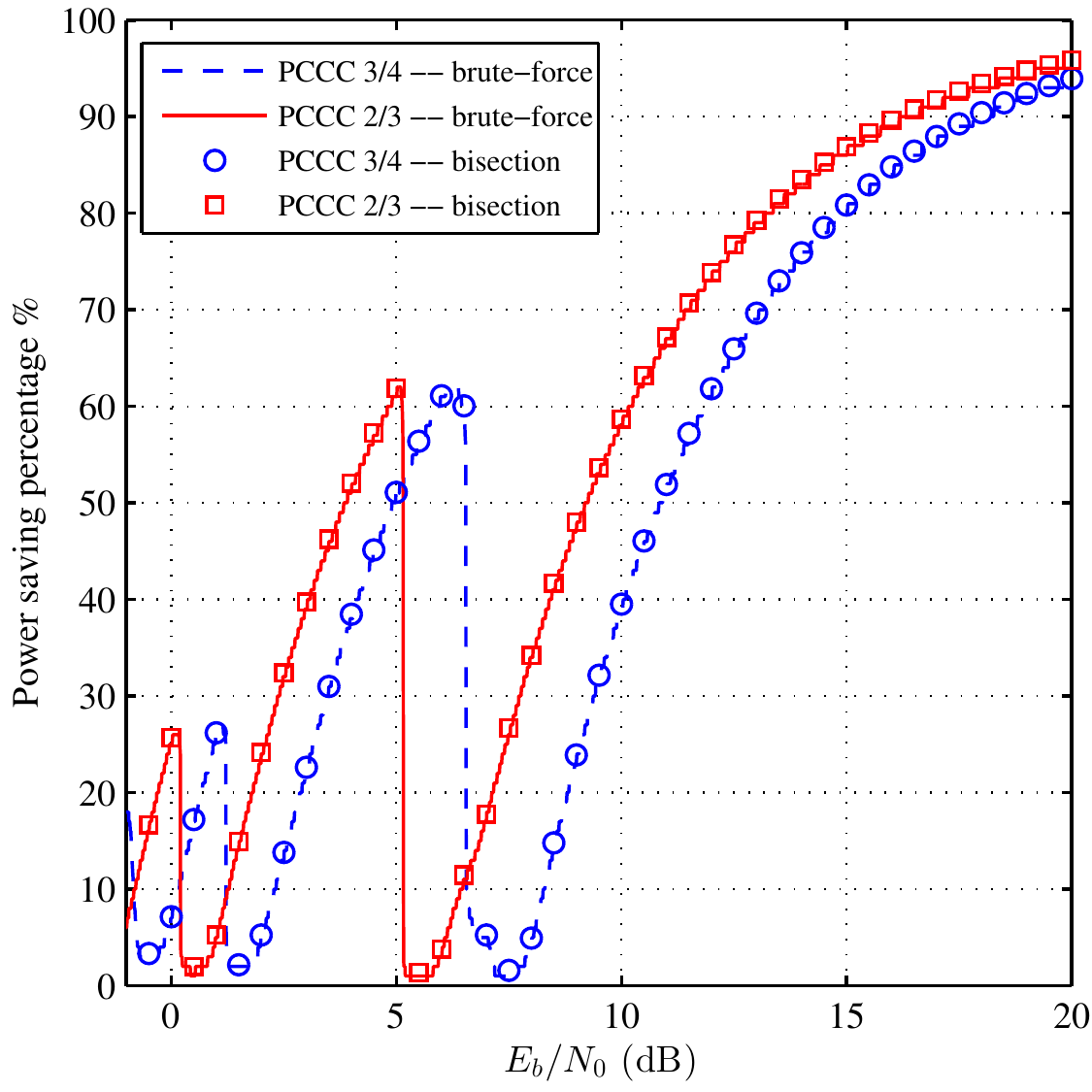}%
\caption{The proposed power optimization using PCCC in Rayleigh fading
channels with $\epsilon=0.01$.}%
\label{F-PCCC-1}%
\end{figure}

Generally speaking, the proposed algorithm can be applied to any HARQ\ system
that exhibits a staircase throughput, which is not exclusive to TPC. For
example, we consider PCCC which is a class of turbo codes implemented using
two identical recursive systematic convolutional encoders
\cite{W-Yafeng-VTC2003}. The constituent encoders used in this example have
parameters $l_\text{i}=1$, $l_\text{o}=2$, $l_\text{k}=4$ where $l_\text{i}$, $l_\text{o}$ and $l_\text{k}$
denote the number of input bits, number of output bits and the constraint
length, respectively. The generator polynomials and the feedback connection
polynomial in octal form are $G_{1}=13$, $G_{2}=15$ and $G_\text{f}=13$,
respectively. The first encoder operates on the input information bits
directly, while the second encoder operates on interleaved information bits.
The PCCC encoder basic code rate is $\zeta=1/3$. The parity bits at the output
of the encoder are punctured to obtain higher code rates. The PCCC is
evaluated with code rates $\zeta=2/3$ and $\zeta=3/4$ using puncturing periods
of $4$ and $6$, respectively. The decoder performs six soft decision decoding
iterations using the BCJR algorithm \cite{berrou1993near,bahl1974optimal}. A fixed packet size of $\mathcal{N}=16,384$
bits is maintained for all code rates and subpacketization is not used. As
depicted in Fig. \ref{F-PCCC-1}, the power saving that can be obtained at low
and moderate $E_\text{b}/N_{0}$ is about $27\%$ and $62\%$, respectively. At high
$E_\text{b}/N_{0}$, the power saving can be as high as $50\%$ for $E_\text{b}%
/N_{0}\gtrsim9.25$ dB and it becomes greater than $80\%$ at $E_\text{b}%
/N_{0}\gtrsim15$ dB. Fig. \ref{F-PCCC-1} also presents the power saving
results using the brute-force and Bisection methods. As it can be seen from
the figure, the difference between the two methods is negligible.

\section{Conclusions}

\label{sec:conclusion5} 

A power optimization algorithm is proposed in this
chapter for wireless multimedia systems which uses truncated HARQ with Chase
combining. The average transmit power per information bit is minimized subject
to a desired throughput value. A semi-analytical solution for the system
throughput is used in the optimization algorithm to find the minimum
transmission power required to achieve the desired throughput. Simulation and
numerical results show that the transmission power can be adjusted to reduce
average transmit power while maintaining the throughput unchanged. Moreover,
the results show that reducing the transmission power per ARQ round does not
necessarily reduce the average transmit power. On the contrary, increasing the
transmission power per ARQ round can reduce the average transmit power. The
proposed power optimization algorithm reduces the average transmit power
significantly with power saving percentage of up to $80\%$.

\chapter{The Performance of Power-Adaptive HARQ with No Channel State Information Feedback}
\label{chap:6}

This chapter presents an adaptive power control
scheme for HARQ systems with no channel state information (CSI)\nomenclature{CSI}{Channel State Information} feedback. The power optimization algorithm is implemented based on ACK/NACK feedback which is an inherent part of the HARQ process. The obtained results show that the proposed
optimization algorithm can achieve a significant power saving
of about $80\%$ for particular scenarios. The system considered in
this work is a truncated HARQ with turbo product codes.
Chase combining is also used to combine the retransmitted
packets with the original transmission. HARQ ACK/NACK feedback is used to
estimate the packet error rate from which the system throughput
is computed.

\section{Related Work}

In Chapter \ref{chap:5}, a power optimization algorithm is proposed to minimize the transmit energy under throughput constraints \cite{mukhtar2014low}. The flat regions in the throughput performance are exploited to optimize the HARQ systems' energy. However, the optimization algorithm requires accurate CSI to obtain the corresponding PER from a lookup table, and subsequently the system throughput. In practice, maintaining perfect CSI is difficult and introduces additional cost to HARQ systems \cite{aggarwal2012joint,zwingelstein2014random}. 

Therefore, in this chapter, we propose a blind power-adaptive HARQ system where no CSI feedback is required. The power optimization algorithm is implemented based on ACK/NACK feedback, which is an essential component of the HARQ process. The transmit power is adjusted iteratively to achieve the minimum average transmit energy while maintaining the throughput unchanged. The throughput is computed at the transmitter side using the packet error rate, which is estimated using the ACK/NACK feedback. The HARQ is implemented using TPC for error correction, and Stop-and-Wait for the ARQ.

\section{Power Optimization with no CSI Feedback}

As discussed in Section \ref{sec:algorithm5}, minimizing the power at the transmitter side for a fixed $\eta$ requires the knowledge of the probability of packet error $P_\text{E}^{(.)}$ which is a function of the channel statistics and the SNR at the receiver side. Hence, the channel condition at the receiver should be sent via a feedback link to the transmitter. However, we propose in this work to estimate $P_\text{E}^{(.)}$ by utilizing the inherent feedback information represented by the ACK/NACK messages \cite{hara1996throughput}. In this sense, the proposed system does not require any specific channel state information feedback.

We implement the Bisection method to find $\mathcal{P}^{\ast}$ as described in the optimization algorithm given in Table~\ref{T-Optbi}.  

\begin{table}[H]%
\caption{Power optimization algorithm based on estimated PER and using the Bisection method.}%
\begin{tabular}
[c]{lllll}%
1. & set $\mathcal{P}_{\max}=\mathcal{P}$ &  &  & \\
2. & set $\mathcal{P}_{\min}=0$ &  &  & \\
3. & estimate $P_\text{E}^{(.)}(\mathcal{P})$ &  &  & \\
4. & set $\eta_{\max}=\mu\eta(\mathcal{P})$ &  &  & \\
5. & set $\mathcal{P}^{\ast}=0.5(\mathcal{P}_{\max}+\mathcal{P}_{\min})$ &  &  & \\
6. & if $0.5(\mathcal{P}_{\max}-\mathcal{P}_{\min})<\epsilon$ &  &  & \\
7. & \ \ \ \ go to 15 &  &  & \\
8. & else: go to 9 &  &  & \\
9. & estimate $P_\text{E}^{(.)}(\mathcal{P}^{\ast})$ &  &  & \\
10. & compute $\eta(\mathcal{P}^{\ast})$ &  &  & \\
11. & if $\eta(\mathcal{P}^{\ast})<\eta_{\max}$ &  &  & \\
12. & $\ \ \ \ \mathcal{P}_{\min}=\mathcal{P}^{\ast}$ &  &  & \\
13. & else: $ \mathcal{P}_{\max}=\mathcal{P}^{\ast}$ &  &  & \\
14. & go to 5 &  &  & \\
15. & end &  &  &
\end{tabular}
\label{T-Optbi}%
\end{table}%

The search process starts by specifying the search bounds. The initial transmit power is assumed to be the maximum $\mathcal{P}_{\max}=\mathcal{P}$ and zero power is the minimum $\mathcal{P}_{\min}=0$. At each iteration, the algorithm divides the search interval into two and sets the transmit power to $\mathcal{P}^{\ast}=0.5(\mathcal{P}_{\max}+\mathcal{P}_{\min})$. The corresponding $P_\text{E}^{(.)}$ is estimated to compute the throughput at the new power value. If the computed throughput is less than the target throughput $\eta_{\max}=\mu\eta(\mathcal{P})$, the midpoint is set as the lower bound $\mathcal{P}_{\min}=\mathcal{P}^{\ast}$; otherwise, it is set as the upper bound $\mathcal{P}_{\max}=\mathcal{P}^{\ast}$ to form a new search subinterval. The process is repeated until the search interval is sufficiently small. It should be noted that the Bisection method is an efficient search algorithm which requires few iterations \cite{mukhtar2014low}.

\section{Numerical and Simulation Results}
\label{sec:Results6}
In this section, simulation and numerical results are
presented to evaluate the performance of the proposed power optimization
algorithm with no CSI feedback. The results are obtained for Rayleigh fading
channels. The interleaving and deinterleaving process are assumed to be ideal;
hence, the envelope of all bits in a particular packet are assumed to be
iid Rayleigh distributed random variables. The HARQ system is constructed
using TPC for error correction and CRC for error detection and the maximum number of allowed ARQ rounds per packet is $M=4$. The considered TPC is the eBCH$(128,120,4)^{2}$ with HIHO decoding. 

The power optimization algorithm described in Section \ref{sec:algorithm5} requires the knowledge of the packet error probability $P_\text{E}^{(.)}$. In HARQ with packet combining, $P_\text{E}^{(.)}$ improves with every additional ARQ round due to combining. Therefore, for each transmission round, $P_\text{E}^{(i)}$ is estimated from the ACK/NACK messages which belong to the $i$th transmission round. 

\begin{figure}[H]
\centering
\includegraphics[]{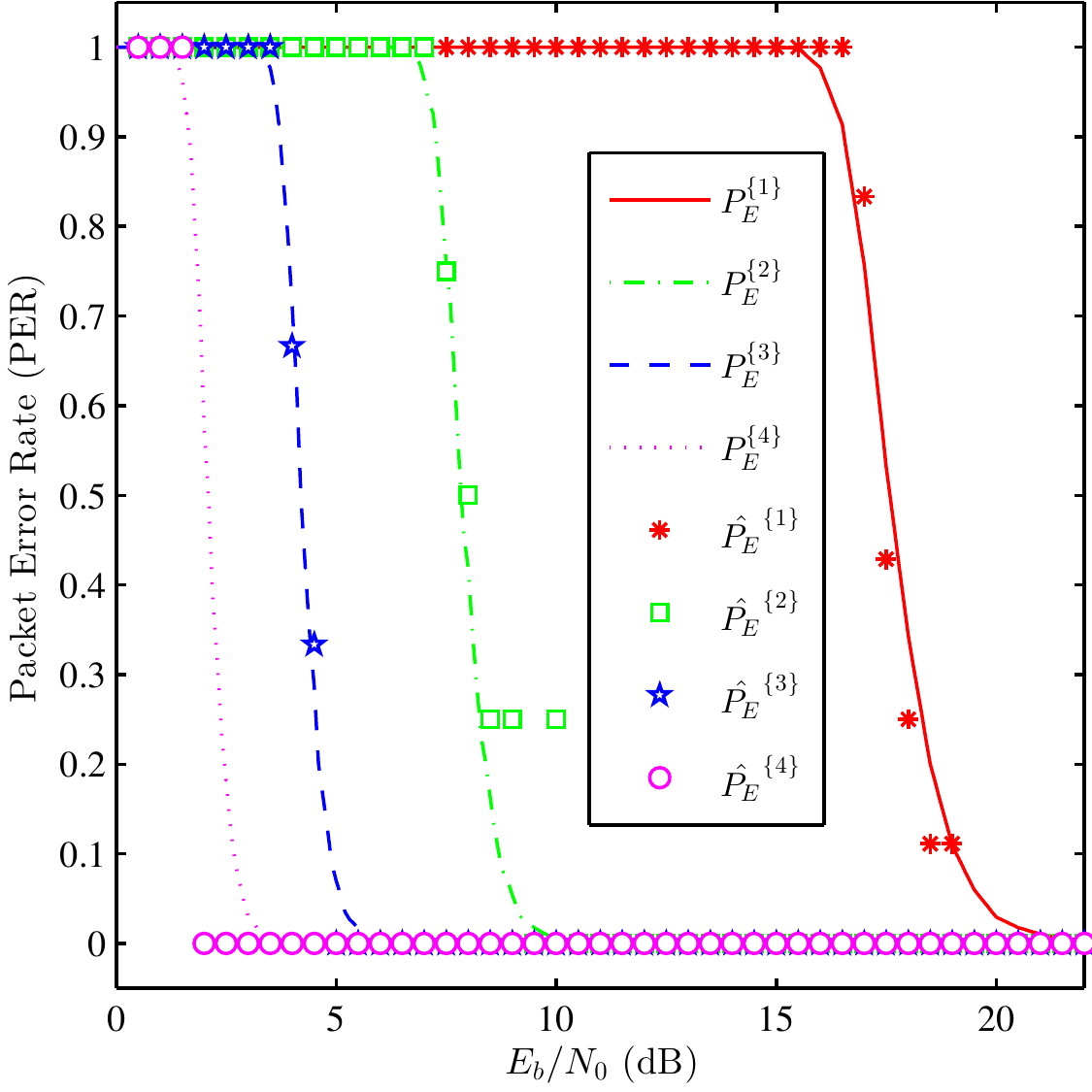}
\caption{PER estimated from 10 ACK/NACK messages.}
\label{fig:Pe_1e1nACKs}
\end{figure}

Fig.~\ref{fig:Pe_1e1nACKs} shows the estimated PER $\hat{P}_\text{E}^{(.)}$ from only 10 ACK/NACK messages. The PER is shown with respect to the SNR per bit $\dfrac{E_\text{b}}{N_0}$. Although the number of ACK/NACK messages is small, a reasonable approximation of the actual $P_\text{E}^{(.)}$ is obtained. The accuracy of the PER estimation improves as the number of considered ACK/NACK messages increases. However, using a small number of ACK/NACK messages makes the adaptation process faster.  

\begin{figure}[H]
\centering
\includegraphics[]{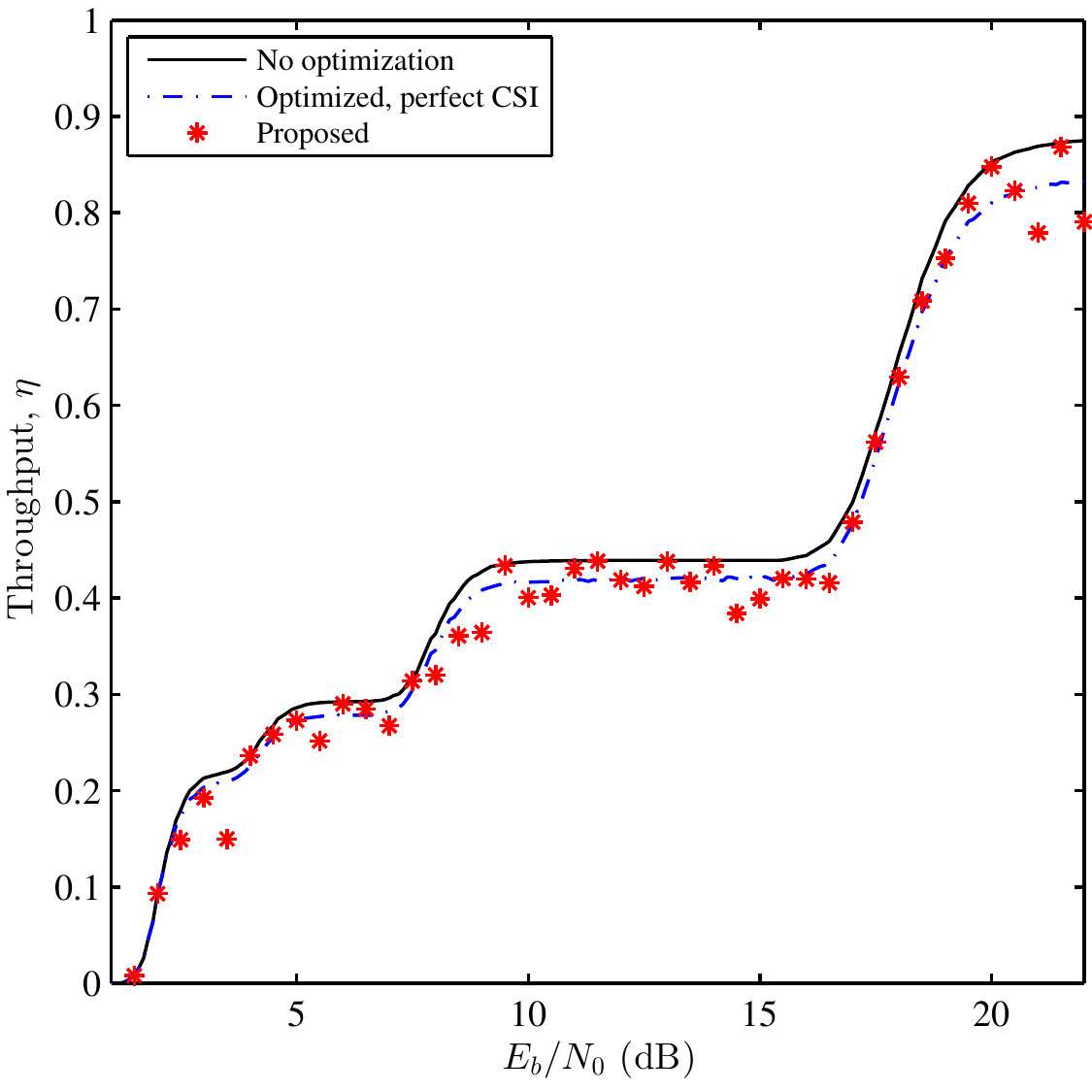}
\caption{Throughput based on PER estimated from 10 ACK/NACK messages.}
\label{fig:eta_1e1nACKs}
\end{figure}

To study the impact of PER estimation inaccuracy on the adaptive HARQ system, we implement the optimization algorithm using only the ACK/NACK feedback. Fig.~\ref{fig:eta_1e1nACKs} shows the HARQ system throughput with and without power optimization using perfect CSI and the estimated PER. We observe that despite considering small number of ACK/NACK messages for PER estimation, the obtained system throughput with optimization using the estimated PER closely matches the throughput values with optimization using perfect CSI.  

\begin{figure}[H]
\centering
\includegraphics[]{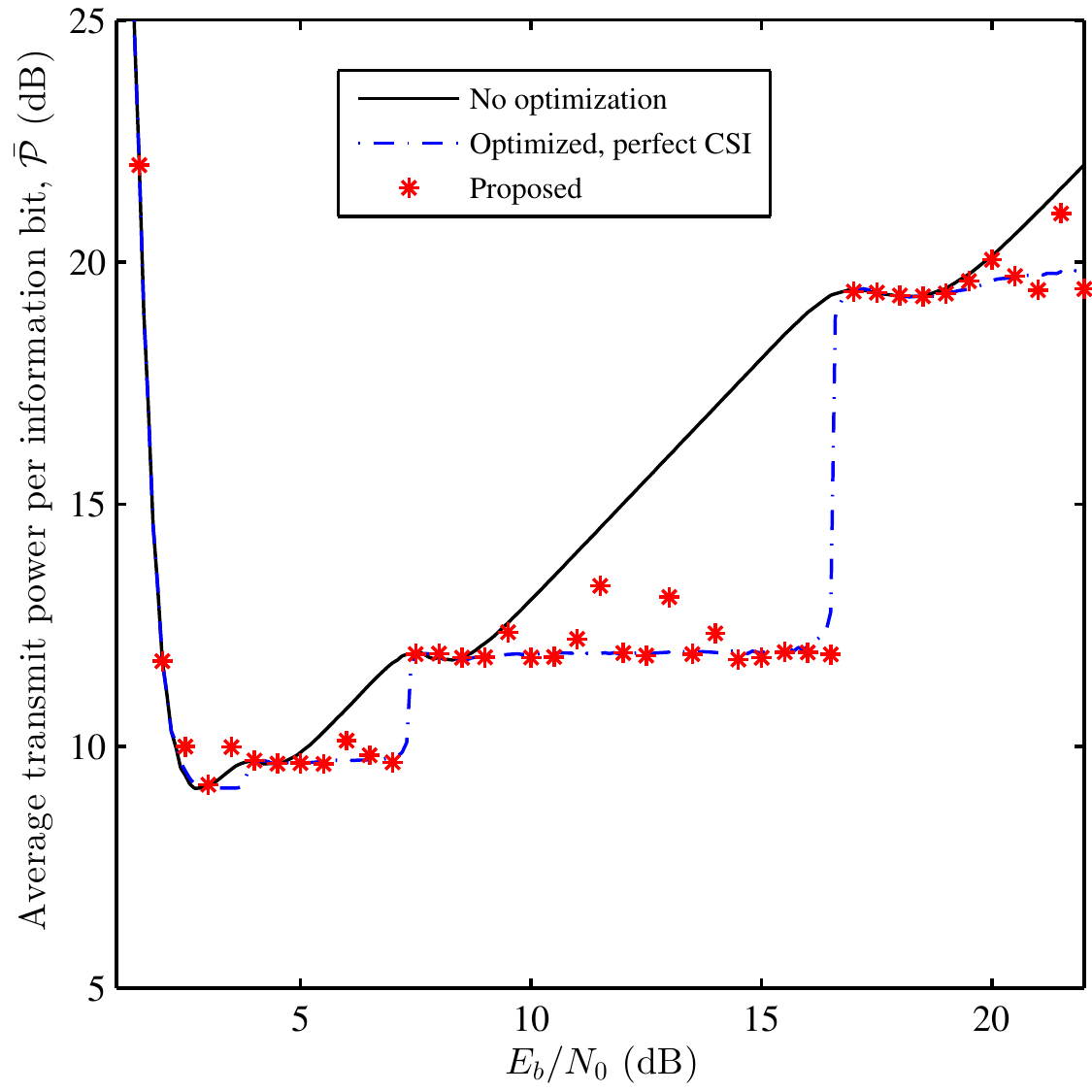}
\caption{Average transmit power based on PER estimated from 10 ACK/NACK messages.}
\label{fig:Pav_1e1nACKs}
\end{figure}

In the proposed optimization algorithm, the objective is to minimize $\mathcal{\bar{P}}$ while maintaining the throughput unchanged. Fig.~\ref{fig:Pav_1e1nACKs} compares $\mathcal{\bar{P}}$ of the TPC-HARQ with and without
the proposed optimization algorithm.  As it can be noted from the figure, the proposed power optimization algorithm reduces $\mathcal{\bar{P}}$ significantly even when we rely solely on the ACK/NACK feedback. The values of $E_\text{b}/N_{0}$ in the figure correspond to the initial channel SNR per bit when the optimization algorithm is not used and peak transmit power is used.

\begin{figure}[H]
\centering
\includegraphics[]{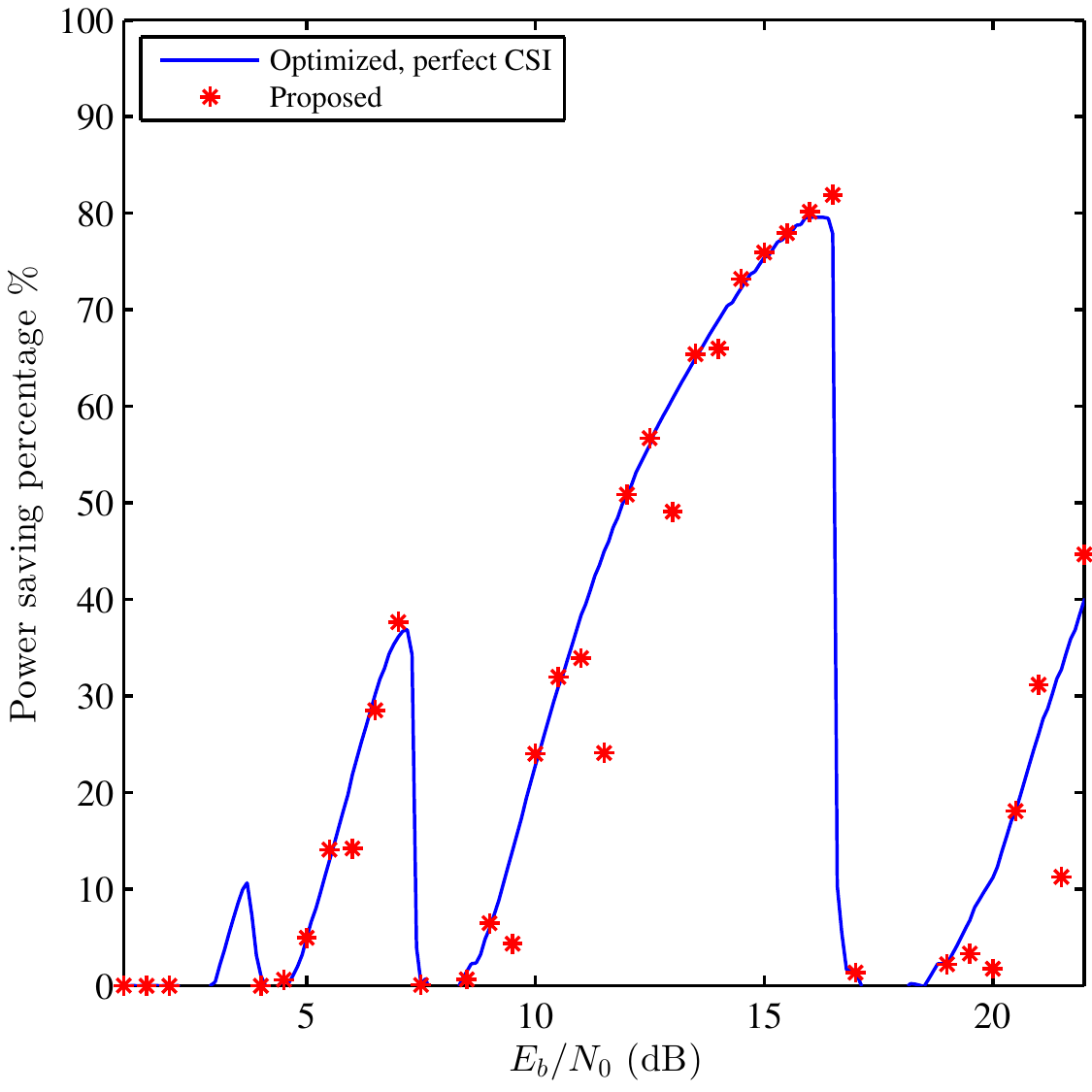}
\caption{Power saving due to power optimization based on PER estimated from 10 ACK/NACK messages.}
\label{fig:pwrSav_1e1nACKs}
\end{figure}

Fig.~\ref{fig:pwrSav_1e1nACKs} shows the power saving percentage as defined in (\ref{eq:pwrSav}). As shown in the figure, a remarkable power saving of about
$80\%$ can be achieved without significant loss in throughput. The algorithm parameters $\mu$ was set to 0.95. The solid line in the figure represents the power saving achieved using the power optimization algorithm with perfect CSI. The algorithm with no CSI feedback also manages to achieve similar power saving where the PER is estimated from only 10 ACK/NACK messages.  

\begin{figure}[H]
\centering
\includegraphics[width=4.3in]{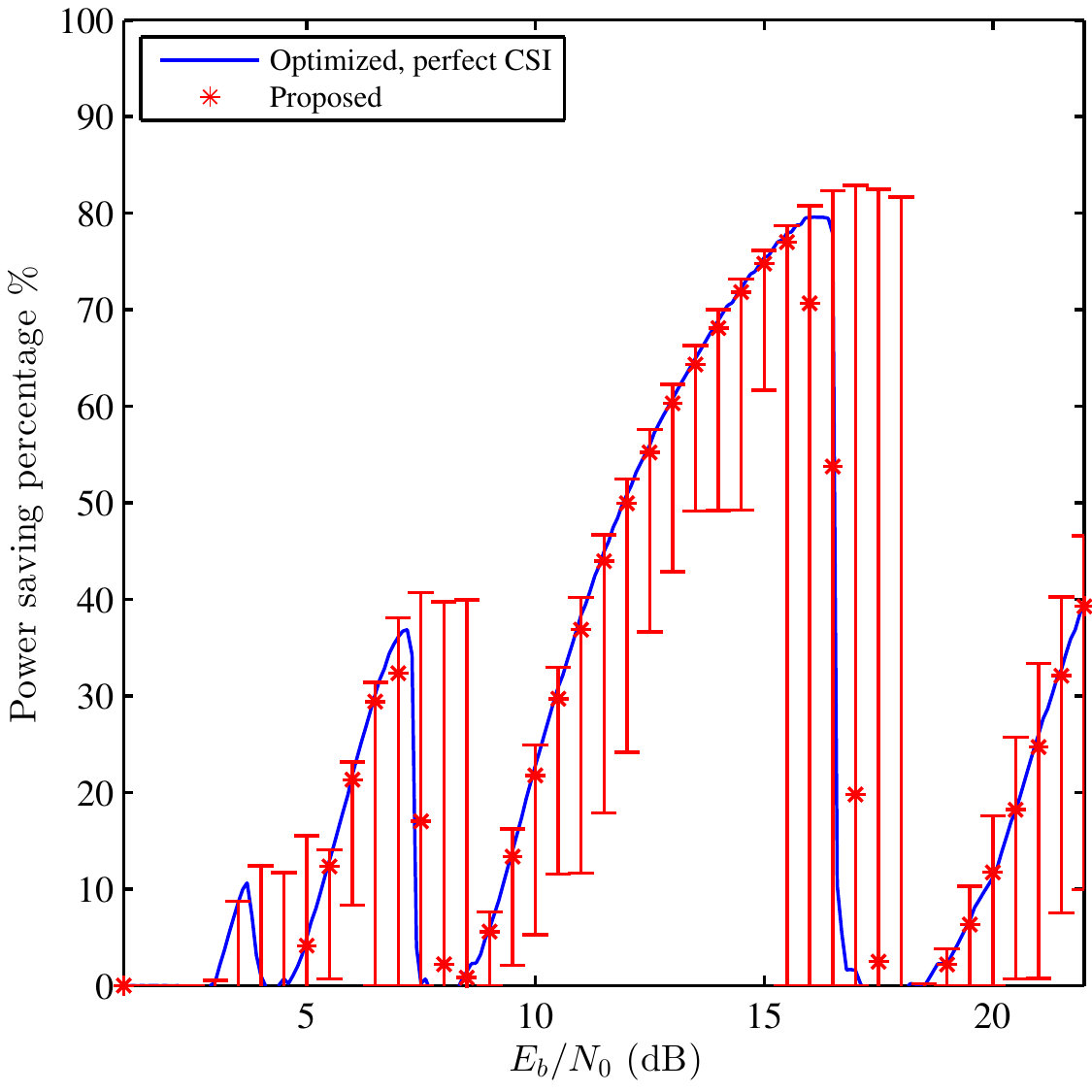}
\caption{Range of power saving values based on PER estimated from 10 ACK/NACK messages.}
\label{fig:pwrSav_1e1nACKs_1e3runs}
\end{figure}

\begin{figure}[H]
\centering
\includegraphics[width=4.3in]{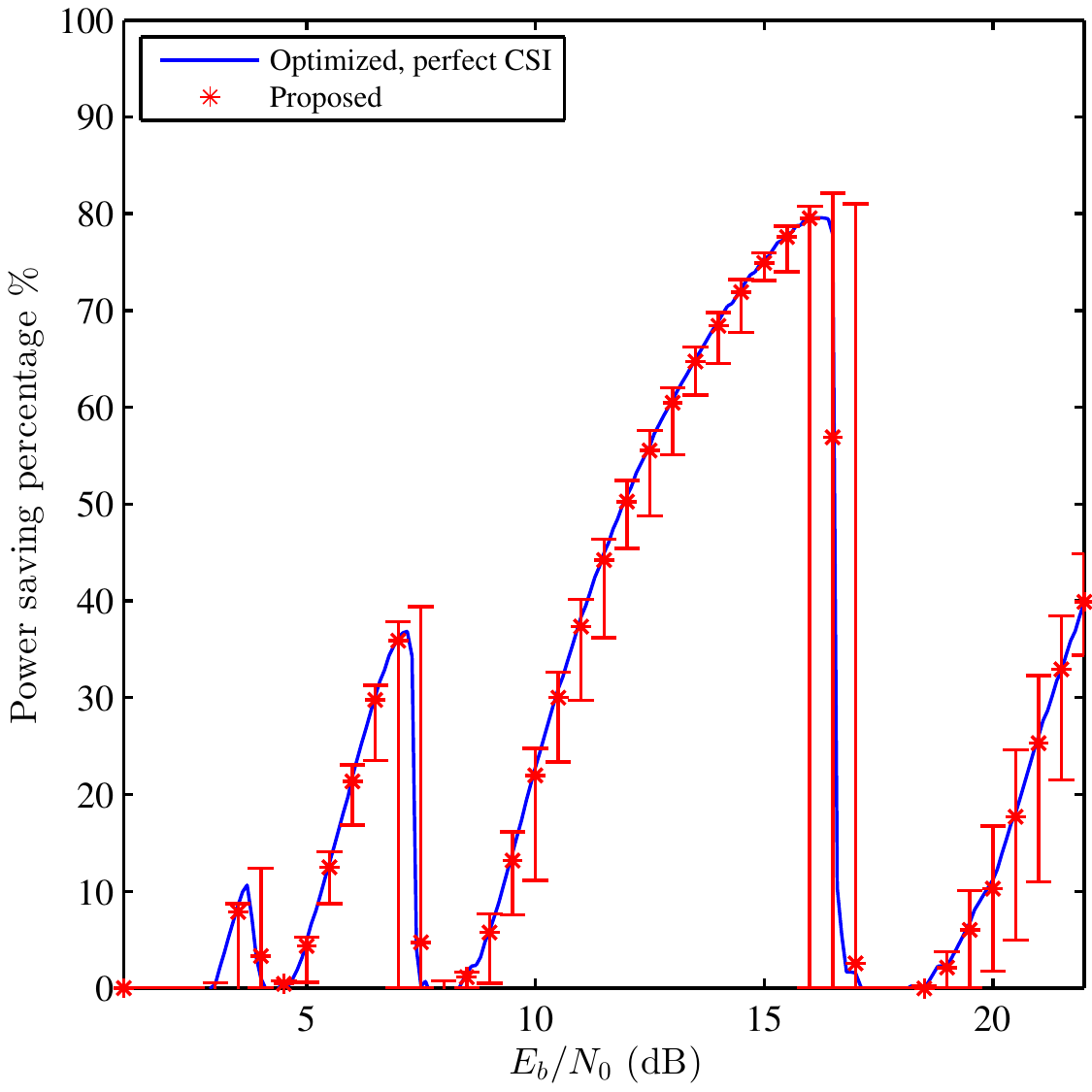}
\caption{Range of power saving values based on PER estimated from 100 ACK/NACK messages.}
\label{fig:pwrSav_1e2nACKs_1e3runs}
\end{figure}

Fig.~\ref{fig:pwrSav_1e1nACKs_1e3runs} and Fig.~\ref{fig:pwrSav_1e2nACKs_1e3runs} show the range of power saving values when the PER is estimated from 10 and 100 ACK/NACK messages, respectively. The accuracy of the optimization algorithm improves as the number of ACK/NACK messages, which are used in PER estimation, increases. Nevertheless, improved accuracy comes at the cost of slower power adaptation.

\section{Conclusions}
\label{sec:conclusion6} A power optimization algorithm is proposed in this
chapter for wireless communication systems, which uses truncated HARQ with Chase
combining. The power control algorithm eliminates the need for channel state information feedback and relies on simple feedback inherent to HARQ. The average transmit power per information bit is minimized subject
to a desired throughput value. The throughput is computed at the transmitter side using the packet error rate, which is estimated using the ACK/NACK feedback. The accuracy of PER estimation improves as the number of considered ACK/NACK messages increases. Simulation and numerical results show that the proposed power optimization algorithm achieves similar power saving when compared to the adaptive system with perfect CSI. Using a small number of ACK/NACK messages for PER estimation provides a practical solution to adjust the transmit power to the optimal value with tolerable inaccuracy. 
\chapter{CRC-Free Hybrid ARQ System using Turbo Product Codes}
\label{chap:7}

This chapter presents an HARQ system using turbo product codes. The inherent word-error detection capability of
TPC is exploited to perform packet error detection as an alternative to CRC codes. Therefore, the
TPC is used for joint bit error correction and packet error detection.
Consequently, the HARQ system efficiency can be improved by increasing the
system throughput when short packets are transmitted, or by reducing the
computational complexity/delay when the packets transmitted are long.
Numerical and simulation results reveal that the
CRC-free TPC-HARQ system consistently provides equivalent or higher
throughput than CRC-based HARQ systems. Moreover, numerical results show
that the TPC self-detection has lower computational complexity than CRC
detection, especially for TPC with high code rates. In particular scenarios,
the relative complexity of the self-detection approach with respect to
popular CRC techniques is about $0.3\%$.
%
%

In conventional HARQ systems, the error correction is performed in two phases.
In the first phase, the FEC codes are applied to correct the bit errors in
the received packets. In the second phase, the receiver verifies if the FEC
process is successful, and then sends a message to the transmitter to
retransmit the packet if the FEC process is unsuccessful; otherwise, the
message informs the transmitter to send a new packet. In general, most
of the work reported in the literature performs the bit error correction and
packet error detection (PED)\nomenclature{PED}{Packet Error Detection} as two independent processes where the FEC is
implemented at the Physical Layer (PHY)\nomenclature{PHY}{Physical Layer} while the packet error detection is
performed at the Data Link Control (DLC)\nomenclature{DLC}{Data Link Control} layer \cite%
{J-Ramis-2011}. Upper layers such as the transport and network layers also provide a limited form of error detection using checksum methods \cite{kurose2003computer}. However, error detection in the DLC layer is usually more sophisticated and is commonly implemented using CRC codes.  

\section{Related Work}

In the literature, the performance of HARQ systems has been evaluated using
various FEC codes such as LDPC \cite%
{linadaptive2012,Sesia04}, TPC \cite%
{H-Mukhtar-2013}, convolutional codes \cite{zhou2006optimum} and turbo
convolutional codes \cite{B-Zhang-2013}. The PED is usually performed using
CRC codes at the DLC layer \cite%
{J-Ramis-2011,H-Mukhtar-2013}. Because the FEC and PED are implemented at
two different layers, the interaction between the two FEC and PED operations
is considered as a cross-layer cooperation \cite{J-Ramis-2011}.

The CRC operation can be realized using different software and hardware
implementations. However, speed requirements usually make software schemes
impractical and hence, dedicated hardware is required \cite%
{M-Grymel-2011,G-Nguyen-2009}. The generic hardware implementation of the
CRC process is based on low complexity linear feedback shift register (LFSR)\nomenclature{LFSR}{Linear Feedback Shift Register}
that performs the polynomial division process of the serial data input. In
the presence of wide data buses, the serial computation can be extended to
parallel versions that process large number of bits simultaneously at the
expense of additional hardware complexity \cite{M.Ayinala-2011,J-Satran-2005}%
. Moreover, the energy consumed by CRC circuits becomes non-negligible for
systems with continuous and high data rate transmission \cite{B-Fu-2009}.

In the literature, the problem of designing CRC-free error detection schemes
has received noticeable attention with the aim of reducing the adverse
effects of the CRC process on the computational complexity, delay,
throughput and energy consumption. For example, Coulton \textit{et al.} \cite%
{coulton2000simple} proposed a simple PED by comparing the ratio of the
standard deviation (STD)\nomenclature{STD}{Standard Deviation} to the mean of the soft information energy at the
output of the decoder. However, the results given in \cite%
{coulton2000simple} indicate that a reliable performance requires a large
sample space to compute the mean and STD accurately. Moreover, the system
requires a substantial number of operations to compute the energy, and then
the mean and STD of the soft information. Buckley and Wicker \cite%
{buckley2000design} proposed to use neural networks for CRC-free error detection. However, training the neural network requires very large
number of samples, which might cause severe performance degradation when the
channel is time variant.

In \cite{zhai2003techniques}, Zhai and Fair proposed an error detection
criterion by continuously monitoring log-likelihood ratio (LLR)\nomenclature{LLR}{Log-Likelihood Ratio} of the soft
information at the decoder output. However, the system performance depends on the SNR, frame size and code rate. Moreover, computing the LLR requires
accurate knowledge of the channel statistics, which are hard to compute and
track. Fricke \textit{et al.} \cite{fricke2009reliability} proposed an error detection technique using
the word and bit error probabilities. However, this technique is unreliable and usually causes throughput inflation.

As it can be noted from the above discussion, the performance of CRC-free techniques is usually limited by the proper
selection of certain thresholds, which are dependent on channel statistics,
codeword size, code rate and sample space. The performance of such
techniques has been evaluated for HARQ systems with various FEC codes such
as LDPC, convolutional, and turbo codes. However, to the best of our
knowledge, there is no work reported in the literature that considers
CRC-free HARQ systems with TPC. TPC have low complexity
convergence detection characteristics that are commonly used for aborting
the iterative decoding process \cite{al2011closed}. However, such
characteristics were never used in the context of HARQ systems.

In this work, we consider a TPC-based HARQ scheme where the error self-detection capability of TPC is exploited. The TPC inherent error
detection capability is used to provide CRC-free PED, which can have a
significant impact on the overall system performance. The performance of the
TPC-HARQ is evaluated in terms of complexity and throughput.

The complexity analysis demonstrated that a significant
complexity reduction can be achieved using the proposed CRC-free system.
However, the achieved complexity reduction is proportional to the TPC code
rate and number of CRC bits. In addition
to the complexity reduction, the CRC-free system offers throughput
enhancement because the CRC bits can be replaced by other redundant bits to
support the error correction process, which can be observed at low SNR. It
is worth noting that the throughput enhancement is more apparent for short
TPC where the number of CRC bits is non-negligible compared to the codeword
size.



\section{TPC Error Detection}

The system model for TPC HARQ with CRC is the same as the one described in Chapter~\ref{chap:3}. However, when TPC self-detection is used, the CRC bits are replaced by information bits such that $\kappa=k^2$. Moreover, to simplify the discussion, we assume that a packet is composed of one TPC codeword. 

At the receiver side, the combiner output $\mathbf{R}_{C}$ is fed directly to the TPC decoder for SISO decoding \cite{pyndiah1998iterative}. If HIHO decoding is desired, the matrix $\mathbf{R}_{C}$ is converted
to a binary matrix $\mathbf{B}$ that is fed to the HIHO decoder, where $\ 
\mathbf{B}=0.5\left(\sign\left[ \mathbf{R}_{C}\right] +1\right) $ and 
$\sign(.)$ is the signum function.

\nomenclature{$\sign(.)$}{signum function}

The TPC decoder takes as input the matrix $\mathbf{R}_{C}$ or $\mathbf{B}$,
and performs a series of soft/hard decoding iterations to ultimately produce
the decoded binary $n\times n$ matrix $\mathbf{\widehat{D}}$. In the
iterative decoding of TPC, all rows of the code matrix are decoded
sequentially followed by column decoding. A full iteration corresponds to
the decoding of all set of rows and columns while a half iteration
corresponds to the decoding of either all the rows or all columns. All
row/column component codewords are decoded independently. The row/column
decoding in every iteration is performed using MLD when HIHO is considered,
and it is performed as described in \cite{pyndiah1998iterative} when SISO is
considered. The decoding process is terminated if the maximum number of
iterations is reached, or if all rows and columns are valid codewords of
their respective elementary codes. In the latter case, the decoding process
is declared successful. Consequently, error detection after the last column
decoding process can be performed by checking the syndromes of the first $k$
rows in $\mathbf{\widehat{D}}$. If all $k$ syndromes are equal to zero, then
the TPC packet is declared error-free; otherwise, the packet is declared
erroneous. There is no need to check the syndromes of the last $n-k$ rows
because they only contain parity bits.

\nomenclature{$\mathbf{\widehat{D}}$}{decoded binary $n\times n$ matrix}

In general, the last column decoding process guarantees that all columns in
the matrix are valid codewords. However, if the row/column decoding process
does not necessarily produce a valid codeword \cite{al2011closed}, then the
last two half iterations should be checked.

The transmission process is repeated until the packet is declared
error-free, or the maximum number of ARQ rounds $M$ is reached after which
the erroneous packet is dropped.

\section{Complexity Analysis}

\label{sec:complexity}

\subsection{Complexity Analysis of TPC Self Detection}

\label{sec:complx7}

\label{sec:complx} In this work, eBCH is used as the elementary code to produce the TPC code matrix. The eBCH
codes are cyclic codes which can be decoded using efficient algebraic
decoding algorithms such as the Berlekamp-Massey algorithm \cite%
{proakis1995digital,berlekamp1968algebraic,massey1969shift}.

Let $\mathbf{\widehat{d}}_{x}=[\widehat{d}_{x,1},\widehat{d}_{x,2},\cdots,%
\widehat{d}_{x,n}]$ refer to the $x$th row in $\mathbf{\widehat{D}}$ where $%
x=1,2,\cdots,n$. Therefore, assuming the last decoding process is
column-wise, then $\mathbf{\widehat{d}}_{x}=\mathbf{c}_{x}+\mathbf{e}_{x}$,
where $\mathbf{c}_{x}$ and $\mathbf{e}_{x}$ are the corresponding rows in
the transmitted binary matrix $\mathbf{C}$ and post-decoding error pattern
matrix $\mathbf{E}$, respectively. For error detection, the syndromes of the
rows in $\mathbf{\widehat{D}}$ can be obtained using:

\nomenclature{$\mathbf{\widehat{d}}_{x}$}{$x$th row in $\mathbf{\widehat{D}}$}
\nomenclature{$\mathbf{c}_{x}$}{$x$th row in $\mathbf{C}$}
\nomenclature{$\mathbf{e}_{x}$}{$x$th row in $\mathbf{E}$}
\nomenclature{$\mathbf{E}$}{post-decoding error pattern matrix}

\begin{itemize}
\item matrix multiplication $\mathbf{S}=\mathbf{\widehat{D}}\mathbf{H}^{\top}$
where $\mathbf{H}$ is $(n-k)\times n$ parity check matrix for the used eBCH
component code and $\{.\}^{\top}$ is the matrix transpose operation. Each row $\mathbf{s}%
_{x}$ in $\mathbf{S}$ corresponds to the syndrome of $\mathbf{\widehat{d}}%
_{x}$; or

\item polynomial division of $\widehat{d}_{x}(X)$ by $g(X)$ where $\widehat{d%
}_{x}(X)$ is the polynomial representation of $\mathbf{\widehat{d}}_{x}$ and 
$g(X)$ is an $(n-k)$-order generator polynomial. The syndrome polynomial $%
s_{x}(X)$ is the remainder of the polynomial division $s_{x}(X)=\Rem%
\left[ {\dfrac{\widehat{d}_{x}(X)}{g(X)}}\right]$. The coefficients of $%
s_{x}(X)$ correspond to the binary vector $\mathbf{s}_{x}$.
\end{itemize}

\nomenclature{$\mathbf{S}$}{syndrome matrix}
\nomenclature{$\mathbf{H}$}{parity check matrix}
\nomenclature{$\{.\}^{\top}$}{matrix transpose operation}
\nomenclature{$\mathbf{s}_{x}$}{$x$th row in $\mathbf{S}$}
\nomenclature{$\widehat{d}_{x}(X)$}{polynomial representation of $\mathbf{\widehat{d}}_{x}$}
\nomenclature{$g(X)$}{generator polynomial}
\nomenclature{$s_{x}(X)$}{syndrome polynomial}

To evaluate the computational complexity of TPC self-detection and CRC
detection, the number of operations carried out by each technique is
counted. For TPC error self-detection, the number of operations $N_\text{S}$
depends on the used eBCH code parameters $n$ and $k$. Moreover, $N_\text{S}$
varies randomly based on the index of the first row that has an error.
Because the row syndromes $\mathbf{s}_{x}$ are computed sequentially, once a
non-zero row syndrome is found the self-detection process is aborted and the
received matrix is declared erroneous.

\nomenclature{$N_\text{S}$}{number of XOR operations in TPC self-detection}
\nomenclature{$N_\text{S}^{\text{mtx}}$}{$N_\text{S}$ using matrix multiplication}
\nomenclature{$\mathbf{h}_{x}$}{$x$th row in $\mathbf{H}$}
\nomenclature{$\left\Vert . \right\Vert _{1}$}{Manhattan norm}
\nomenclature{$\hat{x}$}{index of row in which first error is discovered}
\nomenclature{$\mathbf{g}$}{vector representation of $g(X)$}
\nomenclature{$\nu$}{number of ones in $\mathbf{g}$}
\nomenclature{$N_\text{S}^{\text{lfsr}}$}{$N_\text{S}$ using LFSR}

Binary matrix multiplication can be implemented using only exclusive-OR
(XOR)\nomenclature{XOR}{Exclusive-OR} operations \cite[p. 340]{sklar2001digital}. Therefore, the number of
XOR operations required to perform TPC self-detection using matrix
multiplication is given by 
\begin{equation}
N_\text{S}^{\text{mtx}}=\hat{x}\left[ \sum_{x=1}^{n-k}\left(\left\Vert \mathbf{h}%
_{x}\right\Vert _{1}-1\right)\right]  \label{eq:Ns1}
\end{equation}%
where $\left\Vert \mathbf{h}_{x}\right\Vert _{1}$ is the Manhattan norm of
the $x$th row in $\mathbf{H}$ and $\hat{x}$ is the index of the row in which
the first error is discovered, $\hat{x}\in \{1,2,\cdots ,k\}$. If no error
is detected until the $k$th row, the self-detection process is aborted and
the TPC packet is declared error-free.

Alternatively, TPC self-detection can be performed using polynomial division
which is implemented using a low complexity $(n-k)$-stage LFSR. As reported
in \cite{G-Nguyen-2009}, polynomial division requires $k$ iterations/shifts
per row. Moreover, each shift has an average of $\frac{(2\nu-1)}{2}$ XOR
operations where $\nu=||\mathbf{g}||_{1}$ and $\mathbf{g}$ is the vector
representation of $g(X)$. Therefore, the number of XOR operations required
to perform one TPC self-detection using LFSR is 
\begin{equation}
N_\text{S}^{\text{lfsr}}=\tfrac{1}{2}(2\nu-1)k\hat{x}.  \label{eq:Ns2}
\end{equation}

\begin{figure}[H]
\begin{center}
\includegraphics[
]%
{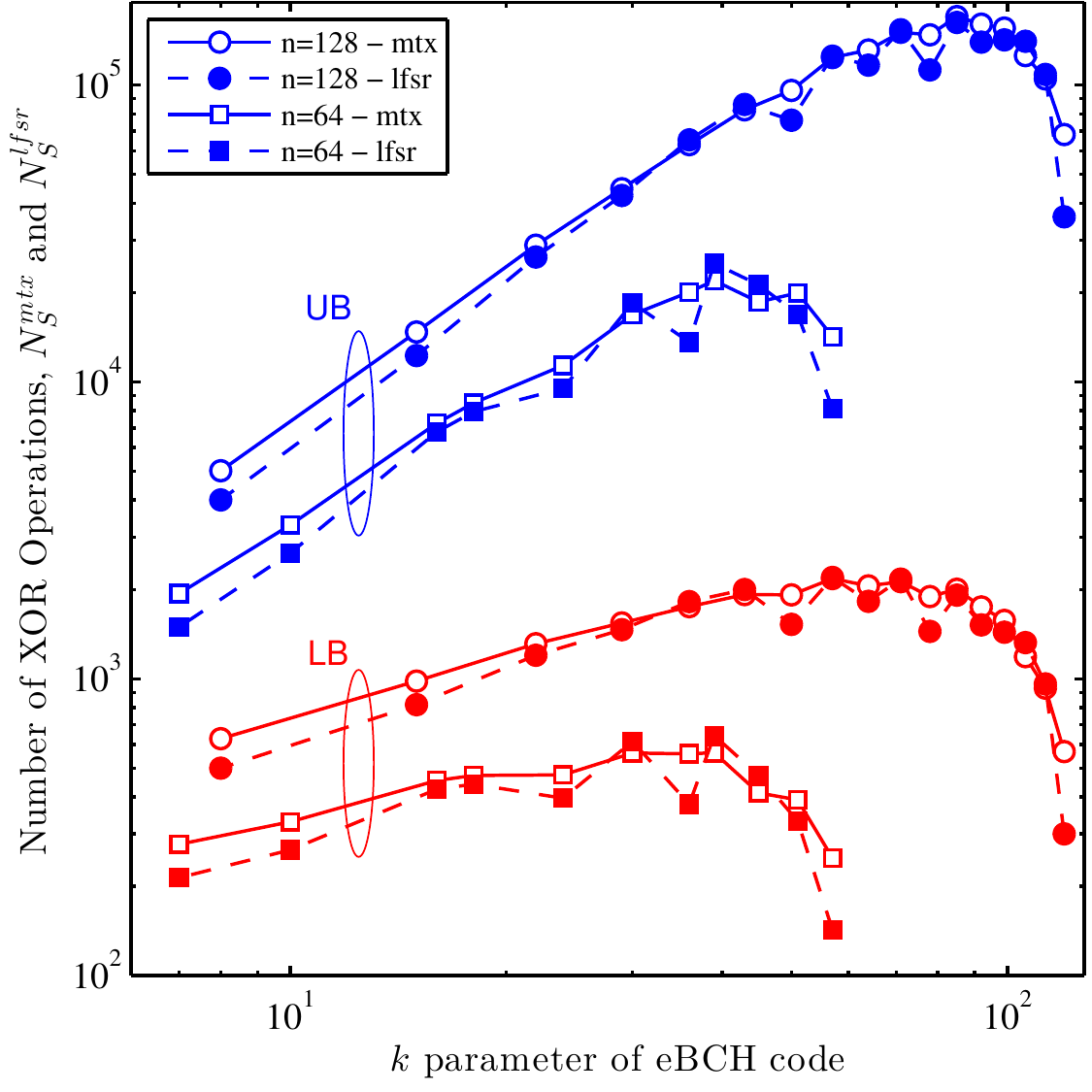}%
\caption{Upper and lower bounds on the complexity of the TPC self-detection
process for the eBCH$(128,120,4)^{2}$ and eBCH$(64,57,4)^{2}$.}%
\label{fig:mtx_vs_lfsr}%
\end{center}
\end{figure}

As it can be noted from (\ref{eq:Ns1}) and (\ref{eq:Ns2}), the complexity of
the TPC self-detection is a random variable. Therefore, expressing (\ref%
{eq:Ns1}) and (\ref{eq:Ns2}) numerically can be achieved using the average
value of $\hat{x}$, which requires the knowledge of the distribution of $\hat{x}$.
In TPC, the errors that cannot be corrected after performing a sufficient
number of iterations are the closed chains of errors \cite{al2011closed}.
The location, number and size of such chains are all random variables that
depend on the TPC code used and the SNR. Consequently, a general expression for the distribution of $\hat{x}$ is not easily obtained, and hence, we
use upper and lower bounds to indicate the system complexity, where 
\begin{equation}
\sum_{x=1}^{n-k}\left(\left\Vert \mathbf{h}_{x}\right\Vert _{1}-1\right)\leq N_\text{S}^{\text{%
mtx}}\leq k\left[ \sum_{x=1}^{n-k}\left(\left\Vert \mathbf{h}_{x}\right\Vert _{1}-1\right)%
\right]  \label{eq:ENs1}
\end{equation}%
and

\begin{equation}
\tfrac{1}{2}(2\nu-1)k\leq N_\text{S}^{\text{lfsr}}\leq\tfrac{1}{2}(2\nu-1)k^{2}\text{.%
}  \label{eq:ENs2}
\end{equation}
Fig.~\ref{fig:mtx_vs_lfsr} compares $N_\text{S}^{\text{mtx}}$ with $N_\text{S}^{\text{%
lfsr}}$ for all possible values of $k$ and $g(X)$ as tabulated in \cite[p. 466]%
{proakis1995digital} for $n$~=~128 and 64. The figure shows that the LFSR
implementation has a slight advantage over matrix multiplication. Therefore,
we adopt the LFSR implementation for TPC self-detection. In the figure, UB
and LB denote the upper and lower bound, respectively.

It is worth noting that matrix multiplication method has
higher computational complexity as compared to the serial implementation of
the LFSR, but it may have lower computational delay due to the fact that
matrix multiplication can naturally achieve high level of parallelism. In
the literature, the computational delay inherent in serial LFSR
implementation can be reduced by using parallel architectures at the cost of
additional complexity \cite{M.Ayinala-2011}.

\subsection{Complexity Analysis of CRC-Based Systems}

Using CRC codes that comprise of $16$ to $32$ bits is
generally sufficient to provide adequate PED accuracy for various
applications. As reported in \cite{K-Witzke-1985}, the misdetection
probability for a CRC code that consists of $l_\text{crc}$ bits approaches $%
2^{-l_\text{crc}}$ over binary symmetric channels (BSC)\nomenclature{BSC}{Binary Symmetric Channel} with large error
probabilities. Therefore, the impact of the CRC bits on the system
throughput might be negligible for packets with large number of bits.
However, the packet size is an important optimization parameter that depends
on the channel conditions \cite{zhou2006optimum}. In bursty traffic
conditions, transmitting small packet sizes is necessary to maximize the
system throughput and energy efficiency. Consequently, the value of $l_\text{%
crc}$ becomes the dominant factor that determines the system efficiency \cite%
{zhou2006optimum}.

The CRC-based TPC-HARQ system requires a number
of operations for error detection, denoted as $N_\text{C}$. CRC encoding and
decoding are also implemented using LFSR. The degree of the CRC generator
polynomial $\bar{g}(X)$ is $l_{\text{crc}}$; therefore, an $l_{\text{crc}}$%
-stage LFSR is used to implement CRC encoding or decoding. Polynomial
division for CRC detection requires $\kappa =k^{2}-l_{\text{crc}}$ shifts.
Moreover, each shift has an average of $(2\bar{\nu}-1)/2$ XOR operations where 
$\bar{\nu}=||\mathbf{\bar{g}}||_{1}$ and $\mathbf{\bar{g}}$ is the vector
representation of $\bar{g}(X)$. Therefore, the total $N_\text{C}$ including CRC
encoding and decoding operations is given by 
\begin{equation}
N_\text{C}=(2\bar{\nu}-1)\kappa .  \label{eq:Nc}
\end{equation}%
Unlike the TPC self-detection, it can be noted from (\ref{eq:Nc}) that $%
N_\text{C} $ is deterministic because it depends only on $\bar{g}(X)$ and $\kappa 
$.

\nomenclature{$N_\text{C}$}{number of XOR operations for CRC detection}
\nomenclature{$\bar{g}(X)$}{CRC generator polynomial}
\nomenclature{$\mathbf{\bar{g}}$}{vector representation of $\bar{g}(X)$}
\nomenclature{$\bar{\nu}$}{number of ones in $\mathbf{\bar{g}}$}
\nomenclature{$\mathcal{C}_\text{S}$}{relative complexity TPC self-detection to CRC detection}

Comparing the computational complexity of the TPC self-detection to other
CRC-based detection standards can be achieved using the relative complexity $%
\mathcal{C}_\text{S}$, where%
\begin{equation}
\mathcal{C}_\text{S}=\frac{N_\text{S}^{\text{lfsr}}}{N_\text{C}}\text{.}
\end{equation}%
Table~\ref{TableKey} presents upper and lower bounds of $\mathcal{C}_\text{S}$
for the codewords $(128,120,4)^{2}$, $(64,57,4)^{2}$, $(32,26,4)^{2}$ and $%
(16,11,4)^{2}$. The corresponding $g(X)$ for each codeword is obtained from 
\cite[p. 466]{proakis1995digital}. Moreover, three commonly used CRC codes,
namely, 16-bit CRC-8005, 16-bit CRC-8BB7, and 32-bit CRC-1EDC6F41 are used
in the comparison. It can be observed that TPC self-detection is remarkably
less complex than CRC detection. The complexity reduction is proportional to
the codeword size and number of CRC bits. For example, the eBCH$%
(128,120,4)^{2}$ TPC self-detection complexity is bounded by $3\times
10^{-3}\leq \mathcal{C}_\text{S}\leq 0.3575$ of the 16-bit CRC 8005 case, and it
is $6\times 10^{-4}\leq \mathcal{C}_\text{S}\leq 0.0715$ of 32-bit CRC 1EDC6F41
case. It is also worth noting that the CRC process requires additional
hardware for encoding and decoding unlike the TPC self-detection which uses
the existing error correction hardware for packet error detection.
Therefore, the elimination of CRC overhead can reduce the system complexity
and improve the system throughput, particularly for small packet sizes.

\begin{table*}[tbp] \centering
\caption{The relative complexity $\mathcal{C}_{S}$ of the TPC error detection versus
CRC error detection using three common CRC standards.}{\small
\begin{tabular}
[c]{|c|c|c|c|c|c|c|c|c|}\hline
{CRC} & \multicolumn{2}{c|}{$(128,120,4)^{2}$} &
\multicolumn{2}{c|}{$(64,57,4)^{2}$} & \multicolumn{2}{c|}{$(32,26,4)^{2}$} &
\multicolumn{2}{c|}{$(16,11,4)^{2}$}\\\cline{2-9}
& LB & UB & LB & UB & LB & UB & LB & UB\\\hline\hline
CRC 8005 & 0.0030 & 0.3575 & 0.0063 & 0.3589 & 0.0141 & 0.3658 & 0.0374 &
0.4116\\
CRC 8BB7 & 0.0010 & 0.1192 & 0.0021 & 0.1196 & 0.0047 & 0.1219 & 0.0125 &
0.1372\\
CRC 1EDC6F41 & 0.0006 & 0.0715 & 0.0013 & 0.0718 & 0.0028 & 0.0732 & 0.0075 &
0.0823\\\hline
\end{tabular}
}\label{TableKey}%
\end{table*}%

Although the obtained results demonstrate that TPC error
detection introduces substantial complexity reduction as compared to
conventional CRC techniques, more comparisons with the overall system
complexity are required to evaluate the significance of the gained
complexity reduction. The complexity of SISO and HIHO TPC decoders is mainly
determined by the number of hard decision decoding (HDD)\nomenclature{HDD}{Hard Decision Decoding} operations
performed \cite{al2011closed}. For SISO TPC, the number of HDD operations
per half iteration used to decode the received matrix $\mathbf{R}$ is $%
n\times 2^{p}$ where $p$ is the number of least reliable elements used to
generate the test patterns in the Chase-II decoder \cite{chase1972class}.
For HIHO decoding, the number of HDD operations per half iteration is $n$.
The complexity of each HDD operation depends on the adopted HDD algorithm.
For example, complexity analysis of BCH-HDD using Burton's algorithm is
given in \cite{Bajoga} where the computational complexity for each HDD
operation $N_{\text{HDD}}$ is bounded by%
\begin{equation}
45\lambda ^{2}n^{2}(\log _{10}n)^{2}<N_{\text{HDD}}<(45\lambda +4)\lambda
n^{2}(\log _{10}n)^{2}
\end{equation}%
where $\lambda =t/n$, and $t$ is the number of errors that can be corrected
by the code. By noting that $t=1$ for all the codes used in this work, the
total HDD complexity of SISO and HIHO TPC per half iteration is bounded by 
\begin{equation}
\left( n\times 2^{p}\right) N_{\text{HDD}}^{\min }<N_{\text{TPC}}^{\text{S}%
}<\left( n\times 2^{p}\right) N_{\text{HDD}}^{\max }
\end{equation}%
and 
\begin{equation}
n\text{ }N_{\text{HDD}}^{\min }<N_{\text{TPC}}^{\text{H}}<n\text{ }N_{\text{%
HDD}}^{\max }
\end{equation}%
for SISO and HIHO TPC, respectively.

\nomenclature{$p$}{number of least reliable elements in Chase-II decoder}
\nomenclature{$N_{\text{HDD}}$}{computational complexity for BCH HDD operation}
\nomenclature{$t$}{number of errors that can be corrected by a code}
\nomenclature{$N_{\text{TPC}}^{\text{S}}$}{HDD complexity of SISO TPC per half iteration}
\nomenclature{$N_{\text{TPC}}^{\text{H}}$}{HDD complexity of HIHO TPC per half iteration}

\begin{figure}[H]\centering
\includegraphics[]
{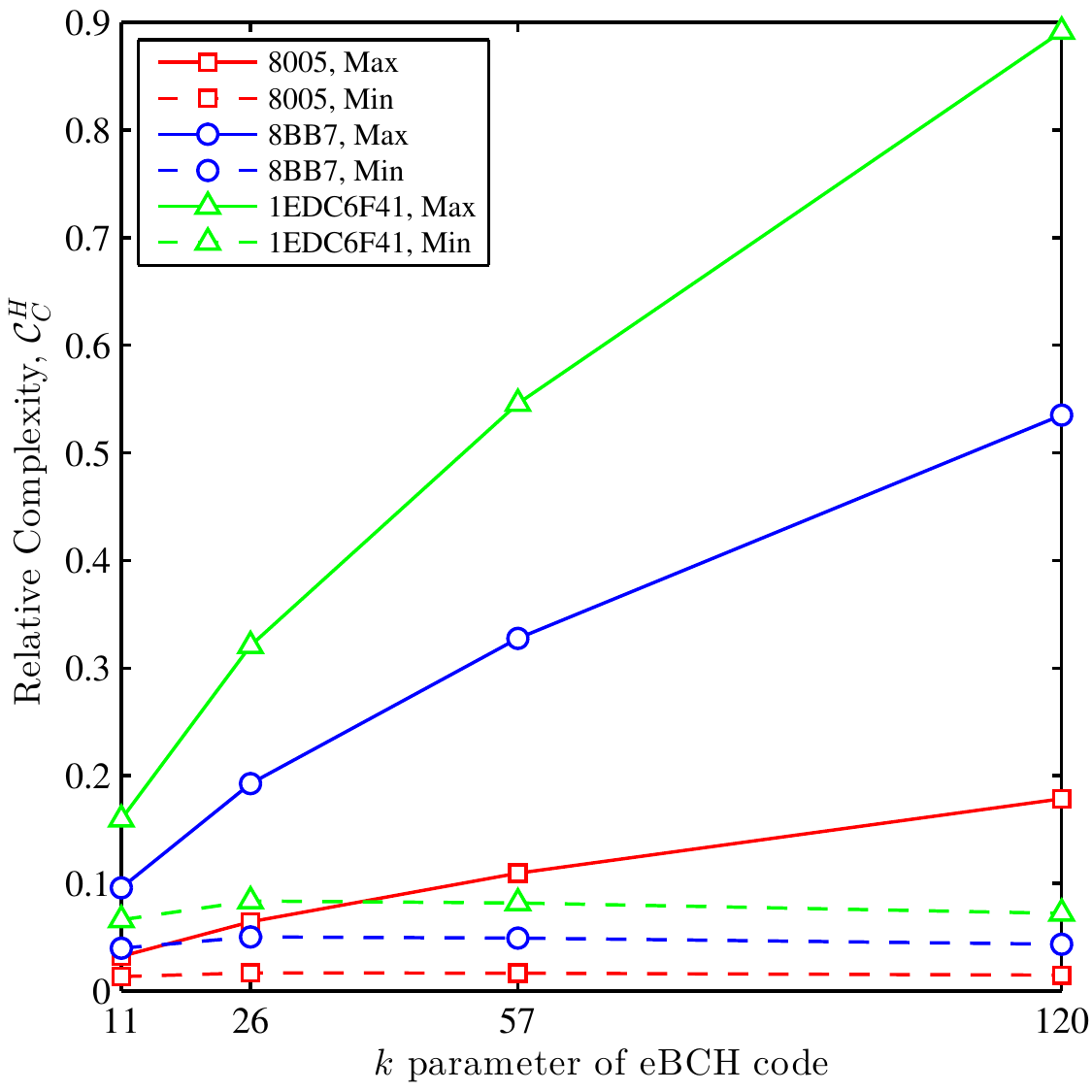}%
\caption{The relative complexity of the CRC error detection to TPC HIHO
decoding.}\label{F-T-comp-HIHO}%
\end{figure}

The error detection process is usually required to stop the iterative
decoding, and hence each half TPC decoding iteration is followed by an error
detection process. Therefore, the number of iterations and transmission
rounds can be ignored and the relative computational complexity of the CRC
detection to TPC decoding per half iteration is given by 
\begin{equation}
\frac{0.5N_\text{C}}{n\/N_\text{HDD}^{\max}}<\mathcal{C}_\text{C}^\text{H}<\frac{%
0.5N_\text{C}}{n\/N_\text{HDD}^{\min}}.  \label{E-T-COM-HIHO}
\end{equation}
Similarly, the relative complexity when SISO is used can be expressed as 
\begin{equation}
\frac{0.5N_\text{C}}{2^{p}\text{\/}n\/N_\text{HDD}^{\max}}<\mathcal{C}_\text{C}^\text{S%
}<\frac{0.5N_\text{C}}{2^{p}\text{\/}n\/N_\text{HDD}^{\min}}.
\end{equation}
The $\mathcal{C}_\text{C}^\text{H}$ given in (\ref{E-T-COM-HIHO}) is depicted in
Fig.~\ref{F-T-comp-HIHO} for the CRC 8005, 8BB7 and 1EDC6F41. The codes
considered are the eBCH $(128,120,4)^{2}$, $(64,57,4)^{2}$, $(32,26,4)^{2}$
and $(16,11,4)^{2}$. The $x$-axis in the figure represents the value of $k$
for these codes. As it can be noted from the figure, the CRC complexity can
have a significant impact on the overall system complexity. For example when
considering eBCH$(128,120,4)^{2}$, the complexity of the CRC-1EDC6F41 can be
as high as $0.89N_\text{TPC}^\text{H}$, and for the CRC-8BB7 it can be $%
0.54N_\text{TPC}^\text{H}$.

For SISO decoding, $\mathcal{C}_{C}^\text{S}=\mathcal{C}_{C}^\text{H}/2^{p}$%
, which implies that the significance of the CRC complexity is scaled by a
factor of $2^{p}$. Since typically $p\leq4$, then $\mathcal{C}_{C}^\text{S}$
will sustain non-negligible values. For example, $\mathcal{C}_{C}^\text{S}$
for the CRC-1EDC6F41 is upper bounded by $0.11$ and $0.055$ for $p=3$ and $4$%
, respectively.

\nomenclature{$\mathcal{C}_\text{C}^\text{H}$}{relative complexity of CRC detection to HIHO decoding}
\nomenclature{$\mathcal{C}_\text{C}^\text{S}$}{relative complexity of CRC detection to SISO decoding}

\section{Throughput using TPC Detection}

In practical HARQ systems, the error detection mechanism is not perfect
where some correct packets are falsely rejected and some erroneous packets
are misdetected. The ratio of falsely rejected packets to the total number
of received packets is defined as the false alarm probability $P_\text{F}$,
whereas, the ratio of misdetected errors to the total number of received
packets is defined as the misdetection probability $P_\text{M}$. The probability
that a packet is declared erroneous during the $i$th transmission round $%
\left( {\normalsize \tilde{P}_\text{E}^{(i)}}\right) ${\normalsize \ by the
receiver is therefore defined as 
\begin{equation}
\tilde{P}_\text{E}^{(i)}=P_\text{E}^{(i)}+P_\text{F}^{(i)}-P_\text{M}^{(i)}\text{.}
\label{E-Pb-1}
\end{equation}%
}As it can be noted from (\ref{E-Pb-1}), false rejections increase the error
probability $\tilde{P}_\text{E}$ and the number of retransmissions which degrades
the system throughput. On the other hand, misdetections reduce $\tilde{P}%
_\text{E} $ and allow erroneous packets to be forwarded to upper layers causing a
throughput inflation. Therefore, the HARQ throughput using TPC error
detection can be obtained by substituting (\ref{E-Pb-1}) in (\ref{E-eta-2}). %

\nomenclature{$P_\text{M}$}{misdetection probability}
\nomenclature{$P_\text{F}$}{false alarm probability}


\section{Numerical and Simulation Results}

\label{sec:Results7}

The performance of the CRC-based and CRC-free error detection
schemes is first evaluated in terms of false alarm rate (FAR)\nomenclature{FAR}{False Alarm Rate} and
misdetection rate (MDR)\nomenclature{MDR}{Misdetection Rate}. The FAR is the number of false alarms divided by
the number of correct packets, whereas, the MDR is the number of
misdetections divided by the number of erroneous packets. The FAR and MDR
are obtained using Monte-Carlo simulation of one-shot transmission to gain
an insight into the basic properties of both error detection schemes.
Different TPC codes are considered in the evaluation of the CRC-free and
CRC-based schemes where TPC self-detection and 16-bit CRC code are used,
respectively.

The FAR and MDR are given in Table~\ref{tab:FA} and~\ref{tab:MD},
respectively. The results are generated for different ${E_\text{b}}/{N_0}$ values in Rayleigh
fading channels. For each SNR value, the simulation is performed using $10^{6}$ transmitted packets. At low SNR values, the number of correctly received packets is not enough to reliably obtain the FAR especially for large packet sizes. On the other hand, for high SNR values, the number of erroneous packets is not enough to reliably obtained the MDR. Therefore, a hyphen `--' is used in the tables to indicate that the FAR or MDR cannot be obtained reliably using $10^{6}$ transmitted packets.

Table~\ref{tab:FA} shows that the FAR rate decreases as the channel SNR
increases. A false alarm happens when the information bits are correct but
the parity bits related to error detection are in error. Therefore, when the
channel SNR increases the probability of having an error in the parity bits
but not in the information bits reduces and the FAR subsequently decreases.
The table also shows that the CRC-based detection has lower FAR as compared
to the TPC self-detection, which is due to the fact that the
number of parity bits of the TPC is large and hence the probability of
error within these bits is large as well. 

\begin{sidewaystable}
\centering%
\renewcommand{\arraystretch}{0.725}
\caption{False alarm rate of TPC self-detection versus 16-bit CRC detection  in Rayleigh fading channels when SISO decoding is used and $M=1$. The hyphen `--' means that the number of error-free packets after transmitting $10^6$ packets is not enough to compute the FAR reliably. The symbol $\left\langle 10^{-5}\right\rangle$ means $\text{FAR} < 10^{-5}$.}
{\normalsize
\begin{tabular}{|c|c|c|c|c|c|c|c|c|}
\hline
{$E_{b}/N_{0}$ (dB)} & \multicolumn{2}{c|}{$(16,11,4)^{2}$} & \multicolumn{2}{c|}{$(32,26,4)^{2}$} & \multicolumn{2}{c|}{$(64,57,4)^{2}$}	& \multicolumn{2}{c|}{$(128,120,4)^{2}$} \\ \cline{2-9}	
	 & TPC 								  & CRC 								 & TPC 								 	& CRC 								 & TPC 								  & CRC 								 & TPC 								 	& CRC \\ \hline
0  & $3.26\times 10^{-1}$ & -- 									 & -- 									& -- 								 	 & -- 									& -- 									 & -- 									& --\\ \hline
1  & $2.85\times 10^{-1}$ & $1.33\times 10^{-1}$ & -- 									& -- 								 	 & -- 								 	& -- 								   & -- 								 	& --\\ \hline
2  & $1.37\times 10^{-1}$ & $8.22\times 10^{-2}$ & -- 									& -- 								 	 & -- 									& -- 									 & -- 									& --\\ \hline
3  & $7.52\times 10^{-2}$ & $4.17\times 10^{-2}$ & -- 									& -- 								 	 & -- 									& -- 									 & -- 									& --\\ \hline
4  & $2.81\times 10^{-2}$ & $1.82\times 10^{-2}$ & -- 									& -- 									 & -- 									& -- 									 & -- 									& -- \\ \hline
5  & $7.40\times 10^{-3}$ & $6.54\times 10^{-3}$ & $6.05\times 10^{-2}$ & $2.26\times 10^{-3}$ & -- 									& -- 									 & -- 									& -- \\ \hline
6  & $9.86\times 10^{-3}$ & $1.41\times 10^{-3}$ & $1.80\times 10^{-2}$ & $8.35\times 10^{-4}$ & -- 									& -- 									 & -- 									& -- \\ \hline
7  & $9.01\times 10^{-5}$ & $1.66\times 10^{-4}$ & $2.93\times 10^{-3}$ & $1.29\times 10^{-4}$ & -- 									& -- 									 & -- 									& -- \\ \hline
8  & $\left\langle 10^{-5}\right\rangle$ 			 	& $\left\langle 10^{-5}\right\rangle$ 			 & $1.40\times 10^{-4}$ & $\left\langle 10^{-5}\right\rangle$ 			 	 & $2.07\times 10^{-2}$ & $9.38\times 10^{-5}$ & -- 									& -- \\ \hline
9  & $\left\langle 10^{-5}\right\rangle$ 			 	& $\left\langle 10^{-5}\right\rangle$ 			 	 & $\left\langle 10^{-5}\right\rangle$ 			 	& $\left\langle 10^{-5}\right\rangle$ 			 	 & $3.95\times 10^{-3}$ & $1.05\times 10^{-5}$ & -- 									& -- \\ \hline
10 & $\left\langle 10^{-5}\right\rangle$ 			 	& $\left\langle 10^{-5}\right\rangle$ 			 	 & $\left\langle 10^{-5}\right\rangle$ 			 	& $\left\langle 10^{-5}\right\rangle$ 			 	 & $5.21\times 10^{-5}$ & $\left\langle 10^{-5}\right\rangle$ 			 	 & -- 									& -- \\ \hline
11 & $\left\langle 10^{-5}\right\rangle$ 			 	& $\left\langle 10^{-5}\right\rangle$ 			 	 & $\left\langle 10^{-5}\right\rangle$ 			 	& $\left\langle 10^{-5}\right\rangle$ 			 	 & $\left\langle 10^{-5}\right\rangle$ 			 	& $\left\langle 10^{-5}\right\rangle$ 			 	 & $1.43\times 10^{-2}$ & -- 			 	\\ \hline
12 & $\left\langle 10^{-5}\right\rangle$ 			 	& $\left\langle 10^{-5}\right\rangle$ 			 	 & $\left\langle 10^{-5}\right\rangle$ 			 	& $\left\langle 10^{-5}\right\rangle$ 			 	 & $\left\langle 10^{-5}\right\rangle$ 			 	& $\left\langle 10^{-5}\right\rangle$ 			 	 & $4.34\times 10^{-4}$ & $\left\langle 10^{-5}\right\rangle$ 			 	\\ \hline
13 & $\left\langle 10^{-5}\right\rangle$ 			 	& $\left\langle 10^{-5}\right\rangle$ 			 	 & $\left\langle 10^{-5}\right\rangle$ 			 	& $\left\langle 10^{-5}\right\rangle$ 			 	 & $\left\langle 10^{-5}\right\rangle$ 			 	& $\left\langle 10^{-5}\right\rangle$ 			 	 & $\left\langle 10^{-5}\right\rangle$ 			 	& $\left\langle 10^{-5}\right\rangle$ 			 	\\ \hline
\end{tabular}
}\label{tab:FA}%
\end{sidewaystable}

Table~\ref{tab:MD} shows that the MDR of TPC self-detection is smaller than
16-bit CRC code for all SNR values when large TPC are used such as $%
(128,120,4)^{2}$. For smaller TPC code sizes, the MDR of TPC self-detection
is smaller than CRC detection only at low SNR values. The MDR of TPC
self-detection deteriorates as the TPC code size is reduced. Small TPC
codewords have less number of component codes involved in TPC self-detection
as compared to larger TPC codewords. Therefore, the joint error detection
capability of these component codes decreases when smaller TPC codewords are
used.

\begin{sidewaystable}
\centering
\renewcommand{\arraystretch}{0.725}
\caption{Misdetection rate of TPC self-detection versus 16-bit CRC detection in Rayleigh fading channels when SISO decoding is used and $M=1$. The hyphen `--' means that the number of erroneous packets after transmitting $10^6$ packets is not enough to compute the MDR reliably. The symbol $\left\langle 10^{-5}\right\rangle$ means $\text{MDR} < 10^{-5}$.}{\normalsize 
\begin{tabular}{|c|c|c|c|c|c|c|c|c|}
\hline
{$E_{b}/N_{0}$ (dB)} & \multicolumn{2}{c|}{$(16,11,4)^{2}$} & \multicolumn{2}{c|}{$(32,26,4)^{2}$} & \multicolumn{2}{c|}{$(64,57,4)^{2}$}& \multicolumn{2}{c|}{$(128,120,4)^{2}$} \\ \cline{2-9}
	 & TPC 									& CRC 								 & TPC 									& CRC 								 & TPC 									& CRC 								 & TPC 									& CRC 								\\ \hline
0  & $1.19\times 10^{-4}$ & $2.30\times 10^{-5}$ & $\left\langle 10^{-5}\right\rangle$ 			 	& $1.30\times 10^{-5}$ & $\left\langle 10^{-5}\right\rangle$ 			 	& $1.50\times 10^{-5}$ & $\left\langle 10^{-5}\right\rangle$ 			 	& $2.00\times 10^{-5}$\\ \hline
1  & $5.62\times 10^{-4}$ & $1.40\times 10^{-5}$ & $\left\langle 10^{-5}\right\rangle$ 			 	& $1.30\times 10^{-5}$ & $\left\langle 10^{-5}\right\rangle$ 			 	& $2.20\times 10^{-5}$ & $\left\langle 10^{-5}\right\rangle$ 			 	& $1.10\times 10^{-5}$ \\ \hline
2  & $4.34\times 10^{-3}$ & $1.11\times 10^{-5}$ & $\left\langle 10^{-5}\right\rangle$ 			 	& $1.70\times 10^{-5}$ & $\left\langle 10^{-5}\right\rangle$ 			 	& $1.00\times 10^{-5}$ & $\left\langle 10^{-5}\right\rangle$ 			 	& $1.10\times 10^{-5}$\\ \hline
3  & $1.71\times 10^{-2}$ & $2.75\times 10^{-5}$ & $\left\langle 10^{-5}\right\rangle$ 			 	& $1.90\times 10^{-5}$ & $\left\langle 10^{-5}\right\rangle$ 			 	& $1.00\times 10^{-5}$ & $\left\langle 10^{-5}\right\rangle$ 			 	& $1.80\times 10^{-5}$ \\ \hline
4  & $4.73\times 10^{-2}$ & $2.35\times 10^{-5}$ & $\left\langle 10^{-5}\right\rangle$ 			 	& $1.90\times 10^{-5}$ & $\left\langle 10^{-5}\right\rangle$ 			 	& $2.10\times 10^{-5}$ & $\left\langle 10^{-5}\right\rangle$ 			 	& $2.10\times 10^{-5}$ \\ \hline
5  & $1.00\times 10^{-1}$ & $2.65\times 10^{-5}$ & $2.04\times 10^{-4}$ & $2.23\times 10^{-5}$ & $\left\langle 10^{-5}\right\rangle$ 			 	& $1.40\times 10^{-5}$ & $\left\langle 10^{-5}\right\rangle$ 			 	& $1.20\times 10^{-5}$ \\ \hline
6  & $1.80\times 10^{-1}$ & $1.78\times 10^{-5}$ & $3.26\times 10^{-3}$ & $1.29\times 10^{-5}$ & $\left\langle 10^{-5}\right\rangle$ 			 	& $1.70\times 10^{-5}$ & $\left\langle 10^{-5}\right\rangle$ 			 	& $1.60\times 10^{-5}$ \\ \hline
7  & $3.74\times 10^{-1}$ & -- 									 & $1.90\times 10^{-2}$ & $1.69\times 10^{-5}$ & $\left\langle 10^{-5}\right\rangle$ 			 	& $1.80\times 10^{-5}$ & $\left\langle 10^{-5}\right\rangle$ 			 	& $2.00\times 10^{-5}$ \\ \hline
8  & $4.63\times 10^{-1}$ & -- 									 & $8.41\times 10^{-2}$ & -- 									 & $\left\langle 10^{-5}\right\rangle$ 			 	& $1.42\times 10^{-5}$ & $\left\langle 10^{-5}\right\rangle$ 			 	& $1.60\times 10^{-5}$ \\ \hline
9  & -- 									& -- 									 & $3.64\times 10^{-1}$ & -- 									 & $4.73\times 10^{-4}$ & $4.17\times 10^{-5}$ & $\left\langle 10^{-5}\right\rangle$ 			 	& $1.70\times 10^{-5}$ \\ \hline
10 & -- 									& -- 									 & -- 									& -- 									 & $1.78\times 10^{-2}$ & -- 									 & $\left\langle 10^{-5}\right\rangle$ 			 	& $1.30\times 10^{-5}$ \\ \hline
11 & -- 									& -- 									 & -- 									& -- 									 & -- 									& -- 									 & $\left\langle 10^{-5}\right\rangle$ 			 	& $1.62\times 10^{-5}$ \\ \hline
12 & -- 									& -- 									 & -- 									& -- 									 & -- 									& -- 									 & -- 									& -- 									 \\ \hline
13 & -- 									& -- 									 & -- 									& -- 									 & -- 									& -- 									 & --  									& -- 									\\ \hline
\end{tabular}
}\label{tab:MD}%
\end{sidewaystable}

Moreover, for small TPC codeword sizes, we observe that the MDR of TPC
self-detection increases as the SNR increases. At low SNR values, the number
of errors in the decoded TPC is high with various patterns; therefore,
the probability of having undetected errors in all component codewords is
small. As the channel SNR increases, the number of errors decreases;
however, the errors tend to have closed chain patterns. These types of
errors are likely to be misdetected in all affected component codewords if
the error pattern is larger than their individual error detection
capability. Hence, the TPC misdetection probability increases. However, it
should be noted that at high SNR the probability of having an erroneous
packet is low which alleviates the effect of high MDR on the system
throughput. On the other hand, the MDR of CRC detection is almost constant
and matches the theoretical approximation $2^{-16}\approx 1.53\times 10^{-5}$%
.

The HARQ system with MRC is simulated for different TPC codeword sizes and code rates. The packet is
TPC encoded with code eBCH$(n,k,d_{\text{min}})^{2}$. For the CRC-based
HARQ, 16-bit CRC 8005 code is used before encoding with TPC for error
correction. Moreover, Rayleigh fading is assumed as described in Chapter~\ref%
{chap:3}. The maximum number of ARQ rounds per packet is $M=4$. For each
simulation run (i.e. for a given SNR value) $1000$ packets are transmitted.
The TPC decoder is configured to perform a maximum of four iterations.
Moreover, the number of reliability bits for the Chase-II decoder \cite%
{chase1972class} is set to 4 in the SISO decoder \cite{pyndiah1998iterative}%
. 

\begin{figure}[H]
\begin{center}
\includegraphics[
]%
{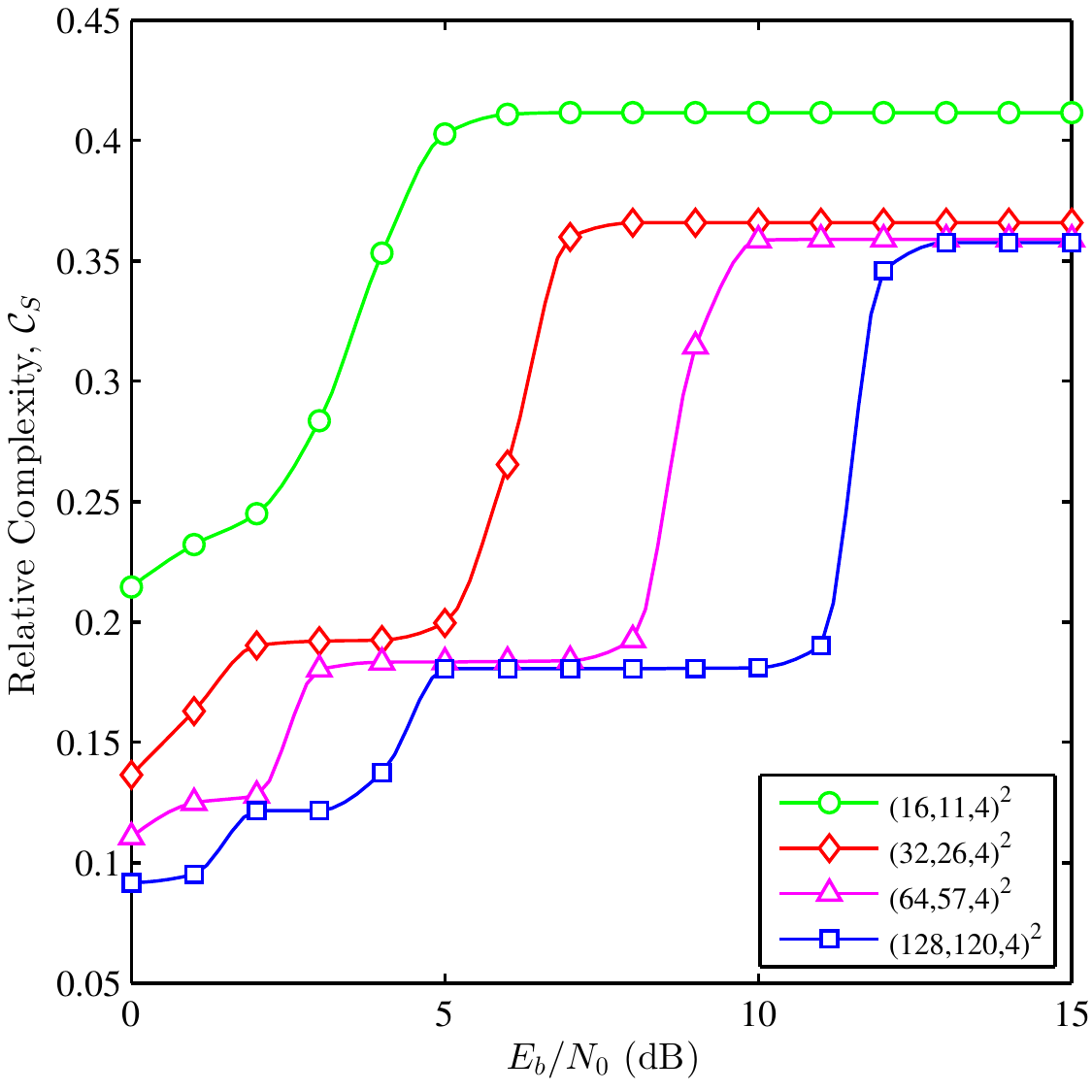}%
\caption{Simulated relative complexity of TPC detection to 8005 CRC detection for HARQ with SISO decoding and $M=4$ in Rayleigh fading channels.}%
\label{fig:Cr_vs_snr}%
\end{center}
\end{figure}

Fig.~\ref{fig:Cr_vs_snr} shows the simulated relative complexity of TPC
detection to CRC 8005 for a range of SNR values in Rayleigh fading channels. The relative complexity
decreases as the SNR decreases. That is because the number of bit errors
becomes higher at low SNR and an error is more likely to be detected at the
beginning of the TPC codeword. The figure also shows that the relative
complexity decreases as the codeword size increases. Fig.~\ref{fig:Cr_vs_snr}
and Table~\ref{TableKey} show the complexity advantage of TPC self-detection
over CRC-based detection.


\begin{figure}[H]
\begin{center}
\includegraphics[
]%
{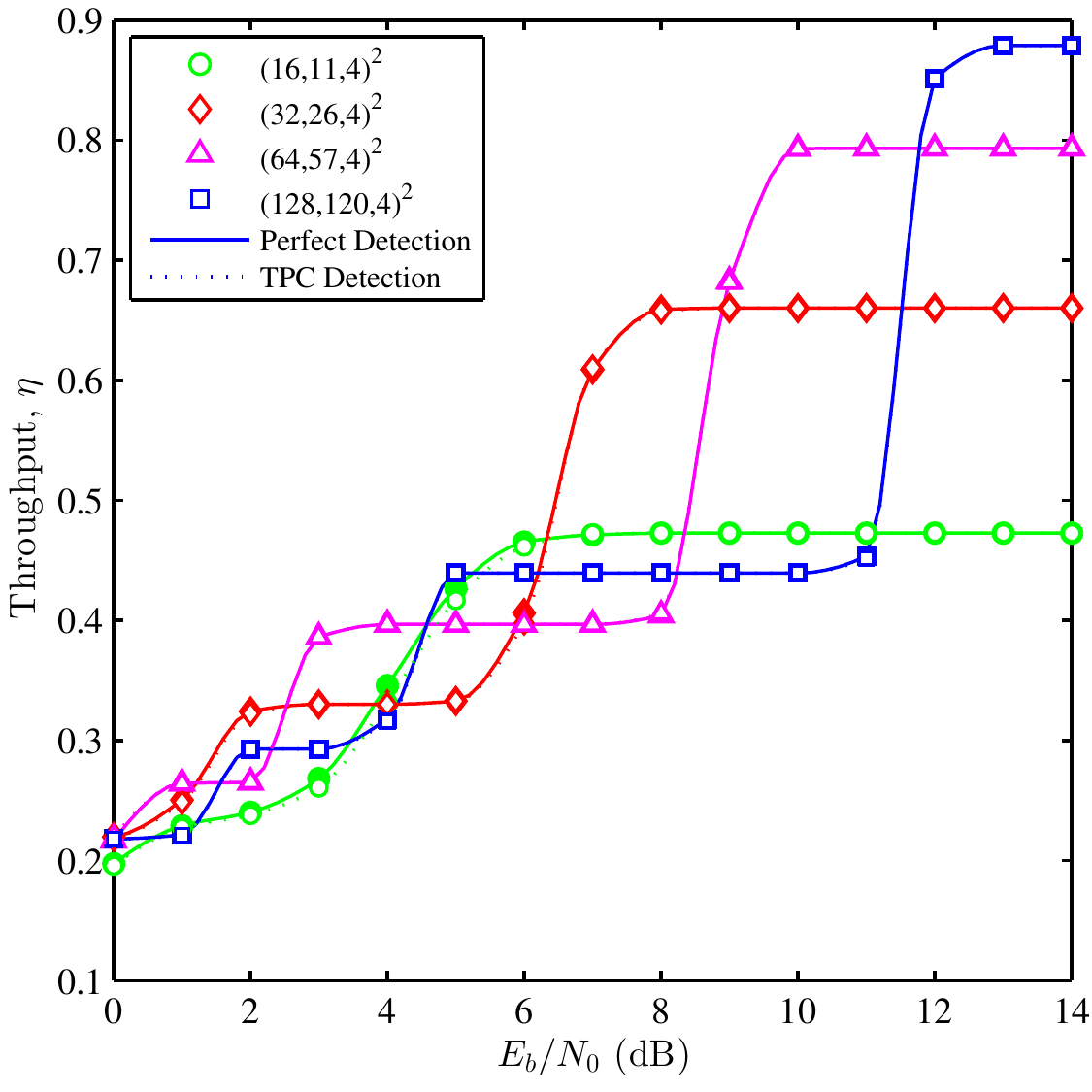}%
\caption{TPC-HARQ throughput using perfect detection and TPC detection with SISO decoding and $M=4$ in Rayleigh fading channels.}%
\label{fig:virt_vs_eff}%
\end{center}
\end{figure}

Fig.~\ref{fig:virt_vs_eff} compares the system throughput when
perfect detection and TPC self-detection are used. The TPC-HARQ system is
simulated for various TPC codes with SISO decoding in Rayleigh fading channels. The figure shows that
the throughput with TPC self-detection is almost the same as the throughput
with perfect detection which means that the misdetections and false alarms
in TPC self-detection has negligible effect on the system throughput.

\begin{figure}[H]
\begin{center}
\includegraphics[
]%
{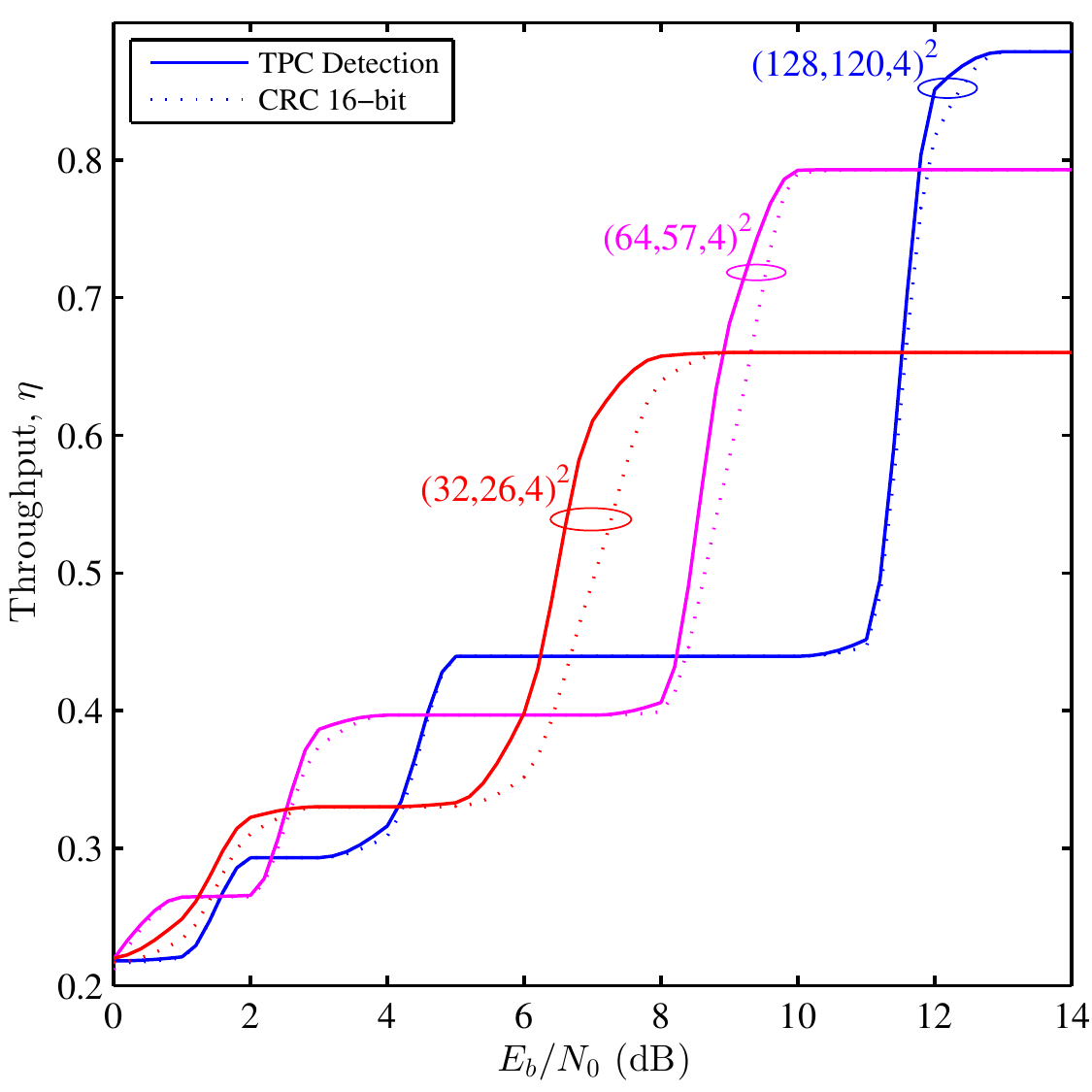}%
\caption{TPC and CRC throughput using equal numbers of redundant bits with SISO decoding and $M=4$ in Rayleigh fading channels.}%
\label{F-Thr_equ_cod_rate}%
\end{center}
\end{figure}

In addition to the complexity advantage of TPC self-detection
over CRC-based detection, TPC self-detection has an advantage in
terms of system throughput. Fig.~\ref{F-Thr_equ_cod_rate} compares the throughput of the
CRC-based and CRC-free HARQ using equal code rates in both systems. In CRC
systems, the redundant bits are composed of the CRC bits and the parity bits
added by the TPC encoder. Therefore, to obtain equal code rates in both systems,
the number of TPC parity bits should be reduced (punctured) by the number of
CRC bit multiplied by the inverse of the code rate. The simulated
throughput using equal code rates is depicted in Fig.~\ref{F-Thr_equ_cod_rate}
for relatively long TPC. The number of bits punctured is $24$, $20$ and $18$
for the eBCH$(32,26,4)^{2}$, eBCH$(64,57,4)^{2}$ and eBCH$(128,120,4)^{2}$,
respectively. As it can be noted from the figure, the throughput with TPC
detection is higher than the CRC-based system. However, the difference
becomes very small as the TPC code size increases and almost vanishes at
high SNRs. Such behavior is expected because the puncturing process reduces
the error correction capability of the TPC.

\begin{figure}[H]
\begin{center}
\includegraphics[
]%
{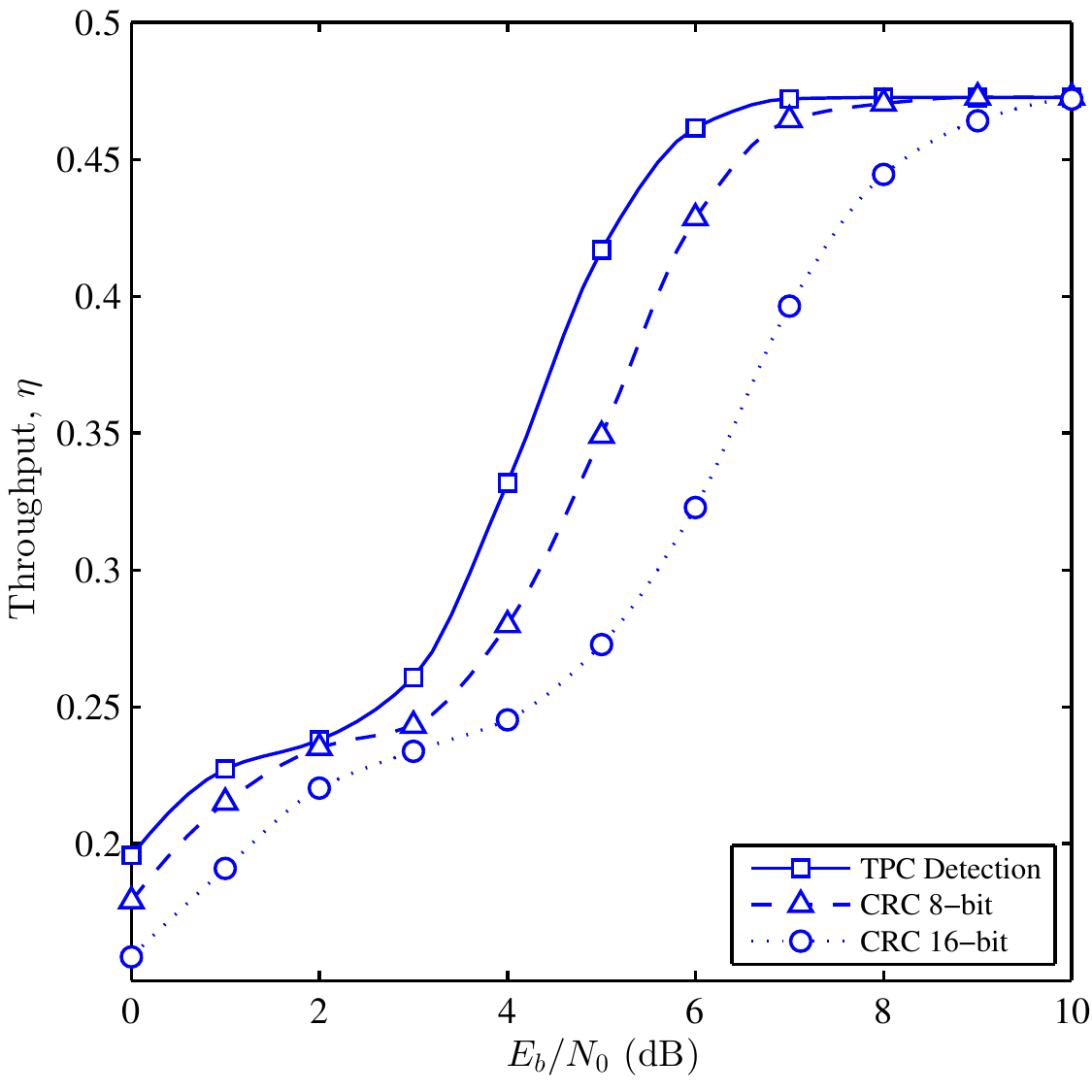}%
\caption{Throughput of eBCH$(16,11,4)^{2}$ with SISO decoding and $M=4$ in Rayleigh fading channels using TPC detection and CRC with 8 and 16 bits. Both systems have equal number of redundant bits.}%
\label{F-Thr_16_11}%
\end{center}
\end{figure}

For short TPC codes, the impact of the CRC bits length is more significant,
and hence, smaller numbers of CRC bits can be employed. However, smaller
number of CRC bits might result in unreliable performance because of the
limited capability of such codes to detect the errors. Moreover, the
puncturing process has a more significant effect on the error correction
capability of the TPC, which is exhibited as throughput reduction at low
SNRs. Fig.~\ref{F-Thr_16_11} shows the throughput of the eBCH$(16$,$11$,$%
4)^{2}$ TPC using $16$ and $8$-bit CRC using equal code rates, where the
number of punctured bits is $34$ and $17$ for the CRC with $16$ and $8$
bits, respectively. As it can be noted from the figure, the throughput results of
the CRC-free and CRC-based systems converge to the same value at high SNRs.
However, the puncturing process reduces the throughput at low SNR because
the decoding process is less effective in terms of error correction.
Overall, using CRC-8 is a reasonable compromise.%

\section{Conclusions}

\label{sec:Conclusions7}
 
A CRC-free HARQ scheme based on TPC error correction and self-detection is proposed and evaluated in this chapter.
The TPC inherent error detection provides additional degrees of freedom by
eliminating the CRC overhead. The false alarm rate, misdetection rate and transmission efficiency are
evaluated using extensive Monte Carlo simulations. The obtained
results show that the CRC-free system consistently provides equivalent or
higher throughput than the CRC-based HARQ with a noticeable advantage when
small packet sizes are used. However, it was observed that the misdetection rate for short codeword lengths increases at high SNR. Although such performance should not be of significant impact for many applications because the probability of packet error is small at high SNR, for applications which do not tolerate error misdetection, CRC-based transmission might be preferred. In addition, the numerical results show that the
TPC self-detection has lower computational complexity than CRC detection
especially for TPC with high code rates. Unlike CRC detection, the
complexity of TPC self-detection scales with the location of error where
significant speedup is achieved when errors are located at the beginning of
the TPC packet which is usually the case at low SNR.

\chapter{Content-Aware and Occupancy-Based Adaptive Hybrid ARQ for Video Transmission}
\label{chap:8}

In this chapter, we propose a cross-layer link adaptation solution where the receiver controls the HARQ ACK messages based on the error detection, video content and playback deadline. The adaptation algorithm utilizes a delay model to estimate the transmission time of a video frame and its delivery budget time. Accordingly, the ARQ limit is adapted based on the playback buffer occupancy and importance of video packets where erroneously received packets may be falsely acknowledged to ensure continuous playback. The performance of the adaptive HARQ is evaluated with and without subpacket fragmentation. Subpacket-based HARQ outperforms packet-based HARQ in terms of delay and continuity of video playback. Moreover, the adaptation algorithm further improves the delay performance in the video transmission system and eliminates buffer starvation with minimum spatial quality distortion.  

In the previous chapters, the performance of HARQ with TPC is evaluated using semi-analytical models and extensive simulations. The obtained results show that TPC-HARQ is a suitable candidate to meet the strict QoS constraints of video transmission in terms of high transmission efficiency, bounded delay, and reduced power consumption while not ignoring the low complexity consideration. However, similar to other link adaptation techniques, TPC-HARQ may result in undesirable QoE if it is used without regard to the characteristics of video bitstreams. Video packets have unequal importance in terms of the impact on video quality distortion. Link adaptation techniques should understand the content of the transmitted stream in order to allocate the available system resources to the most important packets. In addition, the continuity of video playback is another important metric which affects QoE. Therefore, the ultimate objective of link adaptation schemes for video transmission should be the timely delivery of the most important packets within the available playback budget time. 



\section{Related Work}

Many solutions have been proposed in the literature to optimize the transmission of delay-sensitive video. Examples of these solutions are \cite{atzori2007cycle,argyriou2008cross,harsini2009adp,bobarshad2010low,bobarshad2012any}. These solutions consider the bounded delay requirement for video transmission, but ignore the unequal importance of video packets. For example, the authors in~\cite{atzori2007cycle} proposed a rate control approach for video streaming over wireless channels. The wireless channel is characterized by an arguable two-state channel model that provides a coarse approximation of the channel behavior. The source rate and channel code parameters are adaptively computed in a cycle basis subject to a constraint on the probability of starvation at the playback buffer. However, the unequal importance of video content is not considered in their video transmission scheme. In \cite{argyriou2008cross} and \cite{harsini2009adp} adaptive transmission schemes are proposed to minimize the packet loss rate for delay-sensitive application. The proposed schemes consider the loss of packets due to transmission errors and delay-bound violation, but ignores the unequal loss impact of packets on video quality. Moreover, the delay and packet loss rate performance of hybrid ARQ schemes has been studied in \cite{bobarshad2010low} and \cite{bobarshad2012any}, but independent of the video content.

Content-aware link adaptation solutions for video transmission have been proposed in the literature to enhance the performance of wireless networks which traditionally treat transmitted content with equal importance. In \cite{chou2006rate} an optimized rate-distortion packetization scheme is proposed to choose the best policy for what, when and how media packets are transmitted. The paper provides an excellent problem formulation for minimizing the distortion subject to rate constraint in packetized video streaming systems. However, the optimization algorithm requires offline computation of video packets importance in terms of distortion impact. Obtaining such information accurately is a complex task due to the inter-dependency between video packets.

In \cite{hassan2007vso} the authors introduced two channel adaptive rate control schemes for slowly and fast varying channels. Both schemes adjust the source rate by scaling down video frames in size to satisfy playback deadlines. Scaling is only performed on B frames to minimize the distortion impact due to source rate reduction. The authors assumed conventional Stop-and-Wait ARQ in their proposed video streaming system. While this is an acceptable assumption in wireless environments with small RTT, it is typically not a plausible one for wireless networks with large RTT. An adaptive transmission scheme is proposed in \cite{mukhtar2011occupancy} to enhance video streaming over wireless channels by integrating scalable video coding, adaptive channel coding and adaptive modulation. The proposed scheme monitors the playback buffer occupancy through a feedback channel and implements frame-type-dependent scaling at the transmitter to prevent potential playback discontinuities. The amount of scaling is inversely proportional to the importance of video frames.

Moreover, the authors in \cite{khalek2012cross} propose a cross-layer design to maximize video QoE by jointly selecting the optimal source rate and implementing unequal error protection subject to delay and PER constraints. In addition, an energy and content-aware multi-homing video transmission scheme is proposed in \cite{ismail2013energy}. The objective of the proposed scheme is to maximize video quality subject to delay and energy constraints by optimizing the power allocation for different radio interfaces and scheduling the most important packets for transmission while dropping other less important packets if necessary. However, the proposed algorithm did not consider transmission errors and retransmissions. Other examples of content-aware and delay-constrained link adaptation solutions are described in \cite{schaar2003adaptive,schar2007cross,chen2004multi,li2004providing,li2008content,chen2010cross}. 

Most link adaptation solutions for wireless video communication available in the literature assume traditional Type-I HARQ with no subpacketization and packet combining. As discussed in Chapter \ref{chap:3} subpacketization and packet combining provide significant gains in delay and throughput performance. A content-aware packet fragmentation scheme is proposed in \cite{kambhatla2012wireless} to enhance the quality of video transmission. However, the proposed scheme assumes neither retransmission nor channel coding.

In this chapter, we propose a content-aware and occupancy-based adaptive HARQ scheme (A-HARQ) to ensure continuous video playback with minimum quality distortion. The proposed algorithm estimates the transmission time of video frames based on the channel condition and size of video frames. The delay analysis is provided for subpacket-based and packet-based HARQ. The ARQ limit is adapted on a frame-by-frame basis to avoid playback buffer starvation. The receiver falsely acknowledges erroneous video packets if they are expected to be successfully delivered beyond their budget time. In the adaptation algorithm, the delivery budget time is set differently based on the importance of video frames. The adaptive solution is implemented at the receiver which facilitates accurate tracking of the channel condition and playback buffer occupancy.

\section{Adaptive Video Transmission Algorithm}
\label{sec:proposed8}

\begin{figure}[H]
\centering
\includegraphics[angle=270,width=\textwidth]{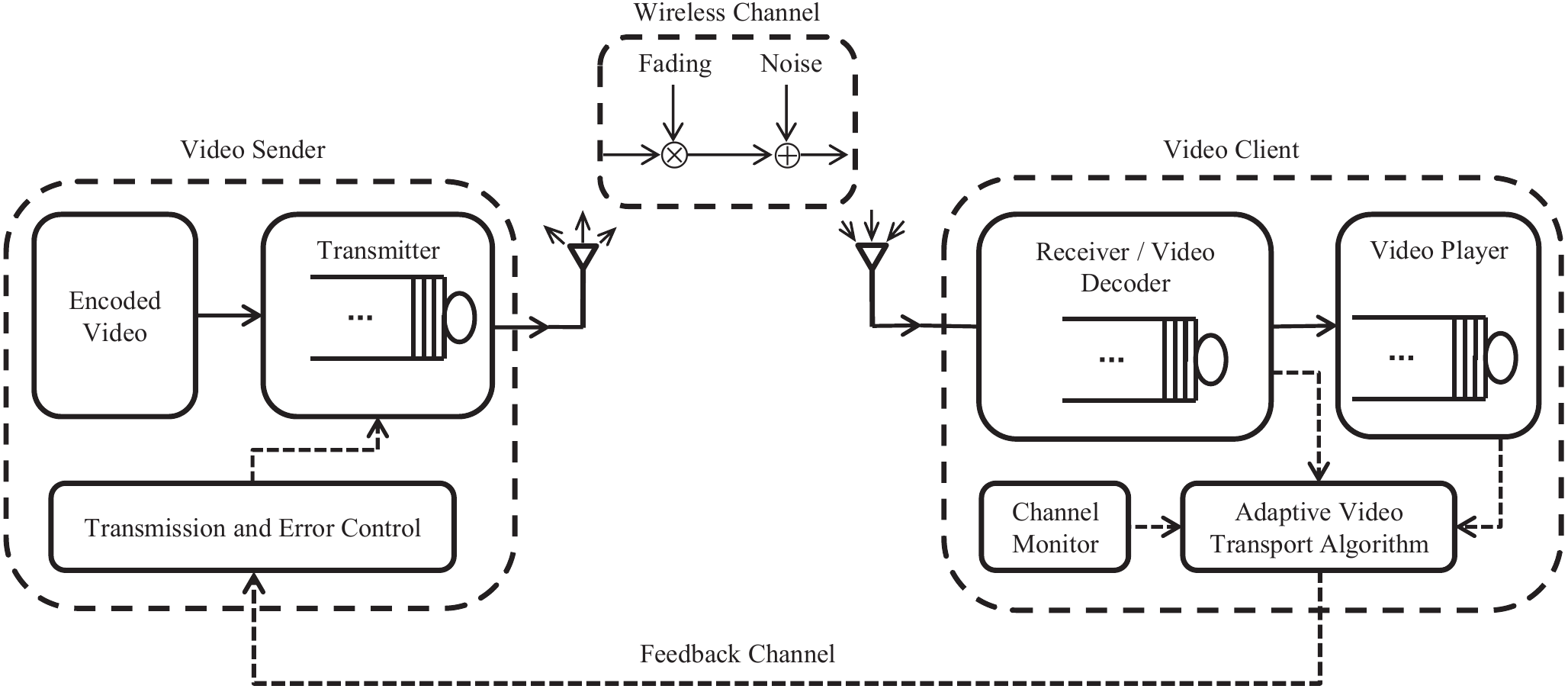}
\caption{Adaptive wireless video transmission system overview.}
\label{fig:sysModel}
\end{figure}

Fig.~\ref{fig:sysModel} provides an overview of the proposed adaptive video transmission system. The receiver monitors the channel condition and the playback buffer occupancy. This information is used by the adaptive video transport algorithm which controls the ARQ feedback messages. The algorithm sends false positive acknowledgments for erroneous video packets which cannot be delivered within their budget time. Hence, the transmitter proceeds to transmit other more important video packets. The less important video frames which are falsely acknowledged are concealed using video error concealment techniques. This way we allocate more budget time for the delivery of important video frames while minimizing the quality distortion due to false acknowledgment. 
     
\begin{figure}[H]
\centering
\includegraphics[width=\textwidth]{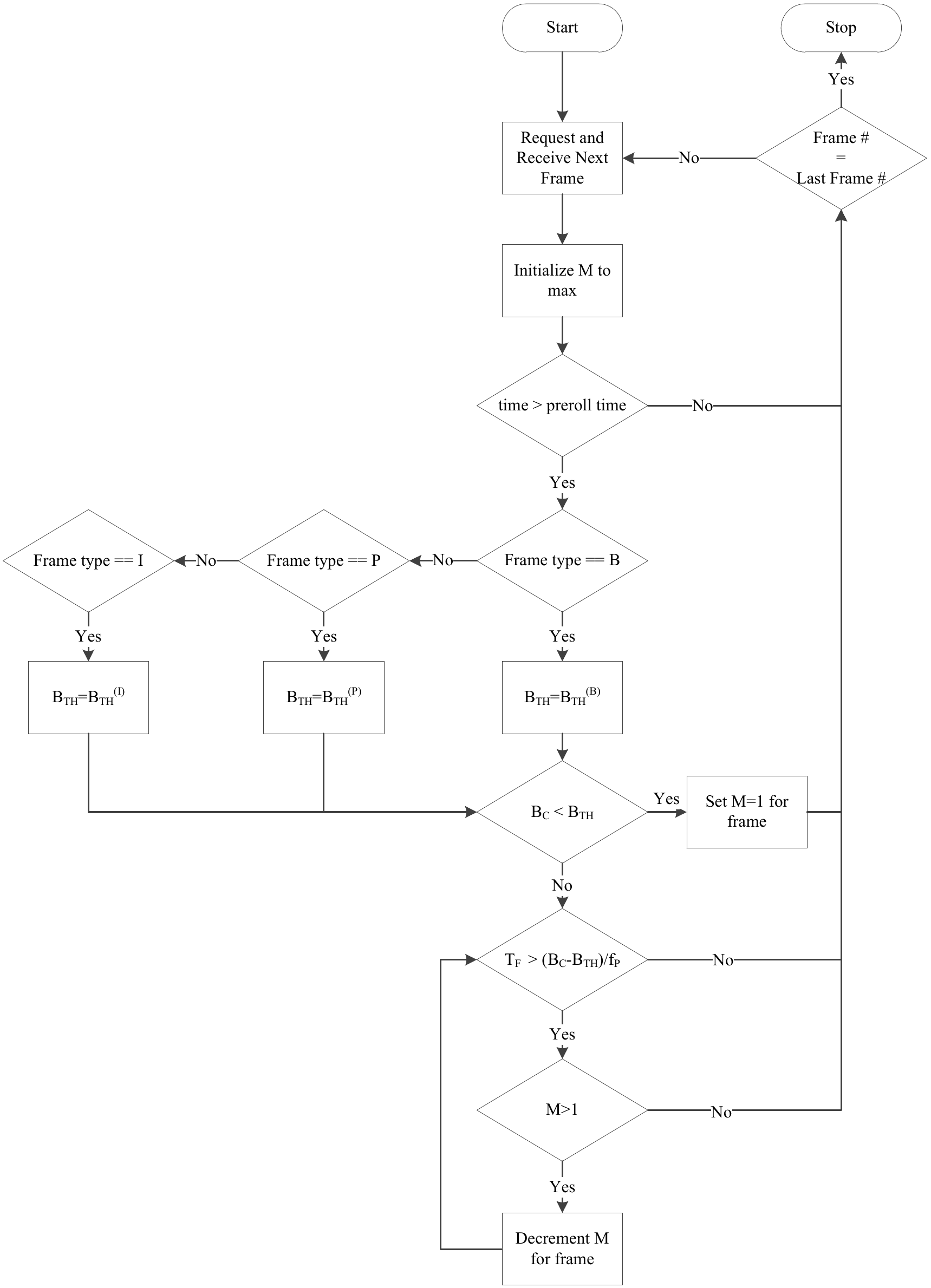}
\caption{Content-aware and occupancy-based transmission algorithm flow chart.}
\label{fig:AdpVidFlowChart}
\end{figure}

The adaptive video transmission algorithm is described using a flow chart in Fig.~\ref{fig:AdpVidFlowChart}. The algorithm operates on a frame-by-frame basis and starts after a preroll delay. The delivery budget time is set as
\begin{equation}
T_\text{B}=\dfrac{B_\text{C}-B_\text{TH}}{f_\text{P}},
\label{eq:budget}
\end{equation}  
where $B_\text{C}$ is the playback buffer occupancy, $B_\text{TH}$ is a buffer occupancy threshold, and $f_\text{P}$ is the playback rate in frames per second (fps)\nomenclature{fps}{frames per second}. The algorithm estimates the delivery time of video frames ($T_\text{F}$) based on the channel condition and video frame sizes using (\ref{eq:taus}). The algorithm adjusts the ARQ limit so that retransmissions are only allowed within the budget time. Video packets which are not received successfully within their budget time are falsely acknowledged. Moreover; false ACKs are sent without computing $T_\text{F}$ when the playback buffer occupancy is below a specified $B_\text{TH}$.  

\nomenclature{$T_\text{B}$}{budget time for delivery of a video frame}
\nomenclature{$B_\text{C}$}{playback buffer occupancy in frames}
\nomenclature{$B_\text{TH}$}{buffer occupancy threshold for adapting ARQ limit}
\nomenclature{$f_\text{P}$}{playback rate in fps}
\nomenclature{$T_\text{F}$}{delivery time of a video frame}
\nomenclature{$B_\text{TH}^\text{(I)}$}{$B_\text{TH}$ for I frame}
\nomenclature{$B_\text{TH}^\text{(P)}$}{$B_\text{TH}$ for P frame}
\nomenclature{$B_\text{TH}^\text{(B)}$}{$B_\text{TH}$ for B frame}

In the algorithm, $B_\text{TH}$ is set differently based on the importance of video frames. We use three thresholds $B_\text{TH}^\text{(I)}$, $B_\text{TH}^\text{(P)}$, and $B_\text{TH}^\text{(B)}$ for I, P, and B frames, respectively, where $B_\text{TH}^\text{(I)}<B_\text{TH}^\text{(P)}<B_\text{TH}^\text{(B)}$. This design translates into more budget time allocation for important frames; therefore, the system resources are effectively utilized to deliver the most important frames before their playback deadline.    

\section{Numerical and Simulation Results}

\label{sec:Results8}

In this section, we evaluate the performance of our proposed A-HARQ video transmission system in terms of playback buffer occupancy and average PSNR. Subpacket-based and packet-based A-HARQ are both numerically evaluated for comparison. Two VBR (variable bit rate)\nomenclature{VBR}{Variable Bit Rate} video sequences are considered in the evaluation, namely, the NBC News (frames 7505 to 8405) and Sony (frames 12305 to 13205) video sequences \cite{seeling2012video}. Both sequences are encoded using H.264 encoder with quantization parameters (28,28,30) for I, P, and B frames, respectively. Moreover, the sequences are encoded using a GoP size of 16 frames with 3 B frames in between I/P key frames. The NBC News is a CIF video sequence with $352\times288$ resolution, whereas, Sony is an HD video with $1920\times1080$ resolution. 

The transmission bit rate $\chi$ is matched with the video sequences source rate $\vartheta$ such that $\chi=\vartheta/\eta$. The video sequences are transmitted over Rayleigh fading channels with ${E_\text{b}}/{N_0}=3$~dB. TPC$(64,57,4)^2$ is used for FEC and the ARQ limit is initially set to $M=4$. In subpacket-based HARQ, 4 TPC codewords are concatenated to form a packet, whereas, in packet-based HARQ, a packet contains a single TPC codeword. It is assumed that a packet may contain information from one video frame only. The packet size is less than or equal to one video frame and hence multiple packets may be needed to accommodate a video frame. 

The video player at the receiver implements a preroll delay before starting the playback. The preroll threshold is set to $B_\text{p}$ frames. After the preroll delay, the adaptation algorithm adjusts the ARQ limit based on the playback buffer occupancy and the channel condition as described in Section~\ref{sec:proposed8}. The change is applied to all packets belonging to the same video frame. ARQ limit adaptation and false ACK messages are only enabled for B frames to minimize the quality distortion impact of false acknowledgment. It is assumed that a video frame is lost if it is falsely acknowledged or its ARQ limit is adjusted by the adaptation algorithm. Lost frames are concealed at the receiver using a frame copy error concealment technique \cite{seeling2012video}. We denote the percentage of concealed frames as $\mathbb{C}$, the average PSNR of the original video as $\mathbb{Q}$, and the average PSNR of the concealed video as $\hat{\mathbb{Q}}$.

\nomenclature{$\mathbb{C}$}{percentage of concealed frames in decoded video}
\nomenclature{$\mathbb{Q}$}{average PSNR of the original transmitted video}
\nomenclature{$B_\text{p}$}{preroll buffer occupancy threshold}

\begin{figure}[H]
\centering
\subfloat[packet-based regular HARQ]{\label{fig:NBC_pkt_regACK_3ms}\includegraphics[width=0.5\textwidth]{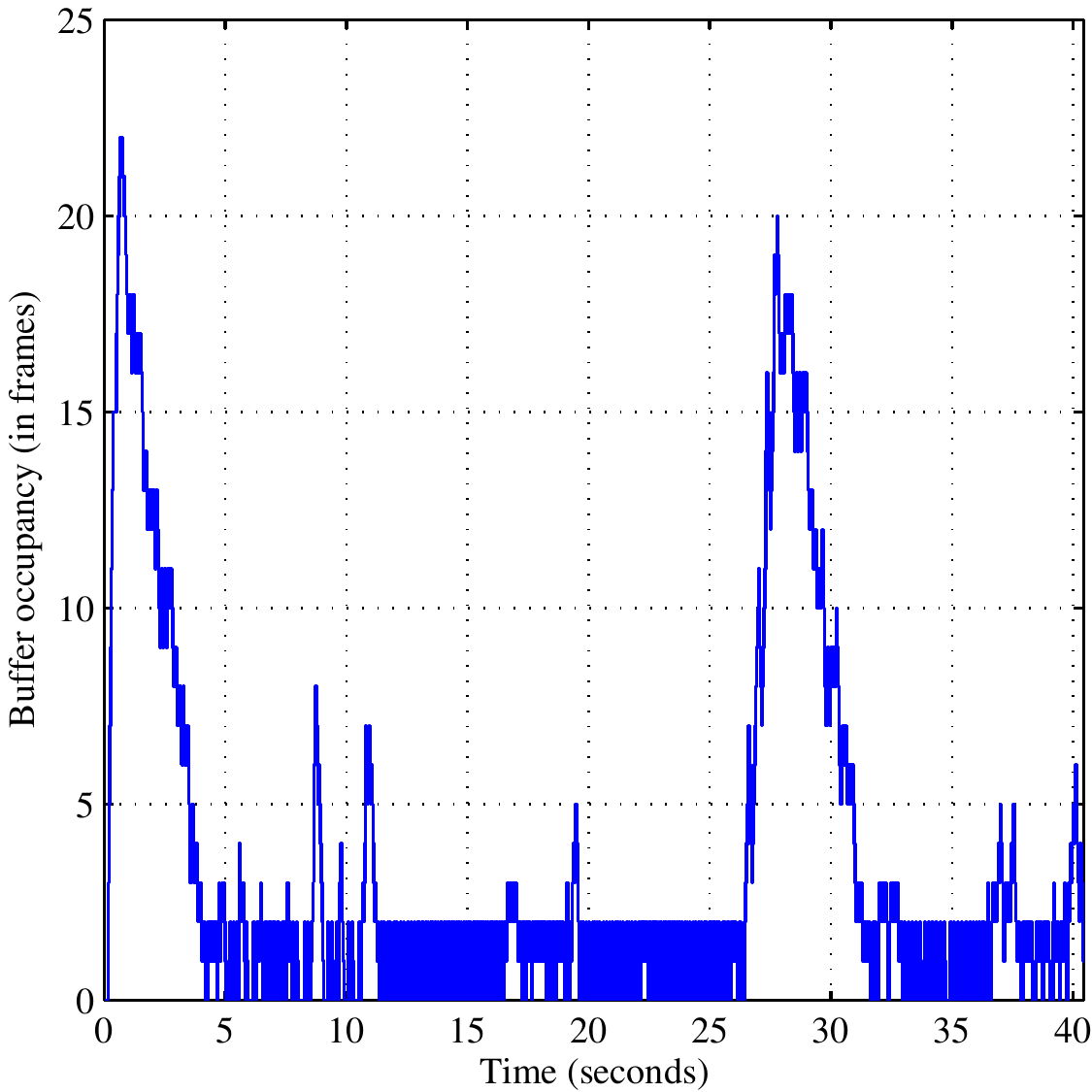}}
\subfloat[packet-based A-HARQ]{\label{fig:NBC_pkt_falseACK_3ms}\includegraphics[width=0.5\textwidth]{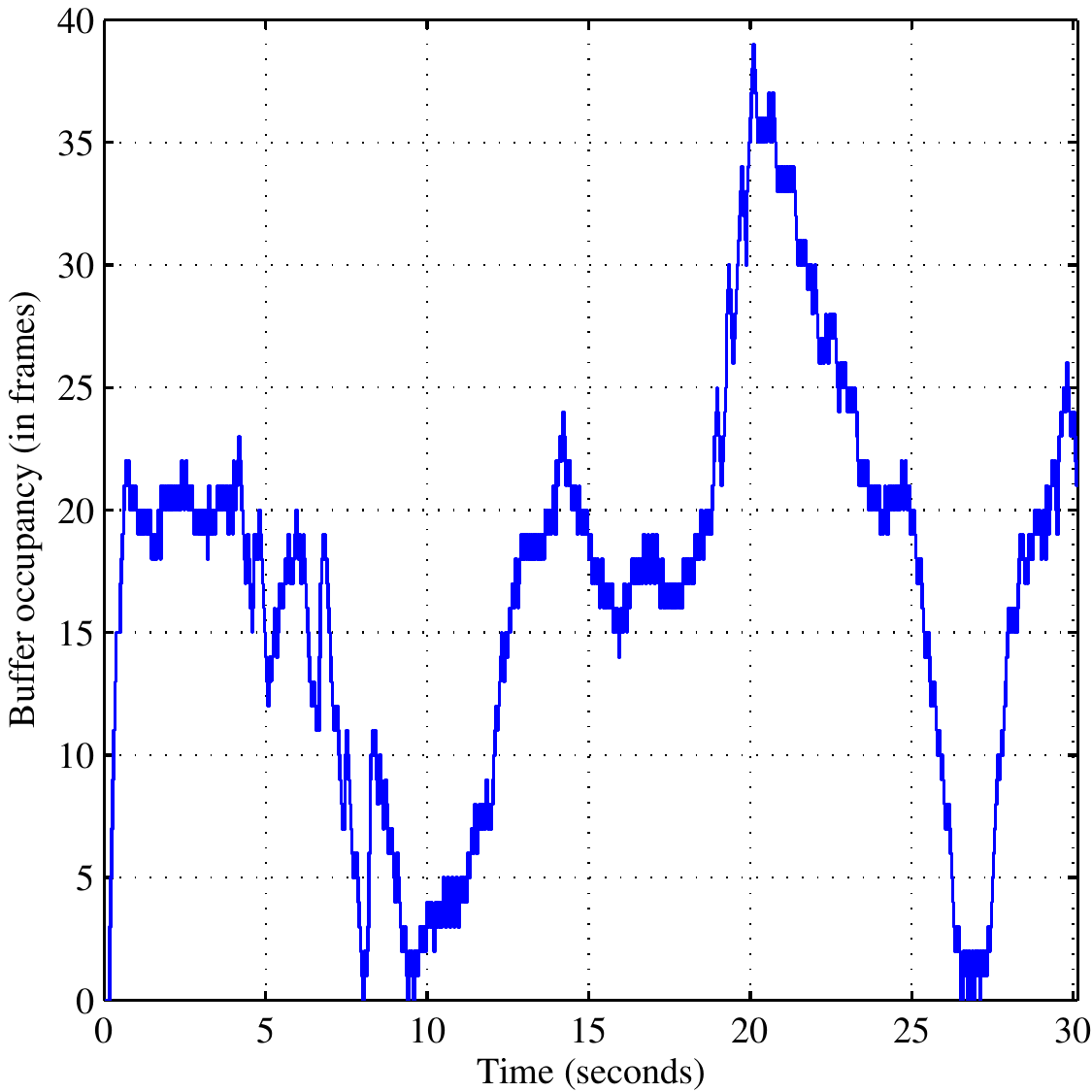}}\\
\subfloat[subpacket-based regular HARQ]{\label{fig:NBC_sub_regACK_3ms}\includegraphics[width=0.5\textwidth]{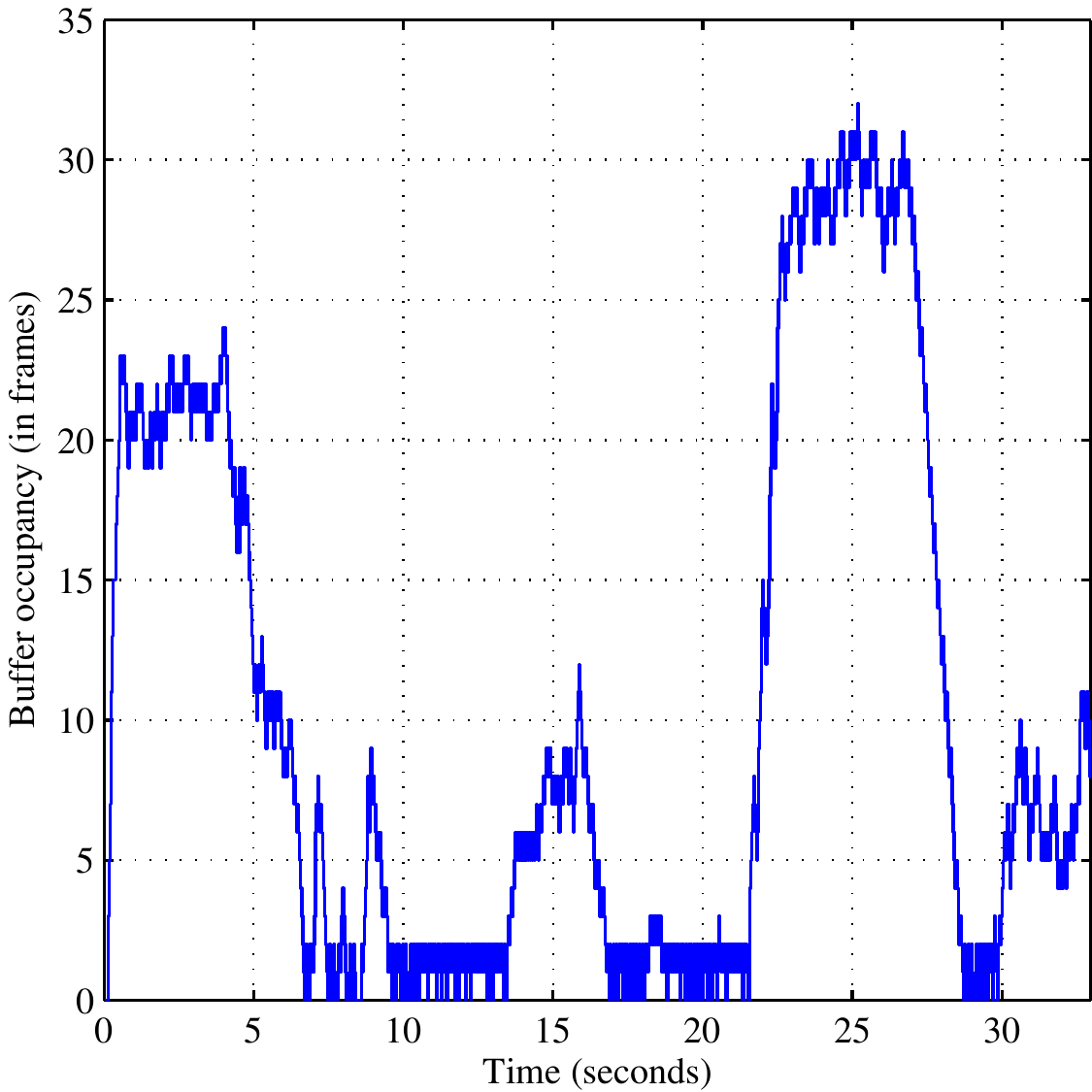}}
\subfloat[subpacket-based A-HARQ]{\label{fig:NBC_sub_falseACK_3ms}\includegraphics[width=0.5\textwidth]{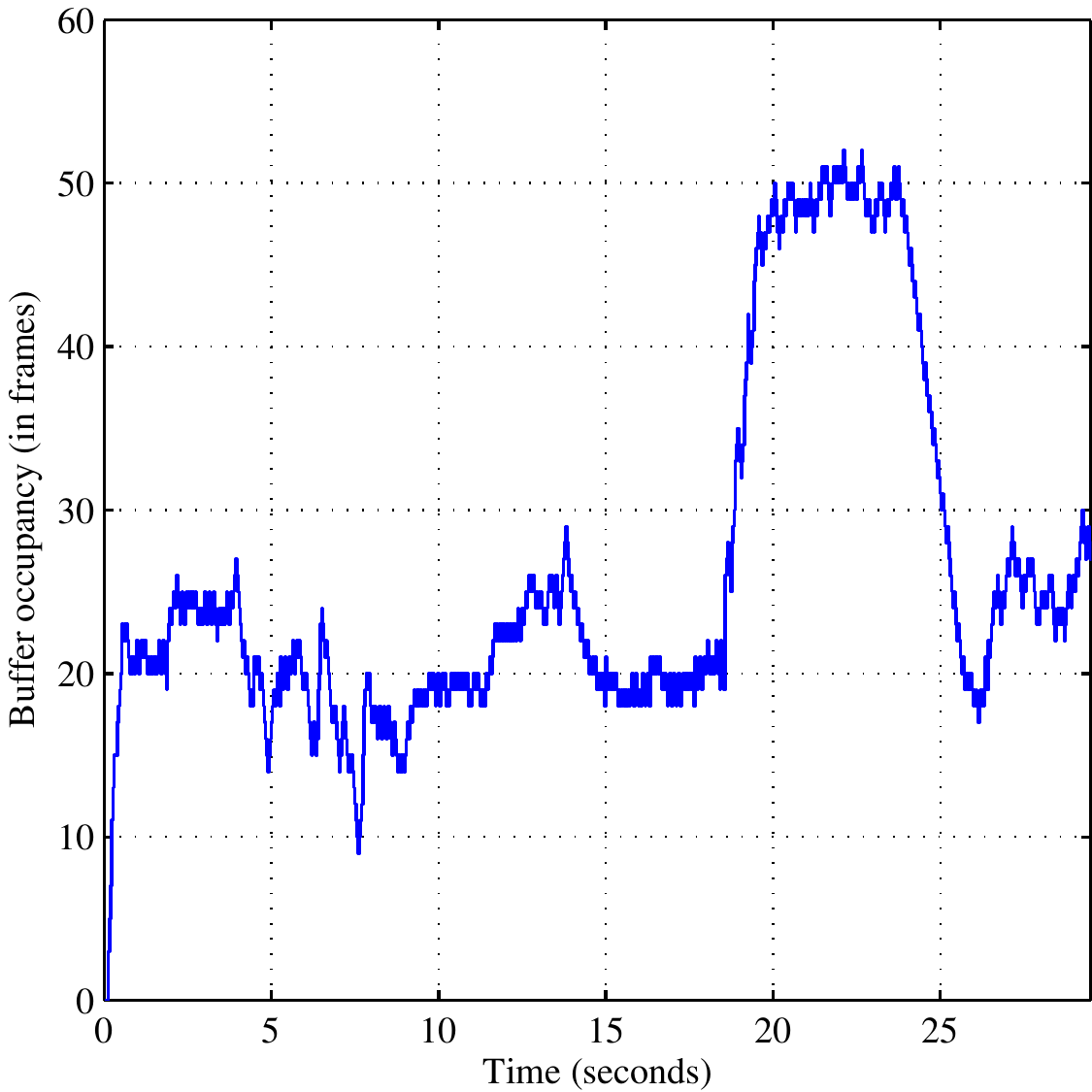}}
\caption{Playback buffer occupancy when transmitting the NBC News using TPC$(64,57,4)^2$ over Rayleigh fading channels with ${E_\text{b}}/{N_0}=3$~dB and $t_\text{p}=3$~ms.}
\label{fig:NBC3ms}
\end{figure}

Fig.~\ref{fig:NBC3ms} compares the playback buffer evolution when transmitting the NBC News sequence using packet-based and subpacket-based systems with regular HARQ and the proposed A-HARQ. The numerical results are obtained assuming a propagation time of $t_\text{p}=3$~ms. The A-HARQ algorithm is implemented with $B_\text{p}=B_\text{TH}^\text{B}=16$ frames. The packet-based regular HARQ provides the worst performance with many starvation instants as shown in Fig.~\ref{fig:NBC_pkt_regACK_3ms}. Playback buffer starvation corresponds to interruption in video playback. The subpacket-based regular HARQ improves the playback buffer evolution but still suffers from considerable playback buffer starvation as shown in Fig.~\ref{fig:NBC_sub_regACK_3ms}. Implementing the proposed A-HARQ significantly improves the playback buffer evolution. The best performance with no starvation is achieved when subpacket-based A-HARQ is used as shown in Fig.~\ref{fig:NBC_sub_falseACK_3ms}.
            
\begin{figure}[H]
\centering
\subfloat[packet-based regular HARQ]{\label{fig:NBC_pkt_regACK_300us}\includegraphics[width=0.5\textwidth]{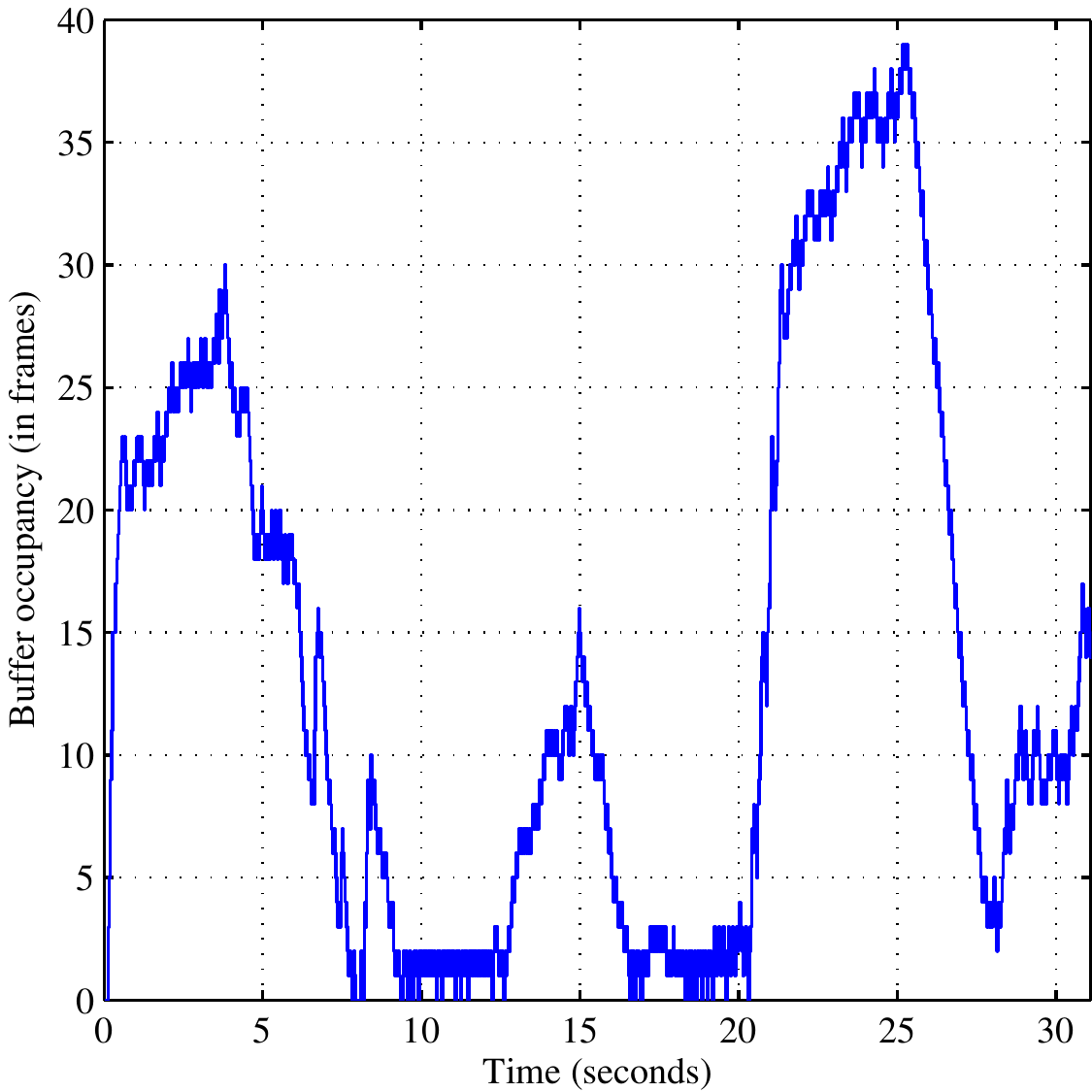}}
\subfloat[packet-based A-HARQ]{\label{fig:NBC_pkt_falseACK_300us}\includegraphics[width=0.5\textwidth]{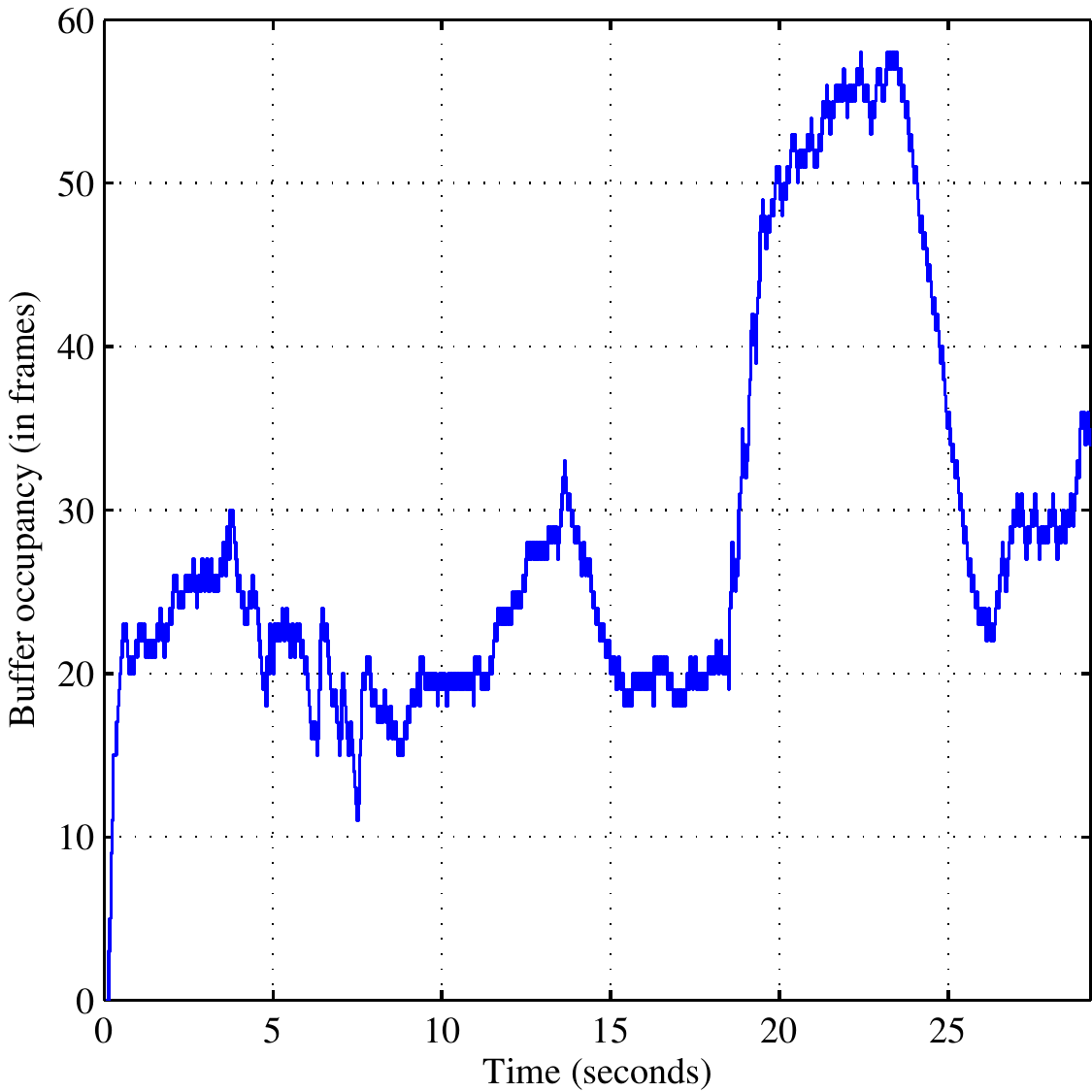}}\\
\subfloat[subpacket-based regular HARQ]{\label{fig:NBC_sub_regACK_300us}\includegraphics[width=0.5\textwidth]{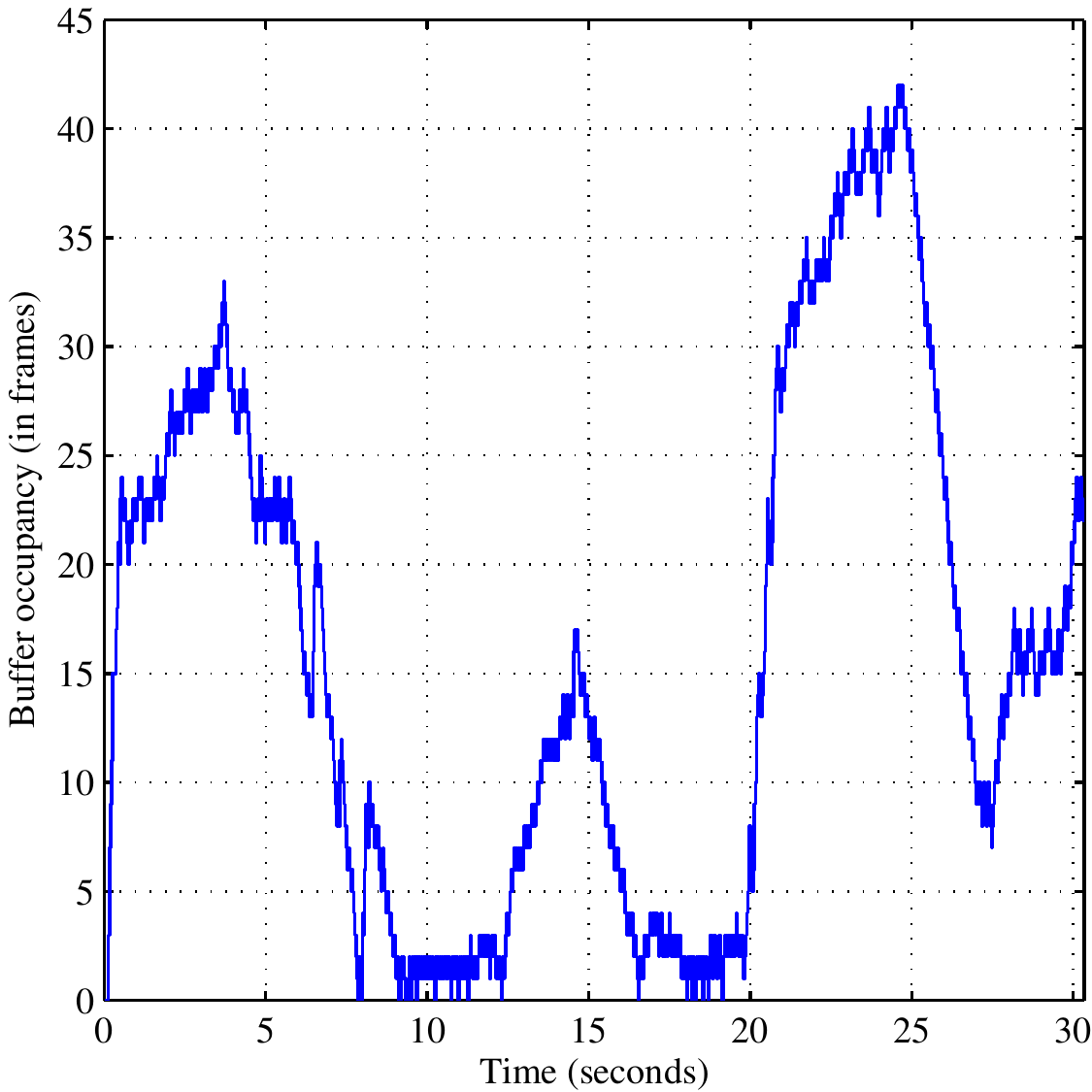}}
\subfloat[subpacket-based A-HARQ]{\label{fig:NBC_sub_falseACK_300us}\includegraphics[width=0.5\textwidth]{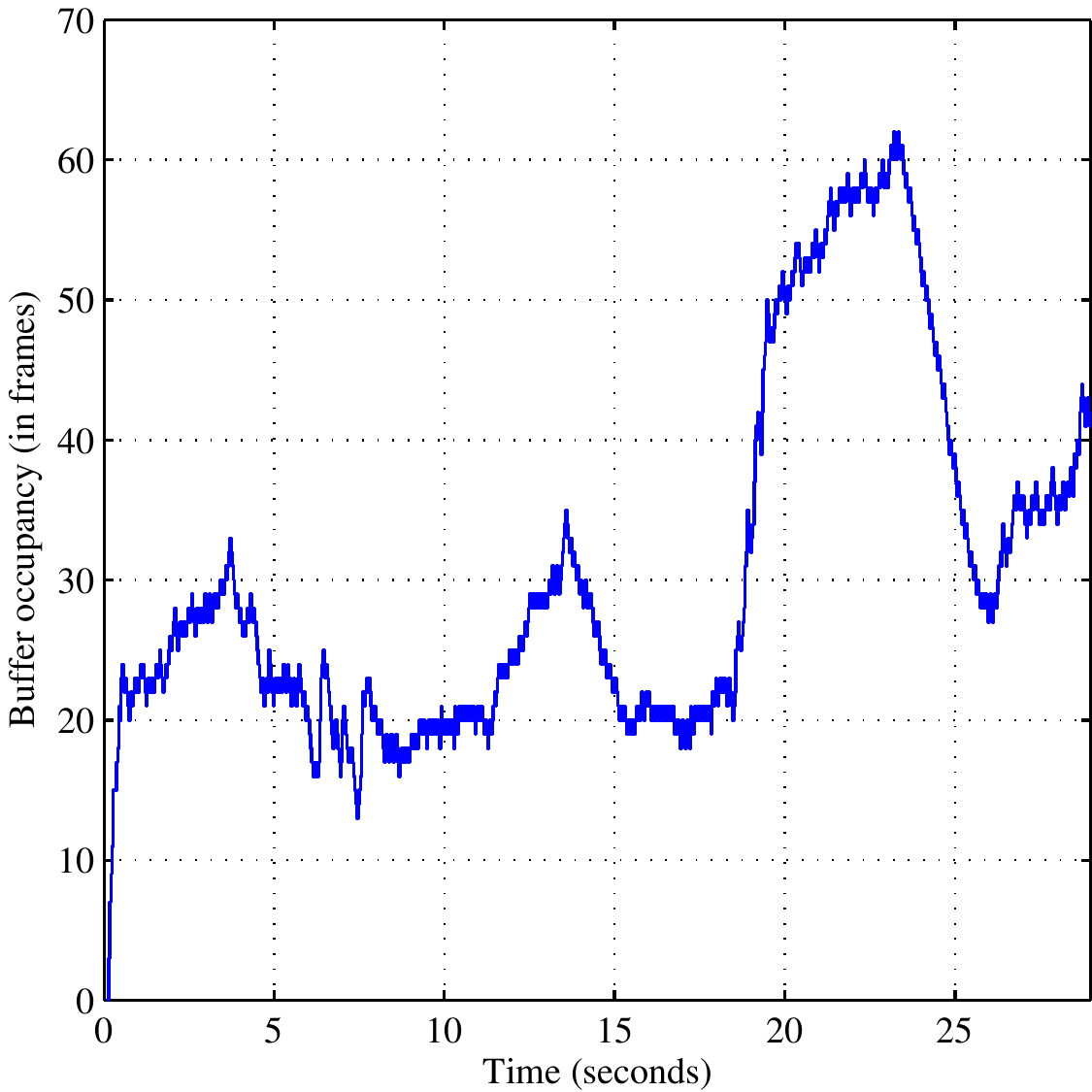}}
\caption{Playback buffer occupancy when transmitting the NBC News using TPC$(64,57,4)^2$ over Rayleigh fading channels with ${E_\text{b}}/{N_0}=3$~dB and $t_\text{p}=300~\mu$s.}
\label{fig:NBC300us}
\end{figure}

Fig.~\ref{fig:NBC300us} also compares the playback buffer evolution when transmitting the NBC News sequence using packet-based and subpacket-based systems; however, with $t_\text{p}=300~\mu$s. The regular HARQ system suffers from lower buffer starvation instants when compared to the corresponding results for $t_\text{p}=3$~ms in Fig.~\ref{fig:NBC3ms}. Comparing Fig.~\ref{fig:NBC_pkt_regACK_300us} with Fig.~\ref{fig:NBC_sub_regACK_300us}, we observe that subpacket-based HARQ provides unnoticeable improvement in the playback buffer evolution when assuming a small value for $t_\text{p}$. However, the proposed A-HARQ eliminates the playback buffer starvation as shown in Fig.~\ref{fig:NBC_pkt_falseACK_300us} and \ref{fig:NBC_sub_falseACK_300us}.

\begin{figure}[H]
\centering
\subfloat[packet-based regular HARQ]{\label{fig:Sony_pkt_regACK_300us}\includegraphics[width=0.5\textwidth]{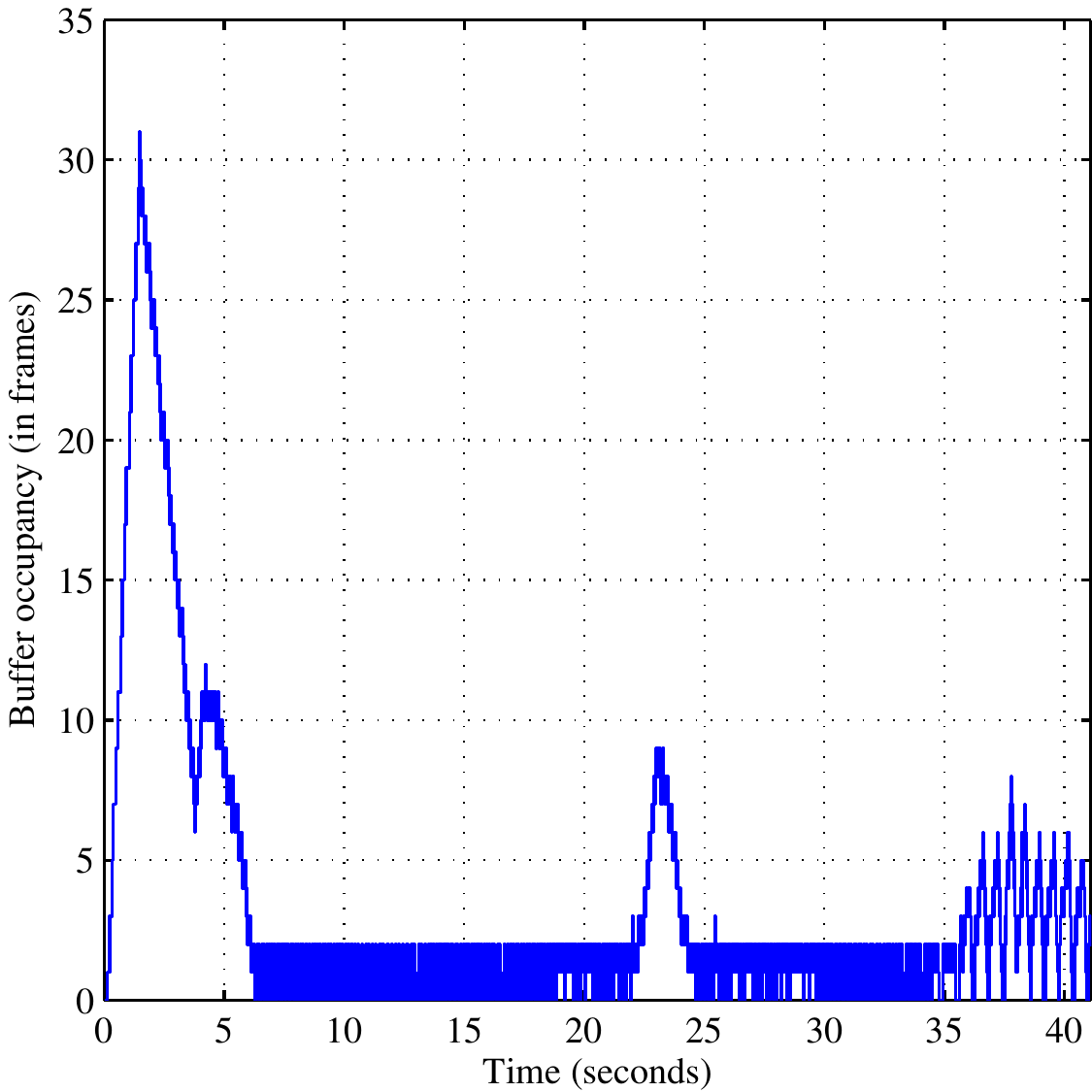}}
\subfloat[packet-based A-HARQ]{\label{fig:Sony_pkt_falseACK_300us}\includegraphics[width=0.5\textwidth]{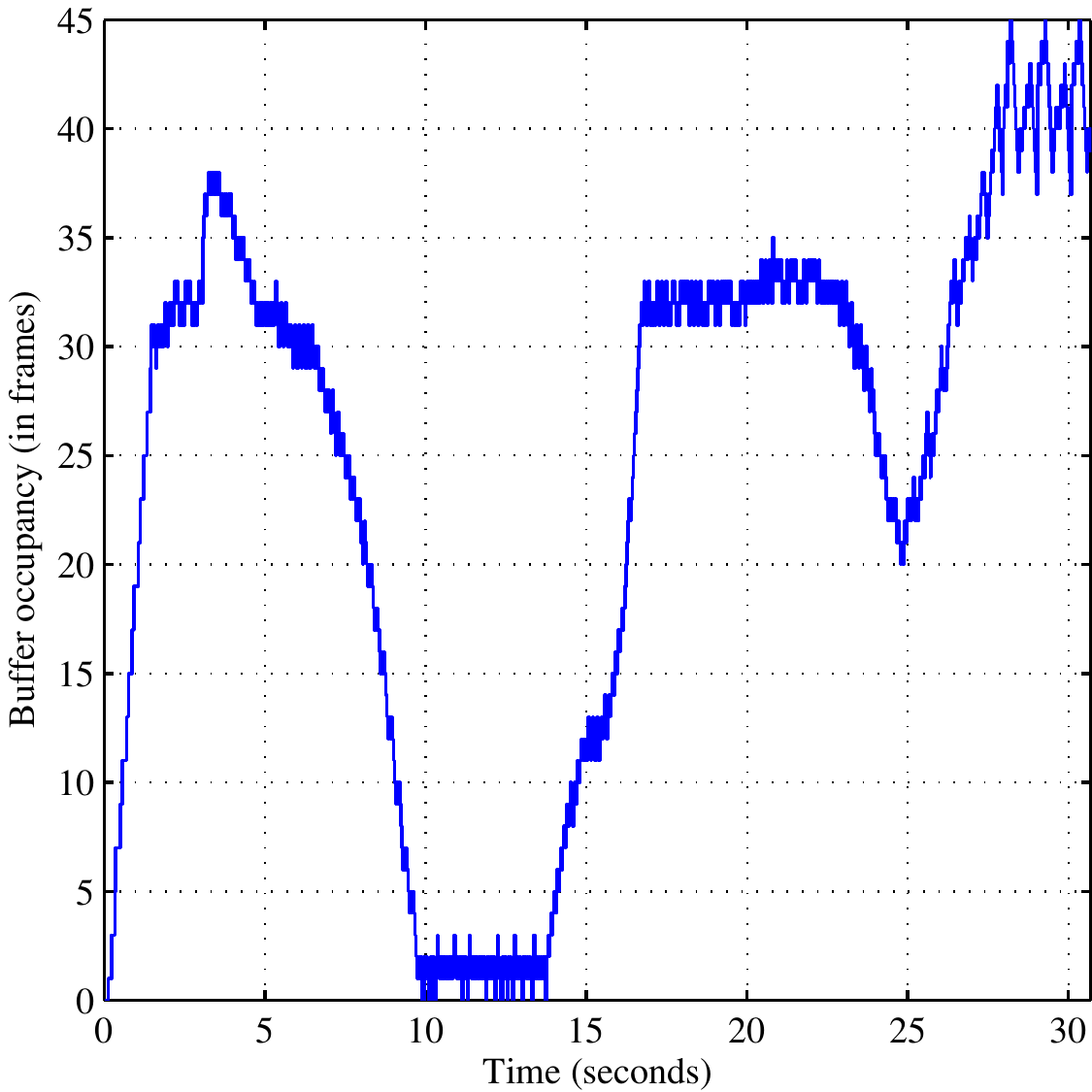}}\\
\subfloat[subpacket-based regular HARQ]{\label{fig:Sony_sub_regACK_300us}\includegraphics[width=0.5\textwidth]{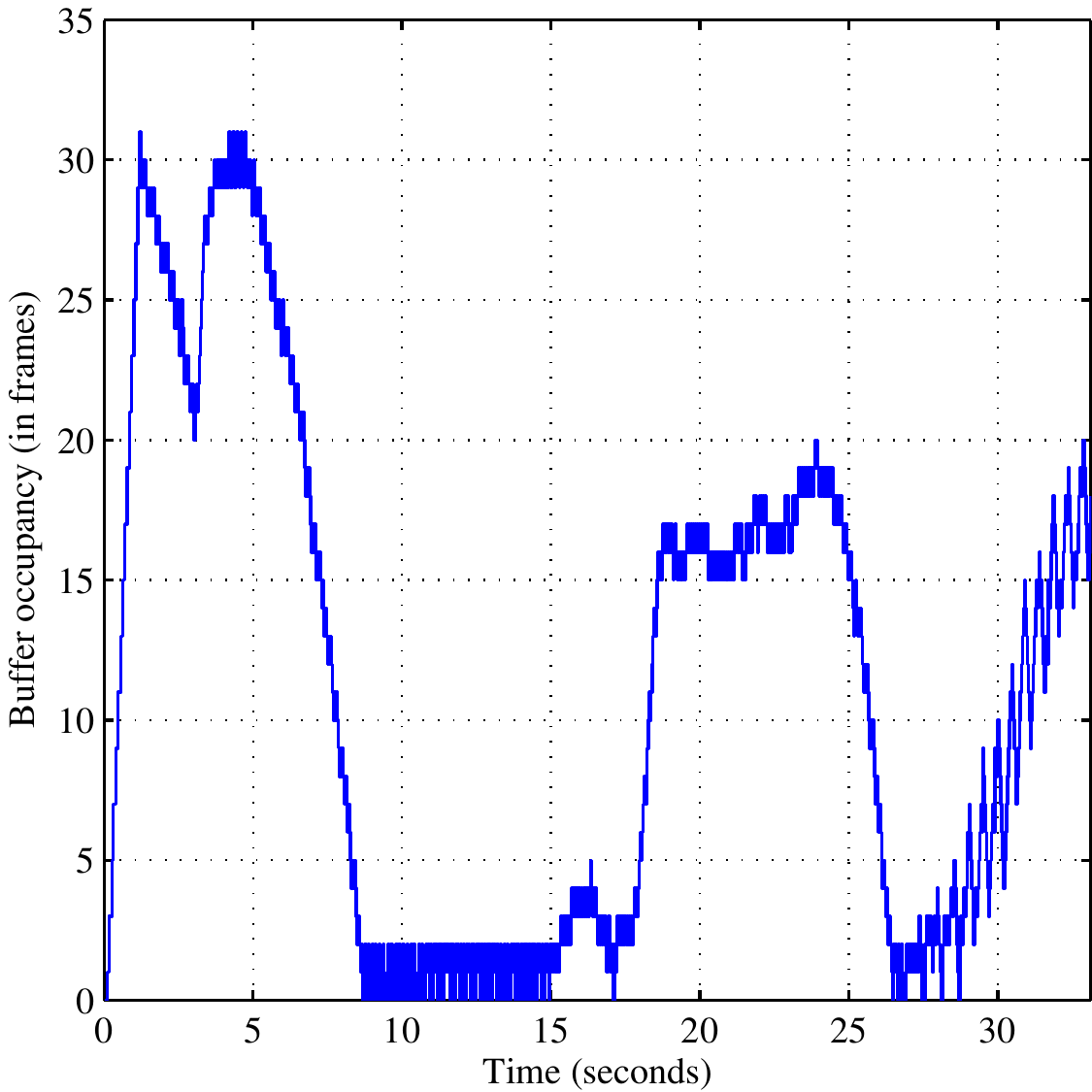}}
\subfloat[subpacket-based A-HARQ]{\label{fig:Sony_sub_falseACK_300us}\includegraphics[width=0.5\textwidth]{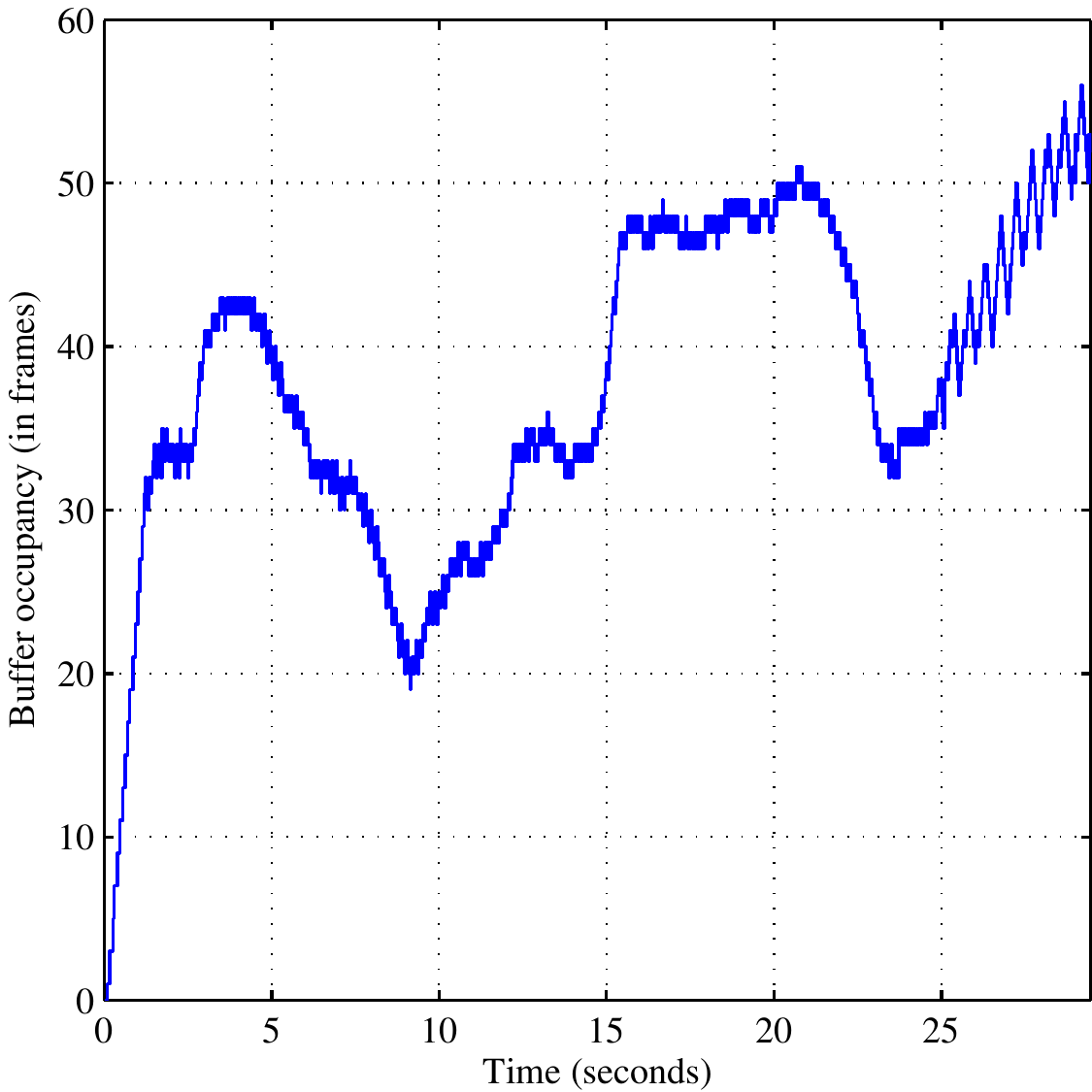}}
\caption{Playback buffer occupancy when transmitting the Sony HD sequence using TPC$(64,57,4)^2$ over Rayleigh fading channels with ${E_\text{b}}/{N_0}=3$~dB and $t_\text{p}=300~\mu$s.}
\label{fig:Sony300us}
\end{figure}

Fig.~\ref{fig:Sony300us} shows the playback buffer occupancy when transmitting the Sony HD video sequence with $t_\text{p}=300~\mu$s. Unlike the NBC News case, small propagation time has significant delay impact which translates into severe playback buffer starvation when regular packet-based HARQ is used as show in Fig.~\ref{fig:Sony_pkt_regACK_300us}. This is due to the large frame sizes of the HD video sequence which increases the number of packets per video frame. Each packet experiences propagation delay and hence the total propagation delay becomes significant when the number of packets per video frame is high. The subpacket-based regular HARQ is more robust to propagation delay with milder playback buffer starvation as shown in Fig.~\ref{fig:Sony_sub_regACK_300us}. The A-HARQ significantly reduces playback buffer starvation for the packet-based system and eliminates it for the subpacket-based system as shown in Fig.~\ref{fig:Sony_pkt_falseACK_300us} and \ref{fig:Sony_sub_falseACK_300us}, respectively. The A-HARQ algorithm is implemented with $B_\text{p}=B_\text{TH}^\text{B}=30$ frames.  

The A-HARQ improves the playback buffer evolution to ensure continuous playback by falsely acknowledging or adjusting the ARQ limit for some video frames. Those affected frames are assumed lost and replaced by redisplaying the last successfully decoded frame. We evaluated the effect of this process in the received video in terms of the average PSNR. Table~\ref{tab:PSNR} shows the average PSNR of the original transmitted video and the received video, and the percentage of concealed frames when using packet-based and subpacket-based A-HARQ. We observe that the subpacket-based A-HARQ provides higher average PSNR when compared to the packet-based A-HARQ. That is because the subpacket-based system has better delay performance where the number of lost frames due to false ACKs or ARQ limit adaptation is lower than the number of affected frames in the packet-based system. This translates into less number of concealed frames and hence higher average PSNR as shown in the table.
     
\begin{table}[H] \centering
\caption{The average PSNR of the original transmitted video and the received video, and the percentage of concealed frames when using packet-based and subpacket-based A-HARQ.}
\begin{tabular}{l|c|c|}
\cline{2-3}
& \multicolumn{2}{|c|}{NBC News (CIF)} \\ \cline{2-3}
& \multicolumn{2}{|c|}{$t_{\text{p}}=3$ ms} \\ \cline{2-3}
& packet-based A-HARQ & \multicolumn{1}{|c|}{subpacket-based A-HARQ} \\ 
\hline
\multicolumn{1}{|l|}{$\mathbb{Q}$ (dB)} & \multicolumn{1}{|c|}{$36.35$} & 
\multicolumn{1}{|c|}{$36.35$} \\ \hline
\multicolumn{1}{|l|}{$\widehat{\mathbb{Q}}$ (dB)} & \multicolumn{1}{|c|}{$%
32.63$} & \multicolumn{1}{|c|}{$35.01$} \\ \hline
\multicolumn{1}{|l|}{$\mathbb{C}$} & \multicolumn{1}{|c|}{$26\%$} & 
\multicolumn{1}{|c|}{$9\%$} \\ \hline \hline
& \multicolumn{2}{|c|}{NBC News (CIF)} \\ \cline{2-3}
& \multicolumn{2}{|c|}{$t_{\text{p}}=300$ $\mu $s} \\ \cline{2-3}
& packet-based A-HARQ & \multicolumn{1}{|l|}{subpacket-based A-HARQ} \\ 
\hline
\multicolumn{1}{|l|}{$\mathbb{Q}$ (dB)} & $36.35$ & \multicolumn{1}{|c|}{$%
36.35$} \\ \hline
\multicolumn{1}{|l|}{$\widehat{\mathbb{Q}}$ (dB)} & $35.60$ & 
\multicolumn{1}{|c|}{$35.75$} \\ \hline
\multicolumn{1}{|l|}{$\mathbb{C}$} & $5\%$ & $4\%$ \\ \hline \hline
& \multicolumn{2}{|c|}{Sony (1080)} \\ \cline{2-3}
& \multicolumn{2}{|c|}{$t_{\text{p}}=300$ $\mu $s} \\ \cline{2-3}
& packet-based A-HARQ & \multicolumn{1}{|c|}{subpacket-based A-HARQ} \\ 
\hline
\multicolumn{1}{|l|}{$\mathbb{Q}$ (dB)} & $40.64$ & \multicolumn{1}{|c|}{$%
40.64$} \\ \hline
\multicolumn{1}{|l|}{$\widehat{\mathbb{Q}}$ (dB)} & $34.98$ & 
\multicolumn{1}{|c|}{$38.44$} \\ \hline
\multicolumn{1}{|l|}{$\mathbb{C}$} & $32\%$ & \multicolumn{1}{|c|}{$12\%$}
\\ \hline
\end{tabular}%
\newline
\label{tab:PSNR}
\end{table}

\section{Conclusions}

\label{sec:Conclusions8} 

A content-aware and occupancy-based adaptive HARQ is proposed for delay-sensitive video transmission. The objective of the proposed scheme is to ensure minimum video quality distortion with continuous playback. The adaptation algorithm uses false acknowledgments and dynamically adjusts the ARQ limit for video packets based on the playback buffer occupancy and the importance of video frames. The performance of the proposed A-HARQ is evaluated with packet-based and subpacket-based systems. The A-HARQ significantly improves the playback buffer evolution of both systems. Moreover, the subpacket-based A-HARQ outperforms the packet-based A-HARQ in terms of delay and packet loss due to false acknowledgment or ARQ limit adaptation. Therefore, the subpacket-based A-HARQ minimizes playback buffer starvation at the expense of a slight reduction in the average PSNR quality of the received video.

\chapter{Conclusions and Future Work}
\label{chap:9}

In this PhD thesis work, turbo product codes have been studied with the objective of improving the state-of-the-art HARQ in wireless video communication systems. The results of this research work on TPC-based HARQ have revealed many interesting observations and achieved significant improvements to HARQ in terms of delay, complexity and power consumption which are the main challenges in wireless video communication. The obtained results were published in leading international journals, namely, the IEEE Communication Letters \cite{H-Mukhtar-2013}, IEEE Journal on Selected Topics in Signal Processing \cite{mukhtar2014low} and IEEE Transactions on Communications \cite{mukhtar2014crc}. The following is a summary of the obtained conclusions.

In Chapter~\ref{chap:4}, the throughput performance of HARQ is considered using TPC with iterative hard and soft decision decoding. The obtained results show that the coding gain advantage of soft decision decoding over hard decision decoding is reduced significantly when HARQ is used with Chase combining, and it actually vanishes completely for some codes. Moreover, an adaptive subpacket TPC HARQ scheme is proposed to maximize the system throughput by changing the subpacket size and code rate based on the channel SNR. In the adaptive system, the obtained throughput using HIHO is only about 1.4 dB less than the obtained throughput using SISO. It is shown that this advantage may not be significant for particular systems such as video transmission.

In Chapter~\ref{chap:5}, the problem of transmit power optimization for HARQ systems with packet combining is considered. The obtained results show that the transmit power can be reduced significantly while the throughput of the HARQ system remains almost unchanged. The proposed power optimization algorithm achieves significant power saving of about 80\% for particular scenarios. Moreover, the results show that reducing the transmission power per ARQ round does not necessarily reduce the average transmit power. On the contrary, increasing the transmission power per ARQ round can reduce the average transmit power.

In Chapter~\ref{chap:6}, an adaptive power control scheme is presented for HARQ systems with no channel state information feedback. The power control algorithm eliminates the need for channel state feedback and relies on simple feedback inherent to HARQ. The average transmit power per information bit is minimized subject to a desired throughput value. The throughput is computed using the packet error rate which is estimated using the ACK/NACK feedback. The accuracy of PER estimation improves as the number of considered ACK/NACK messages increases. Simulation and numerical results show that the proposed power optimization algorithm achieves similar power saving when compared to the adaptive system with perfect CSI. Using a small number of ACK/NACK messages for PER estimation provides a practical solution to adjust the transmit power to the optimal value with tolerable inaccuracy.

In Chapter~\ref{chap:7}, an HARQ system which employs the inherent word-error detection capability of TPC is presented to replace the conventional cyclic redundancy check (CRC). The obtained results show that the CRC-free system consistently provides equivalent or higher throughput than the CRC-based HARQ with a noticeable advantage when small packet sizes are used. In addition, the numerical results show that the TPC self-detection has lower computational complexity than CRC detection
especially for TPC with high code rates. Unlike CRC detection, the complexity of TPC self-detection scales with the location of error where significant speedup is achieved when errors are located at the beginning of the TPC packet which is usually the case at low SNR.

In Chapter~\ref{chap:8}, a content-aware and occupancy-based adaptive HARQ is proposed for delay-sensitive video transmission. The objective of the proposed scheme is to ensure minimum video quality distortion with continuous playback. The adaptation algorithm uses false acknowledgments and dynamically adjusts the ARQ limit for video packets based on the playback buffer occupancy and the importance of video frames. The performance of the proposed A-HARQ is evaluated with packet-based and subpacket-based systems. The A-HARQ significantly improves the playback buffer evolution of both systems. Moreover, the subpacket-based A-HARQ outperforms the packet-based A-HARQ in terms of delay and packet loss due to false acknowledgment or ARQ limit adaptation. Therefore, the subpacket-based A-HARQ minimizes playback buffer starvation at the expense of a slight reduction in the average PSNR quality of the received video.

In conclusion, this PhD thesis work reveals that HARQ performance can be significantly improved to suite delay-sensitive and energy-scarce mobile applications such as wireless video communication. The proposed adaptive HARQ schemes may be incorporated in the link adaptation modules of current and future wireless communication networks to improve their performance in transmitting video. As a future work, we intend to expand the work in this PhD thesis by considering other channel models such as block fading channels and imperfect feedback channels. We also intend to combine the content-aware and occupancy-based HARQ scheme with scalable video coding to achieve graceful degradation in the quality of received video which may result from false ACKs. 

As a general outlook on the road to future wireless networks, in particular 5G networks \cite{osseiran2014scen}, many communication technologies are being developed such as massive MIMO, cognitive radio and visible light communication. At the same time, new video applications and services are being commercialized such as Ultra-HD, 3D, and holographic video streaming. Therefore, video streaming traffic is expected to continuously grow requiring new techniques to optimize wireless networks to satisfy the demand of consumers. We intend to study the HARQ and link adaptation modules in 5G networks with the objective of proposing new optimization techniques which will enable future video applications and achieve responsive and higher quality video streaming.



\backmatter
\renewcommand{\chaptermark}[1]{%
\markboth{#1}{}}

\addcontentsline{toc}{chapter}{Bibliography}
\begin{onehalfspace}
\bibliographystyle{IEEEtran}
\bibliography{IEEEabrv,myref}
\end{onehalfspace}

\end{document}